\documentclass[11pt,a4paper,twoside,openright]{report}

\usepackage{amsfonts,amsmath,amssymb,bold-extra,caption,enumerate,float,graphicx,setspace}
\usepackage{epsfig}
\usepackage{siunitx}
\usepackage{lipsum,blindtext} 
\usepackage{fancyhdr}
\graphicspath{{figures/}}
\usepackage{mathrsfs}
\usepackage{braket}
\usepackage{physics}
\usepackage{doi}

\usepackage{url,array,csquotes,multirow,tabularx}

\usepackage{subcaption} 

\usepackage[usenames,dvipsnames]{xcolor}   
\usepackage{pdfpages}

\usepackage{geometry}
\usepackage[times]{quotchap}
\renewcommand{\chapterheadstartvskip}{\vspace*{-1cm}}

\usepackage[nottoc]{tocbibind}
\usepackage[titles]{tocloft}
\makeatletter
\newlength{\@chapterlength} 
\newcommand\@chapterheadsmark{1}
\ifnum \@chapterheadsmark = 1
\fi

\def\l@section{\@dottedtocline{1}{\@chapterlength}{2.3em}}

\newlength{\@subsectionlength}
\def\l@subsection{\@dottedtocline{2}{\@subsectionlength}{3.2em}}

\newlength{\@subsubsectionlength}
\def\l@subsubsection{\@dottedtocline{3}{\@subsubsectionlength}{0em}}

\makeatother

\renewcommand\cftchappresnum{\chaptername~}
\newlength\mylength
\usepackage[acronym,nomain,toc]{glossaries}
\makeglossaries
\usepackage[intoc,noprefix]{nomencl}
\makenomenclature

\usepackage[toc,page]{appendix}

\definecolor{chaptergrey}{RGB}{166,28,48}
\definecolor{titlecolor}{RGB}{166,28,48}
\definecolor{authorcolor}{RGB}{166,28,48}
\definecolor{refcolor}{rgb}{.0,.23,.45}
\definecolor{linkcolor}{RGB}{148,39,9}

\usepackage{datetime}
\newdateformat{thisdate}{%
	\twodigit{\THEDAY}}
\newdateformat{thismonthdigit}{%
	\twodigit{\THEMONTH}}
\newdateformat{thismonth}{%
	\monthname[\THEMONTH]}
\newdateformat{thisyear}{%
	\THEYEAR}
 
\providecommand{\keywords}[1]
{
    \large
    \textbf{\textit{Keywords---}} #1
}

\usepackage[backend=biber, sorting=none, %
            defernumbers=true, style=nature, doi=true]{biblatex}
\usepackage{hyperref}

\hypersetup{linktoc=all, colorlinks=true, linkcolor=blue, citecolor=teal, urlcolor=linkcolor}

\begin{document}

\pagenumbering{gobble}













\newcommand{\ttitle}{Characterizing the Properties and Constitution
of Compact Objects in Gravitational-Wave Binaries}
\newcommand{\authorname}{Samanwaya Mukherjee}
\newcommand{\cosupname}{Prof. Sukanta Bose}
\newcommand{\supname}{Prof. Gulab Chand Dewangan}
\newcommand{\degreename}{Doctor of Philosophy}

\begin{titlepage}
	\linespread{1.2}
	\begin{center}
		\includegraphics[width=0.17\textwidth]{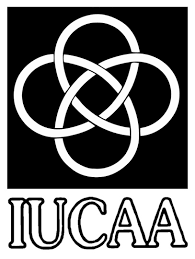}\\
            { \Large \textbf{The Inter-University Centre for\\ Astronomy and Astrophysics}\\ \medskip \large Post Bag 4, Ganeshkhind, Pune 411007, India}\\
            \noindent\rule{\textwidth}{0.4pt}
		{\color{titlecolor} \Huge\textbf{\ttitle}\\}
            \noindent\rule{\textwidth}{0.4pt}
		\vfill
            {\Large by\\}
            \bigskip
		{\huge \textbf{\authorname}\\}
		{\Large \textbf{Supervisor: \supname}\\}
		{\Large \textbf{Co-Supervisor: \cosupname}\\}
            \bigskip
		{\large A thesis to be submitted for the degree of\\}
		\medskip 
		{\LARGE \textit{\degreename \\}}
		\bigskip
		{\large to the\\}
            \bigskip
            \vfill
		{\Large \textbf{Jawaharlal Nehru University}\\ \medskip \large New Mehrauli Road, New Delhi 110067, India\\}
            \medskip	
            \includegraphics[width=0.15\textwidth]{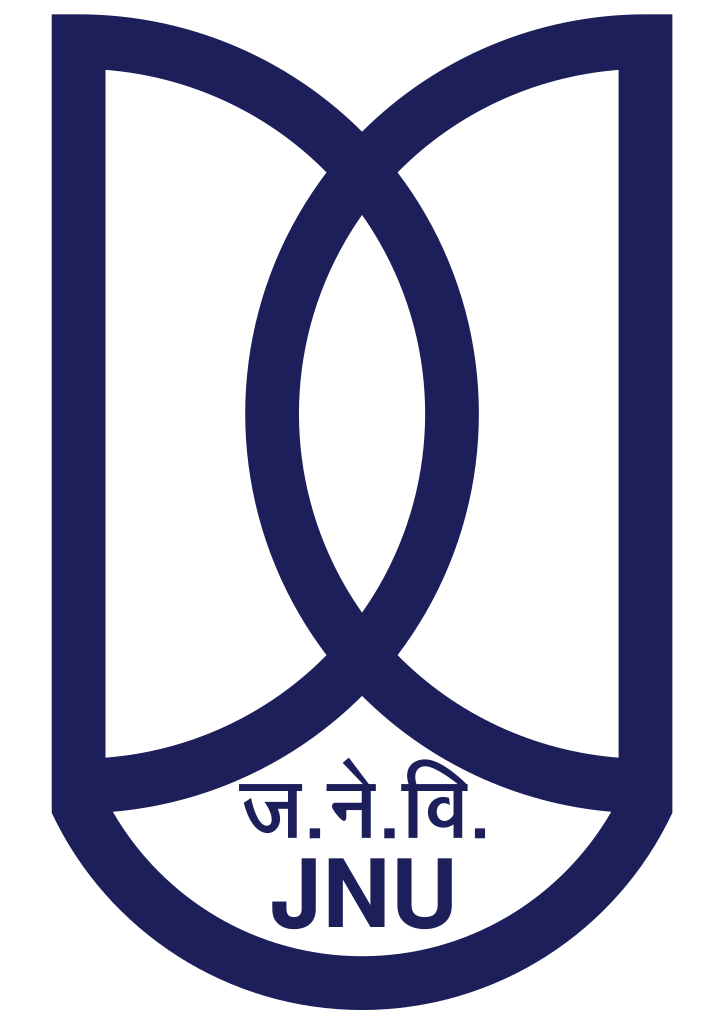}\\
            \end{center}
\end{titlepage}

\pagenumbering{roman}

\chapter*{}

\begin{flushleft}
\textit{\textbf{\Large Dedicated to}}
\end{flushleft}
\vspace*{1em}
\begin{flushleft}
\textit{\textbf{\Large myself -- for the unyielding aspiration;}}
\newline
\textit{\textbf{\Large to my wife Ayantika -- for her unconditional camaraderie;}}
\newline
\textit{\textbf{\Large and to all the young students choosing the hard path to take science forward.}}
\end{flushleft}
\vspace*{\fill}


\newpage
\thispagestyle{empty}
\null\newpage


\chapter*{\centering Declaration by the Student}


I hereby declare that the work reported in this thesis is entirely original. This
thesis is composed independently by me at the Inter-University Centre for Astronomy and Astrophysics, Pune. I further declare that the subject matter presented in the thesis has not previously formed the basis for the award of any degree, diploma, membership, associateship, fellowship, or any other similar title of any University or Institution.

\vspace{4em}
\hfill
\begin{minipage}[t]{0.4\textwidth}
\begin{flushright}
    \bfseries
    \rule{\textwidth}{.5pt}\\
	Samanwaya Mukherjee\\
	(Ph.D. candidate)
\end{flushright}
\end{minipage}

\newpage
\thispagestyle{empty}
\null\newpage


\chapter*{\centering Declaration by the Supervisor}
\vspace{-1cm} 



\par
This is to certify that this thesis titled ``\textbf{\ttitle}" is based on work done by \textbf{Mr. Samanwaya Mukherjee} at the Inter-University Centre for Astronomy and Astrophysics under my supervision. To the best of my knowledge, this thesis is original and has not been published or submitted to any other University for any other Degree or Diploma.

\vspace{4em}
\noindent
\hfill
\noindent
\begin{minipage}[t]{0.4\textwidth}
    \raggedleft
    \hrule\vspace{2ex}
    \textbf{\supname} \\
    IUCAA, Pune, India\\
	\textbf{(Supervisor)}
	
    \vspace{5em}
    \hrule\vspace{2ex}
    \textbf{Prof. Sukanta Bose} \\
    Washington State University, Pullman, USA\\
    \textbf{(Co-Supervisor)}
\end{minipage}


\newpage
\thispagestyle{empty}
\null\newpage


\chapter*{\centering Declaration by the Head of the Institute}
\vspace{-1cm} 



\par
This is to certify that this thesis titled ``\textbf{\ttitle}" is based on work done by \textbf{Mr. Samanwaya Mukherjee} at the Inter-University Centre for Astronomy and Astrophysics. To the best of my knowledge, this thesis is original and has not been published or submitted to any other University for any other Degree or Diploma.

\vspace{4em}
\noindent
\hfill
\noindent
\begin{minipage}[t]{0.4\textwidth}
    \raggedleft
    \hrule\vspace{2ex}
    \textbf{Prof. Raghunathan Srianand} \\
	\textbf{(Director, IUCAA)}
	
\end{minipage}


\newpage
\thispagestyle{empty}
\null\newpage


\chapter*{\centering Acknowledgments}
\addcontentsline{toc}{chapter}{Acknowledgments}

First and foremost, I extend my sincere gratitude to my supervisor Prof. Sukanta Bose for his guidance and mentorship. His continuous support and encouragement have been instrumental in shaping my academic journey. I am thankful to Prof. Gulab Chand Dewangan for his support in academic and administrative matters. I thank my research advisory committee members Prof. Sanjit Mitra and Prof. Debarati Chatterjee for their feedback and suggestions. I thank Prof. K. G. Arun and Prof. Harald Pfeiffer for examining this thesis and providing valuable suggestions and comments.

Since the beginning of my PhD journey, IUCAA has offered me a peaceful and productive atmosphere. I express my heartfelt gratitude to the security guards of IUCAA, maintenance staff at Nalanda and Takshashila for their help, and their pleasant and welcoming behaviour. I thank all the administrative staff at IUCAA, especially Mr. Senith Samuel for always being helpful with a smile, and Ms. Deepika Susainathan for arranging my travel abroad with incredible efficiency. I deeply appreciate the way IUCAA dealt with the COVID situation so aptly, making life easier during the crisis.

I have made so many fond memories over the last few years that looking back, I feel lucky to have been here. I met so many wonderful people, made valuable friendships that I will always cherish. Special thanks go to Anuj for being such a great friend; I will always remember (and continue to have) my long discussions with him on diverse topics. I am glad to have Sayak Datta as a close friend. Sayak and Khun Sang are two of the wonderful collaborators that I worked with during my PhD. I treasure the joyful memories I made with Souvik, Soham, Rahul, Aditya, Atri at NCRA. Cooking dinner with Debu (Debabrata), Souvik, Soham, Rahul, Ayantika; playing chess with Soham; playing the card game twenty-nine with Debu, Souvik, Soham, Ayantika -- all these made our weekends cheerful and lively. I thank Niladri Da and Debajyoti Da for inviting me and Ayantika on numerous occasions and being like a brother to us. Thanks to Chiranjeeb for his jovial company. I am grateful to have enjoyed the colorful and zealous company of Apratim Da, who has motivated me from time to time. I want to thank all my batchmates at IUCAA for their warm friendship; and Tathagata, Suprovo, Bikram, Rajendra, Sayak (Dutta) and so many others for being such wonderful juniors and friends. This list is far from complete. I hope I can be forgiven for not being able to mention the names of all the fabulous people I met and spent time with during the course of my PhD.

I will never forget the times Shalu and Soumendu came to our place by motorcycle all the way from Mumbai, and spent evenings with Ayantika and me in buoyant spirits. Sometimes we returned the favour by going to their place on our scooter, and the scenic beauty of Western Ghats was an additional reward to marvel at.

Above all, I consider myself fortunate to have had my wife, Ayantika, by my side throughout my journey. Her heartfelt support has alleviated countless hardships. On this note, I would like to express my gratitude to IUCAA for providing us with accommodation.


\newpage
\thispagestyle{empty}
\null\newpage

\pagestyle{fancy}
\fancyhf{}
\fancyhead[LE,RO]{\thepage}
 
\setstretch{2}  

\chapter*{\centering Abstract}
\fancyhead[LO,RE]{Abstract}
\addcontentsline{toc}{chapter}{Abstract}

Astrophysical observations point toward strong evidence for the existence of black holes, from stellar-mass ones to the supermassive ones dwelling at the centres of galaxies. Nevertheless, it is yet to be established or ruled out with confidence whether some exotic compact objects, capable of mimicking black holes from an observational point of view, are indeed doing so. In classical General Relativity, a horizon is the defining feature of a black hole, which prevents any event inside from causally affecting the outside Universe. When it is perturbed by an external source, the energy carried by the perturbation is absorbed entirely by a black hole. Any deviation from this scenario would suggest the possibility of an alternative description of the object, and indicative of physics beyond the standard model. The detection of gravitational waves by LIGO and Virgo has offered a unique testbed to probe the nature of dark compact objects. Gravitational waves emitted from the coalescence of compact binaries carry signatures of the presence of horizons. In this thesis, we present ways of improving the detectibility of such signatures.

The quest for distinguishing black holes from horizonless compact objects using gravitational wave signals from coalescing compact binaries can be helped by utilizing the phenomenon of tidal heating, which leaves its imprint on the binary waveforms through the horizon parameters. A black hole absorbs the energy of the gravitational perturbation created by its companion entirely, whereas an exotic black-hole mimicker would only absorb it partially. The horizon parameters quantify this ratio and signify the presence of horizons in a binary. We define two effective parameters to minimize degeneracy and study their measurability by Bayesian and Fisher analyses. Using post-Newtonian waveforms spanning the inspiral part only, we conclude that the effective horizon parameters can be subjected to tight constraints -- especially in future ground-based third-generation detectors Einstein Telescope and Cosmic Explorer where the signal-to-noise ratio is expected to be significantly higher than the current detectors. Analytical inspiral waveforms cease to be sufficient for binaries heavier than $\sim 12M_\odot$ in the second-generation detectors, when the merger-ringdown parts start to contribute significantly to the signal power. We also construct an inspiral-merger-ringdown waveform by using post-Newtonian calculations for the inspiral and numerical relativity data for the merger-ringdown part that incorporates the effects of tidal heating of black holes in the phase and the amplitude. The new model shows a modest yet notable improvement in waveform systematics when compared to numerical relativity data. In the late inspiral phase when the compact objects are closer to each other, the effects of tidal heating are stronger, opening up the possibility of identifying the objects more precisely. Finally, from numerical relativity data of binary black holes, we demonstrate how one can model tidal heating in the late inspiral regime and leverage this knowledge to test for horizonless compact objects mimicking black holes. The studies conducted in this thesis bear significance in determining the nature of compact objects having masses in the entire range that LIGO and future ground-based detectors can detect. Quite possibly, the methods presented in this thesis will pave the way for tidal heating to emerge as a useful discriminator for black holes, especially when the late inspiral regime is accounted for, thereby allowing us a more informed understanding of the constituents of the Universe. 

\keywords{Black Holes, Compact Binary Coalescence, Gravitational Waves, Tidal Heating}

\cleardoublepage\clearpage







\newacronym{gr}{GR}{General Relativity}
\newacronym{gws}{GWs}{Gravitational Waves}
\newacronym{bhs}{BHs}{Black Holes}
\newacronym{sbh}{SBH}{Schwarzschild Black Hole}
\newacronym{kbh}{KBH}{Kerr Black Hole}
\newacronym{smbh}{SMBH}{Supermassive Black Hole}
\newacronym{imbhs}{IMBHs}{Intermediate-Mass Black Holes}
\newacronym{nr}{NR}{Numerical Relativity}
\newacronym{snr}{SNR}{Signal-to-noise ratio}
\newacronym{eob}{EOB}{Effective one-Body}
\newacronym{bbhs}{BBHs}{Binary Black Holes}
\newacronym{pn}{PN}{Post-Newtonian}
\newacronym{nss}{NSs}{Neutron Stars}
\newacronym{ecos}{ECOs}{Exotic Compact Objects}
\newacronym{isco}{ISCO}{Innermost Stable Circular Orbit}
\newacronym{cbc}{CBC}{Compact Binary Coalescence}
\newacronym{th}{TH}{Tidal Heating}


\setstretch{1.5}

\renewcommand{\contentsname}{Table of Contents}
\phantomsection
\fancyhead[LO,RE]{\contentsname}
\tableofcontents
\cleardoublepage\clearpage

\fancyhead[LO,RE]{\listfigurename}
\listoffigures
\cleardoublepage\clearpage


\listoftables
\fancyhead[LO,RE]{List of Tables}

\printglossary[type=\acronymtype, title=Abbreviations, toctitle=List of Abbreviations]
\fancyhead[LO,RE]{List of Abbreviations}

\printnomenclature
\fancyhead[LO,RE]{List of Symbols}
\cleardoublepage\clearpage

\pagenumbering{arabic}
\setstretch{2}
\renewcommand{\chapterheadstartvskip}{\vspace*{-2.5cm}}

\renewcommand{\sectionmark}[1]{\markright{\thesection~#1}{}}
\renewcommand{\chaptermark}[1]{\markboth{\chaptername~\thechapter~-~#1}{}}
\fancyhead[RE]{\leftmark}
\fancyhead[LO]{\rightmark}


\chapter{Introduction}
\label{Chapter1}
\graphicspath{{Chapter_1/Vector/}{Chapter_1/}}


Black holes (\acrshort{bhs}), one of the most astounding predictions of Einstein's \acrfull{gr}, are now widely accepted to be not only existent but quite ubiquitous in our Universe. It is now generally believed that at the centre of every large galaxy there resides a supermassive black hole (\acrshort{smbh})~\cite{Kormendy:1995er}, typically $10^6$ to $10^{10}$ times more massive than the sun. The SMBHs strongly correlate with the dynamics of their host galaxy~\cite{2015pabh.book..427K}, serving as the gravitational nuclei at their core. BHs are one of the simplest yet most fascinating objects in the Universe. To an observer in its vicinity, there is nothing extraordinary about the impact of its presence on the spacetime -- if the sun were to be replaced by a BH of identical mass, barring the eternal darkness and cold, the \textit{gravitational} dynamics of the solar system would remain completely oblivious to that change. Nevertheless, BHs contain within them one of the most puzzling mysteries in the field of astrophysics -- a spacetime singularity, which is enveloped by another spectacular aspect of GR -- an event horizon. In the classical picture, the former feature is the result of an infinite spacetime curvature at the centre of a BH, where any predictive theory, including GR, is expected to break down. In this thesis, we will explore the latter feature of a BH, which bestows on its spacetime some unique properties that are quite demanding for other objects, whether they are observed or hypothetical, to emulate.

BHs were quickly adopted as the only possible dark compact objects setting off violent high-energy phenomena in the Universe, and for good reason. The spacetime pathologies related to the curvature singularity are enclosed by the horizon and thus harmless to the external world. In fact, Penrose's (weak) Cosmic Censorship Conjecture~\cite{Penrose:1969pc} prohibits any singularities to form in the Universe that are not cloaked by horizons, and efforts to produce such naked singularities have failed so far. BHs are strongly believed to be stable against small fluctuations. In addition, BHs satisfy some exceptional uniqueness properties~\cite{Chrusciel:2012jk}. The formation processes of BHs are also well understood, promoting them to possible real physical objects rather than merely a curious mathematical concoction. Neutron stars (\acrshort{nss}), the heaviest stellar remnants within the ambit of standard known physics, can only sustain masses up to $\sim 2M_\odot$, with an uncertainty introduced by an incomplete knowledge of their equation of state. There is no known interaction of matter that can prevent the gravitational collapse of a celestial compact object heavier than $\sim 3M_\odot$, and consequently, any dark compact object having mass beyond this limit is usually classified as a BH.

However, there are still many aspects of BH physics that remain unresolved. Although the exterior of a BH horizon is well-behaved, the interior is not. The Kerr family of BHs contain closed timelike curves inside their horizon, and more generically they feature a Cauchy horizon that indicates a breakdown of predictability. Quantum field theory in a BH background induces a loss of unitarity, a property any predictive theory must conform to. Some possible resolutions of this problem modify the near-horizon structure of spacetime, or eliminate the existence of horizons altogether~\cite{Giddings:2019ujs,Unruh:2017uaw}.
 
From an observational point of view, several exotic compact objects (\acrshort{ecos}) have been hypothesized that are capable of mimicking BHs. As it stands, one can artificially design models of exotic horizonless (and thus non-BH) objects to match all the observational properties of BHs with arbitrary accuracy. Surely, construction of such scenarios necessitate invoking physics beyond the standard model. Even though BHs are very special in nature, these exotic objects are able to mimic their signatures to a distant observer and incite a false impression of the presence of horizons. Within the standard model, any alternative to BH as a massive dark compact object is discarded. However, it has been well known that the standard model fails to be sufficient in describing the Universe at large scales. For example, dark matter, one of the longest standing puzzles in physics~\cite{Barack:2018yly}, propounds the crucial importance of extending the standard model to accommodate new physics. Several proposed solutions to the dark matter problem rely on introducing new fundamental fields minimally or non-minimally coupled to gravity~\cite{Marsh:2015xka}, and can give rise to self-gravitating compact objects. These objects may appear dark to a distant observer if their constituent particles interact with the standard model particles weakly. 

These circumstances motivate one to entertain the existence of BH alternatives, and objectively test for any observational signatures that can support them or rule them out. This is one of the motivations behind this thesis. Actually, BH mimickers may not necessarily replace black holes; instead, they may coexist as species within a larger family of astrophysical compact objects that include both BHs and NSs. Probing the signatures of horizon demands precision tests for the exclusive properties that a BH horizon possesses. Recently, one of the extremely important avenues of such tests has been opened up by the observation of gravitational waves (\acrshort{gws}) from the mergers of compact objects.

Detection of GWs from coalescence of numerous compact binaries by LIGO~\cite{TheLIGOScientific:2014jea} and Virgo~\cite{TheVirgo:2014hva} has opened up a new era of astronomy~\cite{LIGOScientific:2018mvr, LIGOScientific:2020ibl}. Their observations have motivated a series of tests of GR~\cite{LIGOScientific:2019fpa,Abbott:2018lct}. The components of the binaries observed by LIGO and Virgo are mainly inferred to be either BHs or NSs, which is primarily based on the measurements of component masses, population models, and tidal deformability of NSs~\cite{Cardoso:2017cfl}. 
Merger of two NSs was observed in the event GW170817~\cite{gw170817}, and possibly also GW190425~\cite{Abbott:2020uma}. More recently, confirmed detections of events GW200105 and GW200115~\cite{bhns_LIGOScientific:2021qlt} were made where one of the components is believed to be a BH, and the other an NS. However, for the heavier LIGO-Virgo binaries~\cite{LIGOScientific:2018mvr}, it remains to be conclusively proven whether their components are, in fact, BHs of GR or not. In this thesis, we show that it is possible to detect the existence of horizons using GW signals, and build a framework to facilitate such tests using the strong-field regime of binary coalescences where the compact objects are close to each other before they merge. These tests are particularly important for heavier binaries whose contribution to the GW signal power significantly comes from their late inspiral regime.

\section{A Quick Look at Black Holes}\label{sec:quick look at BHs}

A BH is usually conceptualized as a region of spacetime from which nothing, including electromagnetic radiation, can ever escape~\cite{Curiel:2018cbt} -- at least classically. For an isolated BH in an asymptotically flat spacetime, this property is imparted by its event horizon -- a null hypersurface that encloses the singularity at its heart. In the parlance of GR, a horizon is a hypersurface that causally disconnects one region of spacetime from another. An event horizon is a 3D hypersurface in the 4D spacetime, defined globally in the entire spacetime manifold~\cite{Helou:2016xyu,Ashtekar:2004cn}, which represents a boundary beyond which no event can causally affect an outside observer. In its simplest form, a BH is envisaged as an eternal unchanging vacuum object for which one knows the entire global structure of the spacetime. In reality, however, BHs are not eternal and unchanging; they interact with their surroundings and evolve, resulting in their horizons to be dynamical. While the notion of an event horizon has proved to be very convenient in mathematical relativity, its global nature renders the ideas of its ``detection" or ``observation" meaningless, as they associate an instant or a finite duration of time. What can be observed or detected, rather, is a submanifold of the event horizon described at an instant of time -- often referred to as an \textit{apparent horizon}~\cite{Thornburg:2006zb}. The concept of apparent horizons is used in \acrfull{nr} simulations involving BHs, achieved by decomposing the 4D spacetime into ``slices" of the 3D spatial manifold evolving in time.

The fact that BH horizons sever all the causal ties between their interior and the rest of the Universe has profound ramifications. One of the first confrontations of BH physics has been with the second law of thermodynamics, which dictates that any physical process in the Universe must contribute to an increment in its total entropy. The ``one-way" nature of the BH horizon and the second law of thermodynamics enforce a BH to suffer an entropy increase every time an external object is lost inside its horizon~\cite{Wald:1999vt}. This makes a BH an immense reservoir of entropy, accumulated from all the mass-energy it swallows over time. This was realized by Bekenstein in 1972~\cite{Bekenstein:1972tm}. Additionally, according to the \textit{no-hair theorem}, isolated BHs in equilibrium are completely specified by their mass, angular momentum and a net electric charge -- revealing the extreme simplicity of their nature. Mass of a BH has a simple correspondence with the area of its apparent horizon. Hawking showed, by his famous \textit{black hole area theorem}~\cite{Hawking:1971tu}, that under some conditions the area of a BH horizon can never decrease -- a property it shares with the entropy in thermodynamics. This theorem bears a great importance for merging black holes, which is essentially the subject of this thesis, and puts an upper limit on the amount of gravitational radiation that can be emitted in that process. At the same time, it prohibits any BH to split into smaller BHs -- a process manifestly in violation of the theorem.
\subsection{Classification of Black Holes}

BHs are classified in accordance with their formation history, or their three properties -- mass, spin and electric charge. The former category discriminates BHs by their mass ranges, and with far-reaching implications regarding their origin. Among them, stellar-mass BHs are generally formed at the end stage of massive stars as they deplete the nuclear fuel at their core over their lifetime and lose the ability to counteract the relentless pull of gravity. The star may detonate into a supernova explosion at the end of its life, the remnant of which collapses under gravity to form a BH. Subsequently, over cosmic time these BHs may absorb matter-energy from their surroundings and grow heavier; but typically they are believed to have masses ranging from $\sim 3M_\odot$ to $\sim 100M_\odot$. SMBHs, in contrast, are millions to billions of times heavier than the sun, as mentioned before, and therefore extremely unlikely to have emerged out of the stellar graveyard. At present, formation of SMBHs is still a mystery. The simplest possible resolution is that they grow out of ``seed" BHs by feeding on an enormous amount of matter throughout billions of years. However, this proposition faces serious challenges for the SMBHs observed in galaxies at redshift $z\gtrsim 6$~\cite{Valentini:2021cpr,Latif:2016qau}, when the Universe was barely one billion years old. The large gap between the masses of stellar BHs and SMBHs, from $\sim 100M_\odot$ to $\sim 10^5 M_\odot$, is occupied by the so-called \acrfull{imbhs}. IMBHs are believed to be relatively scarce in the Universe, owing to a lack of observational evidence. Nevertheless, they must exist at some point in cosmic time, and are believed to play a pivotal role in the formation of young SMBHs observed at high redshifts. The first confirmed detection of an IMBH came in 2019 form the gravitational-wave observation from a merging BH binary, the remnant of which was a $\sim 150 M_\odot$ BH~\cite{LIGOScientific:2020iuh,LIGOScientific:2020ufj}.

Classification of BHs by their physical properties, on the other hand, introduces four different kinds considering the presence or absence of their angular momentum and electric charge. Apart from their mass, which is a necessary property, the most general BHs can have an angular momentum (spin) and a net electric charge. These BHs are called \textit{Kerr-Newman black holes}. If they are electrically neutral, they are referred to as \textit{Kerr black holes}. If instead they possess an electric charge but are non-spinning, they are known as \textit{Reissner-Nordstr{\"o}m black holes}. Finally, the simplest form of a black hole is one that is neither spinning nor has any electric charge: the \textit{Schwarzschild black hole}.

Among these classes, the current and proposed future ground-based GW detectors are able to detect BHs of stellar masses to intermediate masses. Moreover, charged BHs are unlikely to exist in the Universe since they quickly accrete matter of opposite charge and become neutral~\cite{Gibbons:1975kk}. So in this thesis we focus on Schwarzschild and Kerr family of BHs having stellar to intermediate masses.

Henceforth, we will use the geometric units, setting $c=G=1$, except where mentioned explicitly otherwise.

\subsection{Schwarzschild Black Holes}

The Schwarzschild metric is a static, spherically symmetric solution of Einstein's field equations for the vacuum. A static spacetime requires that

\begin{enumerate}
    \item[(i)] all the metric components are independent of time $t$,
    \item[(ii)] its geometry is impervious to a time reversal $t\to -t$.
\end{enumerate}

In Schwarzschild coordinates $(t,r,\theta,\phi)$, the vacuum solution for a static, spherically symmetric spacetime can be written as
\begin{equation}
    \dd s^2 = -\left(1-\frac{R_S}{r}\right)\dd t^2 + \left(1-\frac{R_S}{r}\right)^{-1}\dd r^2 + r^2 \dd\Omega^2\,.
\end{equation}
$R_S$ is an undetermined constant (scalar). Imposing the condition that GR should reduce to Newtonian gravity in the weak-field limit, and recalling that in that limit, the $tt$ component of the metric becomes (in geometric units)
\begin{equation}
    g_{tt} = -\left(1-\frac{2M}{r}\right)\,,
\end{equation}
one can relate $R_S=2M$ in the asymptotically flat reference frame. $M$ here denotes the ``mass" of the BH. Since $M$ is a scalar, this relation holds generally, in any reference frame.
So, the Schwarzschild metric becomes
\begin{equation}\label{eq:sch metric}
    \dd s^2 = -\left(1-\frac{2M}{r}\right)\dd t^2 + \left(1-\frac{2M}{r}\right)^{-1}\dd r^2 + r^2 \dd\Omega^2\,.
\end{equation}

The Schwarzschild solution is by no means limited to BHs. It represents the \textit{vacuum} solution for \textit{any} static and spherically symmetric system, affirming that the solution is also valid outside, e.g., a spherically symmetric and static star. Furthermore, Birkhoff's theorem states that the solution is unique: it is the \textit{only} solution in GR for any static spacetime with spherical symmetry, with the ``mass" properly defined. 

\subsubsection{The Event Horizon}

The Schwarzschild metric introduces the concept of the \textit{Schwarzschild radius}, $R_S=2M$, which, for BHs, also represents the radius of their horizon. A BH defined by this geometry is called a \acrfull{sbh}. 



There are a plethora of novel features that the surface and the interior of the horizon exhibit. First, let us look at the horizon itself.

Let $P(x)$ be a smooth function of the spacetime coordinates $x^\mu$, and consider the family of hypersurfaces $P(x)=$ constant. The vector fields normal to these hypersurfaces are
\begin{equation}
    l = f(x)\left[g^{\mu\nu}\partial_\nu P(x)\right]\pdv{}{x^\mu}\,,
\end{equation}
where $f(x)$ is an arbitrary non-zero function. This gives
\begin{equation}
    l^2 = f^2 g^{\mu\nu}\partial_\mu P \partial_\nu P\,.
\end{equation}
Now let us consider the hypersurfaces $P=r-2M=$ constant. Since $P$ is a function of $r$ alone, the norm of a normal vector on $P$ becomes
\begin{equation}
    l^2 = f^2g^{rr}\partial_r P \partial_r P\,.
\end{equation}
In Schwarzschild spacetime, $g_{rr}=\left(1-\frac{2M}{r}\right)^{-1}$, and since the Schwarzschild metric is diagonal, $g^{rr}=\left(1-\frac{2M}{r}\right)$. Also, $\partial_r P =1$. Thus,
\begin{equation}
    l^2 = f^2\left(1-\frac{2M}{r}\right)\,. 
\end{equation}
So, for the hypersurface at $P=0$, i.e. $r=2M$, $l^2=0$. A hypersurface with this property is called a \textit{null hypersurface}, of which, as we have demonstrated, the BH horizon is an example. It has a unique property that the normal vectors are also its tangent vectors: if $\mathcal{N}$ is a null hypersurface with normal $l$ and a tangent $t$, then they must satisfy $l\cdot t = 0$. But $l\cdot l=0$, as we have shown, so $l$ itself is a tangent vector on $\mathcal{N}$. Since $l$ is now also a tangent vector, it can be written as
\begin{equation}
    l^\mu = \frac{\dd x^\mu}{\dd \lambda}\,,
\end{equation}
$x^\mu(\lambda)$ being a null curve on $\mathcal{N}$, as a function of some affine parameter $\lambda$. It can further be proved that $x^\mu(\lambda)$ are \textit{null geodesics}. In fact, a null hypersurface can be conceived as a collection of null geodesics, called the generators of the hypersurface.
The latter property is what offers the horizon its distinctive feature: no light ray can leave the horizon, since all the null geodesics are bound to the horizon surface.  



\subsubsection{Orbital Dynamics}\label{sec:orbital dynamics SBH}
It is of relevance to this thesis that we discuss the orbital dynamics in Schwarzschild spacetime, since we will study the GWs from orbiting compact objects to test their nature. Here, we briefly summarize the orbital dynamics in the test-mass limit, i.e. when the orbiting object is small enough not to impact the spacetime curvature itself.

The equation of orbits for a particle of mass $m$ in a Schwarzschild background of mass $M$ can be written as~\cite{Schutz:1985jx}
\begin{equation}
    \left(\dv{r}{\tau}\right)^2=E_m^2 - \left(1-\frac{2M}{r}\right)\left(1+\frac{L_m^2}{r^2}\right)\,,
\end{equation}
where $E_m=-p_t/m$ is the specific energy (energy per unit mass), $L_m=p_\phi/m$ is the specific angular momentum, and $\tau$ is the particle's proper time. $p_\mu\equiv (p_t,p_r,p_\theta,p_\phi)$ is the momentum one-form in Schwarzschild coordinates. For photons, the equation reads
\begin{equation}
    \left(\dv{r}{\lambda}\right)^2=E_p^2 - \left(1-\frac{2M}{r}\right)\frac{L_p^2}{r^2}\,,
\end{equation}
where $E_p=-p_t$, $L_p=p_\phi$, and $\lambda$ is some affine parameter for null geodesics, since proper time is undefined for massless particles.

One can define the ``effective potential" for the massive test particle as
\begin{equation}
    V^2_m(r)=\left(1-\frac{2M}{r}\right)\left(1+\frac{L_m^2}{r^2}\right)\,,
\end{equation}
and for photons,
\begin{equation}\label{eq:v-photon}
    V_p^2(r)=\left(1-\frac{2M}{r}\right)\frac{L_p^2}{r^2}\,.
\end{equation}

$V_m^2(r)$ has one maximum and one minimum, describing one unstable and one stable orbit, given a value of $L_m$. They are located at
\begin{equation}\label{eq:r-testmass}
    r=\frac{L_m^2}{2M}\left[1\pm \sqrt{1-\frac{12M^2}{L_m^2}}\,\right]\,,
\end{equation}
the lower radius being the unstable one. However, this defines a minimum value of $L_m$ for real solutions of $r$ : $L_m^2=12M^2$. Using this relation in Eq.~\eqref{eq:r-testmass}, we get a minimum radius below which stable circular orbits are not possible: 
\begin{equation}
    r=6M\,.
\end{equation}
Referred to as the \textit{innermost stable circular orbit} (\acrshort{isco}), it provides important wisdom in the analytical calculations of GW physics where compact objects orbit each other and spiral inwards by losing orbital radius.

For photons, on the other hand, Eq.~\eqref{eq:v-photon} provides only one unstable radius, irrespective of their momentum:
\begin{equation}
    r=3M\,.
\end{equation}
This is known as the \textit{photon sphere}. If the central object is a BH, the photon sphere plays a crucial role in its stability and is the primary property that casts the \textit{black hole shadow}~\cite{Mishra:2019trb,Shaikh:2018lcc}. The shadow of a BH is expected to be different from that of an ECO featuring a photon sphere, since some photons can pass through the object if it is weakly interacting with photons, or be reflected by the surface of the object due to a finite reflectivity. Recently, the substantial feat of imaging the SMBH candidates at the centre of the M87 galaxy~\cite{EventHorizonTelescope:2019dse} and our own Milky Way~\cite{EventHorizonTelescope:2022wkp} have been achieved by the Event Horizon Telescope collaboration. The observed results are so far consistent with BHs, and it is extremely challenging to tightly constrain the properties of these objects to test for alternate scenarios, due mainly to astrophysical uncertainties.

\subsection{Kerr Black Holes}
\label{sec:Intro_KBH}

BHs are expected to possess a nonzero spin angular momentum, and indeed, some BHs observed by LIGO show the evidence of spin. This makes it important in this context to discuss the geometry around a spinning BH. A spinning, electrically neutral BH is called a Kerr black hole (\acrshort{kbh}). A KBH is clearly not static, but it is \textit{stationary} -- a superset of the static condition. The metric components are still independent of time, but the condition of time reversal symmetry is, in this case, relaxed.
Due to a theorem by Hawking~\cite{Hawking:1971vc}, any stationary metric must be axially symmetric, which a KBH indeed complies with.

In the so-called Boyer-Lindquist coordinates, the Kerr metric can be written as
\begin{align}\label{eq:kbh-metric}
    \dd s^2 = & -\left(1-\frac{R_S r}{\rho^2}\right)\dd t^2 + \frac{\rho^2}{\Delta}\dd r^2 + \rho^2\dd\theta^2 - \frac{2R_S ar\sin\theta}{\rho^2}\dd t\dd \phi \nonumber \\  &  +\left(r^2+a^2+\frac{R_Sra^2}{\rho^2}\sin^2\theta\right)\sin^2\theta\dd\phi^2\,.
\end{align}
Here 
\begin{align}
    R_S&=2M\,,\\
    \rho^2&=r^2+a^2\cos^2\theta\,,\\
    \Delta&=r^2-R_Sr+a^2\,.
\end{align}
$M$ is the mass of the BH, and $a=J/M$, $J$ being its angular momentum.

One major difference between the Schwarzschild solution and the Kerr solution is that there is no analog of the Birkhoff's theorem for an axially symmetric spacetime. It follows that the Kerr solution, which describes a spinning BH in equilibrium, may not represent the spacetime around a spinning material object, e.g. a star.

\subsubsection{The Event Horizon}

The event horizon is the hypersurface inside which the radial coordinate becomes timelike, as discussed in the SBH case. Mathematically, one demands $g^{rr}=0$ on that surface. In the KBH scenario, Eq.~\eqref{eq:kbh-metric} produces two such hypersurfaces as the solutions for $g^{rr}=0$:
\begin{equation}\label{eq:RK}
    R_K^{\pm} = \frac{R_S}{2} \pm \sqrt{\left(\frac{R_S}{2}\right)^2-a^2}\,.
\end{equation}
It can be shown that both are null hypersurfaces. However, from an observational point of view, only the outer solution at $R_K^+=M+\sqrt{M^2-a^2}$ (using $R_S=2M$) presents the attributes of a horizon and its ``one-way" nature to the outside Universe. 

The expression in Eq.~\eqref{eq:RK} also defines the upper limit of the magnitude of angular momentum that a KBH of mass $M$ can rotate with, since $M^2$ has to exceed $a^2$ for real solution for $R_K^\pm$ :
\begin{equation}
    |a|_{\rm max}=M\quad \quad \implies \quad\quad \left|\frac{J}{M^2}\right|_{\rm max}=1\,.
\end{equation}
The quantity $J/M^2$ is referred to as the dimensionless spin $\chi$, its range being $\chi\in[-1,1]$.

\begin{figure}
    \centering
    \includegraphics[width=1\linewidth]{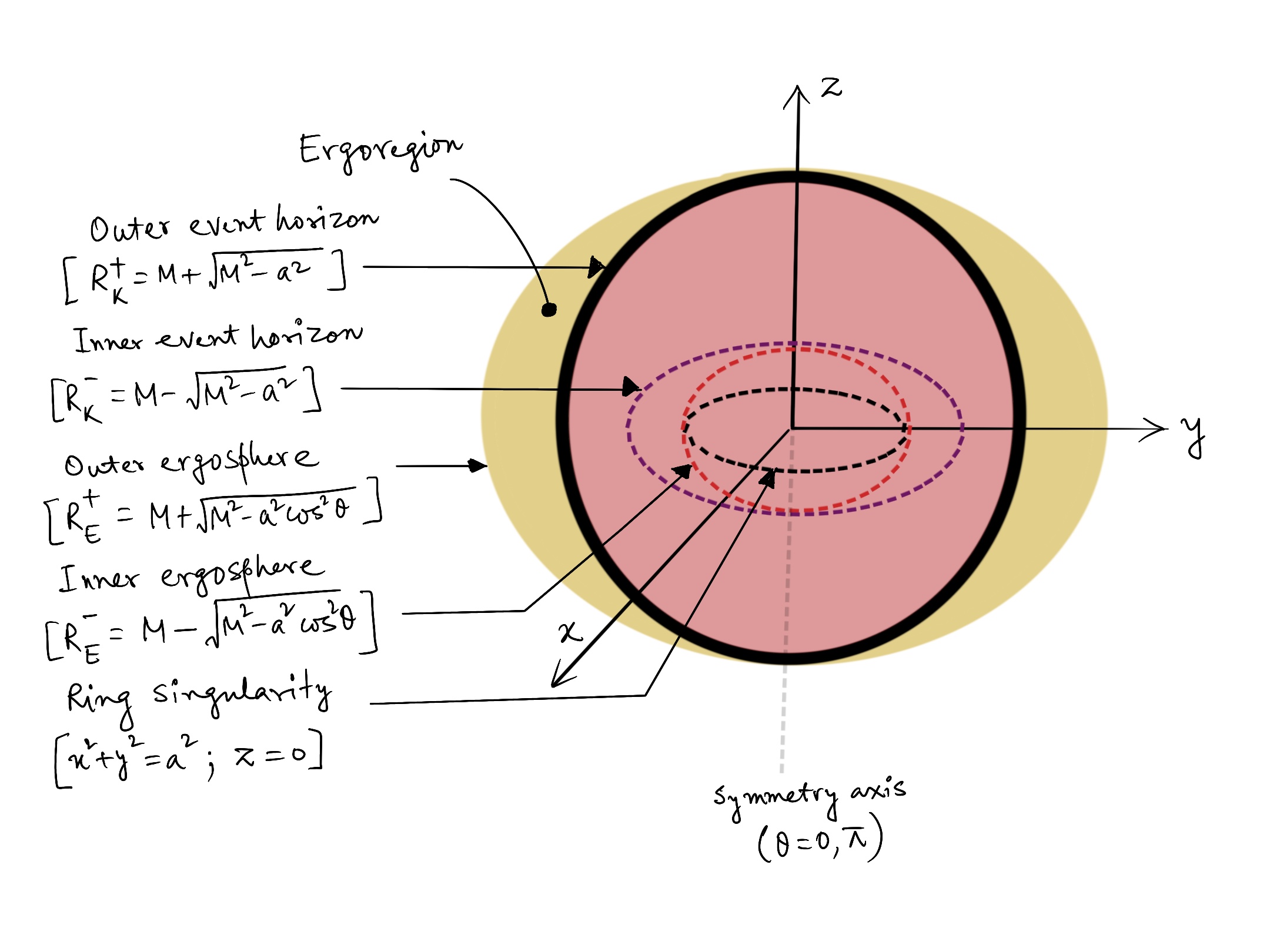}
    \caption{Illustration of the structure of a KBH.}
    \label{fig:kbh}
\end{figure}

\subsubsection{The Ergosphere}

In the Schwarzschild case, the event horizon also represented the ``infinite redshift" surface, i.e. photons emanating from a source reaching the horizon would be infinitely redshifted to an observer far away. This surface appears as the solution of $g_{tt}=0$, which, for an SBH, is also at $r=2M$. For a KBH, however, the horizon and the infinite redshift surfaces are split up; the latter being at
\begin{equation}
    R_E^\pm = M \pm \sqrt{M^2-a^2\cos^2\theta}\,.
\end{equation}
These are called the \textit{ergosurfaces}. Again, to a distant observer, only the outer ergosurface at $R_E^+ = M + \sqrt{M^2-a^2\cos^2\theta}$ has observational consequences.

The region between the ergosurface and the outer horizon is called the \textit{ergoregion}. In the ergoregion, no particle can remain static; moreover, they must rotate along the hole's angular momentum, regardless of the magnitude or direction of its own angular momentum prior to entering the region. In Fig.~\ref{fig:kbh}, different regions and surfaces of a KBH are illustrated. Curiously, the singularity of a KBH defines a ring of radius $a$ instead of a point in space.

One of the most distinct features of a KBH is that it can \textit{lose} energy via interactions with its surroundings. An external agent may cause the spin of a KBH to slow down, thereby leaving the hole with a lesser rotational energy. The energy expense, however, is entirely made by the ergoregion. The horizon area is connected to the entropy, and remains ever-increasing. In 1971, Christodoulou~\cite{Christodoulou:1971pcn} showed that the mass of a rotating BH can be divided into two parts, one is connected to the horizon area and can never decrease, and the other relates to its rotational energy which can be extracted by an external influence. 

In this thesis, we uphold the fact that when subjected to an external perturbation, the response of a KBH has measurable difference than a massive horizonless object. The perturbation, in our case, is the gravitational perturbation caused by the companion of the object in a binary inspiral, whose GW signatures subsequently carry information on the presence of horizons.












\section{The Black-hole Mimickers}

There are several possible species of ECOs hypothesized to exist in the Universe. In our context, we refer to an object as a ``compact object" if it features an ISCO (for the static and spherically symmetric case, this implies that the radius $R<6M$). This property has extremely important consequences in the GWs emitted from a compact binary merger. In this section we will discuss the various possible ECOs that have been proposed.

Within GR, Buchdahl's theorem~\cite{Buchdahl:1959zz} prohibits the compactness of any object to be arbitrarily close to that of a BH. Particularly, the theorem defines an upper limit on the compactness of any object of mass $M$ and radius $R$ under some assumptions: $(M/R)_{\rm max}=4/9$. The assumptions are as follows:
\begin{enumerate}
    \item GR describes the theory of gravity,
    \item The object respects spherical symmetry,
    \item The object is composed of a single perfect fluid,
    \item The fluid is isotropic, or at least the anisotropy is mild,
    \item The radial pressure or the energy density are not negative,
    \item The radial pressure decreases from the centre towards the surface.
\end{enumerate}
Relaxing each of these assumptions, however, opens up possibilities to predict objects that violate the Buchdahl limit and have compactness close to that of BHs. In this section, we discuss a few of such objects.

\subsection{Boson Stars}

Boson stars are one of the earliest and simplest known examples of ECOs. They are the self-gravitating compact configurations of a massive scalar field minimally coupled to gravity~\cite{Kaup:1968zz,Ruffini:1969qy,Khlopov:1985fch,Mukherjee:2014kqa}. For complex scalar fields, static and spherically symmetric solutions are possible. Real scalar fields, on the other hand, produce oscillating geometries with a time-dependent stress-energy tensor.
Both these configurations are possible at the end of a gravitational collapse. Static boson stars correspond to a family of solutions related to a single parameter, namely, the value of the bosonic field at the centre. These solutions predict a maximum mass for the star above which it becomes unstable against radial perturbations. The maximum mass depends heavily on the strength of the self-interactions. A stronger self-interaction allows a higher value of the mass and compactness of the star~\cite{Schunck:2003kk}. They have substantial anisotropies in the stress-energy tensor, allowing them to evade Buchdahl's theorem. However, there are no known solutions for these stars that breach the Buchdahl limit for the compactness. The most compact static boson star theoretically found so far has $R\approx 2.869M$~\cite{Kesden:2004qx}. When spin or nonlinear interactions are introduced, boson star spacetimes can have photonspheres and ergoregions~\cite{Macedo:2013jja,Grandclement:2016eng}, making it possible for these objects to mimic the phenomenological signatures of KBHs related to these spacetime features.

Boson stars are one of the primary candidates of ECOs that can act as BH mimickers. They are also interesting objects in their own right, since the presence of ultralight scalars has the potential to resolve the dark matter puzzle~\cite{Hui:2016ltb}.

\subsection{Anisotropic Stars}

Introduction of fairly large anisotropies can evade the Buchdahl limit~\cite{Andreasson:2007ck}. Various physical scenarios can be responsible for a substantial anisotropy in a compact object~\cite{Kippenhahn:2012qhp,Ruderman:1972aj}. There have been numerous studies on anisotropic stars with static and spherically symmetric configurations~\cite{Bowers:1974tgi,Letelier:1980mxb,Bayin:1982vw,Dev:2000gt,Mak:2001eb,Herrera:2004xc,Silva:2014fca,Yagi:2015hda,Yagi:2016ejg}. In neutron stars, pressure anisotropy can introduce significant modification in their tidal properties~\cite{Biswas:2019gkw}. With increasing anisotropy, highly compact objects arbitrarily close to BHs can be created within a wide range of possible masses.

\subsection{Dark and Hybrid Stars}

Dark matter can play a significant role in producing BH mimickers. Fermionic stars like neutron stars live in dark matter rich environments, and can capture dark matter by gravitational accretion and a nonzero cross-section of interaction with the star material~\cite{Gould:1989gw}. The dark matter material eventually thermalizes with the star and forms a composite object, or a hybrid star. The compactness, however, remains similar to that of the host neutron star and does not violate the Buchdahl limit in general.

Other scenarios include dark matter stars made entirely of bosonic or fermionic dark matter particles~\cite{Giudice:2017dde}. If these particles are light, or they have strong self-interactions, they can form massive and compact stars. These stars can potentially explain the general properties of the GW signals currently inferred as coming from mergers of binary black
holes~\cite{Zhang:2023hxd,Johnson-Mcdaniel:2018cdu}.

\subsection{Gravastars}

The ``gravitational-vacuum stars", or gravastars, are stellar configurations supported by a negative pressure~\cite{Mazur:2001fv,Mazur:2004fk}. The negative pressure leads to violation of some of the energy conditions. Gravastars belong to a wide class of objects, with different models supporting the pressure in different ways. Originally, they were constructed with an interior de Sitter core, and an envelope of a perfect fluid region separated from the interior by a thin shell. Another thin shell separates the perfect fluid region from the Schwarzschild exterior. A simpler model, the thin-shell gravastar, is constructed by separating the exterior and the de Sitter interior by a single thin shell of perfect fluid matter. Gravastars can be obtained as the BH limit of constant-density stars, and violate the Buchdahl limit for compactness. They have been found to be dynamically stable~\cite{Posada:2018agb}, and can appear naturally in an inflationary Universe~\cite{Wang:2018cum}.

\subsection{Wormholes}

Wormholes are one of the very promising and interesting candidates as ECOs and BH mimickers. Originally introduced by Einstein and Rosen~\cite{PhysRev.48.73}, these are two asymptotically flat spacetimes connected by a ``throat". Within GR, they are not vacuum objects and require matter. The structure and theoretical motivation of wormholes are well understood, but their possible formation mechanisms are not known properly. However, recent developments suggest that they can be stabilized and created with possibly reasonable matter content~\cite{Lemos:2003jb,Maldacena:2018gjk}. Wormholes can be arbitrarily compact and massive. Even though their formation is relatively less understood at present, wormholes make a possible alternative to BHs from an observational viewpoint.

\subsection{Other ECOs}

In addition to the ECO candidates discussed above, there are several other exotic alternatives to BHs. They include quasiblack holes~\cite{Lemos:2003gx,Lemos:2008cv}, dark stars produced by a halted collapse~\cite{barcelo2009black,Kawai:2013mda,Baccetti:2017oas}, quantum BHs~\cite{Bekenstein:1995ju,Ashtekar:1997yu} and many others~\cite{Cardoso:2019rvt}.

This diverse zoo of ECOs, capable of masquerading as BHs in the Universe, underscores the importance of designing observational tests to differentiate BHs from their alternatives. In the following chapter, we will concentrate on discerning the signatures of BH properties in the GW signals emitted during compact binary coalescences. Specifically, we will explore one of the potential tests, known as tidal heating of compact objects, to establish a framework for characterizing BHs in LIGO-Virgo binaries.
\clearpage

\chapter{Unmasking the Black-hole Mimickers}
\label{Chapter2}

ECOs, as we discussed in the last chapter, can be arbitrarily similar to BHs, and can show some of the features that are hallmarks of BHs, such as the ISCO or the photon sphere. This introduces severe challenges in identifying a BH unequivocally from the dynamics of the environment around it, or its response to an external influence. Since the telltale signature of a BH is the one-way nature of its horizon, strategies to pinpoint the black-holeness of a compact object should involve null-hypothesis tests by indulging possible deviations in observable signatures and examining the consequences thereof. This chapter devotes itself in discussing some of such deviations. Among them, tidal heating of compact objects is the phenomenon central to the contents of this thesis. 

Electromagnetic (EM) observations, e.g. the images produced by the Event Horizon Telescope discussed in Sec.~\ref{sec:orbital dynamics SBH}, can help towards the goal of classifying compact objects~\cite{Cardoso:2019rvt}. GWs, on the other hand, are produced by the coherent motion of the component objects in a compact binary coalescence (\acrshort{cbc}), and offer a wealth of information about the compact objects from their slow inspiral phase to a highly dynamic merger phase presenting the opportunity to probe the strong-gravity regime. Unlike EM waves, GWs interact very weakly with matter, allowing them to travel through the Universe virtually unimpeded with the information they carry mostly unimpaired. This in turn makes it feasible for simpler models with lesser parameters to describe them, introducing lesser uncertainties. We will retain our attention only to the GW signatures of BH characterization. 


\section{The Distinguishing Properties}

A CBC is broadly divided into three phases, the inspiral, the merger, and the ringdown. The inspiral phase corresponds to the objects being largely separated from each other and moving relatively slowly. Within this regime one can perform, under some approximations, analytical calculations to describe the GWs emitted by the system. The transition from the inspiral to the merger phase is usually chosen to correspond to the ISCO of the binary, following which the objects ``plunge" towards each other. This regime is highly nonlinear, and one needs to solve the Einstein field equations numerically. After the objects merge, the remnant settles down to equilibrium by shedding away the deformities impinged by its perturbed state in the form of GWs. This is known as the ringdown. Here we discuss some viable tests involving these regimes individually or collectively.

\subsection{Quasi-normal Modes}

Studying the ringdown spectra is one of the simplest and most viable tests. The GWs radiated by the merger remnant while it evolves to a Kerr hole from its perturbed state convey unique hallmarks of the identity of the source. Assuming the background metric to be $g_{\mu\nu}=g_{\mu\nu}^{(0)}+h_{\mu\nu}$, where $g_{\mu\nu}^{(0)}$ is the geometry corresponding to the stationary state of the object and $h_{\mu\nu}$ is the small perturbation on it, one arrives at the following equation:
\begin{equation}\label{eq:qnm}
    \pdv[2]{\Psi(t,z)}{z}-\pdv[2]{\Psi(t,z)}{t}-V(r)\Psi(t,z)=S(t,z)\,,
\end{equation}
where $z$ is a suitable coordinate (here we assume $z\to \infty$ at spatial infinity and $z\to -\infty$ at the horizon) and $\Psi(t,z)$ is a master function that captures the metric fluctuations. $S(t,z)$ is the source term, and $V(r)$ is the effective potential defined by the spacetime around the object.

The solution of Eq.~\eqref{eq:qnm} can be obtained by specifying the source term and the boundary conditions. For a BH, the waves can only travel inward at the horizon owing to its unique nature, and only outward at infinity to respect causality. Under these conditions, the solution presents a set of vibrational modes of the BH related to complex frequencies. The frequencies, being complex, reflect both oscillations and damping of the modes. These are called the quasi-normal modes (QNMs).

For a non-BH, the structure of the QNM spectrum can be significantly different due to different boundary conditions. Obeying the no-hair theorem, BH QNMs are completely specified by the mass and the angular momentum of the BH. However, ECOs do not need to adhere to this rule. If the horizon is replaced by a (partially) reflecting surface very close to where the horizon would be, information about the surface becomes encoded in the QNM spectra.

Several studies of the vibrational spectra of ECOs have been performed over the years~\cite{Pani:2009ss,Mazur:2015kia,Macedo:2013jja,Chirenti:2016hzd,Nandi:2016uzg}. Even though currently the GW detectors are not sensitive to multiple QNMs for most of the GW events, the QNMs may be detectable in the near future~\cite{TheLIGOScientific:2016src,Berti:2018vdi}.

\begin{figure}
    \centering
    \includegraphics[width=\textwidth]{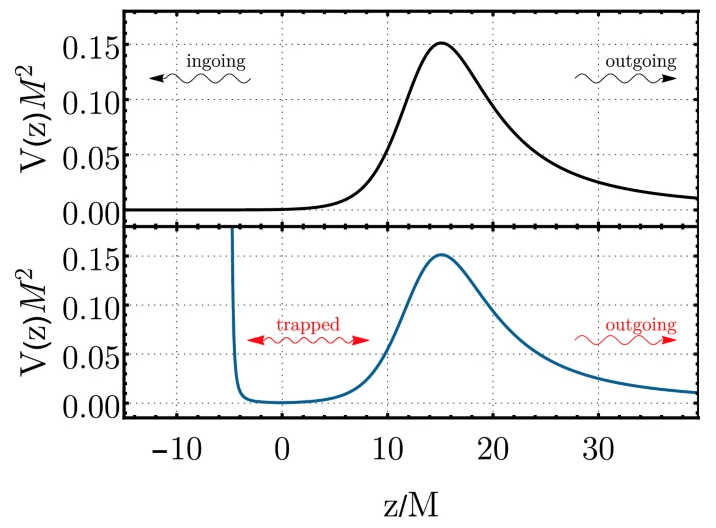}
    \caption{Illustration of the effective potential for a perturbed SBH (top panel) and a horizonless compact object with a partially reflective surface (bottom panel). The peak appears approximately at the photon sphere, $r\approx 3M$. The BH QNMs are only outgoing at infinity ($z\to\infty$) and ingoing at the horizon ($z\to -\infty$). An ECO, on the other hand, introduces a partly reflective surface, causing a part of the waves to be trapped between its surface and the photon sphere. Image source: Cardoso and Pani (2019)~\cite{Cardoso:2019rvt}. }
    \label{fig:echo1}
\end{figure}

\subsection{Echoes}

For an ultracompact object featuring a photon sphere, the latter marks the position of a maximum in the effective potential. For a BH, the ringdown signal is almost entirely described by the lowest QNMs. A fraction of this signal escapes to infinity and has observable properties. The other fraction of the signal generated at the photon sphere travels into the horizon and disappears, leaving no effect on observables at large distances.

In the case of ECOs, the partial reflection at the surface traps a part of the ingoing wave. Under some conditions~\cite{Cardoso:2019rvt}, this results in a series of GW pulses separated by uniform time intervals roughly corresponding to twice the light travel time between the photon sphere and the surface. These are called gravitational-wave echoes. Presence of GW echoes can serve as one of the explicit signatures of a horizonless object~\cite{Cardoso:2017cqb,Maggio:2023fwy,Uchikata:2023zcu,Biswas:2022wah}.

\subsection{Multipole Moments}

So far we have considered only the postmerger part of GW signals from a CBC. This and the remaining sections discuss the inspiral and merger parts, and how one can test for the existence of horizons from a binary system of inspiraling compact objects.

The system of two inspiraling bodies can be analytically solved in a perturbative manner under the post-Newtonian (\acrshort{pn}) approximation~\cite{Blanchet:2013haa} while they are far from each other and slowly moving. 
Starting from the second post-Newtonian (2PN) order, the spin-orbit and spin-spin interactions affect the GW phase evolution of the binary. The dominant effect is the spin-induced quadrupole moment, modifying the GW phase at 2PN order. For an isolated KBH, the quadrupole moment scalar is $Q=-m^3\chi^2$, whereas for a non-BH compact object, a modification is introduced: $Q=-\kappa m^3\chi^2$~\cite{Krishnendu:2017shb}. Evidently, $\kappa=1$ denotes the KBH limit. Depending on the nature of the object, $\kappa$ can have a wide range of possible values~\cite{Psaltis:2008bb,Gair:2012nm}. For example, NSs may have $\kappa$ ranging from $\sim 2$ to $15$~\cite{Laarakkers:1997hb}; for boson stars the range may be from $\sim 10$ to $150$~\cite{Ryan:1996nk}. Constraining the value of $\kappa$, therefore, opens up the possibility to draw inferences regarding the nature of the object(s). Several studies have investigated the measurability of this parameter from GW observations~\cite{Krishnendu:2018nqa,Krishnendu:2019tjp,Saleem:2021vph,}, and with promising prospects. Similar probes can also be performed for higher-order effects like spin-induced octupole moments, appearing at 3.5PN order~\cite{Saini:2023gaw}. Naturally, highly spinning bodies facilitate this test due to the induced multipole moments being higher.

\subsection{Tidal Love Numbers}

Tidal response of the component objects is an important discriminator for BHs, since this response depends largely on the nature of the objects. In a CBC, the orbiting bodies spiral inwards while they are subjected to each other's tidal gravitational field. The tidal field induces deformations in each object; and depending on the nature of the object, these deformations can vary substantially. For objects composed of matter, it is the properties of the matter content that dictate the tidal response. In NSs, for example, the tidal response carries important information about their equation of state~\cite{Flanagan:2007ix,Hinderer:2009ca}, a subject under ardent scrutiny in astrophysical research to understand their composition. In general, an object subjected to a tidal field $\mathcal{E}_{ij}$ develops, to linear order, a tidally induced quadrupole moment
\begin{equation}
    \mathcal{Q}_{ij}=-\lambda\mathcal{E}_{ij}\,,
\end{equation}
with $\lambda$ being a parameter, called the (dimensionful) ``tidal deformability", that reflects the tidal response. $\lambda$ is related to the so-called ``tidal Love number" (TLN) $k_2$ by the relation $k_2=(3/2)\lambda/R^5$, where $R$ is the object's radius. The subscript 2 corresponds to the $l=2$ mode. The contribution of TLNs first appear at 5PN order in the GW phase.

Intriguingly, TLNs for nonspinning BHs identically vanish within GR~\cite{Fang:2005qq,Binnington:2009bb,Gurlebeck:2015xpa}. Significant research has gone into investigating the TLNs for spinning BHs~\cite{LeTiec:2014oez,Chia:2020yla,Bhatt:2023zsy}, but the issue remains unresolved to date. However, TLNs must be nonvanishing for ECOs~\cite{Uchikata:2016qku,Cardoso:2017cfl,Pani:2015hfa}. Constraining TLNs from GW data, therefore, is a viable test for black-holeness~\cite{Sennett:2017etc}.

\subsection{Tidal Heating}

Finally, we arrive at the test that this thesis analyzes primarily -- tidal heating (\acrshort{th}). Broadly, TH occurs in a binary system when the spin of an orbiting body does not match the orbital frequency of its companion, from the reference frame of the former object. This leads to frictional energy dissipation within the object due to its viscosity, leading to what is referred to as tidal heating. Within our solar system, Jupiter's moon Io exhibits rich volcanic activities due to this effect~\cite{tyler2015tidal}. Energy dissipated within the object is drained from the orbit and, over time, the spin and the orbital frequency of the orbiting body tend to coincide, leading to what is known as tidal locking. Most of the large moons of the solar system are tidally locked with their planets, including our own. 

In a CBC radiating GWs, this effect modifies the inspiral rate by draining energy from the orbit. If the orbiting body is a BH, however, the phenomenon is physically different but shows remarkable similarity regarding the consequences~\cite{Hartle:1973zz}. A spinning BH subjected to an external (non-axisymmetric) perturbation suffers a change in its angular momentum so that eventually it matches with the orbital frequency of the perturbing agent. In the case of a CBC, this perturbation is caused by the companion object. The mass of the BH also changes by absorbing energy from the orbit.

A part of the orbital energy in a CBC travels to the future null infinity in the form of GWs. Considering TH, another part is absorbed by the BH horizon. If the BH is replaced by an ECO, however, the presence of a partially reflecting surface instead of the horizon leads to a lesser absorption of the TH fluxes. In a generic binary, this flux can be treated as a fraction of that in a BBH~\cite{Datta:2019epe}. This fraction is represented by the \textit{horizon parameters}. These parameters correspond to unity for BBHs, and are vanishing for binaries with negligible TH compared to BBHs~\cite{datta2020recognizing}. For an ECO with a partially reflecting surface, the horizon parameter satisfies $H\in[0,1]$. Under the post-Newtonian scheme, one can estimate the change in the mass and spin of a KBH influenced by TH~\cite{Alvi:2001mx,Poisson:2004cw,Chatziioannou:2012gq,Chatziioannou:2016kem} and, using the horizon parameters, one can infer about the nature of the compact objects in a generic binary. In the following chapters, we study how well these parameters can be measured, and how one can leverage the strong-gravity regime of a CBC for BH characterization where TH leaves stronger effects in the GW signals.



\clearpage

\chapter{Horizon parameters in Post-Newtonian Waveforms}
\label{Chapter3}
\graphicspath{{30_Chapter_3/figures/}}

\section{Introduction}
\label{intro}


Incorrect inferences about the true nature of the compact objects in a compact binary coalescence (CBC) can have far-reaching implications, such as on population models of compact objects.
In the future, the proposed 3rd generation ground-based GW detectors Einstein Telescope~\cite{maggiore2020science} and Cosmic Explorer~\cite{Reitze:2019iox} are expected to have order-of-magnitude better sensitivity compared to the current ones in estimating the source parameters, which should enable us to probe the nature of these objects more accurately. 
Building separate models for each of these exotic objects is a hard problem, and accurate measurements of their properties are not yet possible with the current detectors. So, a more practical approach would be to devise tests that are generic and
model-independent and are based on our understanding of binary black hole (BBH) dynamics. 

As discussed in Chapter~\ref{Chapter2}, one way to probe the presence of ECOs against BHs is to understand the possible ways in which the characteristics of these objects can differ from those of BHs, and that can be
confirmed or ruled out by introducing appropriate free parameters in the gravitational waveform. In order to develop such model-independent tests of BH mimickers, it is
important to identify the properties that are unique to BHs, and investigate their imprints on the gravitational waveform so that we can measure them from observations. In this chapter, we investigate how the phenomenon of tidal heating can be leveraged in this regard.  


Due to their causal structure, BHs in GR are perfect absorbers that behave as dissipative systems~\cite{MembraneParadigm,Damour_viscous,Poisson:2009di,Cardoso:2012zn}.  A significant feature of a BH is its horizon, which is a null surface and a ``one-way membrane" that does not allow energy to escape outward. These tidal effects cause changes in their mass, angular momentum, and horizon area. This phenomenon is called {\it tidal heating} (TH)~\cite{Hartle:1973zz,Hughes:2001jr,PoissonWill}. If the BH is nonspinning, then energy and angular momentum can only flow into the BH. However, spinning BHs can transfer their rotational energy from the ergoregion out into the orbit due to tidal interactions with their binary companion.
Energy exchange via  TH backreacts on the binary's evolution, resulting in a shift in the phase of the GWs emitted by the system. 
Effect of TH on objects such as NSs or horizonless ECOs is comparatively much less due to their lack of a horizon. So, a careful measurement of this phase shift can be used
 in principle to distinguish BHs from horizonless compact objects~\cite{Maselli:2017cmm, Datta:2019euh, Datta:2019epe, Datta:2020rvo, Agullo:2020hxe, Chakraborty:2021gdf, Sherf:2021ppp, Datta:2021row, Sago:2021iku, Maggio:2021uge, Sago:2022bbj}.

To quantify this effect, two ``horizon parameters", $H_1$ and $H_2$, were introduced in a recent study~\cite{datta2020recognizing} whose utility we will study further in characterizing compact objects. These parameters take the value of 1 when the objects are BHs, and $0\leq H_1,H_2<1$ for other compact objects. The phase shift in GWs due to TH will depend on these parameters. Their accurate measurement, in turn, will indicate the presence or absence of BHs in a binary. It turns out that the covariance of these two parameters is generally finite. We therefore find two combinations of these parameters that are mostly statistically independent. Even then, some covariances between the new parameters can arise due to waveform systematics, non-stationary detector noise, etc. Parameter estimation (PE) exercises for real GW signals widely use Bayesian approaches~\cite{datta2020recognizing, Thrane_2019}, which is a robust method, but computationally expensive. Fisher studies~\cite{Vallisneri:2007ev} can provide reliable estimates for the errors and uncertainties in measuring the source parameters of GW signals with high signal-to-noise ratios, and is much faster and less expensive. In the current work, we will primarily use the latter approach for estimating the errors, and explore Bayesian simulations to corroborate the results.

A CBC consists of three major phases - inspiral, merger, and ringdown. One can model the inspiral phase using post-Newtonian (PN) formalism, whereas numerical relativity (NR) simulations are needed to model the merger regime~\cite{Pretorius:2007nq}. In order to study the ringdown part of the dynamics, one may use BH perturbation theory techniques~\cite{Sasaki:2003xr} or NR. Tidal heating is relevant in the inspiral and is more significant for a binary when the components are closer together so that their tidal interactions are stronger. In the PN regime, TH can be incorporated into the gravitational waveform by adding the phase shift due to this effect into a PN approximant in the time or frequency domain. 

In Sec.~\ref{theory_TH}, we will review the basic framework of tidal heating, describing the waveform parameterization chosen for this work. Sec.~\ref{fisher_basics} will summarize the concepts of the Fisher matrix analysis, discussing various relevant aspects of it, and the corresponding results will be presented in Sec.~\ref{results}. In Sec.~\ref{sec:bayesian} we will present results from  Bayesian simulations to check the consistency of our analyses. In Sec.~\ref{pca}, we will discuss the covariances between the relevant parameters and possibilities of improving the results by diagonalizing the Fisher matrix. We will summarize the results in Sec.~\ref{discussion}, and discuss the relevance of this work to future studies.

Throughout the chapter, we continue use geometric units, assuming $G=c=1$, except when calculating physical quantities.


\section{Theory of Tidal Heating and Waveform Parametrization}
\label{theory_TH}

The PN formalism~\cite{Blanchet:2013haa} describes the gravitational waveform emitted by a stellar-mass compact binary in its early inspiral phase. In this formalism, the evolution of the orbital phase $\Psi(t)$ of a compact binary is computed as a perturbative expansion in a small parameter, typically taken to be the characteristic velocity $v = (\pi M f)^{1/3}$. Here $M$ is the total mass of the binary and $f$ is the instantaneous GW frequency. This analytical procedure demands $v\ll 1$, which makes it useful in the early inspiral phase of a CBC. For building a proper PN waveform, one begins with the gravitational waveform from an inspiraling pair of point particles (PP). Extra terms are added based on the nature of the binary components. If a component has a finite size and inner structure, e.g., an NS, then {\it tidal deformability}  plays an important role \cite{Flanagan:2007ix}. If there is a BH involved, then the effect of its horizon has to be considered. This is where {\it tidal heating} comes into play.

A Kerr black hole (KBH), discussed in Sec.~\ref{sec:Intro_KBH}, is stationary when it is isolated. On the other hand, when a KBH is a member of a binary, it feels its companion's tidal field, which acts as a non-axisymmetric perturbation~\cite{Hartle:1973zz}. This perturbation causes changes in the mass, spin, and horizon area of the KBH over time~\cite{Alvi:2001mx}. Since the KBH experiences the tidal field of its orbiting companion, it absorbs (emits) energy from (into) the orbit. The absorption part is present in non-spinning BHs as well. Additionally, for a KBH, the difference between the spin frequency and the angular frequency of the tidal field causes the spin to slow down, which in turn makes the KBH lose its rotational energy. The slowing down of a rotating BH due to the gravitational dissipation produced by exterior mass is analogous to the slowing down of a rotating planet by viscous dissipation due to tides raised by an exterior moon that increases its internal thermal content - a phenomenon known as tidal heating. Due to this similarity, the energy and angular momentum flux in BBHs is also termed tidal heating~\cite{Poisson:2004cw}.

The gravitational waveform for a specific binary will include contributions from these factors depending on its components. For a generic binary, we can write the frequency domain strain $\Tilde{h}(f)$ as 

\begin{equation}
\label{hf}
    \Tilde{h}(f) =  \Tilde{A}(f) e^{i\left(\Psi_{\rm PP}+\Psi_{\rm TD}+\Psi_{\rm TH}\right)},
\end{equation}
where $\Tilde{A}(f)$ is the frequency-dependent amplitude. The phase terms -- $\Psi_{\rm PP}, \Psi_{\rm TD},$ and $\Psi_{\rm TH}$ -- arise from the point-particle approximation, tidal deformability, and tidal heating, respectively.

Since GW absorption is negligible for matter~\cite{Glampedakis:2013jya},
TH can be a way to discern the existence of horizons~\cite{Datta:2019euh, Maselli:2017cmm}. Reference~\cite{Datta:2019euh} introduced the {\it horizon parameter} $H$ for extreme mass-ratio inspirals (EMRIs) for this purpose. In Ref.~\cite{datta2020recognizing}, the authors extended this to a more general case, introducing horizon parameters for both objects as $H_1$ and $H_2$. Strictly speaking, these two parameters denote the \textit{fraction} of the flux due to TH in any binary to that in a BBH, and they take values $H_{1,2}\in[0,1]$.

In case of circular orbits, the flux of energy at the horizon can be expressed as a PN expansion~\cite{Alvi:2001mx, Poisson:2018qqd, Poisson:2009di, Nagar:2011aa, Bernuzzi:2012ku,Chatziioannou:2016kem, Cardoso:2012zn}. Since TH implies the presence of horizon, the TH energy flux due to each component has to be multiplied with the corresponding $H_i$.

 Let us consider a compact binary with individual masses $m_1$ and $m_2$, dimensionless spins $\chi_1$ and $\chi_2$, total mass $M=m_1+m_2$ and mass-ratio $q=m_1/m_2$ with $m_1\geq m_2$. In the case of partial absorption, one has $0<H_i<1$. Then the absorbed flux can be expressed as~\cite{datta2020recognizing}
\begin{equation}
\label{dedt}
\begin{aligned}
    -\dv{E}{t} = &{}\frac{32}{5}\eta^2 \frac{v^{15}}{4}\sum_{i=1}^{2} H_i\left(\frac{m^{}_i}{M}\right)^3 \left( 1 + 3\chi^2_i\right)\\&
    \times\left\{-(\hat{L}\cdot\hat{S}_i)\chi^{}_i
     + 2 \left[ 1+\sqrt{1 - \chi^2_i}\right]\frac{m^{}_i}{M}v^3\right \}\ ,
\end{aligned}
\end{equation}
where 
$\eta = {m_1 m_2}/{M^2}$ is the symmetric mass-ratio,  $v$ is the characteristic velocity, and
$\hat{S}^{}_i$ and $\hat{L}$ are the unit vectors along the directions of the $i$th object's spin and the orbital angular momentum, respectively.

There are a few things to note from this expression. This is the expression for the rate of energy \textit{absorption} by the compact object. For a spinless binary, i.e. $\chi^{}_1=\chi^{}_2=0$, the right-hand side survives, meaning that tidal heating is still possible; but in that case it is always positive, which means that the energy flux can only be inward and not outward (which is expected for non-spinning BHs). The presence of the term $-(\hat{L}\cdot\hat{S}_i)\chi^{}_i$ contributes to the loss of energy by the BH, which means that energy is being transferred to the orbit. Also, we see that for anti-aligned spins, where ($\hat{L}\cdot\hat{S}_i$) is negative, energy extraction from the BH is not possible.

The horizon parameters $H_{1,2}$ appear in the GW phase in terms that also include the masses and spins. This makes them {\em degenerate} with those parameters, in that it is more practical to measure the following effective observable parameters instead of $H_{1,2}$:
\begin{subequations}
\label{eq:Hparams}
\begin{align}
H_{\rm eff5} \equiv &{} \sum_{i=1}^{2}H^{}_i \left(\frac{m^{}_i}{M}\right)^3 \left(\hat{L}\cdot\hat{S}^{}_i\right)\chi^{} _i \left(3 \chi^{}_i{}^2+1\right)\,,\\
H_{\rm eff8} \equiv &{} ~4 \pi  H_{\rm eff5}+\sum^2_{i=1}H^{}_i \left(\frac{m_i}{M}\right)^4 \left(3 \chi^{}_i{}^2+1\right)\nonumber \\
                &\quad\quad\quad\quad\quad\quad\quad \times \left(\sqrt{1-\chi^{}_i{}^2}+1\right)\,.
\end{align}
\end{subequations}
These are analogous to the effective spin parameter $\chi^{}_{eff}$ that was introduced~\cite{Damour:2001tu, Ajith:2011ec} 
for characterizing spinning compact binary waveforms, where a combination of the spin parameters were introduced as a new parameter that can be measured more precisely. The subscripts here denote the fact that $H_{\rm eff5}$ and $H_{\rm eff8}$ appear in the GW phase in 2.5PN and 4PN order, respectively.

If the system is a BBH, as long as any one of the components has a finite spin, both $H_{\rm eff5}$ and $H_{\rm eff8}$ will be nonzero. On the other hand, when both the components of a BBH have vanishing spins, one has $H_{\rm eff5} \to 0$, but  $H_{\rm eff8}\neq 0$. Therefore, in the low-spin limit, $H_{\rm eff8}$ acts as the discriminator for the presence or absence of horizons. A horizonless binary with negligible tidal heating (e.g. binary neutron star) would have both $H_{\rm eff5}$ and $H_{\rm eff8}$ vanish, regardless of their spin values.

Next, we examine the phase contribution in the gravitational waveforms due to TH. This has been calculated in Ref.~\cite{datta2020recognizing} from Refs.~\cite{Tichy:1999pv, Isoyama:2017tbp} to be

\begin{equation}
\begin{aligned}
\label{eq:phase correction1}
\Psi^{}_{\rm TH} = &{} \frac{3}{128\eta} \left(\frac{1}{v}\right)^5 \left[-  \frac{10 }{9 }v^5 H_{\rm eff5} \left(3 \log \left(v\right)+1\right) \right. \\
&-  \frac{5}{168} v^7 H_{\rm eff5} \left(952 \eta +995\right) \\ 
&\left.+   \frac{5}{9}v^8 \left(3 \log \left(v\right)-1\right)(-4 H_{\rm eff8}+ H_{\rm eff5} \Psi^{}_{\text{SO}} )\right]\,,
\end{aligned}
\end{equation}
where the ``spin-orbit" term $\Psi^{}_{\rm SO}$ is given by

\begin{equation}
\label{spin-orbit}
   \begin{split}
        \Psi^{}_{\text{SO}} =&~ \frac{\big(\hat{L}\cdot\hat{S}^{}_1\big)\chi_1 m_1 (73 m_1 + 45 m_2) + 1\leftrightarrow 2}{3 M^2}\,\\
        =&~\frac{73}{3(1+q)^2}\left\{q^2 \big(\hat{L}\cdot\hat{S}^{}_1\big)\chi_1 + \big(\hat{L}\cdot\hat{S}^{}_2\big)\chi_2\right\}\\
        &+ \frac{15q}{(1+q)^2} \left\{ \big(\hat{L}\cdot\hat{S}^{}_1\big)\chi_1 + \big(\hat{L}\cdot\hat{S}^{}_2\big)\chi_2\right\}\,.
   \end{split}
\end{equation}

Equation~\eqref{eq:phase correction1} gives the total phase contribution in the gravitational waveforms due to TH. We can rewrite this expression in a compact form by identifying the dependence of individual terms on $v$ as
\begin{equation}
\begin{aligned}
    \Psi^{}_{\rm TH} = &\, [C_0 + C_1\log(v) + C_2v^2 + C_3v^3\\
    & + C_4v^3\log(v)] H_{\rm eff5} \\
    & + [D_3v^3 + D_4v^3\log(v)]H_{\rm eff8}\,,
\end{aligned}
\end{equation}
where the coefficients $C_i$ ($i=0,1,2,3,4$) and $D_i$ ($i=3,4$) are functions of the symmetric mass-ratio $\eta$, and $C_3,C_4$ also include $\Psi_{\rm SO}$. The term $C_0 H_{\rm eff5}$ is independent of $v$ and thus independent of $f$. Therefore, this term can be absorbed into the phase of coalescence $\phi_c$, which is also independent of $f$. The terms $C_3 v^3 H_{\rm eff5}$ and $D_3 v^3 H_{\rm eff8}$ have $v^3$ dependence, so they are $\propto f$. These terms can be absorbed into the time of coalescence $t_c$, which appears in the total GW phase as $2\pi f t_c$, and is $\propto f$ as well. For these reasons, we discard these three terms from $\Psi^{}_{\rm TH}$, equivalently redefining $\phi_c$ and $t_c$.

We are then left with the terms containing $C_1, C_2, C_4$ and $D_4$; which give us the GW phase due to tidal heating to be
\begin{equation}
\begin{aligned}
\label{eq:phase correction}
\Psi^{}_{\rm TH} = &{} \frac{3}{128\eta}  \left[-  \frac{10 }{3 } H_{\rm eff5} \,  \log \left(v\right) \right. \\
&-  \frac{5}{168} v^2 H_{\rm eff5} \left(952 \eta +995\right) \\ 
&\left.+   \frac{5}{3}v^3  \log \left(v\right)(-4 H_{\rm eff8}+ H_{\rm eff5} \Psi^{}_{\text{SO}} )\right],
\end{aligned}
\end{equation}
after putting their expressions from Eq.~\eqref{eq:phase correction1}. We will use this expression for $\Psi^{}_{\rm TH}$ here for our analyses. Throughout the chapter, we only consider spins aligned with the orbital angular momentum, so that $\hat{L}\cdot\hat{S}^{}_1=\hat{L}\cdot\hat{S}^{}_2=1$ in Eq.~\eqref{spin-orbit}.

Next, we need the PN approximant to which we will add this phase in order to obtain the complete PN waveform with TH included. For this purpose, we consider the \texttt{TaylorF2} approximant~\cite{tf2PhysRevD.80.084043} upto 3.5PN order ($\sim v^2$), constructed under the ``stationary phase approximation" (SPA)~\cite{cutler_flanagan_PhysRevD.49.2658}. 

\section{Basics of the Fisher Matrix Approach}
\label{fisher_basics}

In this work, we mainly focus on Fisher matrix analysis~\cite{Vallisneri:2007ev, cutler_flanagan_PhysRevD.49.2658, owen_PhysRevD.53.6749, poisson_will_PhysRevD.52.848} for the estimation of errors in the measurement of the horizon parameters in the 3rd generation detectors Einstein Telescope~\cite{maggiore2020science} and Cosmic Explorer~\cite{regimbau2017digging, vitale2017parameter, Reitze:2019iox}. In this section we will briefly summarize the basic concepts of the Fisher matrix approach for parameter estimation. 

A GW signal in the time domain, as emitted by a coalescing compact binary, can be decomposed into two polarization states $h_+(t;\Theta_{\rm GW})$ and $h_{\cross}(t;\Theta_{\rm GW})$, where the parameter vector $\Theta_{\rm GW}$ contains information about the source. For a BBH in PP approximation, $\Theta_{\rm GW}\equiv\{m_1,m_2,\boldsymbol{\chi^{}_1},\boldsymbol{\chi^{}_2},D^{}_L,\iota,t_c,\phi_c\}$, where $m_1,m_2$ are companion masses, $\boldsymbol{\chi^{}_1},\boldsymbol{\chi^{}_2}$ are their dimensionless spin vectors, $D^{}_L$ is the luminosity distance of the binary, $\iota$ is the inclination angle of its orbital plane with respect to the line of sight, and $t_c$ and $\phi_c$ are the time and phase of coalescence, respectively.  We extend this set by including the two parameters $\{H_{\rm eff5}, H_{\rm eff8}\}$, defined in Eq.~\eqref{eq:Hparams}, to incorporate TH. The GW strain in the frequency domain as measured by a detector, $H(f)$, depends on $\Theta_{\rm GW}$, the location of the detector, and three more extrinsic source parameters $\{\alpha, \delta, \psi\}$, which denote right ascension, declination and polarization angle, respectively.

\subsection{The Noise Power Spectral Density} The ability of a GW detector to measure the GW strain depends on its sensitivity, which in turn depends on the \textit{power spectral density} (PSD) of its noise, $n(t)$, and its auto-correlation~\cite{Borhanian:2020ypi} $\kappa=\overline{n(t_1)n(t_2)}$, where the overbar denotes an average over noise realizations. Assuming that the noise is stationary and Gaussian with zero mean, which means $\kappa$ only depends on the time difference $t' =t_1-t_2$, the PSD of the noise (in frequency domain) can be written as,
\begin{equation}
    S_n(f)=\frac{1}{2}\int_{-\infty}^\infty \dd t' \kappa(t')e^{i2\pi ft'}, \quad \text{with}\quad f>0.
\end{equation}
This function denotes the detector sensitivity at different frequencies.

\subsection{The Signal-to-Noise Ratio} The set of all possible detector responses in frequency or time domain forms a vector space. In frequency domain, let us call this space $\mathcal{V}$. We can define, on this space, a \textit{noise-weighted scalar product} of two detector responses $H(f), G(f) \in \mathcal{V}$ as~\cite{Sathyaprakash:2009xs}
\begin{equation}
\label{innerproduct}
   \braket{H}{G}=2\int^\infty_0 \dd f \frac{H^\ast(f) G(f) + G^\ast (f)H(f)}{S_n(f)}.
\end{equation}
Equipped with this definition, we can define the {\it signal-to-noise ratio} (SNR) $\rho$ for a given GW signal $H$ as
\begin{equation}
\label{snr}
    \rho=\sqrt{\braket{H}}=2\sqrt{\int^\infty_0 \dd f \frac{|H(f)|^2}{S_n(f)}},
\end{equation}
where $S_n(f)$ contains information about the sensitivity of the chosen detector. The SNR of a signal characterizes its loudness over a given noise profile.

It is important to mention here that in practical situations, like in this study, we will not cover the entire frequency region (0 to $\infty$), because detectors typically have sensitivity only within a finite frequency band, and the signal band is also finite.
Additionally, since PN expansions fail near the merger phase due to violations of the slow motion and weak gravity conditions, we have to terminate the integration in Eq.~\eqref{innerproduct} or~\eqref{snr} at some point where the binary is still away from the merger. A general choice for such a cut-off frequency is the binary's {\it innermost stable circular orbit} (ISCO), which marks the ``end" of the inspiral phase. ISCO of SBHs, as was discussed in Sec.~\ref{sec:orbital dynamics SBH}, lie at $R=6M$. For a binary of KBHs, location of the ISCO depends on the spin-alignment as well as the component masses and spins. In the case of aligned spins, the ISCO for a KBH is closer to the center of mass of the binary than a Schwarzschild BH of the same mass. In our work, we consider the upper cutoff frequency to be the GW frequency at the ISCO corresponding to the final BH formed after merger, given by (ignoring cosmological redshift)~\cite{Favata:2021vhw} 
\begin{equation}
\label{isco}
    f_{\rm{ISCO}}=\frac{\hat{\Omega}_{\rm{ISCO}}(\chi^{}_f)}{\pi M_f}\,.
\end{equation}
Here $\hat{\Omega}_{\rm{ISCO}}(\chi)=M_{\rm{Kerr}}\Omega_{\rm{ISCO}}$ is the dimensionless angular frequency for a circular equatorial orbit around a KBH with mass $M_{\rm{Kerr}}$ and spin $\chi$~\cite{Bardeen:1972fi}. For the upper cutoff frequency, we choose $M_{\rm{Kerr}}=M_f$ and $\chi=\chi^{}_f$, the final mass and spin of the merger remnant BH, which are obtained by using fitting formulas from NR simulations~\cite{PhysRevD.93.044006}. Explicit expressions for $\hat{\Omega}_{\rm{ISCO}}$, and $M_f,\chi^{}_f$ in terms of initial masses and spins are mentioned in Appendix C of Ref.~\cite{Favata:2021vhw}.

So, Eq.~\eqref{snr} will be replaced by
\begin{equation}
\label{snr2}
    \rho=2\sqrt{\int^{f_{\rm max}}_{f_{\rm min}} \dd f \frac{|H(f)|^2}{S_n(f)}}.
\end{equation}
Here $f_{\rm min}$ will be determined by the detector band's lower frequency cut-off, and $f_{\rm max}$ will correspond to the ISCO (Eq.~\eqref{isco}).

\subsection{The Fisher Information Matrix}
\label{fisher_matrix}
The detector output $S(f)$ in frequency domain is related to the GW strain $H(f)$ and the noise $N(f)$ as $S(f)=H(f)+N(f)$. Since we have assumed a Gaussian profile for the noise, we can write the probability function for $N(f)$~\cite{Borhanian:2020ypi} as
\begin{align}
\label{ptheta}
     p(\Theta)=&p_0(\Theta)e^{-\frac{1}{2}\braket{N}}\nonumber\\
     =&p_0(\Theta)e^{-\frac{1}{2}\braket{S-H(\Theta)}},
\end{align}
where $\Theta$ is the parameter vector and $p_0$ is the prior on these parameters.

Let us denote $E(\Theta)=\braket{S-H(\Theta)}$, and expand this quantity around the ``true" value ($\Theta^\ast$) of the parameters :
\begin{equation}
\label{etheta}
    E(\Theta)=E(\Theta^\ast)+\frac{1}{2}\pdv{E}{\Theta_i}{\Theta_j}\bigg|_{\Theta=\Theta^\ast}\Delta\Theta^i\Delta\Theta^j+\cdots,
\end{equation}
where $\Delta\Theta = (\Theta-\Theta^\ast)$, and we use Einstein summation convention over repeated indices. Also, using the expression for $E(\Theta)$, we can write
\begin{align}
    \pdv{E(\Theta)}{\Theta_i}{\Theta_j} = &~ 2\braket{\partial_{\Theta_i}H(\Theta)}{\partial_{\Theta_j}H(\Theta)}+\braket{\partial_{\Theta_i}\partial_{\Theta_j}H(\Theta)}{N}\nonumber\\
    \approx &~ 2\braket{\partial_{\Theta_i}H(\Theta)}{\partial_{\Theta_j}H(\Theta)},
\label{dedtheta}
\end{align}
where in the second step we have assumed that the SNR value is high enough for the first order derivatives of $H$ to dominate over the second order ones~\cite{cutler_flanagan_PhysRevD.49.2658}.

We now define the {\it Fisher information matrix} $\Gamma$, the elements of which are given as
\begin{equation}
\label{gamma}
    \Gamma_{ij}=\braket{\partial_{\Theta_i}H(\Theta)}{\partial_{\Theta_j}H(\Theta)}.
\end{equation}
Using this in Eq.~\eqref{dedtheta}, and assuming that $\Delta\Theta$ is small, we can infer from Eq.~\eqref{ptheta} that
\begin{equation}
    p(\Theta)\propto\exp{-\frac{1}{2}\Gamma_{ij}\Delta\Theta^i\Delta\Theta^j}.
\end{equation}
The inverse of the Fisher matrix is the \textit{covariance matrix}, $C=\Gamma^{-1}$. Along the diagonal of $C$, one gets the variances of the concerned parameters, from which one can get the 1-$\sigma$ errors in those parameters as $\sigma_{\Theta_i}=\sqrt{C_{ii}}$. The off-diagonal elements are the covariances between the parameters, defined as
\begin{equation}
    C_{ij}={\rm cov}(\Theta_i,\Theta_j)=\overline{(\Theta_i-\overline{\Theta}_i)(\Theta_j-\overline{\Theta}_j)},
\end{equation}
where the bar denotes mean value. For $i=j$, one gets $C_{ii}=\overline{(\Theta_i-\overline{\Theta}_i)^2}$, called the ``variance" of the distribution of $\Theta_i$, which is the square of its standard deviation $\sigma_{\Theta_i}$.

For the Fisher matrix approach to work, not only the SNR has to be high, but the matrix also has to be {\it well-conditioned}~\cite{Borhanian:2020ypi}. This criterion is quantified by the {\it condition number}, which is defined as the ratio of the largest and the smallest eigenvalues of the matrix. If this quantity is too large, then the inversion of $\Gamma$ is not trustworthy. Here, we ensured that it is well within the numerical precision available for our computations~\cite{condnum_PhysRevD.85.062002}.





\section{Results of Fisher Analyses}
\label{results}

In this section we will apply the Fisher matrix approach to estimate the errors in the TH parameters in the three-detector network comprising the Advanced LIGO~\cite{LIGOScientific:2014pky} and Advanced Virgo~\cite{VIRGO:2014yos} detectors, and the proposed 3rd generation detectors Einstein Telescope~\cite{maggiore2020science} and Cosmic Explorer~\cite{Reitze:2019iox}. We have used the package \texttt{GWBENCH}~\cite{Borhanian:2020ypi} for our Fisher matrix calculations.
In order to estimate the errors in certain parameters using this approach, we first inject a gravitational waveform into the relevant detector, then using Eq.~\eqref{gamma}, calculate the Fisher matrix $\Gamma$ and the covariance matrix by inverting it. As mentioned in Sec.~\ref{theory_TH}, in this work we take \texttt{TaylorF2} as the PN approximant and incorporate in it the phase contribution ($\Psi^{}_{\rm TH}$) due to TH, given by Eq.~\eqref{eq:phase correction}. 

In our study, we estimate the projected errors in five parameters, namely $\Theta \equiv\{\mathcal{M}_c,\eta,D_L,H_{\rm eff5},H_{\rm eff8}\}$, where $\mathcal{M}_c$ is the \textit{chirp mass} defined as $\mathcal{M}_c=(m_1m_2)^{3/5}/M^{1/5}$. When we discuss the variation of the errors with component spins in Sec.~\ref{sec:spin}, we extend the parameter space with the two component spins $\chi_1,\chi_2$. As the lower cutoff frequencies, we have used 10 Hz (4 Hz) for LIGO-Virgo (ET, CE), and the upper cutoff frequencies are determined by the spin-dependent ISCO frequencies given by Eq.~\eqref{isco}.

From Eqs.~\eqref{eq:Hparams} we see that $H_{\rm eff5}$ and $H_{\rm eff8}$ are functions of the component masses only through the ratios $m_1/M=q/(1+q)$ and $m_2/M=1/(1+q)$, both of which have values always lying between 0 and 1. Also, $-1\leq\chi^{}_i\leq 1$. This enables one to define a range in the values of these two parameters that can occur physically, for all possible values of $q$. This turns out to be approximately~\cite{datta2020recognizing}
\begin{equation}
\label{range}
    -4\leq H_{\rm eff5}\leq 4\,,  \quad {\rm and} \quad -46.3\lesssim H_{\rm eff8}\lesssim 54.3\,.
\end{equation}
Even though we will treat $H_{\rm eff5}$ and $H_{\rm eff8}$ as free parameters here, we have to keep in mind that this is the physical range of values they can have.

\subsection{LIGO \& Virgo}
\label{ligo-results}


\begin{figure}
     \centering
     \begin{subfigure}[b]{0.49\textwidth}
         \centering
         \includegraphics[width=\textwidth]{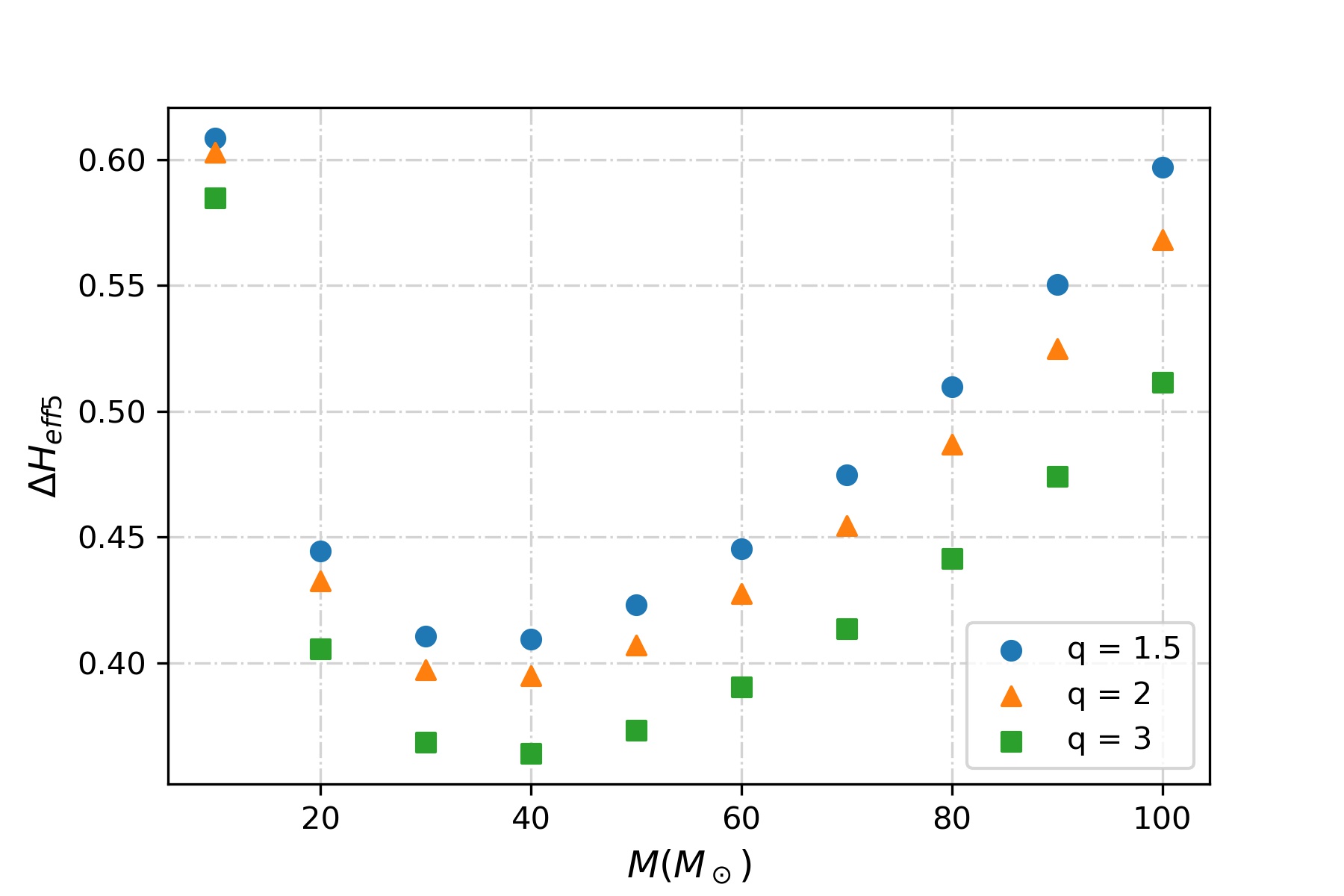}
         \caption{~Errors in $H_{\rm eff5}$ in LIGO-Virgo}
         \label{h5-ligo-m}
     \end{subfigure}
     \hfill
     \begin{subfigure}[b]{0.49\textwidth}
         \centering
         \includegraphics[width=\textwidth]{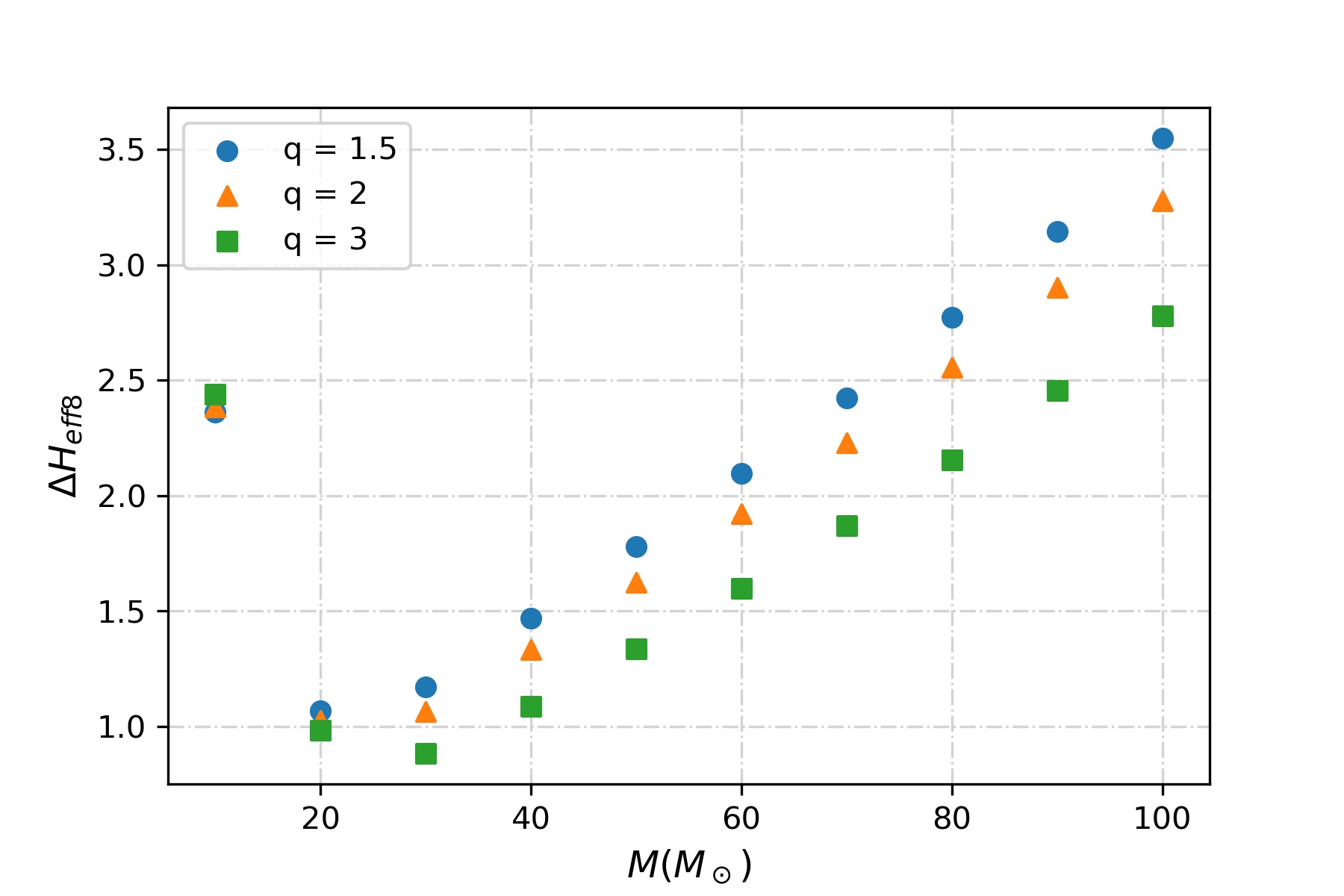}
         \caption{~Errors in $H_{\rm eff8}$ in LIGO-Virgo}
         \label{h8-ligo-m}
     \end{subfigure}
        \caption{Error values in the TH parameters $H_{\rm eff5}$ (top) and $H_{\rm eff8}$ (bottom) as a function of total mass, when measured by the three detector network of LIGO (Hanford, Livingston) and Virgo.  mass-ratio ($q$) has been varied from 1.5 to 3 for getting different curves. We consider aligned spins here, so that $\hat{L}\cdot\hat{S}_i=1$. $H_{\rm eff5}=0.6, H_{\rm eff8}=12, D^{}_L=200$ Mpc, $\chi^{}_1=\chi^{}_2=0.8$ have been taken.}
        \label{ligo-m}
\end{figure}

Figure~\ref{ligo-m} shows the variation of 1-$\sigma$ errors with the total binary mass in the noise spectrum of the three-detector network of LIGO (Hanford, Livingston) and Virgo. The Y-axes report the 1-$\sigma$ errors in $H_{\rm eff5}$ (Fig.~\ref{h5-ligo-m}) and $H_{\rm eff8}$ (Fig.~\ref{h8-ligo-m}), denoted by $\Delta H_{\rm eff5}$ and $\Delta H_{\rm eff8}$, respectively. In our analysis, we have used the most recent design sensitivity curves of Advanced LIGO~\cite{LIGO:design-sensitivity} and Advanced Virgo~\cite{Virgo:design-sensitivity} detectors. Binaries in the range of total mass $10-100M_\odot$ have been considered, situated at a distance of 200 Mpc. The errors initially fall with total binary mass in the range $M<30M_\odot$, and we find that there is a region around $30 M_\odot$ where the errors are minimum. Thereafter, the errors rise rapidly with increasing $M$. An increase in $M$ causes the SNR to rise, which provides a better estimation for $H_{\rm eff5}, H_{\rm eff8}$. This causes the dip in the errors for $M = 10-30 M_\odot$. Further increase in $M$ shrinks the signal band. This is because it lowers the value of the ISCO  frequency while $f_{\rm min}$ remains fixed. This interplay between the SNR and the effective frequency interval creates an optimal region, which turns out to be about $30-40 M_\odot$. We also note that the exact minima in the errors are slightly different for $H_{\rm eff5}$ ($\sim 40 M_\odot$) and $H_{\rm eff8}$ ($\sim 30 M_\odot$).

$H_{\rm eff5}$ and $H_{\rm eff8}$ for binaries with more asymmetric masses appear to be more precisely measurable. This is expected because mass asymmetry lowers the value of the symmetric mass-ratio $\eta$, and the TH phase has a prefactor of $1/\eta$ (see Eq.~\eqref{eq:phase correction}), making the phase contribution due to TH higher for more asymmetric masses, consequently adding more GW cycles into the signal band.  For a binary at 200 Mpc with $M=30M_\odot$ and $q=1.5$, Fisher estimates show an error value of $\sim 0.4 (1.2)$ for $H_{\rm eff5} (H_{\rm eff8})$, which amounts to a relative percentage error of $\sim 67\%(10\%)$. In LIGO, then, the detection of a so-called {\it golden binary} at a distance $\leq 50$ Mpc will make it possible to estimate these TH parameters with better than 17\% precision.

\subsection{3rd Generation Detectors}
\label{3g-results}

The proposed 3rd generation (3G) GW detectors, Einstein Telescope (ET) and Cosmic Explorer (CE) will have a higher sensitivity than current detectors, which will result in higher SNR for CBCs. This makes Fisher error projections quite trustworthy. In this section, we explore the measurement precision of $H_{\rm eff5},H_{\rm eff8}$ in ET and CE. In addition to the variations of the errors with $M$, we will also look at the variations with luminosity distance $D^{}_L$ and the spin values $\chi^{}_{1,2}$.

For our study, we have used the sensitivity curves for the ET-D configuration \cite{Hild:2010id} of Einstein Telescope, and 40 km long CE configuration \cite{CE:sensitivity, LIGOScientific:2016wof} of Cosmic Explorer, optimized for the low fequencies of CBC.

\subsubsection{Dependence on the Total Binary Mass}





\begin{figure}
     \centering
     \begin{subfigure}[b]{0.49\textwidth}
         \centering
         \includegraphics[width=\textwidth]{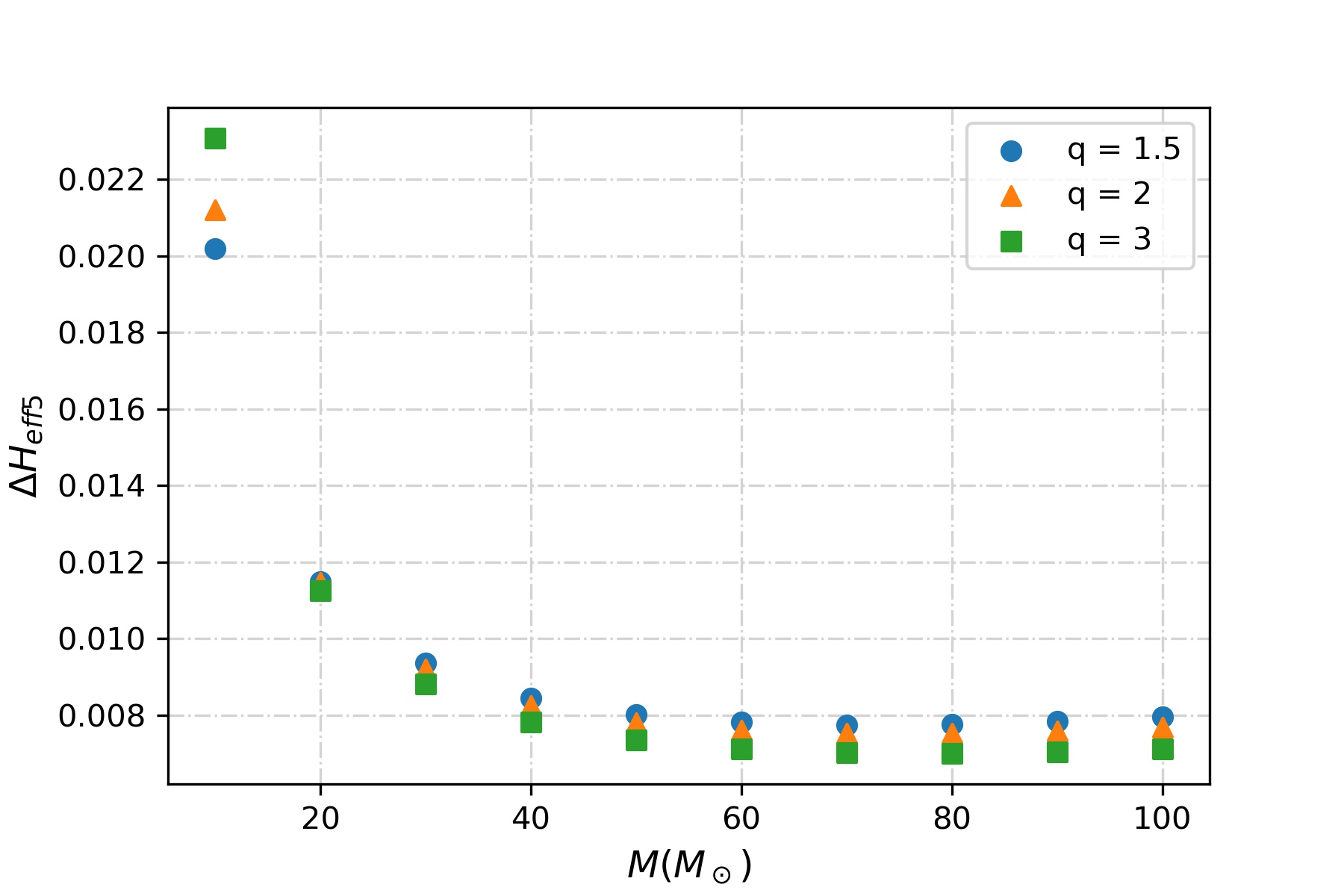}
         \caption{Errors in $H_{\rm eff5}$ in Einstein Telescope}
         \label{h5-et-m}
     \end{subfigure}
     \hfill
     \begin{subfigure}[b]{0.49\textwidth}
         \centering
         \includegraphics[width=\textwidth]{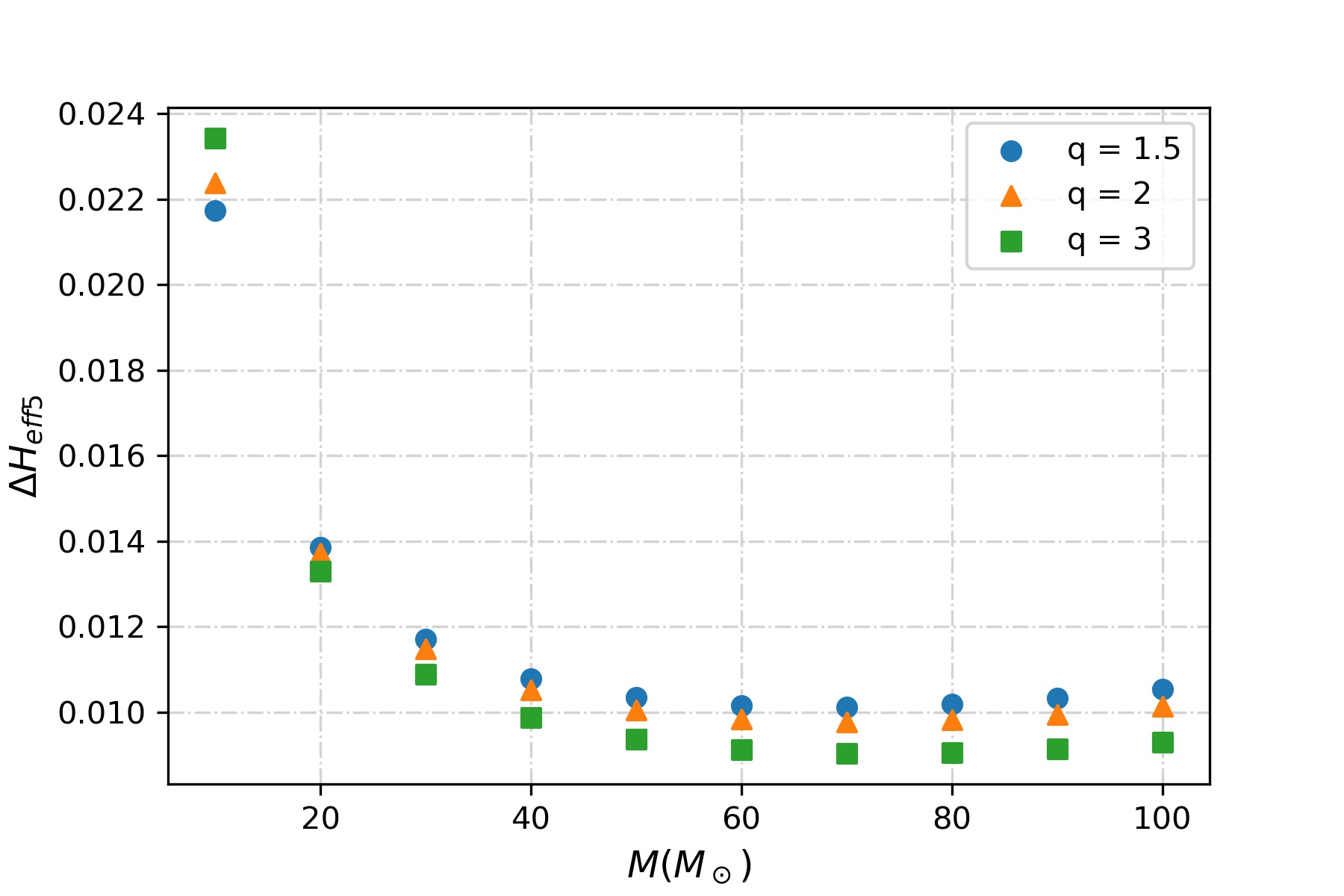}
         \caption{Errors in $H_{\rm eff5}$ in Cosmic Explorer}
         \label{h5-ce-m}
     \end{subfigure}
     \hfill
     \begin{subfigure}[b]{0.49\textwidth}
         \centering
         \includegraphics[width=\textwidth]{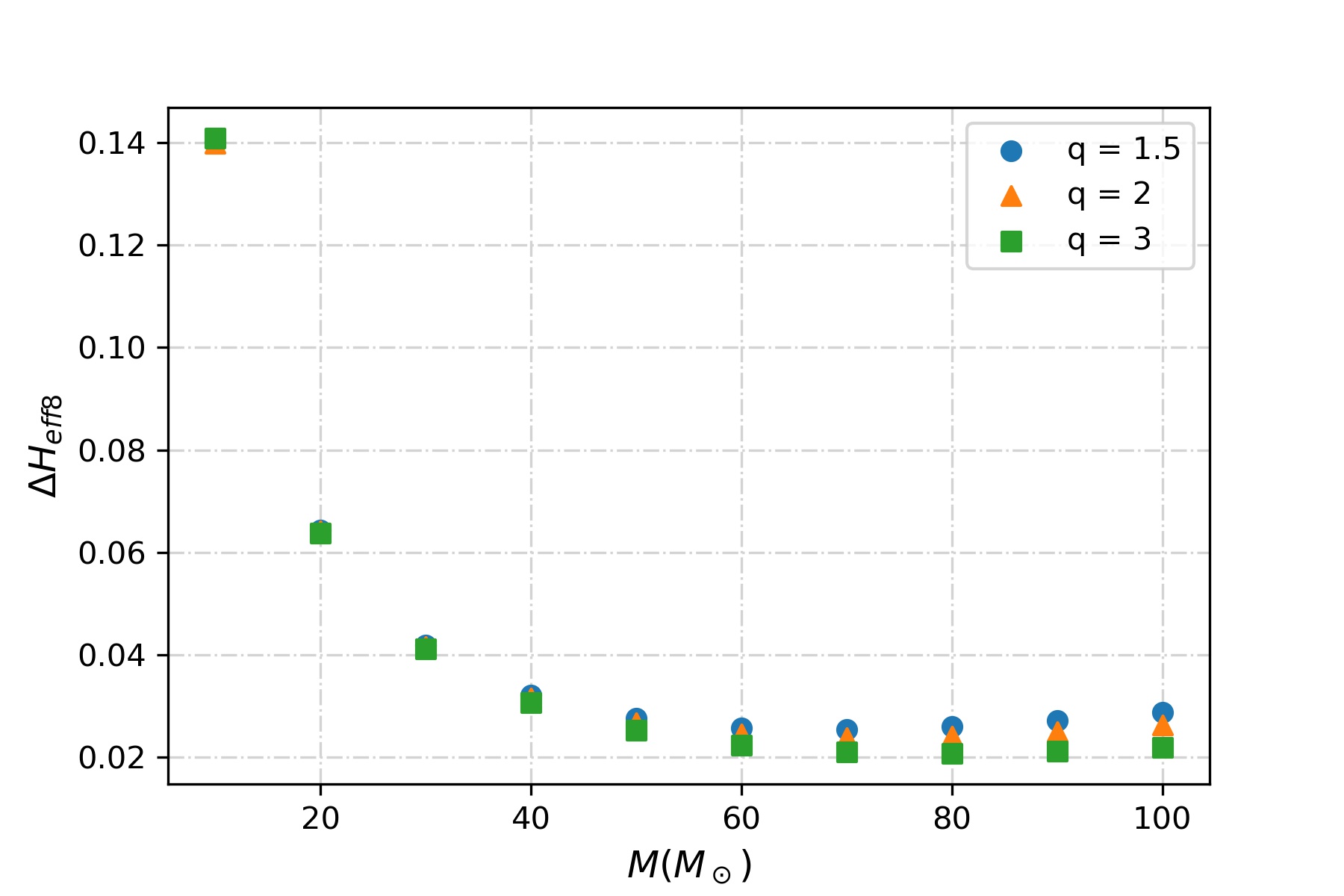}
         \caption{Errors in $H_{\rm eff8}$ in Einstein Telescope}
         \label{h8-et-m}
     \end{subfigure}
     \hfill
     \begin{subfigure}[b]{0.49\textwidth}
         \centering
         \includegraphics[width=\textwidth]{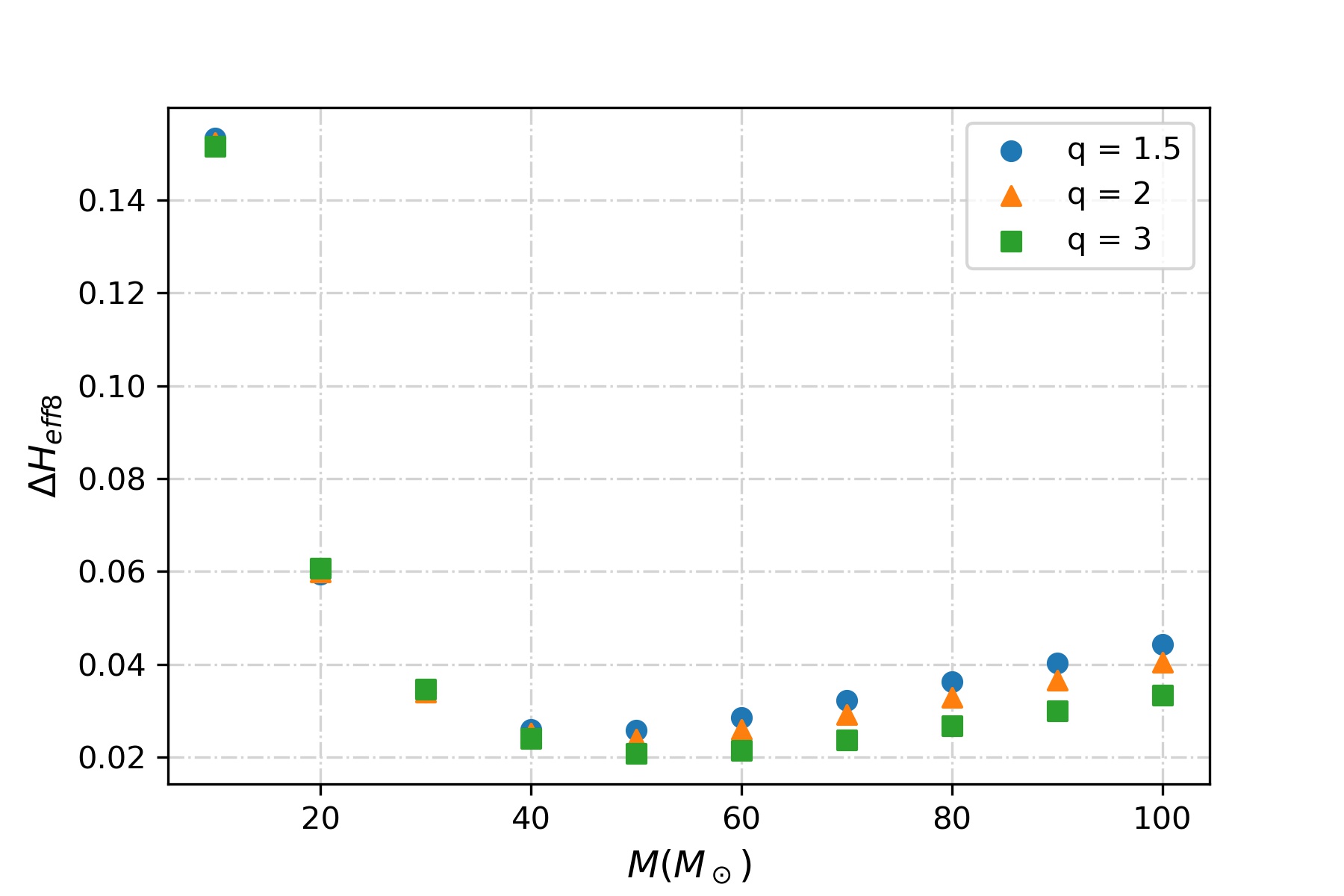}
         \caption{Errors in $H_{\rm eff8}$ in Cosmic Explorer}
         \label{h8-ce-m}
     \end{subfigure}
        \caption{Errors in $H_{\rm eff5}$ (top row) and $H_{\rm eff8}$ (bottom row) as a function of total mass $M$, when measured in ET (first column) and CE (second column). Injection parameters are the same as in Fig.~\ref{ligo-m}. }
        \label{et-ce-m}
\end{figure}


\begin{figure}
    \centering
    \includegraphics[width=85mm]{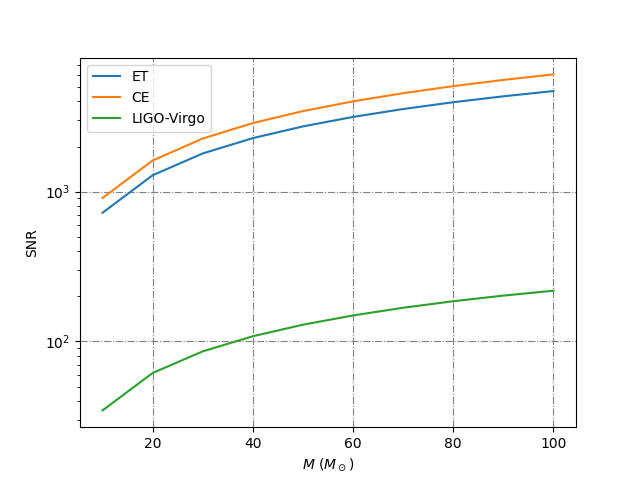}
    \caption{ Variation of SNR with the total mass $M$ in LIGO-Virgo, ET and CE, as calculated from Eq.~\eqref{snr2}. We consider binaries at $D_L = 200$ Mpc with mass-ratio $q=1.5$. }
    \label{snr-m}
\end{figure}

Figure~\ref{et-ce-m} shows the variation of errors with total mass.  As expected, the errors are smaller compared to LIGO-Virgo, due to the high SNR values. In 3G detectors (Fig.~\ref{snr-m}), typically SNR $\sim\mathcal{O}(10^3)$ whereas in LIGO-Virgo, SNR $\sim\mathcal{O}(10^2)$ for the chosen parameter space. Also, SNR increases more rapidly with $M$ in ET and CE than in LIGO-Virgo,  making the rise in errors due to the shortening of the frequency range much slower after $M\sim 60M_\odot$, as seen from Fig.~\ref{et-ce-m}. Comparing Fig.~\ref{ligo-m} and Fig.~\ref{et-ce-m}, we confirm that the precision of measurement in 3G detectors has substantial improvement over LIGO-Virgo.

 For binaries with $M\gtrsim 60M_\odot$ and $q=1.5$, estimation of $H_{\rm eff5}$ and $H_{\rm eff8}$ can be made with 1-$\sigma$ errors $\Delta H_{\rm eff5}\sim$ 0.008 (1.2\%) and $\Delta H_{\rm eff8} \sim 0.02$  (0.22\%) respectively, for BBHs at a distance of 200 Mpc. The error values fall as more component mass asymmetry is introduced. However, for binaries with low masses, we see that this trend is reversed for $H_{\rm eff5}$ -- as seen in Figs.~\ref{h5-et-m} and~\ref{h5-ce-m} for $M=10 M_\odot$. This behaviour possibly follows from the fact that the SNR is not the only factor behind the uncertainty benchmarks provided by Fisher analyses, the latter also depend on how the waveform model produces the Fisher matrix elements across different regions of the parameter space.  We also note that the variation in the error values with changing $q$ is less pronounced in ET, CE than  LIGO-Virgo.

\subsubsection{Dependence on the Luminosity Distance}
\label{sec:distance}

\begin{figure}
     \centering
     \begin{subfigure}[b]{0.49\textwidth}
         \centering
         \includegraphics[width=\textwidth]{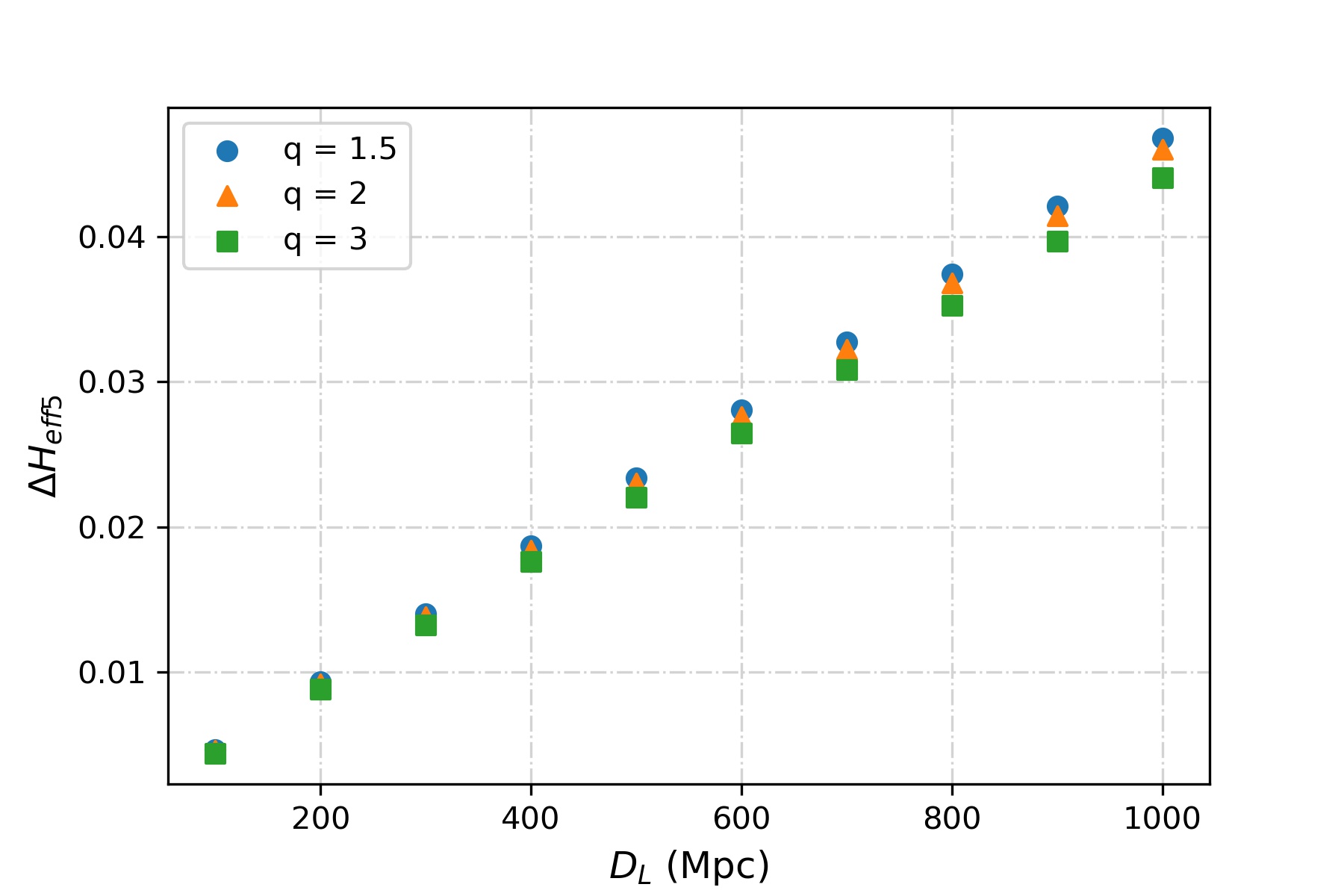}
         \caption{Errors in $H_{\rm eff5}$ in Einstein Telescope}
         \label{h5-et-dl}
     \end{subfigure}
     \hfill
     \begin{subfigure}[b]{0.49\textwidth}
         \centering
         \includegraphics[width=\textwidth]{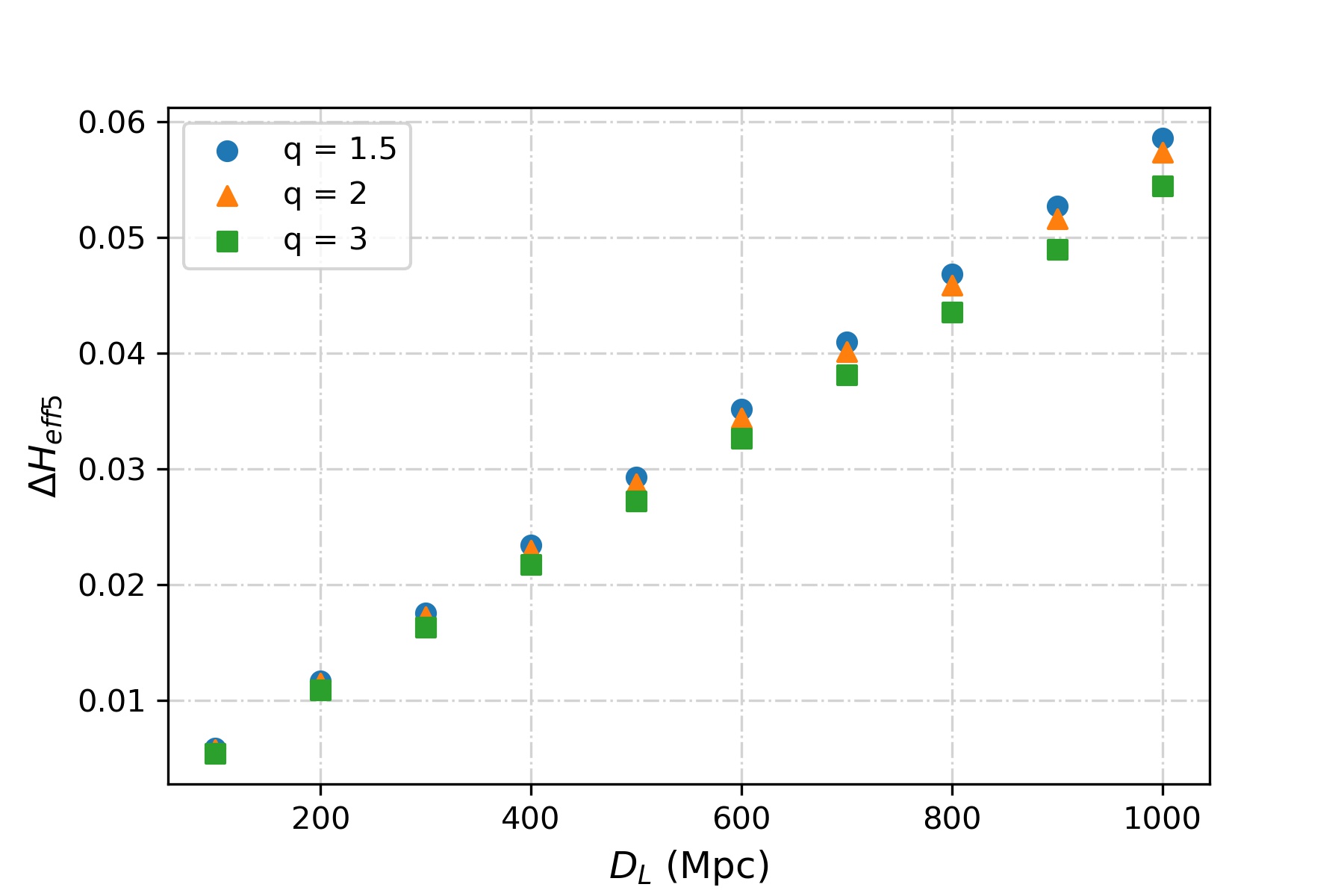}
         \caption{Errors in $H_{\rm eff5}$ in Cosmic Explorer}
         \label{h5-ce-dl}
     \end{subfigure}
     \hfill
     \begin{subfigure}[b]{0.49\textwidth}
         \centering
         \includegraphics[width=\textwidth]{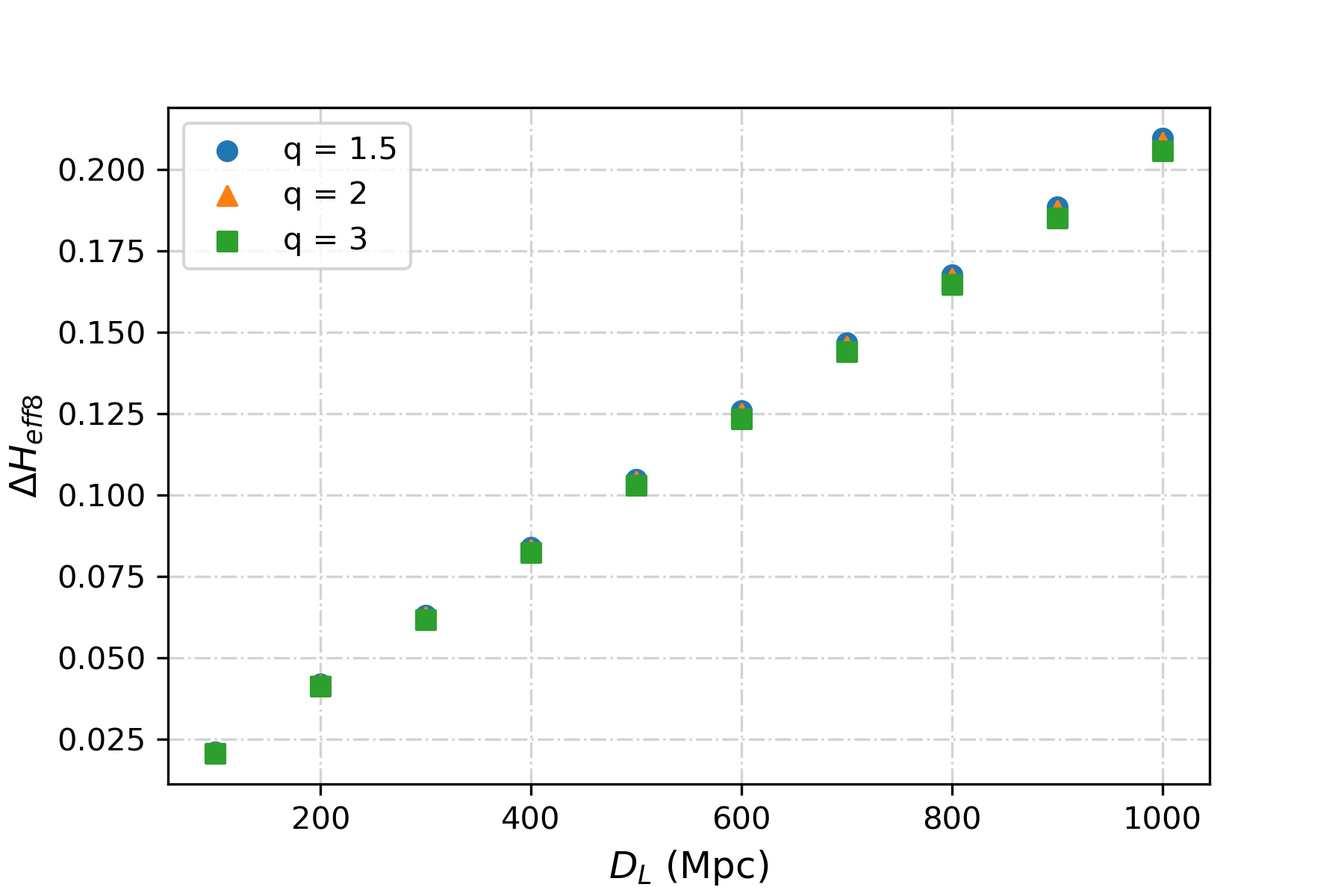}
         \caption{Errors in $H_{\rm eff8}$ in Einstein Telescope}
         \label{h8-et-dl}
     \end{subfigure}
     \hfill
     \begin{subfigure}[b]{0.49\textwidth}
         \centering
         \includegraphics[width=\textwidth]{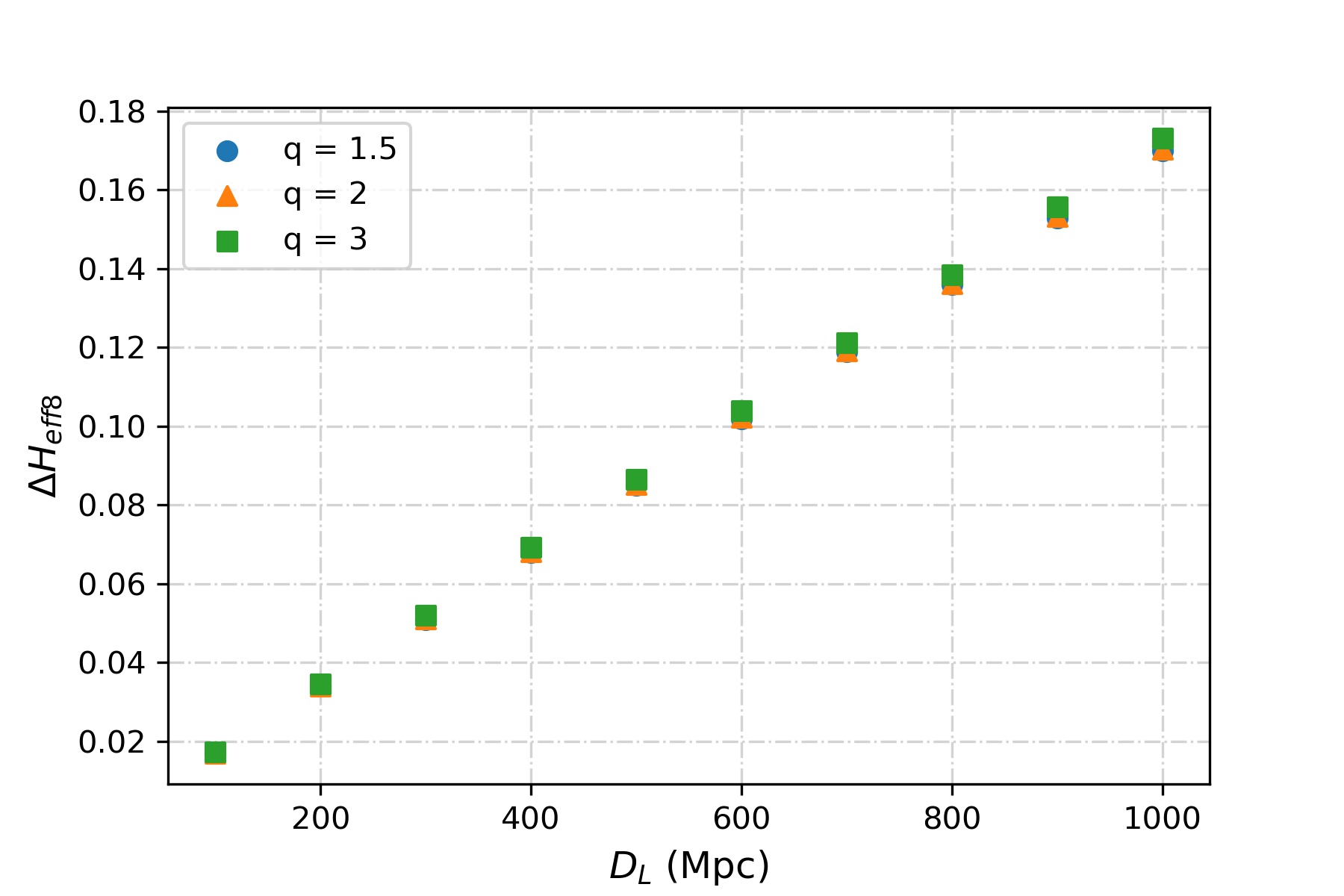}
         \caption{Errors in $H_{\rm eff8}$ in Cosmic Explorer}
         \label{h8-ce-dl}
     \end{subfigure}
    \caption{Errors in $H_{\rm eff5}$ (top row) and $H_{\rm eff8}$ (bottom row) as a function of luminosity distance, when measured in ET (first column) and CE (second column). Along the X-axis, $D^{}_L$ varies from 100 Mpc to 1 Gpc. Other parameters are fixed at $H_{\rm eff5}=0.6, H_{\rm eff8}=12, M=30M_\odot$, $\chi^{}_1=\chi^{}_2=0.8$.}
    \label{et-ce-dl}
\end{figure}

Figure~\ref{et-ce-dl} shows the variation of errors with luminosity distance $D^{}_L$. We keep the total mass fixed at $M=30M_\odot$. In this case only the fall in SNR with increasing distance affects the errors. As expected, the errors rise linearly with $D^{}_L$, with the slope being greater for more symmetric masses. From Fig.~\ref{et-ce-dl}, we see that for binaries as far as 1 Gpc away, the 1-$\sigma$ error in $H_{\rm eff5}$ is $\sim$ 0.06 (10\%) for BH binaries with $q=3$, whereas for $H_{\rm eff8}$ it is $\sim$ 0.2 (1.6\%). Owing to the linear variation in errors, the presence of horizons for all the sources within this range can be tested with linearly increasing precision.




\begin{figure}
    \centering
    \includegraphics[width=86mm]{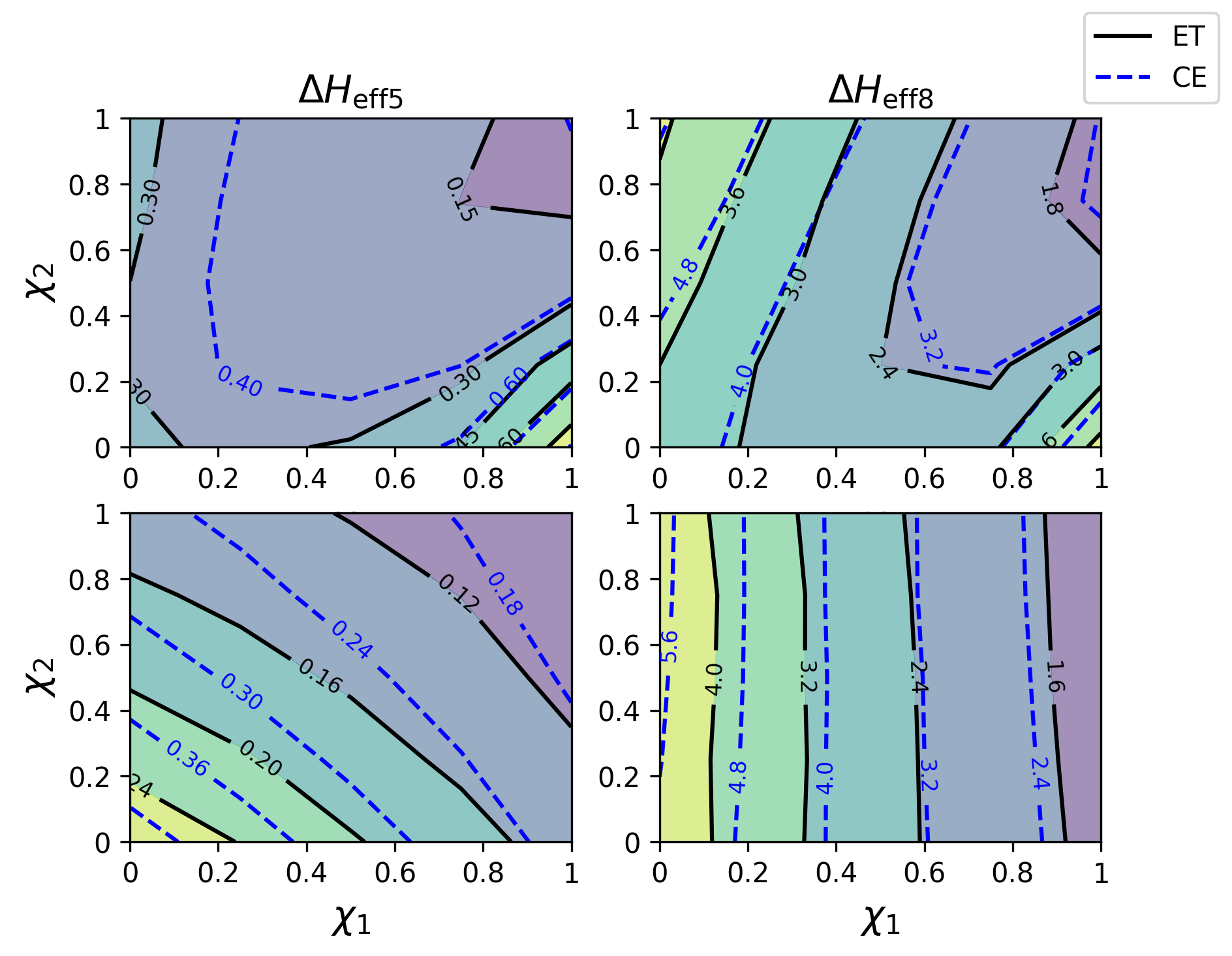}
    \caption{ Variation of errors in $H_{\rm eff5}$ (left column) and $H_{\rm eff8}$ (right column) in ET (solid lines), CE (dashed lines) with dimensionless spins. $\chi_1, \chi_2$ are varied from 0 to 1 along the $X$ and $Y$-axes respectively. Total binary mass is $M=40 M_\odot$, and mass-ratios are $q=1.1$ (top panel) and $q=3$ (bottom panel). We consider optimally oriented binaries at $D_L = 200$ Mpc, with $H_{\rm eff5}=0.6$, and $H_{\rm eff8}=12$.}
\label{spinvar}
\end{figure}

\begin{figure}
    \centering
    \includegraphics[width=86mm]{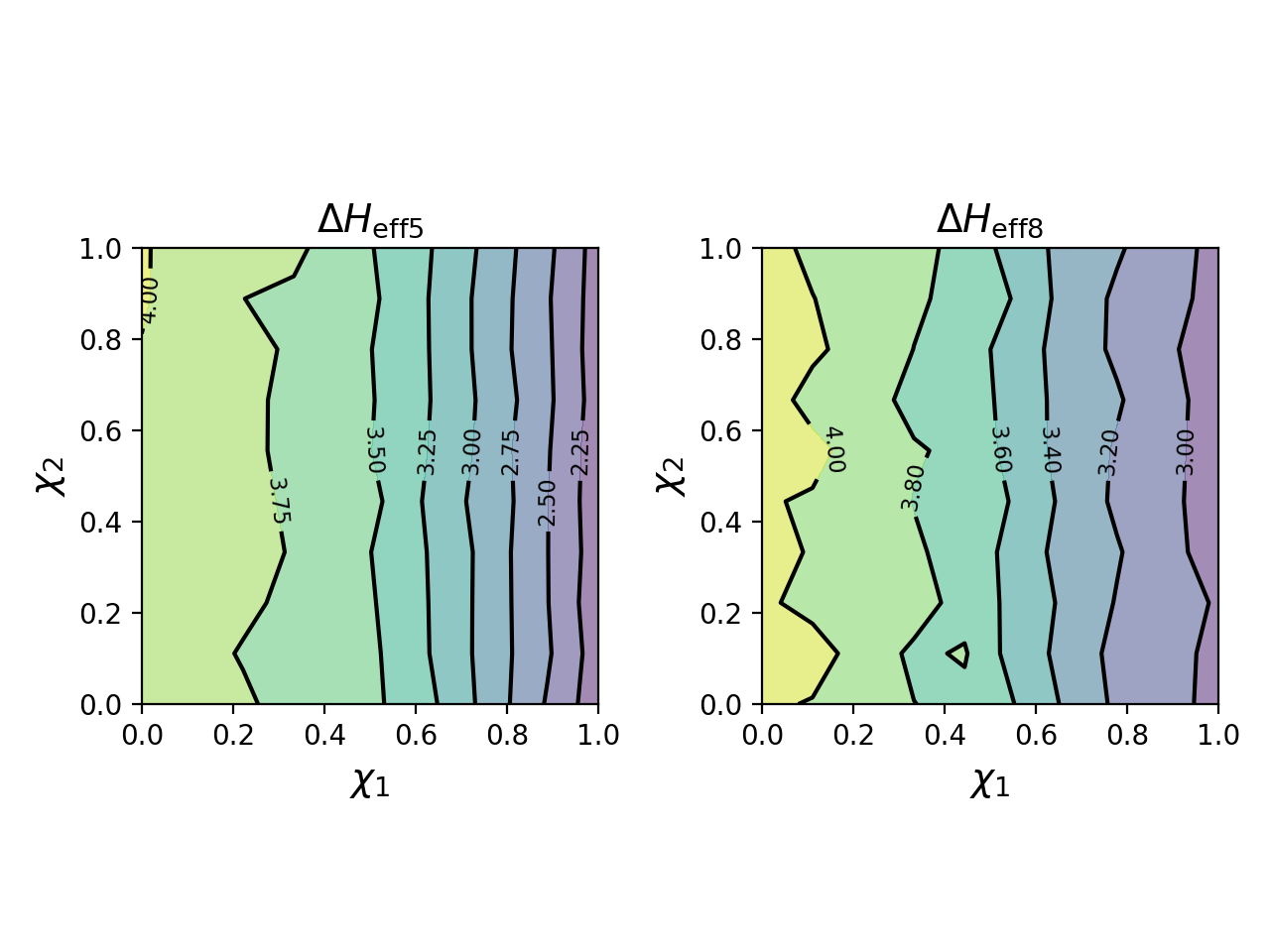}
    \caption{ Variation of errors in $H_{\rm eff5}$ (left column) and $H_{\rm eff8}$ (right column) in CE, with the parameter space $\Theta \equiv\{\mathcal{M}_c,\eta,D_L,\chi_1,\chi_2,H_{\rm eff5},H_{\rm eff8},t_c,\phi_c\}$. $\chi_1, \chi_2$ are varied from 0 to 1 along the $X$ and $Y$-axes respectively. Optimally oriented binaries are considered at $D_L = 200$ Mpc, with $M=40 M_\odot$, $q=3$, $H_{\rm eff5}=0.6$, and $H_{\rm eff8}=12$.}
\label{spinvar2}
\end{figure}

\subsubsection{Dependence on the Spins}
\label{sec:spin}

Measurements of $H_{\rm eff5}, H_{\rm eff8}$ are expected to depend on the spins of the binary components $\chi^{}_1,\chi^{}_2$ due to the presence of the spin-orbit term $\Psi^{}_{\rm SO}$ (Eq.~\eqref{spin-orbit}), and the fact that the upper cutoff frequency $f_{\rm ISCO}$ depends on the component spins (Eq.~\eqref{isco}).

Figure~\ref{spinvar} shows contours of the error values in $H_{\rm eff5}$ and $H_{\rm eff8}$ in ET and CE detectors when the dimensionless (aligned) spins $\chi_1, \chi_2$ are varied from 0 to 1. $\Delta H_{\rm eff5}$ and $\Delta H_{\rm eff8}$ contours are shown in the plots in the left and the right columns, respectively. The parameter space considered for this analysis is $\Theta \equiv\{\mathcal{M}_c,\eta,D_L,\chi_1,\chi_2,H_{\rm eff5},H_{\rm eff8}\}$.

We demonstrate the spin dependence of the errors for total binary mass $M=40M_\odot$ and two values of the mass-ratios, $q=1.1$ (top panel) and $q=3$ (bottom panel). Let us consider one of these binaries in CE, with $q=1.1$ and low values of component spins, $\chi_1=\chi_2=0.2$. For the 7-dimensional parameter space mentioned above, the errors in $H_{\rm eff5}$ ($H_{\rm eff8}$) are $\sim 0.4$ (3), which amounts to a percentage error of $\sim 67\%$ ($25\%$) for this binary. The contours have lower error values as they approach the point $\chi_1=\chi_2=1$, indicating that the errors decrease with increasing spins. This can be attributed to the fact that $f_{\rm ISCO}$ increases with $\chi_1$ and/or $\chi_2$, making the effective frequency range larger, consequently adding more GW cycles in the frequency band. We note here that the values of the parameters themselves increase with the (aligned) spins substantially (Eq.~\eqref{eq:Hparams}; Fig.~1 and Fig.~2 of Ref.~\cite{datta2020recognizing}), implying that for highly spinning compact objects one can put more stringent constraints on them. Owing to the higher dimensionality of the parameter space considered in Fig.~\ref{spinvar} as compared to the analyses presented in Figs.~\ref{et-ce-m} and~\ref{et-ce-dl}, the former reports higher errors in the effective horizon parameters than the latter ones, with an identical set of input parameters. In a real astrophysical scenario where the spins have to be estimated, Fig.~\ref{spinvar} represents more relevant benchmarks for the measurabilities of $H_{\rm eff5}$ and $H_{\rm eff8}$. Unfortunately, it is difficult to explore the uncertainties in measuring the two TH parameters across arbitrary regions of the physical parameter space when all the parameters are measured together, mainly because the Fisher matrices must be invertible. This was the primary reason for choosing a subset of the parameter space in the earlier sections. However, since the results depend notably on the parameter space considered, as seen in this section, we choose another region of the parameter space where $\Gamma$ is still well-conditioned if $t_c$ and $\phi_c$ are also measured. The results are presented in Fig.~\ref{spinvar2}, and these report more realistic estimates within this region. The errors increase roughly by an order of magnitude compared to Fig~\ref{spinvar}. These results from Fisher analyses, however, only provide a benchmark of the errors, and only full Bayesian studies can provide more accurate and realistic estimates. In the next section, we perform some Bayesian studies to corroborate the Fisher analysis results, in the parameter space of $\{\mathcal{M}_c,\eta,D_L,H_{\rm eff5},H_{\rm eff8}\}$. In a future study, more thorough Bayesian analyses can be performed with a larger set of parameters, e.g. including $t_c$ and $\phi_c$.

\section{Comparison with Bayesian Analyses}
\label{sec:bayesian}

We carried out Bayesian parameter estimation with  \texttt{Bilby}~\cite{Ashton:2018jfp} to compare the results with the ones gotten from the Fisher analyses. For each of the detector networks (LIGO-Virgo, ET, CE), we chose one point from the region of the parameter space that is expected to produce the best results according to the Fisher studies above. This was partly to ensure the robustness of the best estimates found by the latter method. Although these regions are different for LIGO-Virgo and 3G detectors, as noted earlier, we choose the values of total mass ($M=40M_\odot$) and mass-ratio ($q=3$) same for all three detector networks for the sake of comparison. We first inject \texttt{TaylorF2} waveforms with the TH phase, then run the parameter estimation to obtain posteriors from the simulations. The starting frequency is 10Hz(4Hz) for LIGO-Virgo(ET, CE), and the upper cutoff frequency is taken to be  the corresponding ISCO frequency. 




\begin{figure}
     \centering
     \begin{subfigure}[b]{0.49\textwidth}
         \centering
         \includegraphics[width=\textwidth]{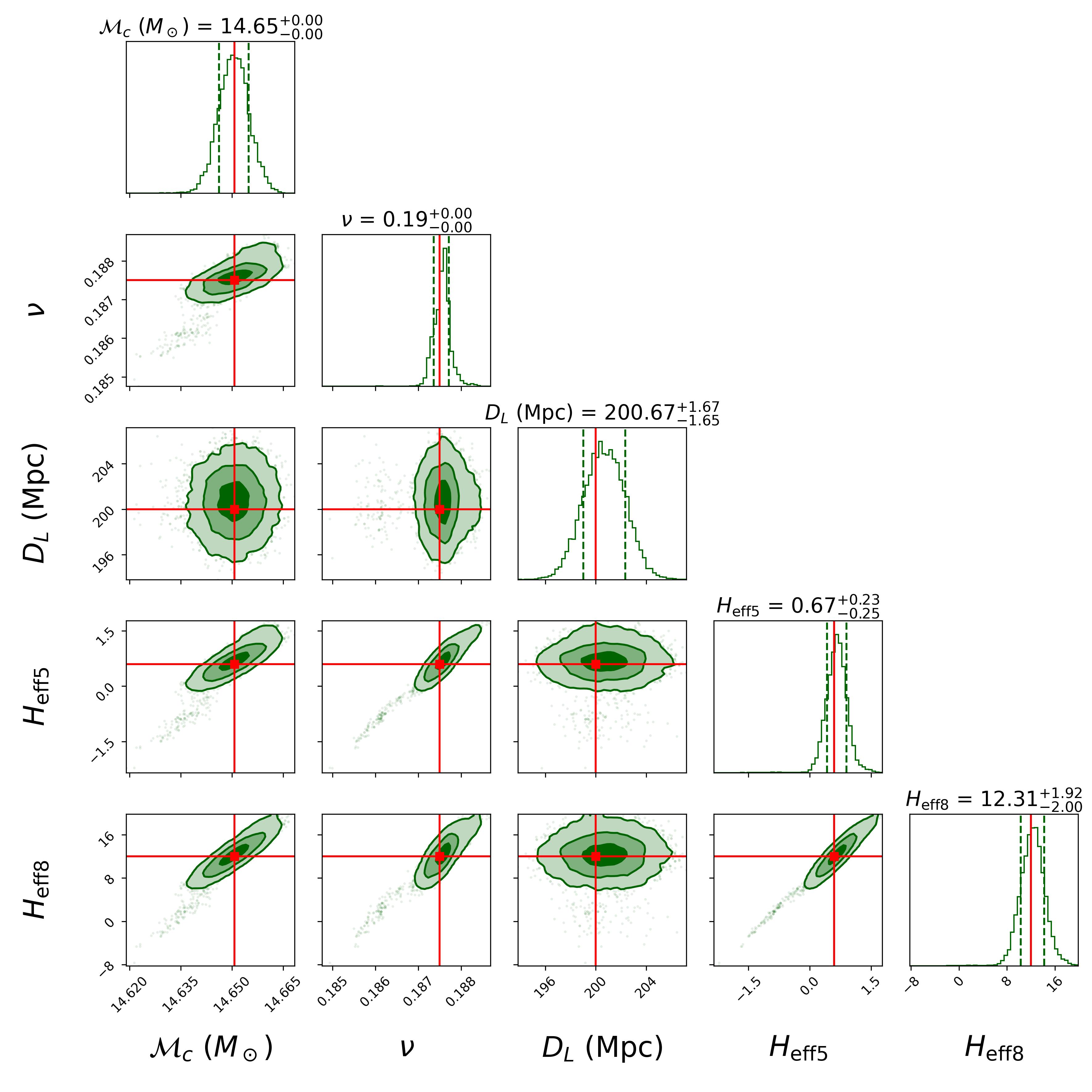}
         \caption{LIGO-Virgo}
         \label{40-ligo}
     \end{subfigure}
     \hfill
     \begin{subfigure}[b]{0.49\textwidth}
         \centering
         \includegraphics[width=\textwidth]{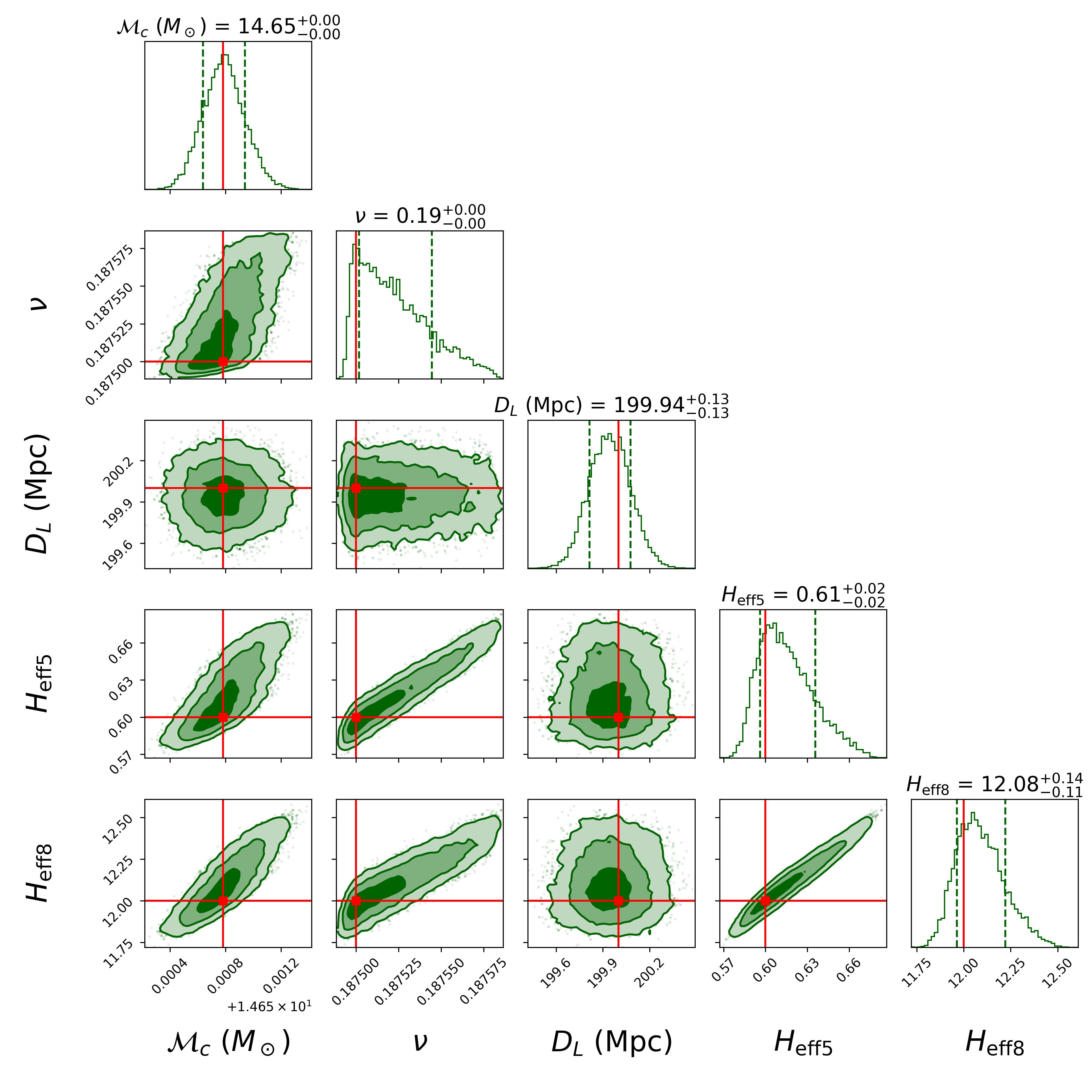}
         \caption{Einstein Telescope}
         \label{40-et}
     \end{subfigure}
     \hfill
     \begin{subfigure}[b]{0.49\textwidth}
         \centering
         \includegraphics[width=\textwidth]{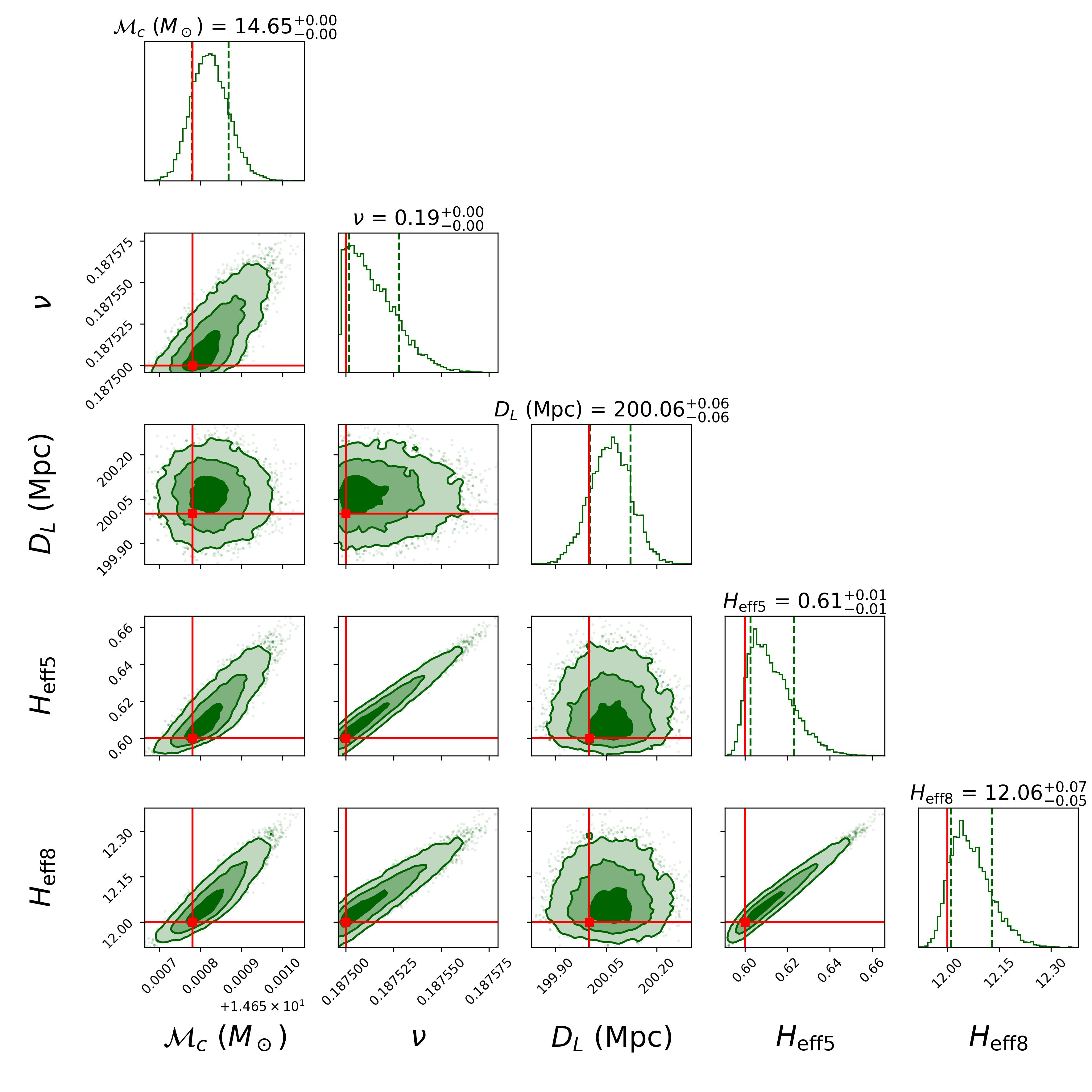}
         \caption{Cosmic explorer}
         \label{40-ce}
     \end{subfigure}
    \caption{Posterior plots from Bayesian parameter estimation. Injection parameters are $M=40 M_\odot$, $q=3$, $D^{}_L=200$ Mpc, $\chi^{}_1=\chi^{}_2=0.8$, $H_{\rm eff5}=0.6$, $H_{\rm eff8}=12$. The solid red lines denote the injected values of the corresponding parameters and the green dashed lines show the standard deviations (for the figures~\ref{40-et} and \ref{40-ce}, 14.65 is to be added to the tick labels of $\mathcal{M}_c$, which is mentioned below the x-axes of the subplots). The priors chosen for these simulations are listed in Table~\ref{priors}.}
    \label{bayes}
\end{figure}


In Table~\ref{priors}, we list the distribution and ranges of priors used for the chosen parameter space.


\begin{table}[h]
\centering
\begin{tabular}{|c c c c|} 
 \hline
 Parameter & Distribution & Range & Units \\ [0.5ex] 
 \hline\hline
 \small{Chirp mass} ($\mathcal{M}_c$) & Uniform & (10, 20) & $M_\odot$ \\ 
 \hline
 \small{Symmetric mass-ratio} & & & 
 \\ ($\eta$) & Uniform & (0.01, 0.25) & $\cdots$ \\
 \hline
 \small{Luminosity distance} & & & \\ 
 ($D_L$) & Uniform & (100, 500) & Mpc \\
 \hline
 $H_{\rm eff5}$ & Uniform & (-4, 4) & $\cdots$ \\
 \hline
 $H_{\rm eff8}$ & Uniform & (-20, 20) & $\cdots$ \\ [1ex] 
 \hline
\end{tabular}
\caption{ Choice of priors for the Bayesian posteriors presented in Fig.~\ref{bayes}.}
\label{priors}
\end{table}

Figure~\ref{bayes} shows the corner plots generated form the posteriors, for the three detector-networks, LIGO-Virgo (Fig.~\ref{40-ligo}), ET (Fig.~\ref{40-et}), and CE (Fig.~\ref{40-ce}). As expected, estimation of $H_{\rm eff5}$ and $H_{\rm eff8}$ are much better in ET and CE than in LIGO-Virgo, with the errors broadly agreeing with their Fisher counterparts for similar systems studied in Figs.~\ref{h5-ligo-m},~\ref{h5-et-m},~\ref{h5-ce-m} for $H_{\rm eff5}$, and Figs.~\ref{h8-ligo-m},~\ref{h8-et-m},~\ref{h8-ce-m} for $H_{\rm eff8}$. The 2-dimensional covariance plots in the Bayesian posteriors highlight the contours of 1-$\sigma$, 2-$\sigma$ and 3-$\sigma$ confidence levels from the smallest to the largest ones, respectively. In some cases (e.g. Fig.~\ref{40-ligo}), thin trails of points are seen in the plots, which are sampled beyond the 3-$\sigma$ confidence level.

\begin{figure}[]
    \centering
    \includegraphics[width=86mm]{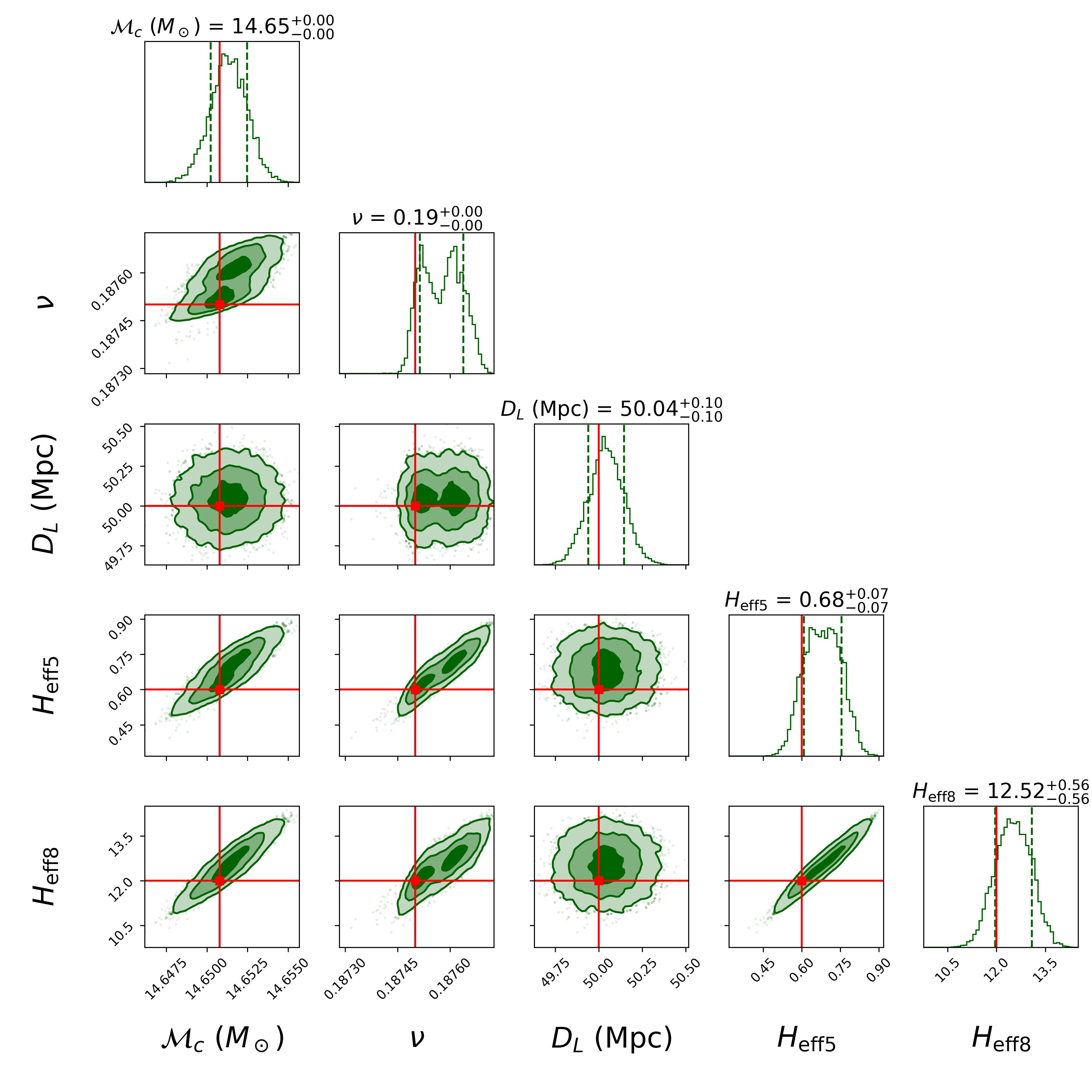}
    \caption{ {Bayesian posterior plot in LIGO-Virgo \\ with $M=40 M_\odot, q=3, D_L=50$ Mpc, $\chi^{}_1=\chi^{}_2=0.8$, $H_{\rm eff5}=0.6$, $H_{\rm eff8}=12$.}}
    \label{ligo-golden}
\end{figure}

We chose a fourth point for the LIGO-Virgo network, which is significantly close -- at $D_L=50$ Mpc, with all the other parameters the same as in Fig.~\ref{40-ligo}. Prior for the luminosity distance is taken to be uniform in the range (10, 100) Mpc. All the other parameters have the same priors as in Table~\ref{priors}. Figure~\ref{ligo-golden} shows the corresponding posterior plot. Comparing 
the two, 
we see the expected improvement in accuracy and precision, arising from the binary being four times closer. The errors in this case are $\sim 11.7\%(\sim 4.7\%)$ for $H_{\rm eff5} (H_{\rm eff8})$ for a BBH.


\begin{figure*}[ht]
    \centering
    \hspace{-1cm}
    \includegraphics[width=\linewidth]{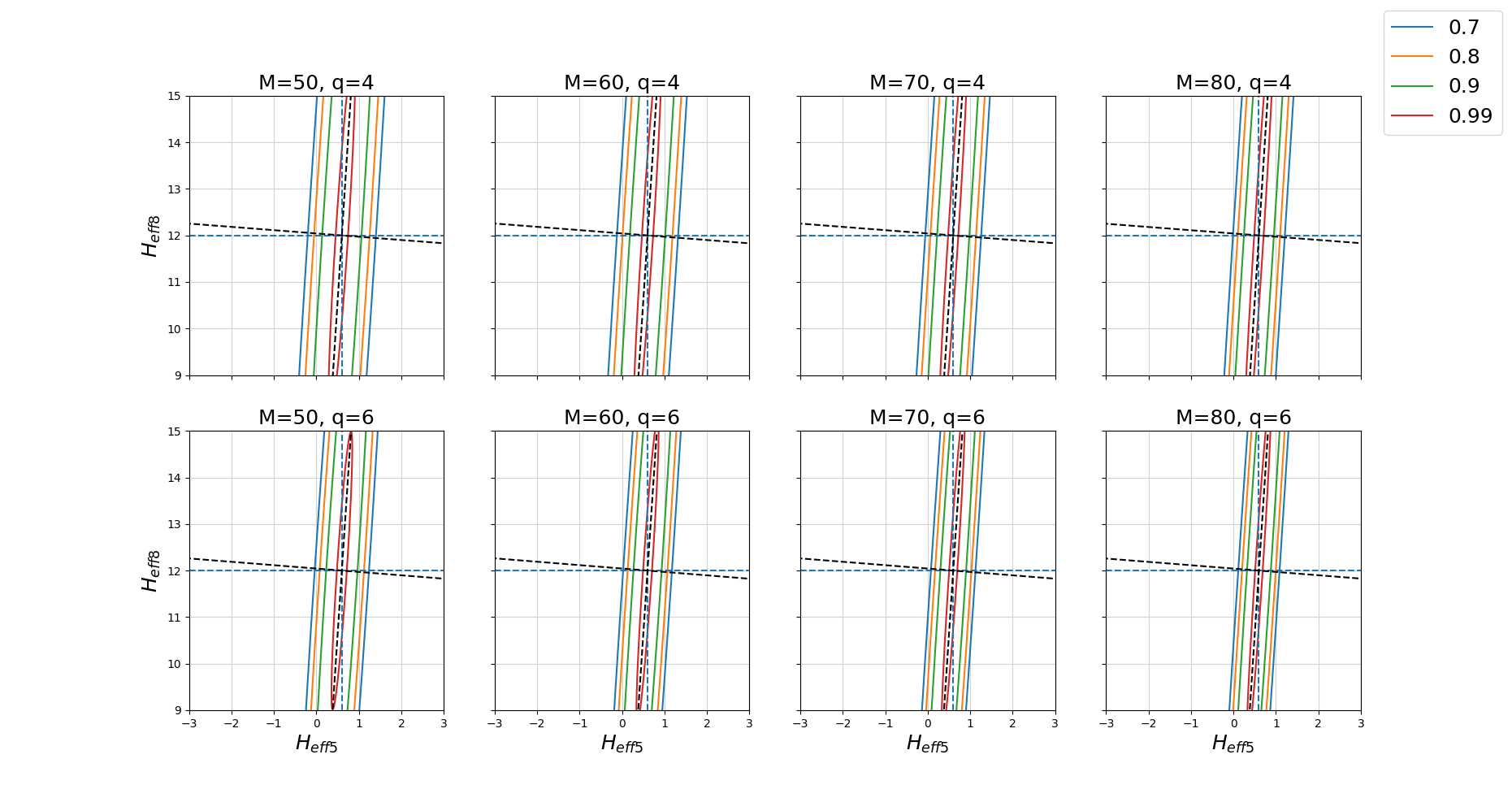}
    \caption{ Fisher ellipses in LIGO-Virgo with normalized templates, implying SNR = 1. The spin values taken here are $\chi_1=\chi_2=0.8$. The ``target waveform" corresponds to the intersection point of the eigenvector directions (the dotted black lines), which is $H_{\rm eff5}=0.6, H_{\rm eff8}=12$. First and second row correspond to mass-ratio $q=4$ and $q=6$ respectively, and the four columns are for total mass values $M=50,60,70,80M_\odot$ from left to right. Match values for different ellipses are shown in the common legend at the upper right corner.}
    \label{ellipse_ligo}
\end{figure*}


    

\begin{figure}
     \centering
     \begin{subfigure}[b]{\textwidth}
         \centering
         \includegraphics[width=\textwidth]{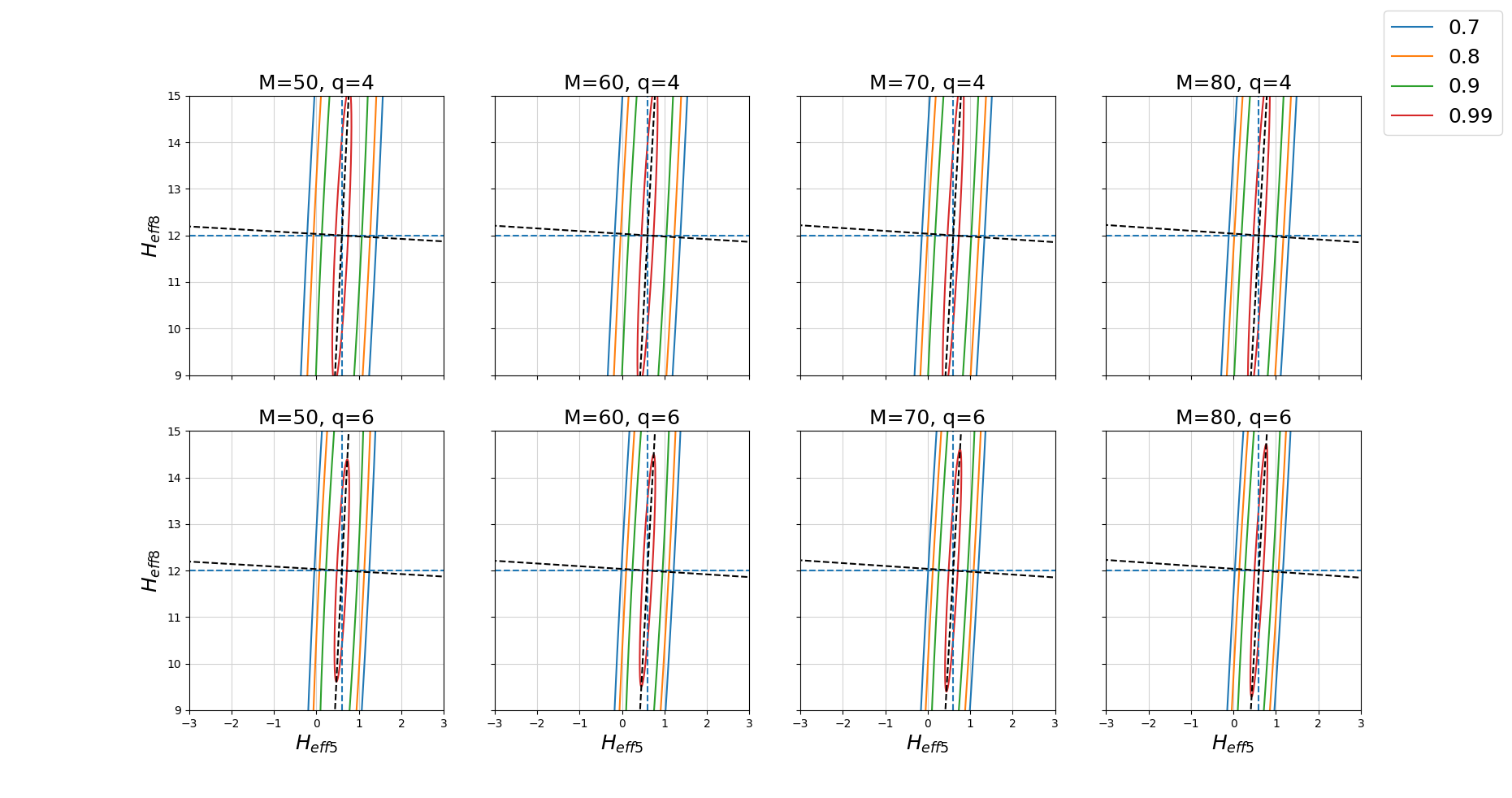}
         \caption{}
         \label{ellipse-et}
     \end{subfigure}
     \hfill
     \begin{subfigure}[b]{\textwidth}
         \centering
         \includegraphics[width=\textwidth]{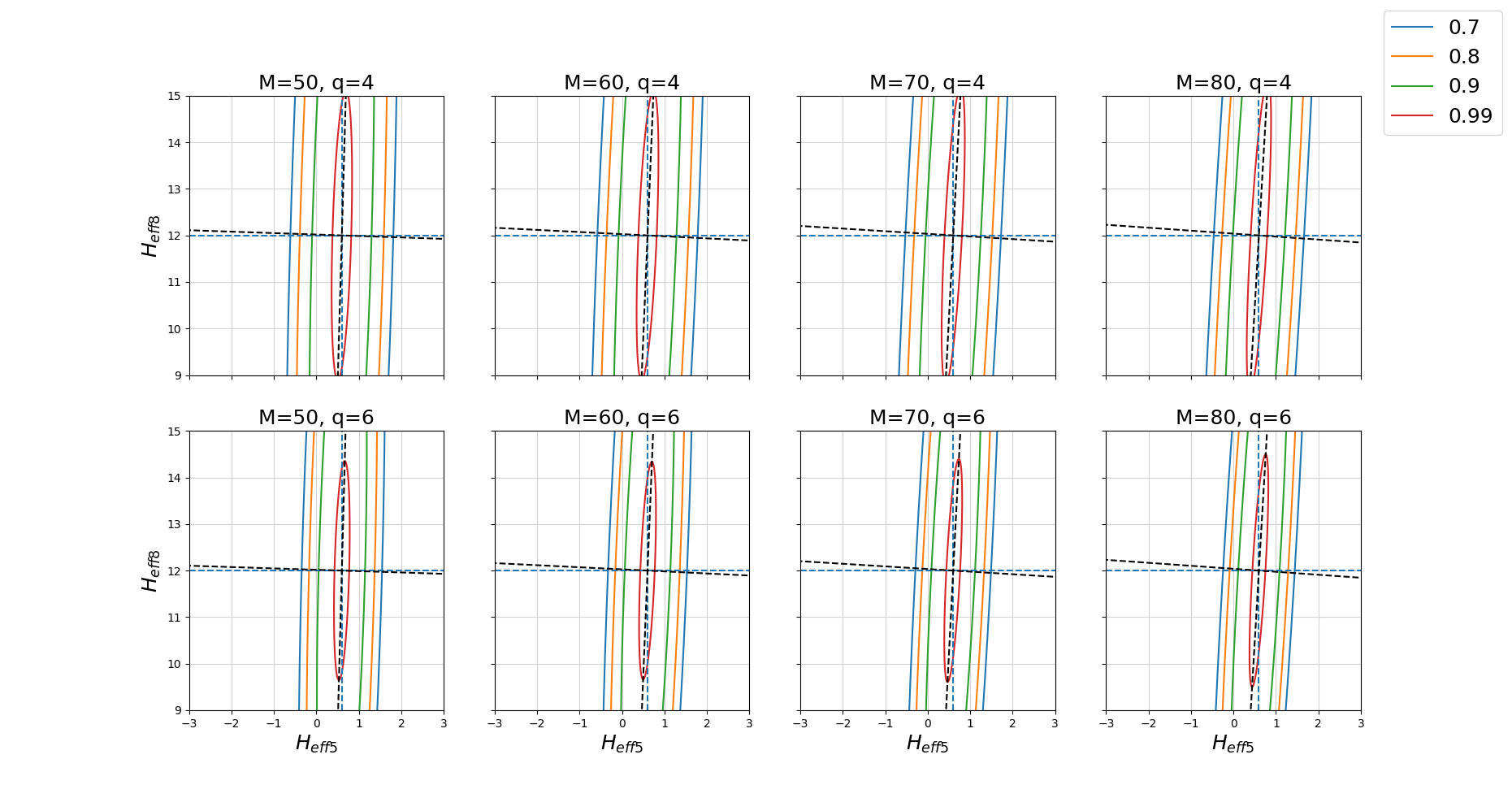}
         \caption{}
         \label{ellipse-ce}
     \end{subfigure}

    \caption{ Same as in Fig.~\ref{ellipse_ligo}, but in the detectors ET (Fig.~\ref{ellipse-et}) and CE (Fig.~\ref{ellipse-ce}).}
    \label{ellipse_et_ce}
\end{figure}


\section{Principal Component Analysis}
\label{pca}

So far we have considered the diagonal elements of the error
covariance matrix $C$, which are the variances. In this section we will consider the covariances between $H_{\rm eff5}$ and $H_{\rm eff8}$, which are the off-diagonal elements of $C$. Diagonalizing $C$ will  yield its eigenvalues and eigenvectors. The eigenvectors will provide the new coordinates (i.e., horizon parameters) that have a vanishing  covariance, and the eigenvalues will correspond to the errors in those new coordinates. Keeping this in mind, we choose a particular waveform as a ``target" waveform, with which we calculate the matches (defined below) of several neighboring ``template" waveforms. If we make a contour plot of these match values, we get ellipses -- with their principal axes along the eigenvectors of $C$. Since $C=\Gamma^{-1}$, the eigenvectors of $C$ and $\Gamma$ are the same, and their eigenvalues are related by $\lambda_i=1/\alpha_i$, where $\lambda_i(\alpha_i)$ are the eigenvalues of $\Gamma(C)$.

In this quest, we take the manifold $\mathcal{V}$ of dimensionality same as that of $\Theta$, with every point on $\mathcal{V}$ corresponding to a template waveform. The distance between two templates on $\mathcal{V}$ corresponding to the parameter vectors $\Theta$ and $(\Theta+\Delta\Theta)$ can be calculated as~\cite{Toubiana:2020cqv}:

\begin{align}
\label{temp_distance}
    |h(\Theta+\Delta\Theta)&-h(\Theta)|^2 \nonumber\\
    =&\braket{h(\Theta+\Delta\Theta)-h(\Theta)}\nonumber\\
    =&\braket{\pdv{h}{\Theta_i}}{\pdv{h}{\Theta_j}}\Delta\Theta^i\Delta\Theta^j\nonumber\\
    =&\Gamma_{ij}\Delta\Theta^i\Delta\Theta^j.
\end{align}
The \textit{match} ($\mathcal{M}$), also known as the \textit{overlap function}~\cite{Ajith:2009fz,owen_PhysRevD.53.6749}, between two template waveforms $h(\Theta)$ and $h(\Theta+\Delta\Theta)$ can be defined as the inner product between them, maximized over the extrinsic parameters $t_c$ and $\phi_c$~\cite{owen_PhysRevD.53.6749} :
\begin{equation}
    \mathcal{M} = \underset{t_c,\phi_c}{\rm max}\braket{h(\Theta+\Delta\Theta)}{h(\Theta)}.
\end{equation}
If all the templates $h(\Theta)$ on $\mathcal{V}$ are normalized by $\hat{h}(\Theta) = h(\Theta)/\braket{h(\Theta)}{h(\Theta)}$, so that $\braket{\hat{h}(\Theta)}{\hat{h}(\Theta)}=1\, \forall\, \hat{h}(\Theta)\in \mathcal{V}$, then the maximum value of $\mathcal{M}$ can be 1, which corresponds to $\Delta\Theta=0$. Then, one can define the \textit{mismatch} between two templates $\hat{h}(\Theta)$ and $\hat{h}(\Theta+\Delta\Theta)$ as $1-\mathcal{M}$, which geometrically denotes the ``distance" between them on the manifold $\mathcal{V}$. One can relate them, using Eq.~\eqref{temp_distance}, as
\begin{equation}
\label{match}
    1-\mathcal{M}=\Gamma^{(n)}_{ij}\Delta\Theta^i\Delta\Theta^j.
\end{equation}

Here $\Gamma^{(n)}_{ij}$ is the Fisher matrix for normalized templates ($\hat{h}(\Theta)$), related to the Fisher matrix for unnormalized templates ($h(\Theta)$) as 

\begin{align}
    \Gamma_{ij}=&\braket{\pdv{h}{\Theta_i}}{\pdv{h}{\Theta_j}}\nonumber\\
    =&\braket{h}{h}\braket{\pdv{\hat{h}}{\Theta_i}}{\pdv{\hat{h}}{\Theta_j}}\nonumber\\
    =&\,\rho^2\, \Gamma^{(n)}_{ij}.
\end{align}
The last expression follows from the fact that $\sqrt{\braket{h}{h}}$ is the SNR $\rho$, given by Eq.~\eqref{snr2}.
Eq.~\eqref{match} motivates one to define a metric $g_{ij}$ on $\mathcal{V}$ to express the distance between two templates $\hat{h}(\Theta)$ and $\hat{h}(\Theta+\Delta\Theta)$ as $g_{ij}\Delta\Theta^i\Delta\Theta^j$, and identify the relation of the metric with the Fisher matrix as $g_{ij}=\Gamma^{(n)}_{ij}=(1/\rho^2)\Gamma_{ij}$.

In our analysis, we consider a 2D manifold with only $H_{\rm eff5}$ and $H_{\rm eff8}$ as parameters, which is a submanifold of $\mathcal{V}$ with all the other parameters fixed. On this submanifold, Eq.~\eqref{match} can be expanded as
\begin{equation}
\begin{split}
    1-\mathcal{M}=\Gamma^{(n)}_{00}(H_{\rm eff5}-H_{\rm eff5}^\ast)^2+\Gamma^{(n)}_{11}(H_{\rm eff8}-H_{\rm eff8}^\ast)^2\\+2\Gamma^{(n)}_{01}(H_{\rm eff5}-H_{\rm eff5}^\ast)(H_{\rm eff8}-H_{\rm eff8}^\ast)\,,
\end{split}
\end{equation}
with $H_{\rm eff5}^\ast(H_{\rm eff8}^\ast)$ being the value of $H_{\rm eff5}(H_{\rm eff8})$ corresponding to the target waveform.
Thereby, the contours of constant values of $\mathcal{M}$ represent ellipses in the space of $H_{\rm eff5}$ and $H_{\rm eff8}$, centered at ($H_{\rm eff5}^\ast$,$H_{\rm eff8}^\ast$), given that the Fisher matrix components are constant.
For a Fisher matrix with only $H_{\rm eff5}$ and $H_{\rm eff8}$ as parameters, none of its components depends on the values of $H_{\rm eff5}$ or $H_{\rm eff8}$. This implies that the metric is flat on this submanifold, and the contours of constant $\mathcal{M}$ are all perfect ellipses. If we diagonalize $\Gamma$, then in the eigen-coordinates, all the covariances will vanish. Let us call the corresponding eigenvectors $(X,Y)$, known as the {\it principal components}~\cite{Ohme:2013nsa}. Then the equation of the ellipses with respect to the eigen-coordinates $(X,Y)$ with corresponding eigenvalues ($\lambda_1, \lambda_2$) becomes
    \begin{equation}
        \lambda_1 (X-X^\ast)^2+\lambda_2 (Y-Y^\ast)^2=(1-\mathcal{M}).
    \end{equation}
These ellipses are centered at $(X^\ast,Y^\ast)$, and have principal axes ($a,b$) given by,
\begin{equation}
    a=\sqrt{(1-\mathcal{M})/\lambda_1}\,, \quad\quad b=\sqrt{(1-\mathcal{M})/\lambda_2}\,.
\label{a,b}
\end{equation}
 
Figure~\ref{ellipse_ligo} shows such ellipses in the sensitivity of Advanced LIGO and Virgo for a target waveform with source parameters $H_{\rm eff5}^\ast=0.6, H_{\rm eff8}^\ast=12$. Figures~\ref{ellipse-et} and \ref{ellipse-ce} show similar ellipses in ET and CE, respectively. We show eight different plots for a combination of different values of the total mass and mass-ratio. Since $\Gamma$ is a symmetric matrix, its eigenvectors, lying along the dotted lines shown in the plots, are orthogonal to each other. 


\begin{figure}[]
    \centering
    \includegraphics[width=85mm]{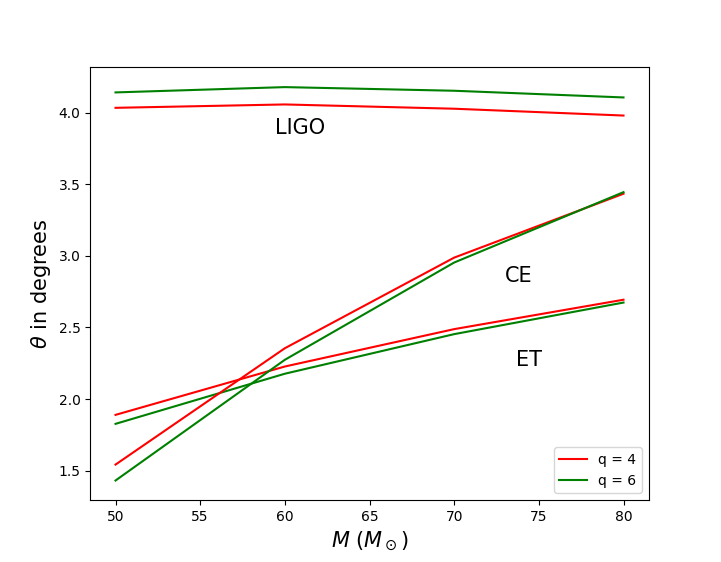}
    \caption{ Variation of the rotation angle of $X-Y$ coordinate axes with respect to the $H_{\rm eff5}-H_{\rm eff8}$ axes with $M$ for $q=4,6$ in Fig.~\ref{ellipse_ligo} and Fig.~\ref{ellipse_et_ce}. }
\label{theta}
\end{figure}


 Covariances between the two parameters cause the ellipses to tilt, with higher tilt angles implying higher covariances between $H_{\rm eff5}$ and $H_{\rm eff8}$.  In Fig.~\ref{theta}, we show the variation of the tilt angles ($\theta$) between the $X-Y$ coordinate axes and the $H_{\rm eff5}-H_{\rm eff8}$ axes with total mass $M$, for $q=4$ and $q=6$. In LIGO-Virgo, the tilt angles vary negligibly with $M$, but their values are higher compared to the 3G detectors. In ET and CE, The tilts of the ellipses increase slowly with $M$, implying that the covarinaces between $H_{\rm eff5}$ and $H_{\rm eff8}$ are higher for higher-mass systems. We also note that CE shows a faster growth in the covariances for higher mass systems than ET. The effect of $q$ on the covariances appears to be different in LIGO-Virgo than in ET, CE -- in the former, they increase with increasing $q$, but the latter two follow the opposite trend. The small tilt angles of the eigen-coordinates, especially in 3G detectors, imply negligible covariances between $H_{\rm eff5}$ and $H_{\rm eff8}$.
 
\begin{figure}[]
    \centering
    \includegraphics[width=85mm]{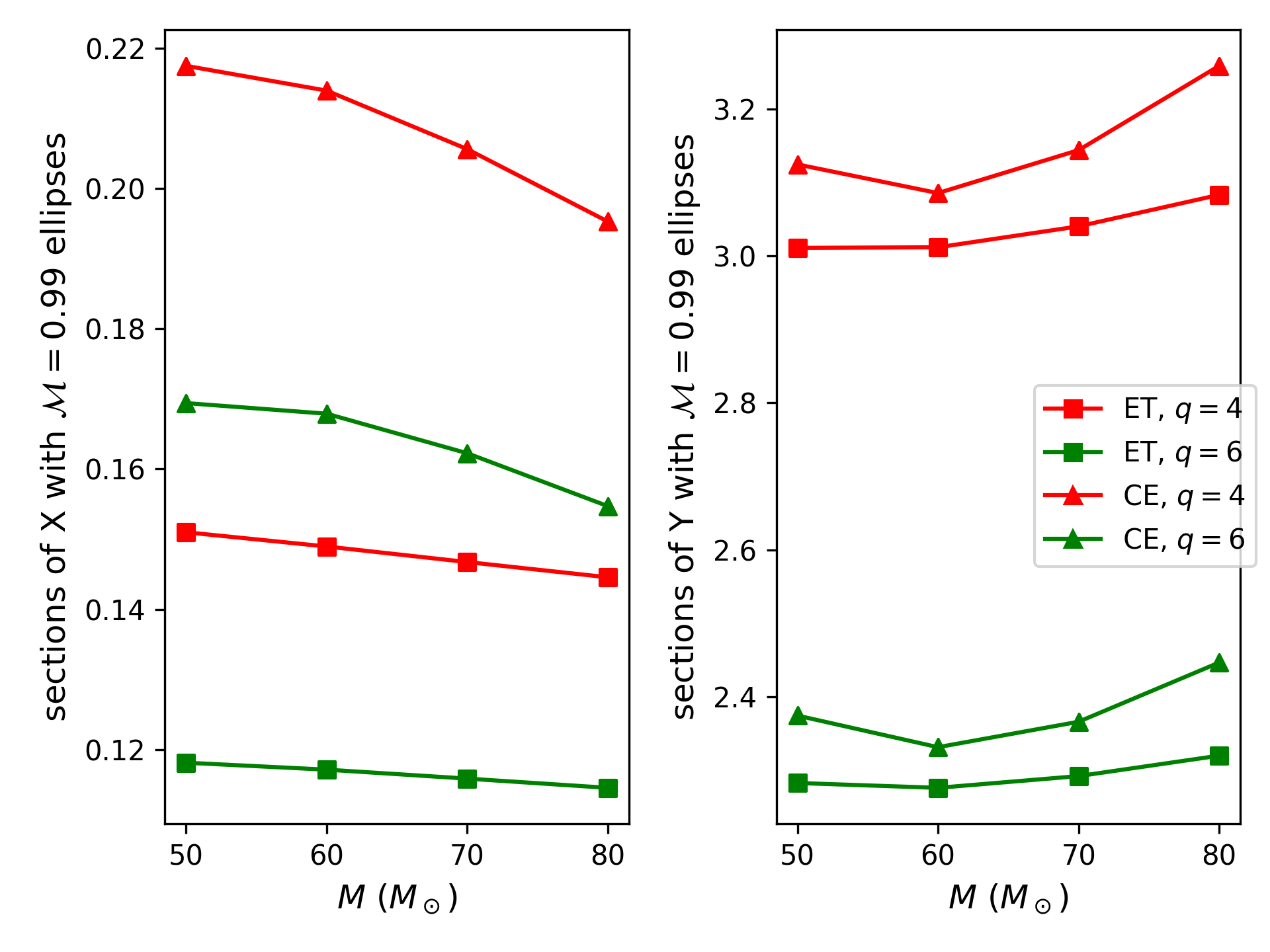}
    \caption{ Sections of the $\mathcal{M}=0.99$ ellipses with their principal axes in Fig.~\ref{ellipse_et_ce}, plotted against $M$ for $q=4$ (red), $q=6$ (green) in ET (square points) and CE (triangular points).}
\label{eigval}
\end{figure}

 The measurability of a certain parameter can be inferred from these ellipses by studying how closely spaced they are along the direction of that parameter, which denotes how rapidly the match values change with small displacements along that direction. Rapid change of match values implies that two different waveforms can be distinguished better; consequently, the errors are smaller. Since we are considering only normalized waveforms for this analysis, effects of the SNR on the statistical errors are absent here, in contrast to Sec.~\ref{results} where the results depend largely on SNR. This enables us to study the variations of the errors in the eigen-coordinates in an SNR-independent way. To demonstrate how the shapes of the ellipses vary with total mass and mass-ratio, in Fig.~\ref{eigval} we plot the principal axes $a$ and $b$ of the ellipses, defined in Eq.~\eqref{a,b}, for the match value $\mathcal{M}=0.99$ (the red ellipses in Fig.~\ref{ellipse_ligo} and Fig.~\ref{ellipse_et_ce}). In this figure only the 3G detectors are considered. The ellipses get stretched out along the eigen-coordinate $Y$ (the semi-major axes) with increasing $M$, implying that the error in that coordinate increases with $M$ for $M> 60M_\odot$. This follows the behavior of $\Delta H_{\rm eff8}$, which increases with $M$ in this region (Figs.~\ref{h8-et-m},~\ref{h8-ce-m}). Along the $X$ direction (semi-minor axes), sections of the ellipses are smaller with increasing $M$,
 implying lesser errors.  Increasing $q$ makes the ellipses more squeezed along both $X$ and $Y$, implying better measurabilities  in both the eigen-coordinates.

Covariance between two parameters is a measure of the degeneracy between them. A vanishing covariance between two parameters allows one to probe the dependence of the model (in our case, the gravitational waveform) on them separately. This possibility is attained by 3G detectors with the introduction of more GW cycles in the early inspiral, and the accompanying fact that these parameters are much more precisely measurable than in LIGO and Virgo. In the case of ET and CE, as we see, in the considered region of the parameter space, the errors and the covariances are not large enough for diagonalization to make any significant difference.

\section{Discussion and Summary}
\label{discussion}

We have explored how well one can measure the two tidal heating parameters $H_{\rm eff5}$ and $H_{\rm eff8}$ in the future ground-based GW detectors Einstein Telescope and Cosmic Explorer as well as in the  2nd generation detectors Advanced LIGO and Advanced Virgo. These parameters account for the flux of energy and angular momentum into 
or out of a (spinning) BH, which is different for other compact objects -- even those that mimic a BH. They appear at 2.5PN and 4PN orders, respectively, in the expression for the phase shift in the gravitational waveform due to tidal heating. 
The prospect of proper estimation of these parameters results in a viable method for distinguishing BHs (in binaries) from other compact objects that do not have horizons, but may otherwise  resemble them. We  chose \texttt{TaylorF2} as the PN approximant and added the tidal-heating phase shift to it. We used primarily the Fisher matrix approach  for  estimating the errors. 

In 3G detectors, we showed that for a total binary mass of $M\gtrsim 50M_\odot$, estimation of the aforementioned parameters is the most precise, whereas for  2G detectors there is a specific region around $20M_\odot\lesssim M \lesssim 40M_\odot$ where we expect the best results with the current waveform. Increasing mass asymmetry results in lesser errors. The errors rise linearly with the luminosity distance. In 3G detectors, we can constrain $H_{\rm eff5}$ ($H_{\rm eff8}$) with a 1-$\sigma$ error value of $\sim 0.05$ ($\sim 0.2$) for a binary with $M=30M_\odot, q=1.5, H_{\rm eff5}=0.6, H_{\rm eff8}=12$, at 1 Gpc distance. This error value amounts to a relative percentage error of approximately 8.3\% (2\%) for $H_{\rm eff5}$ ($H_{\rm eff8}$). In LIGO-Virgo, the errors are higher ($\sim 300\%$ for $H_{\rm eff5}$, $\sim 50\%$ for $H_{\rm eff8}$), as expected, mainly due to the lower SNRs. The measurements can be improved by using coherent mode stacking, by which one can combine observations of $N$ number of GW events and effectively scale the SNR by a factor of $\sqrt{N}$~\cite{Yang:2017zxs,Bose:2017jvk}. Spins of the binary components affect the measurabilities due to the spin-orbit term, and the fact that the upper cutoff frequency used in this work is spin-dependent. Increasing spin makes the considered frequency range wider, which in turn lowers the error values.

We have also demonstrated that in the sensitivity of 3G detectors, covariances between these parameters are not significant, meaning that we do not expect to improve the results any further by introducing any new combination of them by diagonalizing the covariance matrix. However, in 2G detectors, covariances are high enough for this method to produce better results, and we show how we can define a new set of coordinates with less errors from the tilts of the Fisher ellipses. 

This work will be useful in future studies when more complete and accurate tidal-heating waveforms are available that extend deeper into the merger phase. We make inroads in this pursuit in Chapters~\ref{Chapter4} and~\ref{Chapter5}. Our study has identified the regions in the parameter space where one can expect the best results in estimating the tidal heating parameters. We have shown that these results are consistent with Bayesian analyses.

\clearpage
\chapter{Phenomenological Waveforms for Binary Black Holes}
\label{Chapter4}

\section{Introduction}
\label{sec:intro}

The launch of GW astronomy has had a stellar
start, with the detection of over 90 compact binary coalescences (CBCs) so far~\cite{LIGOScientific:2021djp,Nitz:2021zwj,Olsen:2022pin}.
The fourth (O4) observation run of the
ground-based GW detectors LIGO~\cite{LIGOScientific:2014pky}, Virgo~\cite{AdvVIRGO} and KAGRA~\cite{KAGRA:2018plz}
are expected to detect many more CBCs, which enables us to subject
GR to unprecedented tests. Such tests demand accurate CBC waveforms. 
In the field of GW data analysis, waveform models serve as templates against which the real data are compared and estimates of the source parameters are made by the matched filtering technique~\cite{Owen:1998dk} and Bayesian inference~\cite{Veitch:2014wba,Ashton:2018jfp,Romero-Shaw:2020owr}. 
Since a sizable subset of CBCs are high-mass
binaries, which have a good fraction of their signal power arriving
from the late-inspiral and merger parts, these tests benefit from
employing inspiral-merger-ringdown (IMR) waveforms. One needs to solve the full Einstein equations numerically to extract gravitational waveforms predicted by GR to meet the accuracy standard imperative for describing the merger-ringdown parts of a CBC. However, such simulations of numerical relativity (NR) are of extreme computational cost, impeding the usage of this scheme for creating long waveforms that span the entire frequency range of GW detectors for intermediate to stellar-mass binaries. On the other hand, post-Newtonian (PN) approximation (see, e.g., Ref.~\cite{Blanchet:2013haa} and the references therein) or the effective-one-body (EOB) framework~\cite{Buonanno:1998gg,Bohe:2016gbl,Nagar:2018gnk} describes the inspiral part in analytical forms, but fail to be reliably accurate in the late inspiral to the merger regime.  

Efforts to construct accurate IMR models have been based on the EOB formalism with calibrations to NR data (\texttt{SEOBNR}~\cite{Bohe:2016gbl,Taracchini:2012ig,Taracchini:2013rva} and \texttt{TEOBResumS}~\cite{Nagar:2018zoe,Gamba:2021ydi,Nagar:2020pcj} models), phenomenological waveform models built by combining PN and NR waveforms~\cite{Ajith:2007kx,Santamaria:2010yb,Hannam:2013oca,Pratten:2020fqn,Husa:2015iqa,Khan:2015jqa}, and more recently, surrogate models~\cite{Tiglio:2021ysj,Field:2013cfa,Blackman:2017pcm}. The computational cost of EOB models, however, is burdened by the need to tackle the orbital dynamics through solving a complex system of ordinary differential equations. Phenomenological models, in comparison, are equipped with closed-form expressions for the phase and amplitude of the waveforms for a given set of binary parameters, and are much faster to evaluate. The latter models contain PN-inspired phase and amplitude behaviors augmented with higher-order terms calibrated against a set of ``hybrid waveforms". The hybrids are constructed by stitching either PN (e.g., in \texttt{IMRPhenomC}~\cite{Santamaria:2010yb}) or EOB (e.g. \texttt{IMRPhenomD}~\cite{Khan:2015jqa,Husa:2015iqa}, \texttt{IMRPhenomXAS}~\cite{Pratten:2020fqn} for aligned-spin cases) waveforms with NR ones at a suitable frequency, to allow smooth transitions of phase and amplitude from the inspiral to the merger-ringdown regime. Utmost care needs to be borne when producing complete IMR waveforms by combining these parts, to limit introducing waveform systematics, which can reduce the effectiveness of tests of GR. The increasing modeling errors and computational cost of gravitational waveforms with increasing mass ratio and spins~\cite{Kumar:2016dhh} have inspired the advent of reduced-order models (ROMs) and surrogate models.
By fitting interpolated decomposed waveform data pieces over the binary parameter space, surrogate
models can significantly accelerate NR (e.g.~\cite{Blackman:2015pia,Blackman:2017pcm,Varma:2018mmi}) or EOB waveforms (e.g.~\cite{Purrer:2015tud,Lackey:2016krb,Lackey:2018zvw}), while maintaining high accuracy within their parameter space of validity.

In phenomenological models for binary black holes (BBHs), the inspiral waveform is usually constructed with the ``point-particle" approximation, which also acts as a baseline for creating generic waveforms with tidal effects arising from the finite size of the components, e.g. for binary neutron stars (BNSs)~\cite{Barkett:2015wia,Dietrich:2018uni}. For black holes (BHs), 
the absorption of energy and angular momentum by their horizons needs to be consistently incorporated within a complete BBH waveform model. 
This effect is weak but important in many respects. In Chapter~\ref{Chapter3} we have shown that the presence of TH can be measurable, especially in the future era of GW detectors with orders of magnitude higher signal-to-noise ratios (SNRs) compared to the current ones. Combined with the fact that TH is much more significant for BHs than horizonless compact objects, this makes TH a viable discriminator for horizons to identify BHs against theoretically possible exotic compact objects (ECOs)~\cite{Cardoso:2016oxy}, which can mimic BHs in their GW signals~\cite{Datta:2019epe,Datta:2019euh}.

In NR simulations, TH arises naturally due to the presence of the BH apparent horizons. While evolving the binary, BH singularities are expunged from the computational domain by excision techniques or by the ``moving puncture" method. The properties of the apparent horizons -- the masses, spins, and the horizon areas -- continue to get impacted by the energy and angular momentum fluxes throughout the binary evolution up to the merger. Changes in BH masses and spins impact the inspiral rate of the binary, leaving its imprint in the phase of GW signals. Complete IMR waveforms created with NR data for late inspiral and merger, then, would have incomplete information about BBH systems if they are joined with inspiral approximants devoid of the imprints of horizon fluxes. Ignoring this effect may also lead to ambiguities in the tests of GR from GW observations of CBCs, where small deviations from GR predictions are probed under the strong gravity conditions of a CBC. Lesser systematics will facilitate a more accurate evaluation of GR's validity in such cases.

In the EOB formalism, progress has been made to include the effects of horizon absorption for nonspinning binaries~\cite{Nagar:2011aa,Bernuzzi:2012ku,Damour:2012ky} and for aligned spins~\cite{Damour:2014sva}. These works use numerical evaluation of horizon fluxes using a frequency-domain perturbative approach along stable and unstable orbits in the test-mass limit. This test-mass knowledge is then hybridized with lower-order analytical information that is valid for comparable masses. These calculations were implemented in the construction of \texttt{TEOBResumS}~\cite{Nagar:2018zoe}, an NR-informed EOB waveform model which includes the horizon-absorption effects in a PN-inspired resummed form.  Alvi~\cite{Alvi:2001mx} has calculated, under PN approximation, leading-order expressions of mass and spin evolution of BBH systems. Alvi's expressions are explicitly valid for any mass ratio. A resummed version of these expressions was used by Damour and Nagar~\cite{Damour:2014sva} to describe them under the EOB framework.

In Phenomenological models, horizon absorption was taken into account in the flux calculations of \texttt{IMRPhenomC}~\cite{Santamaria:2010yb}, up to 2.5PN order. However, later models with more accurate calibrations spanning larger regions of the parameter space (e.g. \texttt{IMRPhenomD} or \texttt{IMRPhenomXAS}) do not account for this effect explicitly in their inspiral parts. 

In this chapter, we construct a phenomenological BBH waveform model with a consistent inclusion of TH effects, and study its contribution to an improvement in waveform systematics. A complete BBH model, which includes the phase and amplitude modifications for TH from the early inspiral to the merger, not only provides a comprehensive foundation but also helps to lay the groundwork for a more general waveform model with the \textit{horizon parameters}, which can act as discriminators for BHs from ECOs or NSs~\cite{Datta:2020gem,Mukherjee:2022wws}.

We first describe some preliminaries and notations of gravitational waveforms in Sec.~\ref{sec:convention}. Then in Sec.~\ref{sec:inspiral}, we build an inspiral model containing explicit corrections due to TH at 2.5PN, 3.5PN, and 4PN orders in their phase and amplitude. The point-particle baseline for the inspiral waveform is based on an aligned-spin EOB approximant. In Sec.~\ref{sec:NR}, we describe the publicly available NR waveforms of the SXS catalog~\cite{Boyle:2019kee} used for merger and ringdown phases. In Sec.~\ref{sec:hyb}, we describe the construction of new hybrid waveforms by stitching together the TH-corrected inspiral and merger-ringdown parts. Following \texttt{IMRPhenomD} (henceforth referred to as \texttt{PhenomD}), we model the hybrids in the frequency domain in Sec.~\ref{sec:model} and calibrate the inspiral part with augmented PN expressions, but against the new set of hybrid waveforms. Section~\ref{sec:model validation} discusses the faithfulness of the new model, and in Sec.~\ref{sec:comparison}, we compare it with \texttt{PhenomD} and NR data within the parameter range of calibration. We conclude in Sec.~\ref{sec:discussion} with discussions and future directions of this work.

Throughout the chapter we use geometric units $G=c=1$, except while calculating physical quantities.

\section{Waveform Conventions and Notation}\label{sec:convention}

We consider $l=|m|=2$ spherical harmonic modes of gravitational waves from coalescing binary black holes. In the time domain, the complex GW strain can be written as,
\begin{equation}\label{eq:strain}
    h_{2,2}(\Theta,t)=A(\Theta,t)e^{-i\phi(\Theta,t)},
\end{equation}
where $\Theta$ is the parameter vector containing the parameters of the binary. The intrinsic parameters are the dimensionless spin vectors $\vb*{\chi_1},\vb*{\chi_2}$ and the two masses $m_1$ and $m_2$. Since we are concerned with non-precessing binaries only, the spin vectors have nonzero components only along (or opposite to) the direction of the orbital angular momentum. In terms of the normalized orbital angular momentum $\hat{L}$ and the dimensionful spin vectors $\vb*{S_i}$ ($i=1,2$ for the two  objects), the components of dimensionless spins can be expressed as
\begin{equation}\label{eq:spins_def}
    \chi^{}_i=\frac{\vb*{S_i}\cdot \hat{L}}{m_i^2}\,.
\end{equation}
We define the mass ratio as $q=m_1/m_2\geqslant 1$, total mass $M=m_1+m_2$, and the symmetric mass ratio $\eta=m_1m_2/M^2$. 

The complex strain in Eqn.~\eqref{eq:strain} can be constructed from the plus and cross polarization states as
\begin{equation}
    h(\Theta,t)=h_{+}(\Theta,t)-ih_{\cross}(\Theta,t)\,.
\end{equation}
The Fourier transform of the complex strain is defined as
\begin{equation}
    \Tilde{h}(f)=\int_{-\infty}^\infty h(t)e^{-i2\pi ft}\dd t\,,
\end{equation}
where $f$ denotes frequency. For the non-eccentric waveforms that we use, $f$ can be written as a monotonically increasing function of time $t$ until the merger. 

In gravitational waveform modeling, the dimensionless frequency $Mf$ plays an important role, since the total mass $M$ acts as a scaling factor. For better readability, we define $F\equiv Mf$.

\section{Input Waveforms}

\subsection{Inspiral Waveforms}\label{sec:inspiral}
\subsubsection{Effective-one-body Description of a Compact Binary Coalescence}

In the effective-one-body approach to the two-body problem in GR, originally developed by Buonanno and Damour~\cite{Buonanno:1998gg}, the dynamics of two compact objects of masses $m_1$ and $m_2$ and spins $\vb*{\chi^{}_1}$ and $\vb*{\chi^{}_2}$ is mapped onto the dynamics of an effective particle of mass $\mu = m_1m_2/(m_1+m_2)$ and spin $\vb*{\chi_\ast}$ moving in the background of a deformed Kerr geometry with mass $M=m_1+m_2$ and spin $\vb*{\chi}^{}_{\rm Kerr}$. The spin mapping $\{\vb*{\chi^{}_1},\vb*{\chi^{}_2}\}\to \vb*{\chi_\ast}$ and the deformation of the Kerr background, parametrized by the symmetric mass ratio $\eta$, implies that the test-particle dynamics reproduces the PN-expanded dynamics of the original two-body system. Free parameters are introduced into the models that represent unknown, higher-order PN terms, or additional physical effects like corrections due to eccentricity. Such free parameters are calibrated to NR simulations. With the EOB system specified, its conservative dynamics can be described by an EOB Hamiltonian~\cite{Taracchini:2013rva,Taracchini:2012ig}, while the non-conservative dynamics is contained in a parametrized radiation-reaction term that is inserted in the equations of motion. This term sums over the outgoing GW modes and is calibrated to reproduce NR simulations. The combination of these two pieces describes the binary inspiral through to the merger, at which point a ringdown waveform is attached to the inspiral-merger waveform. The ringdown waveform is constructed as a linear superposition of the dominant quasinormal modes (QNMs) of the Kerr BH formed at merger~\cite{Buonanno:2009qa,Damour:2014sva}, with the amplitude and phase of each QNM mode determined by the process of stitching the ringdown part with the inspiral-merger parts.

For the purpose of creating hybrid waveforms in our work, we choose \texttt{SEOBNRv2}~\cite{Taracchini:2013rva}, an aligned-spin EOB approximant, as the point-particle baseline. More recent versions of the \texttt{SEOBNR} family of waveforms are available~\cite{Bohe:2016gbl} at present, but we create our model based on the former one for a direct comparison with \texttt{PhenomD}. We also mention here that for the construction of \texttt{PhenomD}, NR calibrations of \texttt{SEOBNRv2} were removed before creating hybrid waveforms, referred to as \texttt{SEOBv2}. In this work, however, we use the original calibrated version of the EOB model since in the hybrids, stitched at a frequency much lower than the innermost stable circular orbit (ISCO), we expect little to no NR information from the inspiral approximant to be present.

\subsubsection{Corrections for Tidal Heating of Black Holes}

The early inspiral part of a CBC can be described by the analytical formalism of post-Newtonian (PN) expansions~\cite{Blanchet:2013haa}, under the approximation that the source is slowly moving and weakly gravitating. In this formalism, the evolution of the orbital phase $\Psi(t)$ of a compact binary is computed as a perturbative expansion in a small parameter, typically taken to be the characteristic velocity $v = (\pi M f)^{1/3}$. This analytical procedure demands $v\ll 1$, which makes it useful in the early inspiral phase of a CBC.

An electrically neutral spinning black hole in GR -- the Kerr black hole (KBH) -- is stationary when it is isolated. On the other hand, when a KBH is a member of a binary, it feels its companion's tidal field, which acts as a non-axisymmetric perturbation~\cite{Hartle:1973zz}. This perturbation causes changes in the mass, spin, and horizon area of the KBH over time~\cite{Alvi:2001mx}. Since the KBH experiences the tidal field of its orbiting companion, it absorbs (emits) energy from (into) the orbit. The absorption part is present in non-spinning BHs as well. Additionally, for a KBH, a spin higher than the angular frequency of the tidal field decays due to tidal interactions, which in turn makes the KBH lose its rotational energy. The slowing down of a rotating BH due to the gravitational dissipation produced by exterior mass is analogous to the slowing down of a rotating planet by viscous dissipation due to tides raised by an exterior moon that increases its internal thermal content - a phenomenon known as tidal heating. Due to this similarity, the energy and angular momentum flux in BBHs is also termed tidal heating~\cite{Poisson:2004cw}.

During the early stages of the binary evolution, the change in the orbital frequency due to emission of GWs is much smaller than the orbital frequency itself. In this \textit{adiabatic inspiral} regime, loss of binding energy $E(v)$ of the two-body system with time equals the GW flux emitted to future null infinity ($\mathcal{F}^{\infty}(v)$) plus the energy flux absorbed by the two BH horizons ($\mathcal{F}^{\rm H}(v)$). So the energy balance condition becomes
\begin{equation}
    -\dv{E(v)}{t}=\mathcal{F}^{\infty}(v)+\mathcal{F}^{\rm H}(v).
\end{equation}

Evolution of the orbital phase $\phi$ and the characteristic velocity $v$, obtained from this equation, read
\begin{equation}
    \dv{\phi}{t} = \frac{v^3}{M}\,,\quad\quad\quad \dv{v}{t}=-\frac{\mathcal{F}(v)}{E'(v)},
\end{equation}
where $\mathcal{F}(v)=\mathcal{F}^{\infty}(v)+\mathcal{F}^{\rm H}(v)$. These equations yield a solution for the phase $\Psi(f)$ of the frequency-domain waveform $\Tilde{h}(f) =  \Tilde{A}(f) e^{-i\Psi(f)}$~\cite{Tichy:1999pv}:
\begin{equation}
\label{psi}
    \Psi(f)=2(t_c/M)v^3 - 2\phi_c - \pi/4 - \frac{2}{M}\int (v^3-\Bar{v}^3)\frac{E'(\Bar{v})}{\mathcal{F}(\Bar{v})}\dd\Bar{v},
\end{equation}
where $E'(v)=\dv*{E(v)}{v}$. 

The frequency-domain amplitude $\Tilde{A}(f)$ can be written as\footnote{Note that we define $E(v)$ as the total binding energy. Ref.~\cite{Ajith:2011ec} treats $E(v)$ as the specific binding energy (binding energy per unit mass), so the power of $M$ in our treatment is different from theirs in this particular equation.}~\cite{Ajith:2011ec}
\begin{equation}
\label{amp}
    \Tilde{A}(f)=\mathcal{C}\frac{2\eta M^{3/2}}{d_L}v\sqrt{\frac{\pi}{3}}\left[-\frac{E'(v)}{\mathcal{F}(v)}\right]^{1/2},
\end{equation}
where $\mathcal{C}$ is a constant that depends on the inclination of the orbital plane with the line of sight, and $d_L$ denotes the distance to the binary.

Splitting $\mathcal{F}$ into $\mathcal{F}^\infty$ and $\mathcal{F}^{\rm H}$ in Eq.~\eqref{psi} enables one to write the frequency-domain phase in the form
\begin{equation}
\label{eq:psi2}
    \Psi(f)=\Psi_{\rm PP}(f)+\Psi_{\rm TH}(f),
\end{equation}
with $\Psi_{\rm PP}(f)$ being the GW phase under point particle (PP) approximation that corresponds to $\mathcal{F}^\infty$, and $\Psi_{\rm TH}(f)$ being the phase correction due to TH. The absorbed flux can be expressed as the sum of the rates of mass increment~\cite{Alvi:2001mx} for the two BHs, given by Eq.~\eqref{dedt}. 


Using this expression, the phase contribution of tidal heating in gravitational waveforms of BBHs was described in Eq.~\eqref{eq:phase correction}. However, in this chapter, we are considering BH binaries only, so the effective horizon parameters are given by
\begin{subequations}
\label{eq:Hparams_chapter3}
\begin{align}
H^{(\rm BBH)}_{\rm eff5} \equiv &{} \sum_{i=1}^{2} \left(\frac{m^{}_i}{M}\right)^3 \left(\hat{L}\cdot\hat{S}^{}_i\right)\chi^{} _i \left(3 \chi^{}_i{}^2+1\right)\,,\\
H^{(\rm BBH)}_{\rm eff8} \equiv &{} ~4 \pi  H^{(\rm BBH)}_{\rm eff5}+\sum^2_{i=1} \left(\frac{m_i}{M}\right)^4 \left(3 \chi^{}_i{}^2+1\right)\nonumber \\
                &\quad\quad\quad\quad\quad\quad\quad \times \left(\sqrt{1-\chi^{}_i{}^2}+1\right)\,.
\end{align}
\end{subequations}
%

We find the PN expansion of the total amplitude by expanding $\left[-E'(v)/\mathcal{F}(v)\right]^{1/2}$ in Eq.~\eqref{amp} in powers of $v$ up to $v^7$, and separate the contribution due to TH:
\begin{equation}\label{eq:af}
    \Tilde{A}(f)=\Tilde{A}_{\rm PP}(f)+\Tilde{A}_{\rm TH}(f).
\end{equation}
$\Tilde{A}_{\rm PP}(f)$ is mentioned in Eq.~(5.7) of Ref.~\cite{Ajith:2011ec}, and we get the expression for $\Tilde{A}_{\rm TH}(f)$ as
\begin{equation}\label{eq:amp correction}
\begin{aligned}
    \Tilde{A}_{\rm TH}(f)= ~ \mathcal{C}&\frac{M^{5/6}}{d_L\pi^{2/3}}\sqrt{\frac{5\eta}{24}}f^{-7/6}\left[\frac{1}{8}H^{(\rm BBH)}_{\rm eff5}v^5 \right.\\ 
    &\left.~+\left(\frac{1079}{1792}+\frac{103}{192}\eta\right)H^{(\rm BBH)}_{\rm eff5}v^7 \right. \\ &\left. + \frac{1}{32}\left\{\Tilde{A}_{\rm SO}H^{(\rm BBH)}_{\rm eff5} \right.\right.\\ &\left.\left. - 8\left(H^{(\rm BBH)}_{\rm eff8}-\pi H^{(\rm BBH)}_{\rm eff5}\right)\right\} v^8\right],
\end{aligned}
\end{equation}
where $\Tilde{A}_{\rm SO}$ is the spin-orbit term 
\begin{equation}
\begin{aligned}  
    \Tilde{A}_{\rm SO} = &~ \frac{179}{6(1+q)^2}\left\{q^2(\hat{L}\cdot\hat{S}^{}_1)\chi_1 + (\hat{L}\cdot\hat{S}^{}_2)\chi_2\right\} \\ & 
                    +\frac{35q}{2(1+q)^2}\left\{(\hat{L}\cdot\hat{S}^{}_1)\chi_1 + (\hat{L}\cdot\hat{S}^{}_2)\chi_2\right\}.
\end{aligned}
\end{equation}


\subsection{Numerical Relativity Waveforms for the Merger and Ringdown}\label{sec:NR}

As a BBH progresses towards the merger phase, the BHs come so close to each other that the system becomes too compact for the weak-gravity condition to hold. In addition, their orbital velocities culminate to values comparable to $c$, breaching the slow-motion approximation. The analytical PN results, with their highest order of expansion available currently, become increasingly inaccurate to describe the system evolving under such extreme conditions. 
Starting from this \textit{late inspiral} phase to the merger and ringdown phase where the two BHs merge and the final remnant BH settles to equilibrium, one needs to solve the full Einstein equations numerically for extracting GWs to conform to the accuracy standards of the waveforms required for searches of GW signals or their parameter estimation (PE). This is the field of NR, which has seen exciting breakthroughs ~\cite{Pretorius:2005gq,Campanelli:2005dd,Baker:2005vv} in the early 2000s, and great strides towards simulating merging black holes with different masses and spins~\cite{Pfeiffer:2012pc,Centrella:2010mx,LeTiec:2014oez,Hannam:2013pra}.

We use NR data from the publicly available SXS catalog~\cite{Boyle:2019kee}, computed using the Spectral Einstein Code (SpEC)~\cite{Scheel:2008rj,Ossokine:2013zga}. SpEC uses the excision techniques to remove the BH singularity for extracting gravitational waves from binary black hole systems. Unlike the PN framework where the fluxes of energy and angular momentum absorbed by the BHs have to be considered explicitly in the energy balance equation, numerical simulations of full Einstein equations capture this effect inherently. Scheel \textit{et al.}~\cite{Scheel:2014ina} have demonstrated the slow change in the BH masses and spins through the binaries' temporal evolution in NR simulations with SpEC. They track the apparent horizons as a function of time, and at frequent time intervals, they measure both the surface area and the spin of the horizons. The spin computation has been carried out using the approximate Killing vector formalism~\cite{Cook:2007wr}. The mass of the black hole is then computed using Christodoulou's formula~\cite{Christodoulou:1971pcn},
\begin{equation}
    M^2=M_{\rm irr}^2 + \frac{S^2}{4M_{\rm irr}^2},
\end{equation}
where $M_{\rm irr}$ is the irreducible mass of the Kerr BH,
\begin{equation}
    M_{\rm irr}^2=\frac{1}{16\pi}\int_\mathcal{H}\dd A.
\end{equation}
Here $S$ is the (dimensionful) spin of the KBH and the integration covers the total horizon area $A$.

\begin{table*}[t]
    \centering
    \begin{tabular}{p{0.05\linewidth}p{0.22\linewidth}p{0.08\linewidth}p{0.08\linewidth}p{0.08\linewidth}p{0.08\linewidth}p{0.08\linewidth}p{0.08\linewidth}}
        \hline
        \# & Simulation label & $q$ & $\chi^{}_1$ & $\chi^{}_2$ & $\chi^{}_{\rm PN}$ & $N_{\rm orb}$ & $F_{\rm stitch}$\\
        \vspace{0.5mm}\\
        \hline
        \hline
        \vspace{0.5mm}\\
        1 & SXS:BBH:0156 & 1 & $-$0.95 & $-$0.95 & $-$0.79 & 13 & 0.0057 \\
        2 & SXS:BBH:0151 & 1 & $-$0.6 & $-$0.6 & $-$0.5 & 15 & 0.0054 \\
        3 & SXS:BBH:0001 & 1 & 0 & 0 & 0 & 28 & 0.0041 \\
        4 & SXS:BBH:0152 & 1 & 0.6 & 0.6 & 0.5 & 23 & 0.0052 \\
        5 & SXS:BBH:0153 & 1 & 0.85 & 0.85 & 0.7 & 25 & 0.0052 \\
        6 & SXS:BBH:0234 & 2 & $-$0.85 & $-$0.85 & $-$0.72 & 28 & 0.0037 \\
        7 & SXS:BBH:0238 & 2 & $-$0.5 & $-$0.5 & $-$0.43 & 32 & 0.0037 \\
        8 & SXS:BBH:0169 & 2 & 0 & 0 & 0 & 16 & 0.0062 \\
        9 & SXS:BBH:0253 & 2 & 0.5 & 0.5 & 0.43 & 29 & 0.0046 \\
        10 & SXS:BBH:2131 & 2 & 0.85 & 0.85 & 0.72 & 25 & 0.0056 \\
        11 & SXS:BBH:1936 & 4 & $-$0.8 & $-$0.8 & $-$0.71 & 17 & 0.0057 \\
        12 & SXS:BBH:1418 & 4 & $-$0.4 & $-$0.5 & $-$0.37 & 67 & 0.005 \\
        13 & SXS:BBH:0167 & 4 & 0 & 0 & 0 & 16 & 0.0071 \\
        14 & SXS:BBH:1417 & 4 & 0.4 & 0.5 & 0.37 & 80 & 0.0057 \\
        15 & SXS:BBH:1907 & 4 & 0 & 0.8 & 0.12 & 21 & 0.0062 \\
        16 & SXS:BBH:1423 & 8 & $-$0.6 & $-$0.75 & $-$0.57 & 18 & 0.0066 \\
        17 & SXS:BBH:0064 & 8 & $-$0.5 & 0 & $-$0.43 & 19 & 0.0066 \\
        18 & SXS:BBH:0063 & 8 & 0 & 0 & 0 & 26 & 0.0065 \\
        19 & SXS:BBH:0065 & 8 & 0.5 & 0 & 0.43 & 34 & 0.0063 \\
        20 & SXS:BBH:1426 & 8 & 0.48 & 0.75 & 0.47 & 26 & 0.0078 \\
        \vspace{0.5mm}\\
        \hline
    
    \end{tabular}
    \caption{Hybrid waveforms used to calibrate the model. The first column lists the simulation IDs of the corresponding NR data in the SXS catalog. The last column reports the dimensionless frequency at the midpoint of the stitching region. For $M=20M_\odot$, $F=0.005$ corresponds to $f\approx 51$ Hz.}
    \label{tab:hyb}
\end{table*}


\begin{figure}
     \centering
     \begin{subfigure}[b]{\textwidth}
         \centering
         \includegraphics[width=\textwidth]{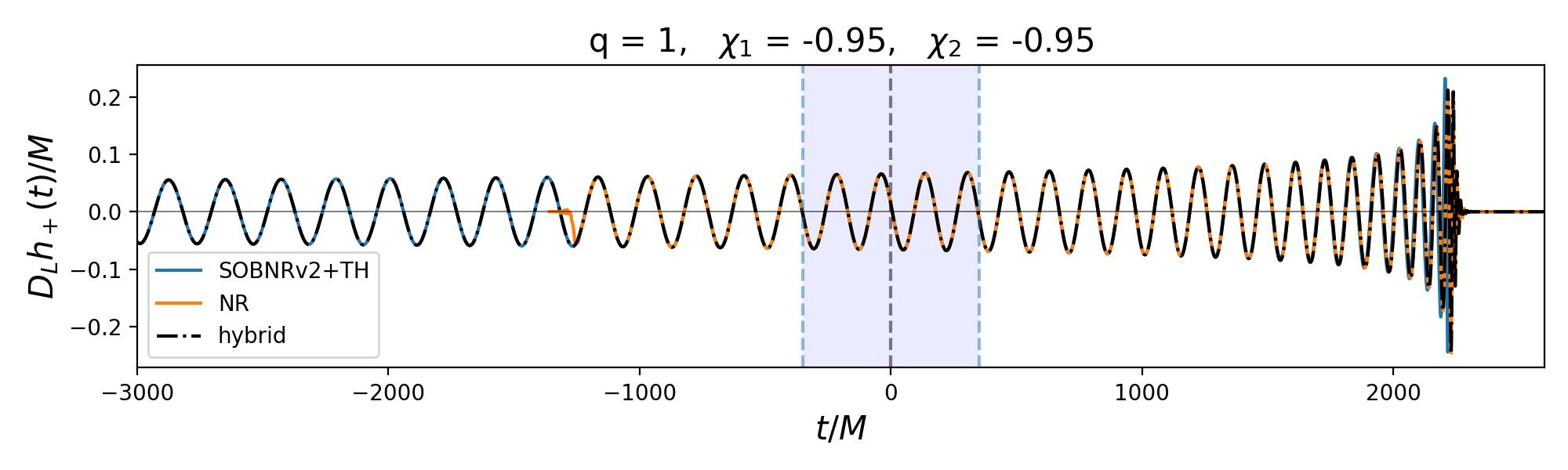}
         \caption{}
     \end{subfigure}
     \hfill
     \begin{subfigure}[b]{\textwidth}
         \centering
         \includegraphics[width=\textwidth]{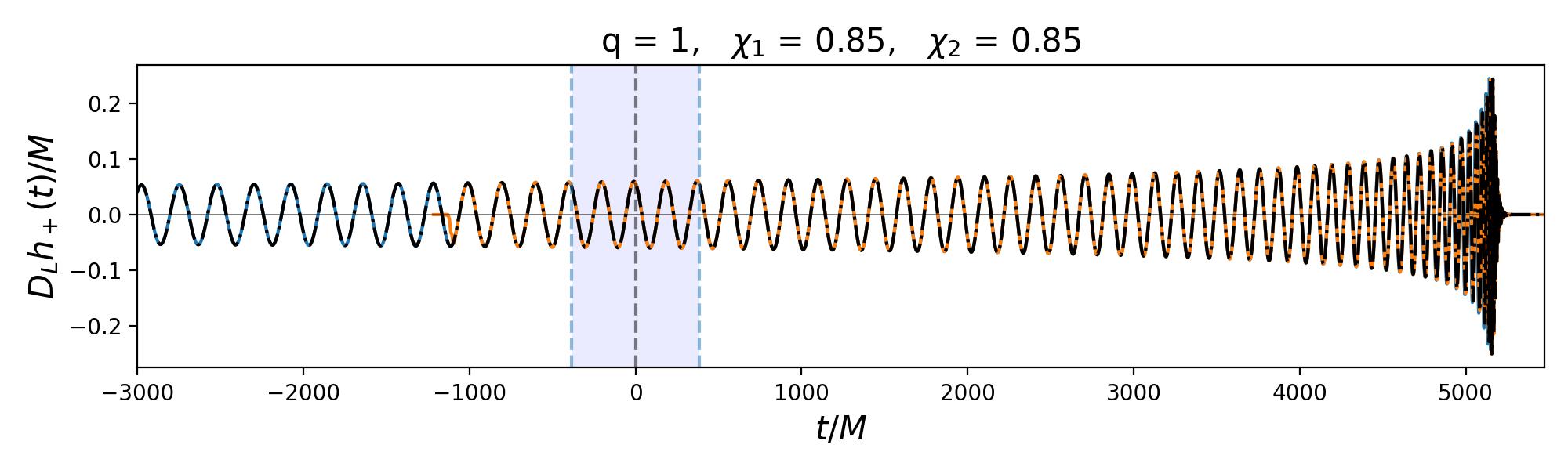}
         \caption{}
     \end{subfigure}
     \hfill
     \begin{subfigure}[b]{\textwidth}
         \centering
         \includegraphics[width=\textwidth]{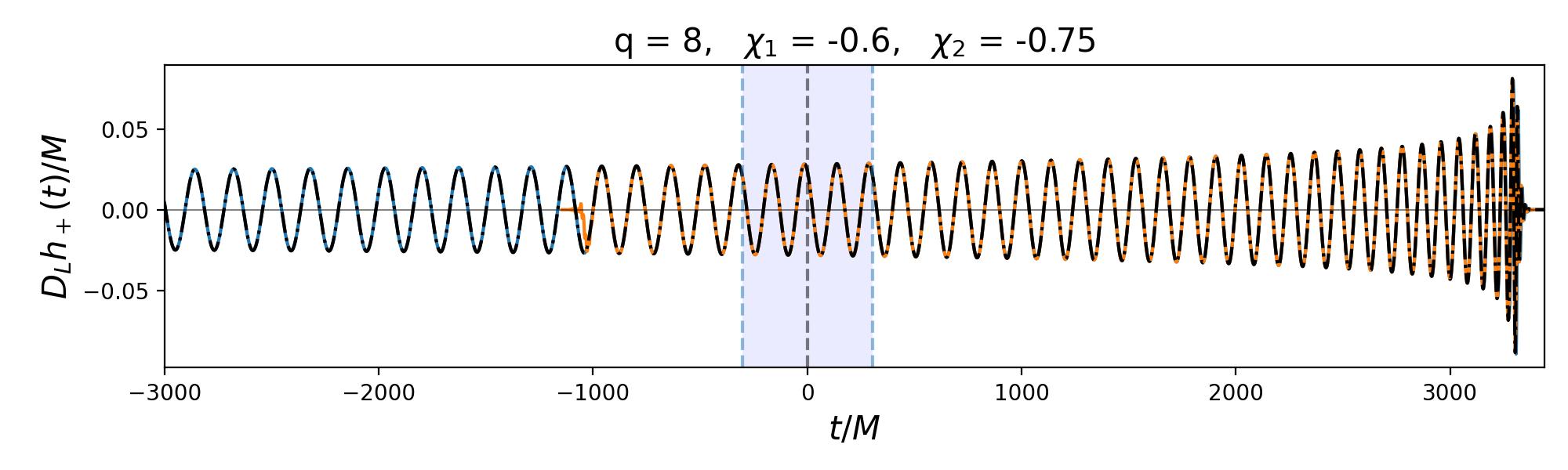}
         \caption{}
     \end{subfigure}
     \hfill
     \begin{subfigure}[b]{\textwidth}
         \centering
         \includegraphics[width=\textwidth]{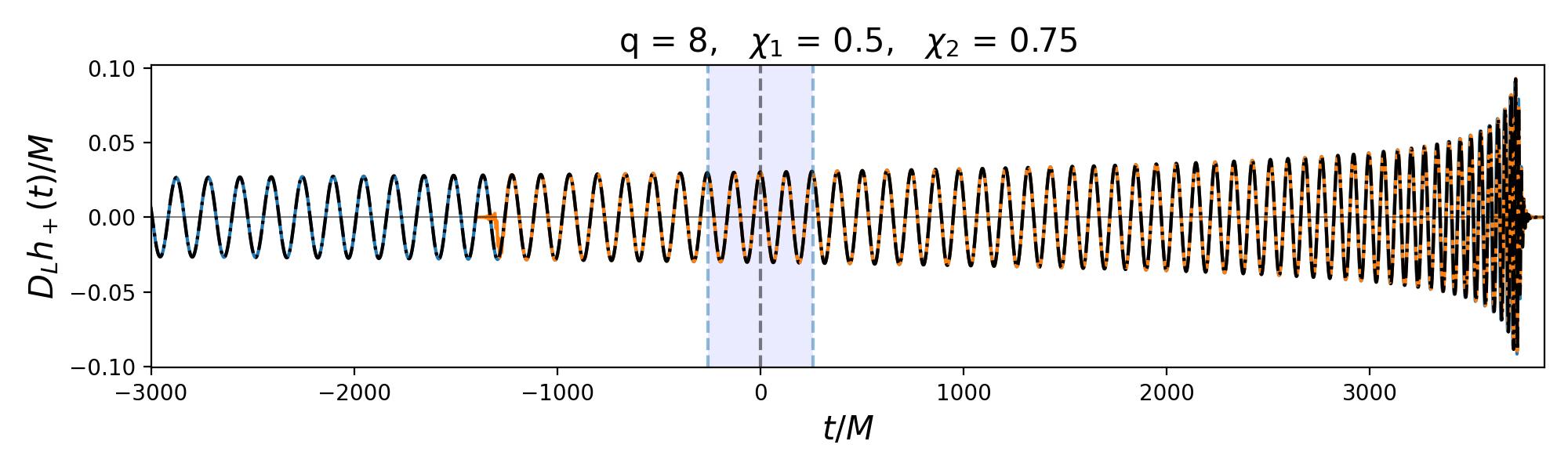}
         \caption{}
     \end{subfigure}
        \caption{Hybrid waveforms for four different configurations in the parameter space of $q,\chi^{}_1,\chi^{}_2.$ Inspiral waveforms are generated by adding the phase and amplitude corrections due to TH to the \texttt{SEOBNRv2} model, shown in blue. NR waveforms are shown in orange, and the hybrid waveforms are shown in black dashed-dotted lines. The $x$ and $y$ axes denote time (in units of total mass) and the real part of time-domain strain, $h_+(t)$ (in units of $M/D_L$), respectively. The blue shaded areas denote the stitching regions.}
        \label{fig:hybrids}
\end{figure}


During the evolution, each SXS NR waveform is extracted at a series of times on a set of concentric coordinate spheres surrounding the binary, decomposed in modes of spin-weighted spherical harmonic functions. Then the waveforms are extrapolated to future null infinity $\mathscr{I}^+$. The dataset of each SXS waveform contains several gravitational waveform modes $(l,m)$, and the orders of extrapolation $N=2,3,4$. A higher order of extrapolation is preferred when accurate waveforms are needed in the inspiral, and lower order extrapolation is preferred for more accuracy at the merger-ringdown phase~\cite{Boyle:2019kee}. Since we use these waveforms for constructing hybrids, we need more accuracy for the merger-ringdown part due to the fact that the inspiral part is described by PN waveforms in the hybrids. We use 20 aligned-spin non-eccentric BBH waveforms from SXS with $1\leqslant q \leqslant 8$ and $-0.95\leqslant \chi^{}_1,\chi^{}_2\leqslant 0.85$ for calibrating our model. We choose $(2,2)$ modes, and the extrapolation order $N=2$ for our purpose.

\section{Hybrid Construction}\label{sec:hyb}

For constructing a hybrid, the NR waveform has to be `stitched' with the analytical inspiral waveform in a frequency region where both the waveforms have sufficient accuracy, 
which also demands that the morphology of these waveforms do not deviate significantly from one another. For PN waveforms, one conventionally chooses ISCO of the binary to be the endmost point of validity. The corresponding GW frequency $f_{\rm ISCO}$ is considered to be the highest frequency for PN expansions. In our work, we ensured that the frequency in the stitching region is below this threshold for all the hybrids, which also ensures that the NR calibrations in \texttt{SEOBNRv2} are minimally present before the stitching starts.

While stitching at a frequency as low as possible (considering ISCO) should enable one to incorporate the maximum number of NR cycles into the hybrid, there are two more factors that affect the choice of the stitching region:

\begin{itemize}
    \item The initial NR data do not perfectly describe two black holes in quasi-equilibrium. At the start of each simulation, the geometry relaxes to equilibrium on the dynamical time-scale of the individual BHs, changing the mass and spin of each BH by a fractional amount of order $10^{-5}$, and emitting a spurious pulse of gravitational radiation (often referred to as `junk radiation’). 
    \item A robust hybrid should depend weakly on the small changes to the stitching region. A monochromatic signal would be completely degenerate under a shift in the coalescence time $t_c$ and the coalescence phase $\phi_c$; this degeneracy is broken by the increase in frequency with time. The start and end of the stitching interval should reflect enough change in the frequency to break this degeneracy. 
\end{itemize}

The first issue is addressed in our work by choosing the stitching region beyond the specified \textit{relaxation time} for each NR simulation in the SXS catalog, 
the time interval (in units of $M$) by which the junk radiation dies out. To address the second requirement, MacDonald \textit{et al.}~\cite{MacDonald:2011ne} have recommended the stitching interval to satisfy $\delta\omega/\omega_m\geqslant 0.1$, where $\omega_m$ is the GW frequency at the midpoint of the stitching interval, and $\delta\omega$ is the change in frequency over the interval. This choice ensures that the residual oscillations in $t_c$ with changing $\omega_m$ are below $1M$. In our work, we perform the stitching over 4 GW cycles, and we place the stitching region in a way to satisfy this condition. In the literature, 
the
construction of hybrids has been performed both in time ~\cite{Ajith:2007kx,Husa:2015iqa,Pratten:2020fqn} and  frequency domains~\cite{Santamaria:2010yb,Damour:2010zb}. In our work, we choose the former,
primarily because the NR waveforms are available in the time domain, and typically they have too few GW cycles for a reliable Fourier transform.

 In time domain, any two non-precessing binary waveforms $h^A(\Theta,t)$ and $h^B(\Theta,t)$, with the same set of intrinsic parameters $\Theta$, only differ by a relative time shift ($\Delta t$) and an overall constant phase difference ($\phi_0$):
\begin{equation*}
    h^A(\Theta,t) = e^{i\phi_0}h^B(\Theta,t+\Delta t).
\end{equation*}
In the region where we expect the PN and NR waveforms to be morphologically similar, we can write
\begin{equation}
 \label{pnnr}
    h^{\rm PN}(\Theta,t) = e^{i\phi_0}h^{\rm NR}(\Theta,t+\Delta t).
\end{equation}
However, in reality, $h^{\rm PN}(\Theta,t)$ and $h^{\rm NR}(\Theta,t)$ differ from each other morphologically due to waveform systematics, so Eq.~\eqref{pnnr} does not hold in general~\cite{MacDonald:2011ne}. To generate a hybrid waveform, then, one needs to find a suitable region where PN and NR waveforms are approximately the same and minimize the square of the difference between the left-hand and right-hand sides in Eq.~\eqref{pnnr}, by varying $\Delta t$ and $\phi_0$. Since the GW frequency of a CBC increases with time, one can alternatively write the angular GW frequency as a function of time ($\omega(t)$) and minimize the quantity
\begin{equation}
\label{delta}
    \delta=\int_{t_1}^{t_2}\left(\omega^{(\rm NR)}(t)-\omega^{(\rm PN)}(t+\Delta t)\right)^2 \dd t
\end{equation}
with respect to $\Delta t$. Here $t_1$ and $t_2$ denote the start and end of the stitching region.

We list the steps taken to construct the hybrids as follows:

\begin{enumerate}
    \item  \texttt{SEOBNRv2} is a time-domain waveform, and the phase and amplitude corrections for TH are calculated in the frequency domain. So, we have to first convert the time-domain data to the Fourier domain to apply the corrections. Since the time-domain data is finite, we use a tapered window function to minimize Gibbs artifacts, called the Planck-taper window~\cite{MacDonald:2011ne}:
        
        \begin{equation}
            \begin{aligned}
                w(x) =
            \left\{
        	    \begin{array}{ll}
        		    0 , &  x \leqslant x_1 \\
        		    \left[e^{y(x)}+1\right]^{-1}, &  x_1 < x < x_2 \\
        		    1 , &  x_2 \leqslant x \leqslant x_3 \\
        		    \left[e^{z(x)}+1\right]^{-1}, &  x_3 < x < x_4 \\
        		    0, &  x_4 \leqslant x,
        	    \end{array}
            \right.
            \end{aligned}
        \end{equation}
        where $y(x) = (x_2-x_1)/(x-x_1)+(x_2-x_1)/(x-x_2)$, and $z(x)=(x_3-x_4)/(x-x_3)+(x_3-x_4)/(x-x_4)$. 

        The frequency-domain waveform, after correcting for TH, is then converted back to the time domain for hybridization.
    
    \item NR data from the SXS catalog sample the GW strain $h^{(\rm NR)}(t)$ in non-uniform timesteps, with a higher sampling rate in regions of higher GW frequency. We resample the data with the coarsest sampling rate by using cubic interpolation. 
    
    \item Having obtained $h^{(\rm PN)}(t)$ and $h^{(\rm NR)}(t)$, we minimize the quantity $\delta$ defined in Eq.~\eqref{delta} with respect to $\Delta t$. Phase alignments are done at the temporal midpoint of the stitching region, $t=(t_1+t_2)/2$.
    
    \item The hybrid waveform is then constructed as

    \begin{equation}
    \begin{aligned}
          h^{(\rm Hyb)}(t) =&~ \mathcal{F}(t)h^{(\rm PN)}(t)\\ & +\qty[1-\mathcal{F}(t)]e^{i\phi'_0}h^{(\rm NR)}(t+\Delta t'),
    \end{aligned}
    \end{equation}
    where $\phi'_0$ and $\Delta t'$ are the phase offset at the midpoint of the stitching region and the value of $\Delta t$ that minimizes $\delta$, respectively.  $\mathcal{F}(t)$ is a blending function defined by
    
    \begin{equation}
        \begin{aligned}
            \mathcal{F}(t) =
        \left\{
    	    \begin{array}{ll}
    		    1 , &  t < t_1 \\
    		    \frac{1}{2}\qty(1+\cos\frac{\pi(t-t_1)}{t_2-t_1}), & t_1 \leqslant t \leqslant t_2 \\
    		    0 , & t_2 < t\,.
    	    \end{array}
        \right.
        \end{aligned}
    \end{equation}
\end{enumerate}

In the time domain hybrids, parts of the waveforms at $t<t_1$ are purely post-Newtonain, and for $t>t_2$ the waveforms contain purely NR data. In between, the blending function $\mathcal{F}(t)$ smoothly stitches the two waveforms together after proper alignment in phase and time. 

Figure~\ref{fig:hybrids} shows four of the total twenty hybrids made for calibrating the waveform model, at the extreme points of the calibration parameter space. The blue shaded regions correspond to the stitching intervals, $t_1\leqslant t \leqslant t_2$. In Table~\ref{tab:hyb}, we list the NR data used to create the set of hybrids for calibration. We report the simulation labels of aligned-spin NR data from SXS, dimensionless spin magnitudes for the two component BHs, the effective spin parameter used for modeling the inspiral (discussed in Sec.~\ref{sec:correspondence}), the number of orbits present in the NR data, and the value of dimensionless frequency $Mf$ at the midpoint of the stitching regions.

\section{Modeling the Hybrids in the Frequency Domain}\label{sec:model}

We perform phenomenological parametrization of the hybrid waveforms in the frequency domain. Template waveforms in the frequency domain are of particular preference since the parametrization is easier, and also a search using frequency-domain templates is computationally inexpensive compared to one using time-domain templates. Our parametrization follows the construction of \texttt{PhenomD}~\cite{Husa:2015iqa,Khan:2015jqa} for modeling the phase and amplitude of the frequency-domain hybrids. 

\texttt{PhenomD} used uncalibrated \texttt{SEOBNRv2} as the inspiral approximant and 19 NR waveforms, extracted by SpEC and BAM~\cite{Bruegmann:2006ulg} codes, for hybrid construction and calibration. The model was constructed in the frequency domain by dividing the entire waveform into three parts -- inspiral, intermediate, and merger-ringdown. 
This modular nature allows one to use a different inspiral model while keeping the merger and ringdown intact. In that model, inspiral is defined to be the region below  $F=0.018$. Merger and ringdown phases are separated by identifying the ringdown frequencies of different hybrids that are used to calibrate the model. In \texttt{PhenomD}, the end of inspiral ($F=0.018$) was chosen in such a way that beyond that frequency, each of the hybrids contained purely NR data. The merger-ringdown model of \texttt{PhenomD}, then, is a model of purely NR data, which we do not need to recalibrate in this work. In fact, as discussed in Ref.~\cite{Husa:2015iqa}, the modular nature of \texttt{PhenomD} allows independent modeling of the inspiral, intermediate, and merger-ringdown parts separately, without even needing any hybrid waveform. The hybrid dataset, however, serves as a benchmark to compare the final model with and provide long time-domain waveforms for better frequency-domain modeling. Additionally, the hybrids are created with a smooth transition of phase and amplitude by the stitching process, producing accurate values of the pseudo-PN parameters capturing higher-order terms in frequency.

\subsection{Inspiral Phase Model}\label{sec:inspiral phase model}

The inspiral approximant used in the hybrid data is \texttt{SEOBNRv2} with corrections in phase and amplitude due to tidal heating, as discussed earlier. Since the final model is a phenomenological one, it requires ready-to-use analytical forms for the phase and amplitude with extra coefficients calibrated to the hybrid data.  
We write the frequency-domain phase of a waveform in the inspiral as 
\begin{equation}\label{eq:phiins}
\begin{aligned}
     \Psi_{\rm \tiny{INS}}(F;\Theta) =&~ \Psi_{\rm \tiny{TF2}}(F;\Theta) + \Psi_{\rm TH}(F;\Theta) \\&+ \frac{1}{\eta}\left[\sigma_0 + \sigma_1 F + \frac{3}{4}\sigma_2 F^{4/3}\right. \\& \left. \quad\quad + \frac{3}{5}\sigma_3 F^{5/3} + \frac{1}{2}\sigma_4 F^2 \right],
\end{aligned}
\end{equation}
where $\Psi_{\rm \tiny{TF2}}(F;\Theta)$ and $\Psi_{\rm TH}(F;\Theta)$ are 3.5PN phase of \texttt{TaylorF2} (described in Appendix~\ref{AppendixA}) and the TH correction (Eq.~\eqref{eq:phase correction} with Eq.~\eqref{eq:Hparams_chapter3}), respectively. $\sigma_i\>\> (i=0-4)$ are 5 phenomenological pseudo-PN parameters. The analytical ansatz in Eq.~\eqref{eq:phiins} is the same as in \texttt{PhenomD} inspiral model. 
Since we want to leverage the modular nature of \texttt{PhenomD} to concatenate its merger-ringdown model with our own inspiral model, the end of the inspiral phase in the current work needs to remain unchanged, which is $F=0.018$. We also note here that one can, in principle, directly add the correction $\Psi_{\rm TH}(F;\Theta)$ to the \texttt{PhenomD} inspiral phase (as is done for the amplitude, discussed in Sec.~\ref{sec:amplitude}), without any recalibration. However, since we intend to avoid adding the correction to the NR information present in the last few cycles of inspiral, recalibrating the phase allows a smooth transition from PN to NR. To find the parameter values corresponding to each hybrid waveform, we fit the first derivative of the phase with respect to $F$:
\begin{equation}\label{eq:phiprimeins}
\begin{aligned}
     \Psi'_{\rm\tiny INS}(F;\Theta) =&~ \Psi'_{\rm \tiny{TF2}}(F;\Theta) + \Psi'_{\rm TH}(F;\Theta) \\&+ \frac{1}{\eta}\left[\sigma_1 + \sigma_2 F^{1/3}\right. \\& \left. \quad\quad + \sigma_3 F^{2/3} + \sigma_4 F \right],
\end{aligned}
\end{equation}
where $\Psi' = \partial\Psi/\partial F$. $\sigma_0$ is determined by imposing $C^{(1)}$ continuity in phase at the boundary between inspiral and merger. The fits are performed over the frequency range $F\in [0.0035,0.019]$, ending at a slightly higher frequency, to reduce boundary effects at the interface and find robust fits for $\sigma_i$. For all the 20 hybrids used for calibration, the stitching regions are placed within this frequency range, which warrants that there is purely NR data beyond $F=0.018$.

\subsection{Inspiral and Intermediate Amplitude Model}\label{sec:amplitude}

To model the amplitude within the frequency range of the inspiral, we build the inspiral amplitude model by adding the correction due to TH, given by Eq.~\eqref{eq:amp correction}, to the \texttt{PhenomD} amplitude: 
\begin{equation}
\label{eq:ampmodel}
    \Tilde{A}_{\rm INS}(F) = \Tilde{A}_{\rm D}(F)+\Tilde{A}_{\rm TH}(F)\,.
\end{equation}
We do not recalibrate the amplitude pseudo-PN parameters of \texttt{PhenomD}, since the inspiral amplitude model ends (at $F=0.014$) before the phase model does (at $F=0.018$), and the amount of NR information is less than that of phase. Moreover, we find that the effect of the amplitude correction is significantly less than the phase correction, diminishing the necessity of a recalibration. 

At the interface of two frequency intervals modeled separately, $C^{(1)}$ continuity on the amplitude cannot be imposed in a straightforward manner. For the frequency-domain phase, one has the freedom to tune the coalescence time $t_c$ and the coalescence phase $\phi_c$ to impose continuity of the phase and its derivative across an interface. For amplitude, however, there is no such freedom. If the amplitude is modeled separately in different frequency regimes, a different strategy has to be followed. To ensure that the amplitude and its derivative are continuous throughout the three phases, one needs a frequency interval to fit a polynomial that satisfies these conditions at the beginning and the end. This `intermediate' interval is defined to be $F\in [0.014,F_{\rm peak}]$, where $F_{\rm peak}\equiv Mf_{\rm peak}$ is the frequency corresponding to the peak amplitude. For $F<0.014$, the amplitude is given by Eq.~\eqref{eq:ampmodel}. In the intermediate region, the amplitude is approximated as a polynomial in $f$:

\begin{equation}
\begin{aligned}
\label{eq:ampint}
     \Tilde{A}_{\rm int} = A_0 & \left[\delta_0 + \delta_1 f + \delta_2 f^2 \right. \\ 
                     & \left. +\delta_3 f^3 + \delta_4 f^4\right]. 
\end{aligned}
\end{equation}
Here $A_0$ includes the leading order $f^{-7/6}$ behaviour. Evaluation of $\delta_i$ follows the steps of \texttt{PhenomD}, described in Sec. V(C) of Ref~\cite{Khan:2015jqa}. We briefly summarize it here:
\begin{itemize}
    \item Equation~\eqref{eq:ampint} has five parameters, requiring five independent equations for unique solutions. Two of them come from the $C^{(1)}$ continuity of $\Tilde{A}_{\rm int}$ with $\Tilde{A}_{\rm INS}(f)$ at the beginning ($F=0.014$), and two from $C^{(1)}$ continuity with \texttt{PhenomD} amplitude at $F_{\rm peak}$. The fifth equation appears from an additional condition that the polynomial coincides with \texttt{PhenomD} amplitude at the mid-frequency (chosen as a collocation point), $F=(0.014+F_{\rm peak})/2$.
    \item Solving the aforementioned set of equations, one finds the form of $\Tilde{A}_{\rm int}$ which smoothly connects the intermediate region with inspiral and merger.
    \item It is worth mentioning that while choosing more collocation points would improve the model accuracy, it would also encumber the model with a larger set of equations to solve. Keeping in mind that amplitude errors are less consequential than phase errors, we refrain from considering more collocation points.
\end{itemize}

\begin{figure}
     \centering
     \begin{subfigure}[b]{0.49\textwidth}
         \centering
         \includegraphics[width=\textwidth]{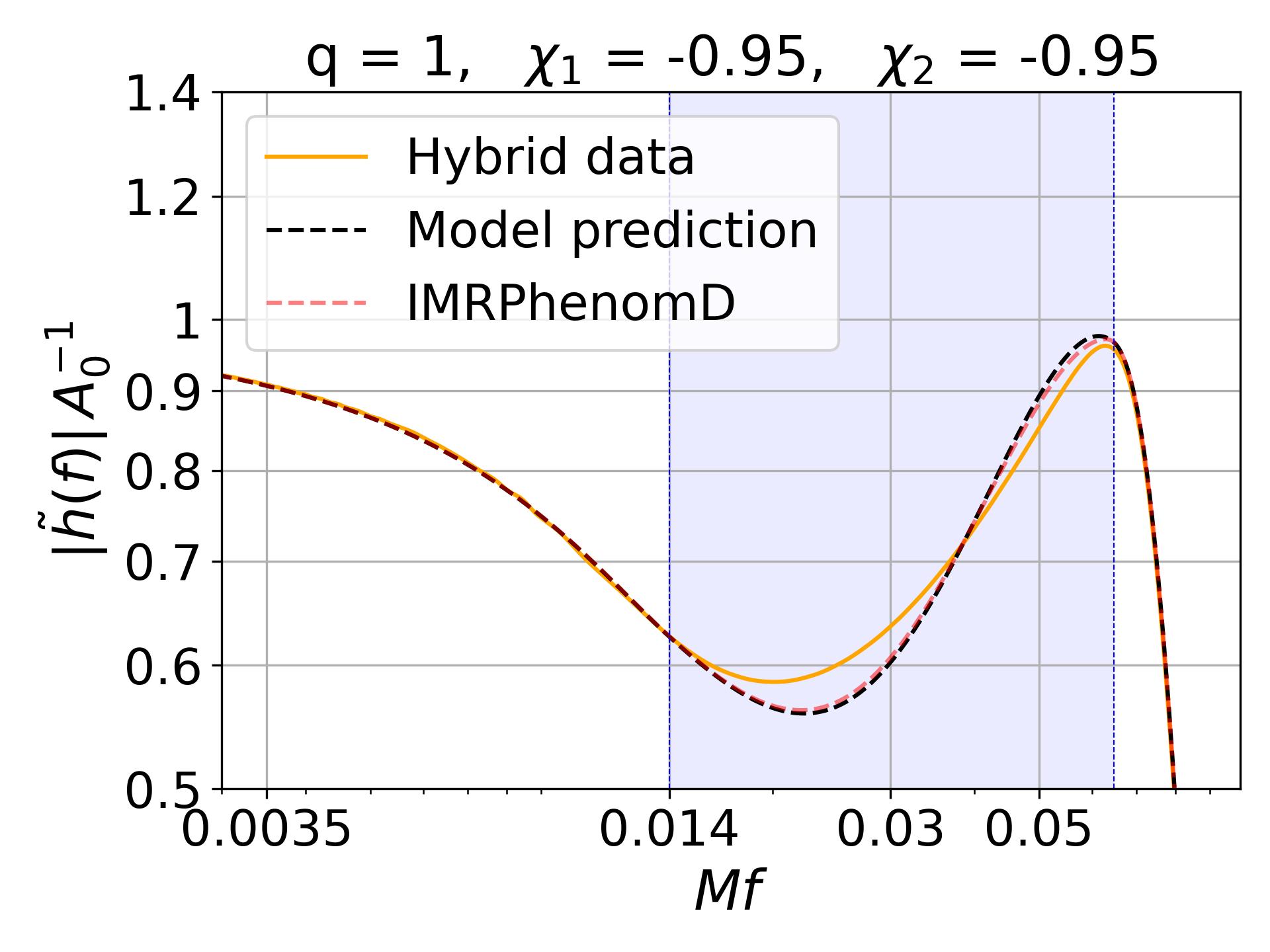}
     \end{subfigure}
     \hfill
     \begin{subfigure}[b]{0.49\textwidth}
         \centering
         \includegraphics[width=\textwidth]{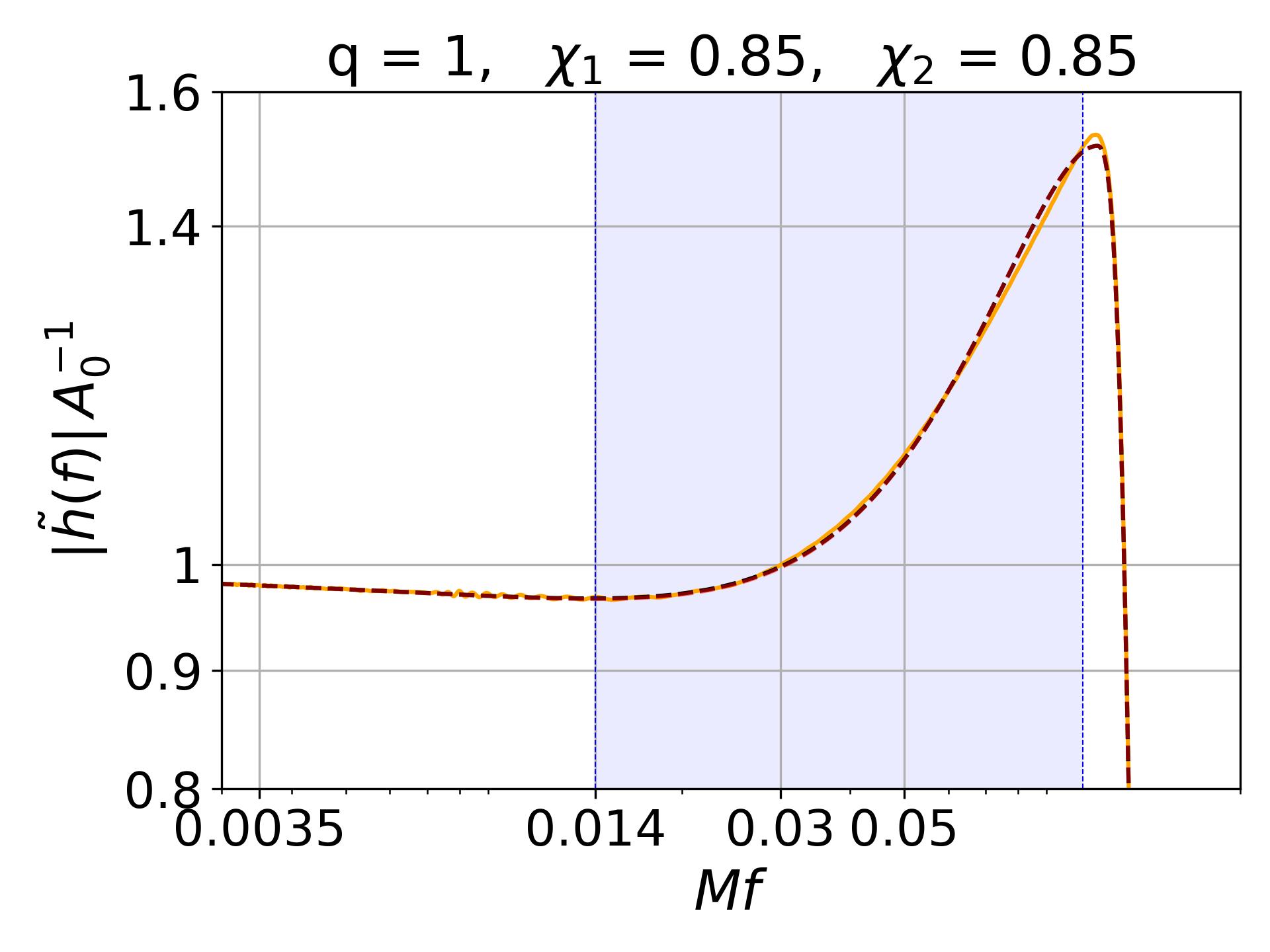}
     \end{subfigure}
     \hfill
     \begin{subfigure}[b]{0.49\textwidth}
         \centering
         \includegraphics[width=\textwidth]{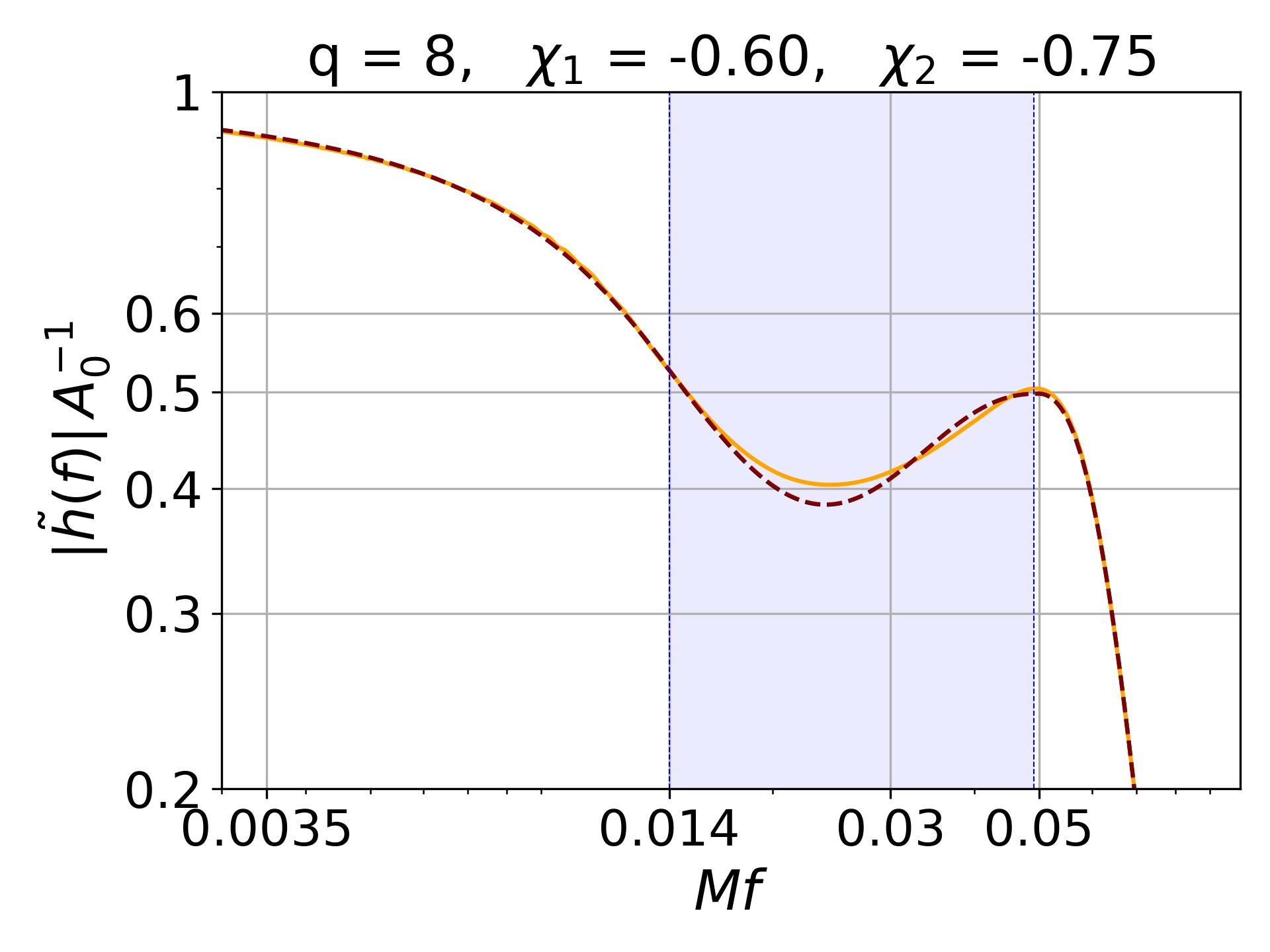}
     \end{subfigure}
     \hfill
     \begin{subfigure}[b]{0.49\textwidth}
         \centering
         \includegraphics[width=\textwidth]{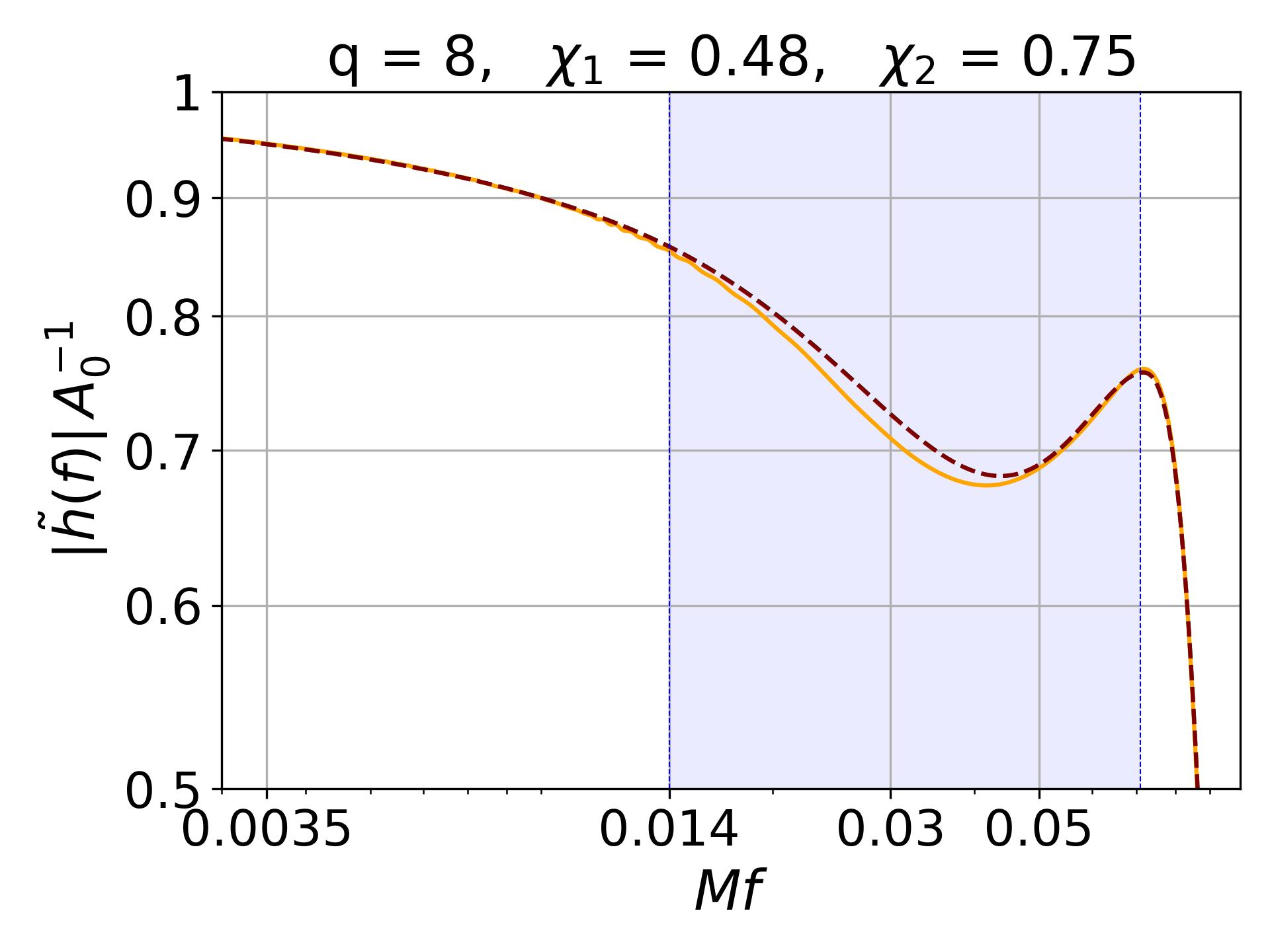}
     \end{subfigure}
        \caption{Rescaled amplitude of the full IMR waveforms as functions of $Mf$. The blue shaded regions denote the intermediate frequency range, $0.014\leqslant Mf \leqslant F_{\rm peak}$.}
        \label{fig:imramp}
\end{figure}

In Appendix~\ref{AppendixB}, we describe the evaluation of the peak frequency $f_{\rm peak}$ and the ringdown frequency $f_{\rm RD}$ as functions of the BH masses and spins.

\subsection{Correspondence between the Phenomenological and Physical Parameters}\label{sec:correspondence}

\begin{figure}
\centering
  \includegraphics[width=\textwidth]{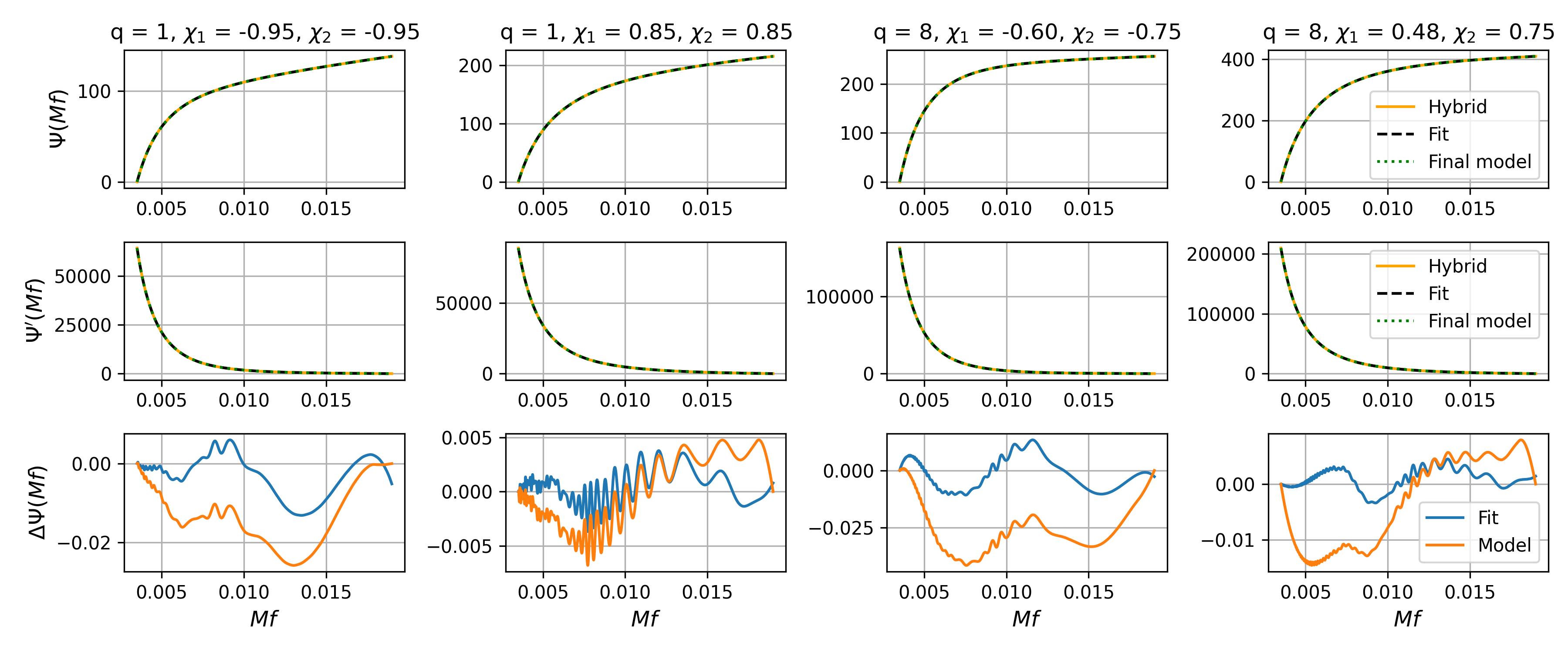}
  \caption{Phase, phase derivative, and phase difference between the hybrids (same as in Fig.~\ref{fig:hybrids}), individual fits obtained from Eq.~\eqref{eq:phiprimeins}, and the final model prediction. \textit{Top panels}: Inspiral phase $\Psi (Mf)$. \textit{Middle panels}: $\Psi^\prime (Mf)=\partial\Psi/\partial (Mf)$ plotted against $Mf$.  \textit{Bottom panels}: $\Delta \Psi_{\rm Fit} = (\Psi_{\rm Hybrid} - \Psi_{\rm Fit})$, and $\Delta \Psi_{\rm Model} = (\Psi_{\rm Hybrid} - \Psi_{\rm Model})$. All the plots are shown within the inspiral frequency range, $0.0035\leqslant Mf \leqslant 0.018$.}
\label{fig:phase}
 \end{figure}


Our model has 4 phenomenological parameters corresponding to the inspiral phase. These parameters play an intermediate role in generating gravitational waveforms of aligned-spin compact binaries. To generate waveforms for arbitrary values of the masses and spins, one needs to build a correspondence between the set of physical parameters $\{\eta,\chi^{}_1,\chi^{}_2\}$ and the set of phenomenological parameters $\{\sigma_i\}$. The total mass $M$ of the binary works as a trivial scaling factor.


\begin{figure}[t]
     \centering
     \begin{subfigure}[b]{0.49\textwidth}
         \centering
         \includegraphics[width=\textwidth]{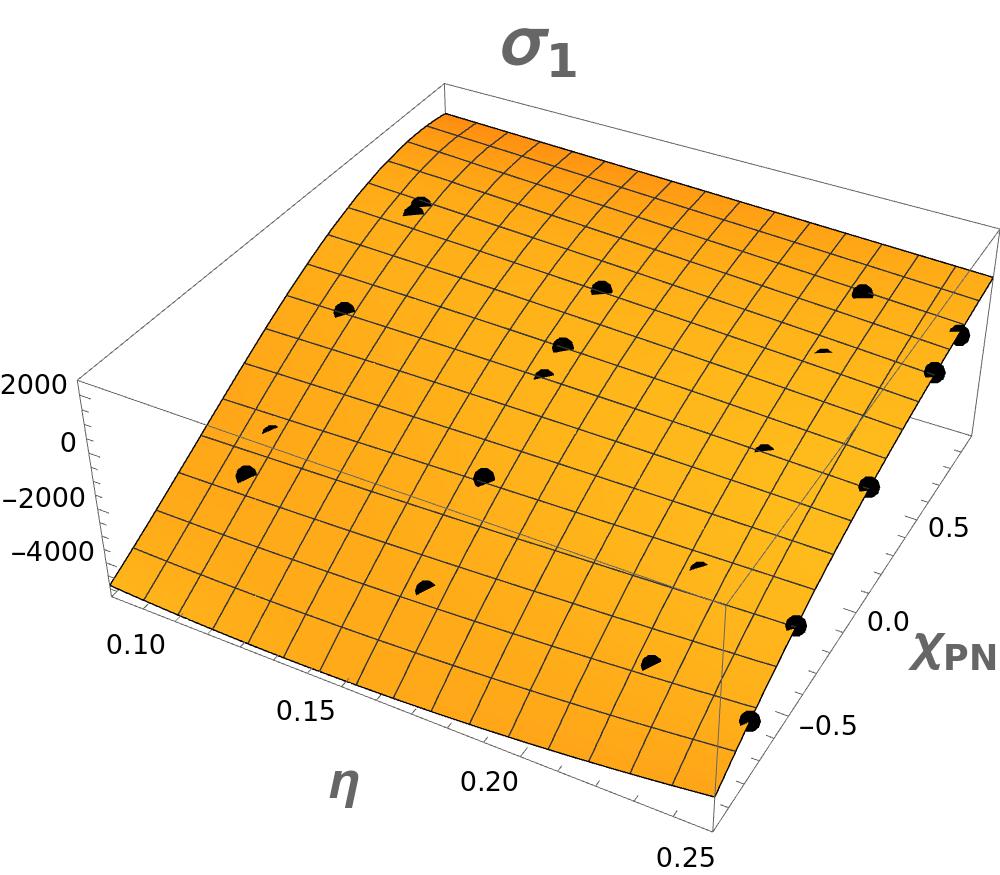}
     \end{subfigure}
     \hfill
     \begin{subfigure}[b]{0.49\textwidth}
         \centering
         \includegraphics[width=\textwidth]{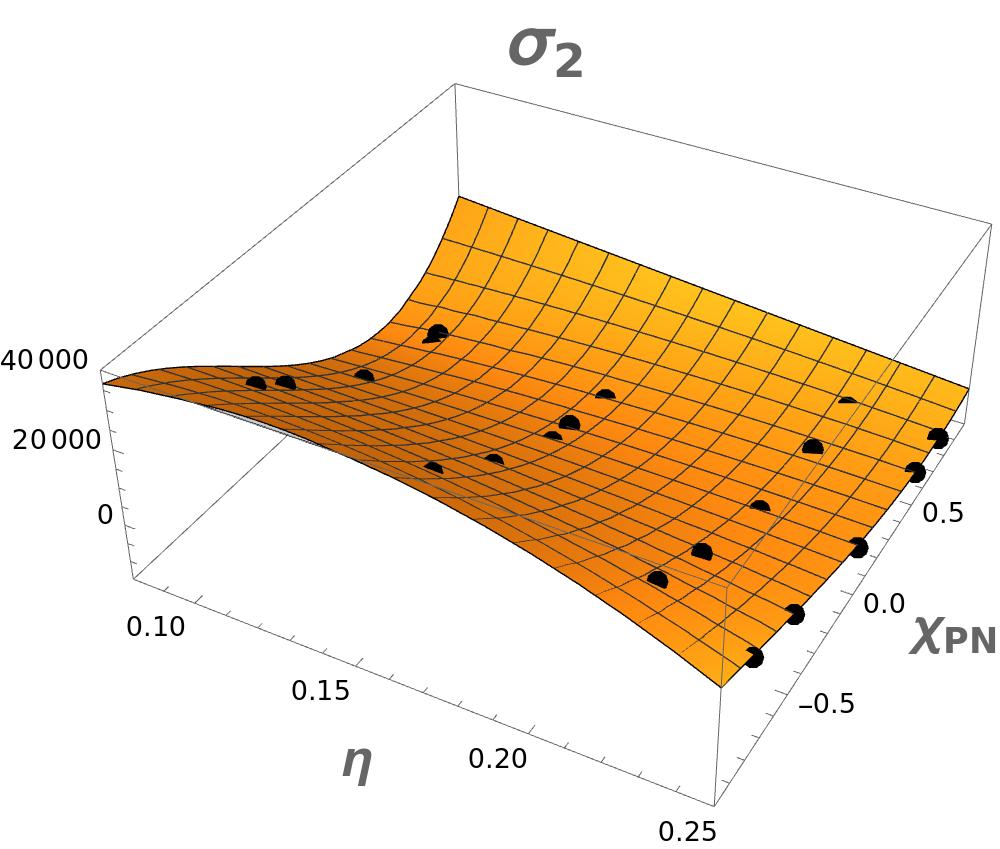}
     \end{subfigure}
     \hfill
     \begin{subfigure}[b]{0.49\textwidth}
         \centering
         \includegraphics[width=\textwidth]{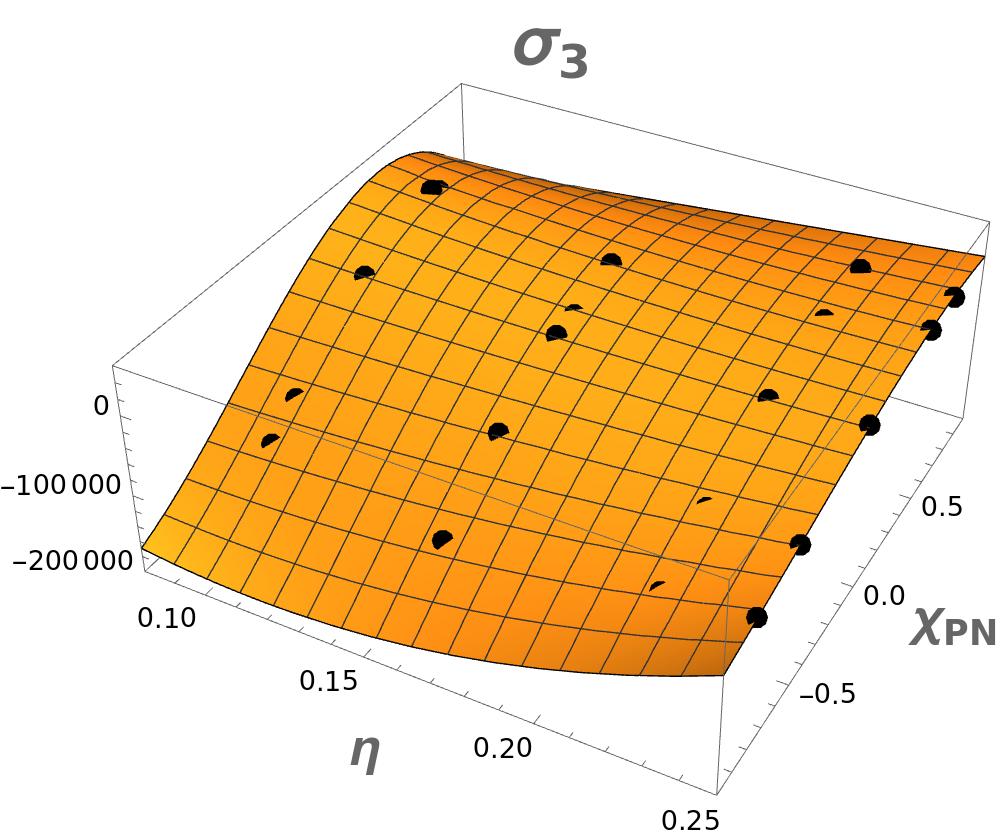}
     \end{subfigure}
     \hfill
     \begin{subfigure}[b]{0.49\textwidth}
         \centering
         \includegraphics[width=\textwidth]{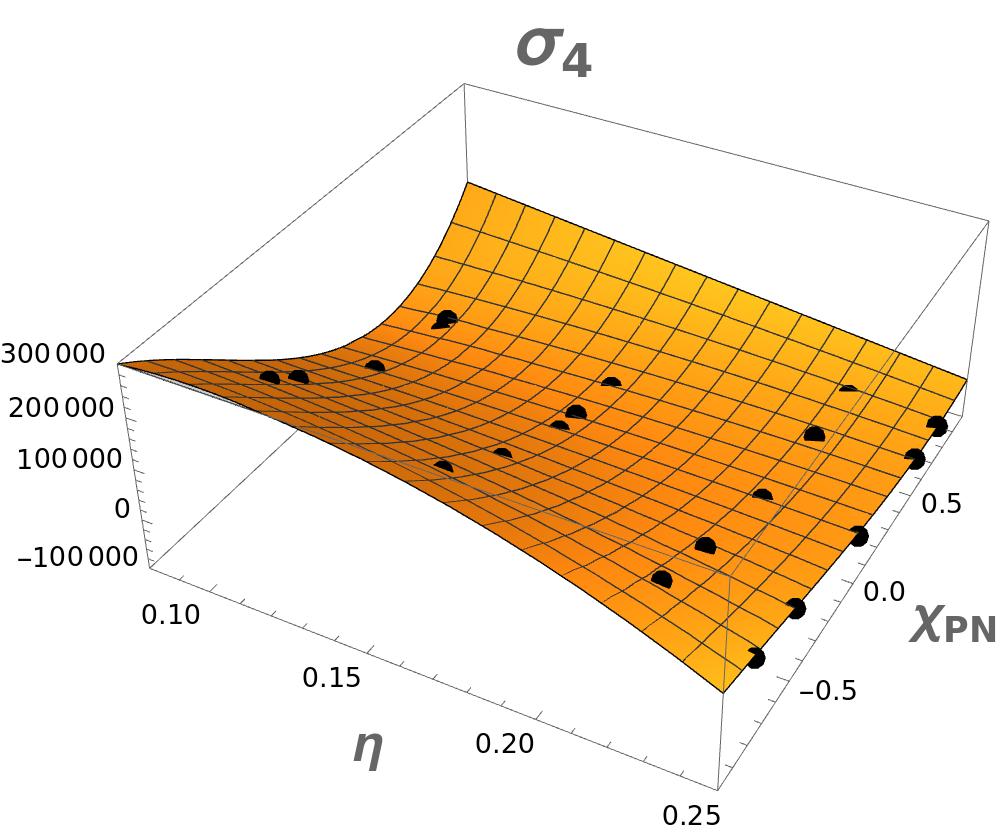}
     \end{subfigure}
        \caption{Fits of $\sigma_i$ calculated by the \texttt{NonLinearModelFit} module of Mathematica. The surfaces correspond to the 2D fits with $\eta$ and $\chi^{}_{\rm PN}$ as described in Eq.~\eqref{eq:Lambda}.}
        \label{fig:sigma}
\end{figure}

In PN expansions, the leading order contribution of spin in the phase of the waveform appears as a function of the combination
\begin{equation}
    \chi^{}_{\rm PN} = \chi^{}_{\rm eff} - \frac{38\eta}{113}(\chi^{}_1+\chi^{}_2),
\end{equation}
where
\begin{equation}
    \chi^{}_{\rm eff}=\frac{m_1\chi^{}_1+m_2\chi^{}_2}{M}.
\end{equation}

In this model, $\chi^{}_{\rm PN}$ is used as a single spin parameter to generate the phenomenological parameters. This ``effective spin approximation" works well for aligned-spin binaries, especially in the inspiral regime~\cite{Khan:2015jqa}.

To establish the correspondence between physical and phenomenological parameters, we fit $\sigma_i$ obtained from different hybrids with a polynomial of $\eta$ (up to second order) and the effective spin parameter $\chi^{}_{\rm PN}$ (up to third order):
\begin{equation}\label{eq:Lambda}
\begin{aligned}   
    \sigma_i = ~ \lambda_{00}^i+\lambda_{10}^i\eta ~& +(\chi^{}_{\rm PN}-1)(\lambda_{01}^i+\lambda_{11}^i\eta+\lambda_{21}^i\eta^2)\\
                & +(\chi^{}_{\rm PN}-1)^2(\lambda_{02}^i+\lambda_{12}^i\eta+\lambda_{22}^i\eta^2)\\
                & +(\chi^{}_{\rm PN}-1)^3(\lambda_{03}^i+\lambda_{13}^i\eta+\lambda_{23}^i\eta^2).
\end{aligned}
\end{equation}
For each $\sigma_i$, this polynomial generates a set $\{\lambda^i_{jk}\}$ to build the correspondence between physical and phenomenological parameter sets. Figure~\ref{fig:sigma} shows the surfaces of $\sigma_i$ as functions of $\{\eta,\chi^{}_{\rm PN}\}$ with the ansatz in Eq.~\eqref{eq:Lambda}. These best-fit surfaces generate a constant set of $\{\lambda^i_{jk}\}$ which define the inspiral model, along with the \texttt{TaylorF2} phase and the TH correction. In Appendix~\ref{AppendixC}, we list the coefficient values corresponding to each $\sigma_i$.
A schematic flowchart for generating the inspiral phase model starting from the intrinsic binary parameters can be described as:
\begin{equation*}
    {\{\eta,\chi^{}_1,\chi^{}_2\}}\xrightarrow[\text{spin}]{\text{Effective}}\{\eta,\chi^{}_{\rm PN}\}\xrightarrow{\lambda^i}{\{\sigma_i\}}\xrightarrow{}{\text{Inspiral model}}.
\end{equation*}

\subsection{Full IMR Model}\label{sec:imr}

For the phase, we have separately modeled the frequency region $0.0035\leqslant F \leqslant 0.018$, and we have the phase model of \texttt{PhenomD} thereafter. For the amplitude, we have a corrected model in the range $F \leqslant 0.014$, and a polynomial function for $0.014\leqslant F \leqslant F_{\rm peak}$. This gives us one interface for the phase and two interfaces for the amplitude to impose $C^{(1)}$ continuity. For the amplitude, however, $C^{(1)}$ continuity is ensured by the intermediate polynomial itself. For the phase, we can vary $t_c$ and $\phi_c$ of the inspiral phase. After imposing $C^{(1)}$ continuity, we can generate the complete IMR phase and amplitude by defining a step function

\begin{equation}
        \begin{aligned}
            \theta (f-f_0) =
        \left\{
    	    \begin{array}{ll}
    		    -1 , &  f < f_0 \\
    		    +1 , & f\geqslant f_0\,.
    	    \end{array}
        \right.
        \end{aligned}
\end{equation}
Using this function, one can define
\begin{equation}
    \theta ^\pm (f;f_0)=\frac{1}{2}\qty[1\pm \theta (f-f_0)],
\end{equation}
so that the IMR phase can be written as
\begin{equation}\label{eq:phiimr1}
    \Phi_{\rm IMR}(F)=\Phi_{\rm INS}(F)\theta^-(F;F_2) + \theta^+(F;F_2)\Phi_{\rm D}(F),
\end{equation}
where $F_2=0.018$, and $\Phi_{\rm D}(F)$ corresponds to the phase model of \texttt{PhenomD}. $\Phi_{\rm INS}(F)$ is given by Eq.~\eqref{eq:phiins}.

The IMR amplitude model follows a similar treatment, which can be expressed as
\begin{equation}\label{eq:ampimr}
\begin{aligned}
    \Tilde{A}_{\rm IMR}(F) & = \Tilde{A}_{\rm INS}(F)\theta^-(F;F_3) \\ + & \theta^+(F;F_3)\Tilde{A}_{\rm int}(F)\theta^-(F;F_4) + \theta^+(F;F_4)\Tilde{A}_{\rm D}(F),
\end{aligned}
\end{equation}
where $F_3=0.014$, and $F_4=F_{\rm peak}$. $\Tilde{A}_{\rm INS}(F)$ is the inspiral amplitude in Eq.~\eqref{eq:ampmodel}, $\Tilde{A}_{\rm int}(F)$ is the intermediate amplitude defined in Eq.~\eqref{eq:ampint}, and $\Tilde{A}_{\rm D}(F)$ is the \texttt{PhenomD} amplitude.

Finally, the plus and cross polarization states of the complete frequency-domain waveform read,
\begin{equation}
\begin{aligned}
    \Tilde{h}_+(f) = & \Tilde{A}_{\rm IMR}(f)\qty(\frac{1+\cos^2\iota}{2}) \\ & \times\exp{-i(\Phi_{\rm IMR}(f)-\phi_0-2\pi ft_0)},
\end{aligned}
\end{equation}
and
\begin{equation}
\begin{aligned}
    \Tilde{h}_{\cross}(f) = & -i\Tilde{A}_{\rm IMR}(f)\cos\iota \\ & \times\exp{-i(\Phi_{\rm IMR}(f)-\phi_0-2\pi ft_0)}.
\end{aligned}
\end{equation}
Here $\iota$ is the angle of inclination of the binary plane to the line of sight, $\phi_0$ and $t_0$ carry the overall phase and timeshift freedom of the complete waveform. We call the final waveform model \texttt{IMRPhenomD\_Horizon}, and abbreviate it to \texttt{PhenomD\_Horizon}.

In Fig.~\ref{fig:phase}, we show the inspiral phase $\Psi(Mf)$, its derivative with respect to $Mf$, and the difference between the phase of the hybrid data and the model, for the same configurations as in Fig.~\ref{fig:hybrids}. We show both the individual fits from the ansatz in Eq.~\eqref{eq:phiins} and the final model prediction obtained from the best fit values of ${\lambda^i_{jk}}$ from Eq.~\eqref{eq:Lambda}. In Fig.~\ref{fig:imramp}, we show the inspiral and intermediate amplitude, rescaled by the leading order factor $A_0=\frac{1}{\pi^{2/3}}\sqrt{5\eta/24}f^{-7/6}$. The intermediate regions are shown in a blue shaded colour. We also show the amplitude of \texttt{PhenomD} on the same plots.

\section{Faithfulness of the Model}\label{sec:model validation}

\begin{table}[t]
\vspace{4.5mm}

    \centering
    \begin{tabular}{p{0.05\textwidth}p{0.25\textwidth}p{0.1\textwidth}p{0.08\textwidth}p{0.08\textwidth}}
        \hline
        \# & Simulation label & $q$ & $\chi^{}_1$ & $\chi^{}_2$ \\
        \vspace{0.5mm}\\
        \hline
        \hline
        \vspace{0.5mm}\\
        1 & SXS:BBH:0159 & 1 & $-$0.9 & $-$0.9 \\
        2 & SXS:BBH:0154 & 1 & $-$0.8 & $-$0.8 \\
        3 & SXS:BBH:0148 & 1 & $-$0.44 & $-$0.44 \\
        4 & SXS:BBH:0150 & 1 & 0.2 & 0.2 \\
        5 & SXS:BBH:0170 & 1 & 0.44 & 0.44 \\
        6 & SXS:BBH:0155 & 1 & 0.8 & 0.8 \\
        7 & SXS:BBH:0160 & 1 & 0.9 & 0.9 \\
        8 & SXS:BBH:0157 & 1 & 0.95 & 0.95 \\
        9 & SXS:BBH:0014 & 1.5 & $-$0.5 & 0 \\
        10 & SXS:BBH:0008 & 1.5 & 0 & 0 \\
        11 & SXS:BBH:0013 & 1.5 & 0.5 & 0 \\
        12 & SXS:BBH:0046 & 3 & $-$0.5 & $-$0.5 \\
        13 & SXS:BBH:0036 & 3 & $-$0.5 & 0 \\
        14 & SXS:BBH:0168 & 3 & 0 & 0\\
        15 & SXS:BBH:0031 & 3 & 0.5 & 0 \\
        16 & SXS:BBH:0047 & 3 & 0.5 & 0.5 \\
        17 & SXS:BBH:0056 & 5 & 0 & 0 \\
        18 & SXS:BBH:0181 & 6 & 0 & 0 \\
        19 & SXS:BBH:1424 & 6.465 & $-$0.66 & $-$0.8 \\
        20 & SXS:BBH:0298 & 7 & 0 & 0 \\
        \vspace{0.5mm}\\
        \hline
    
    \end{tabular}
    \caption{List of the extra hybrids created for validating the model.}
    \label{tab:hyb_nc}
\end{table}

\begin{figure}
    \centering
    \includegraphics[width=\textwidth]{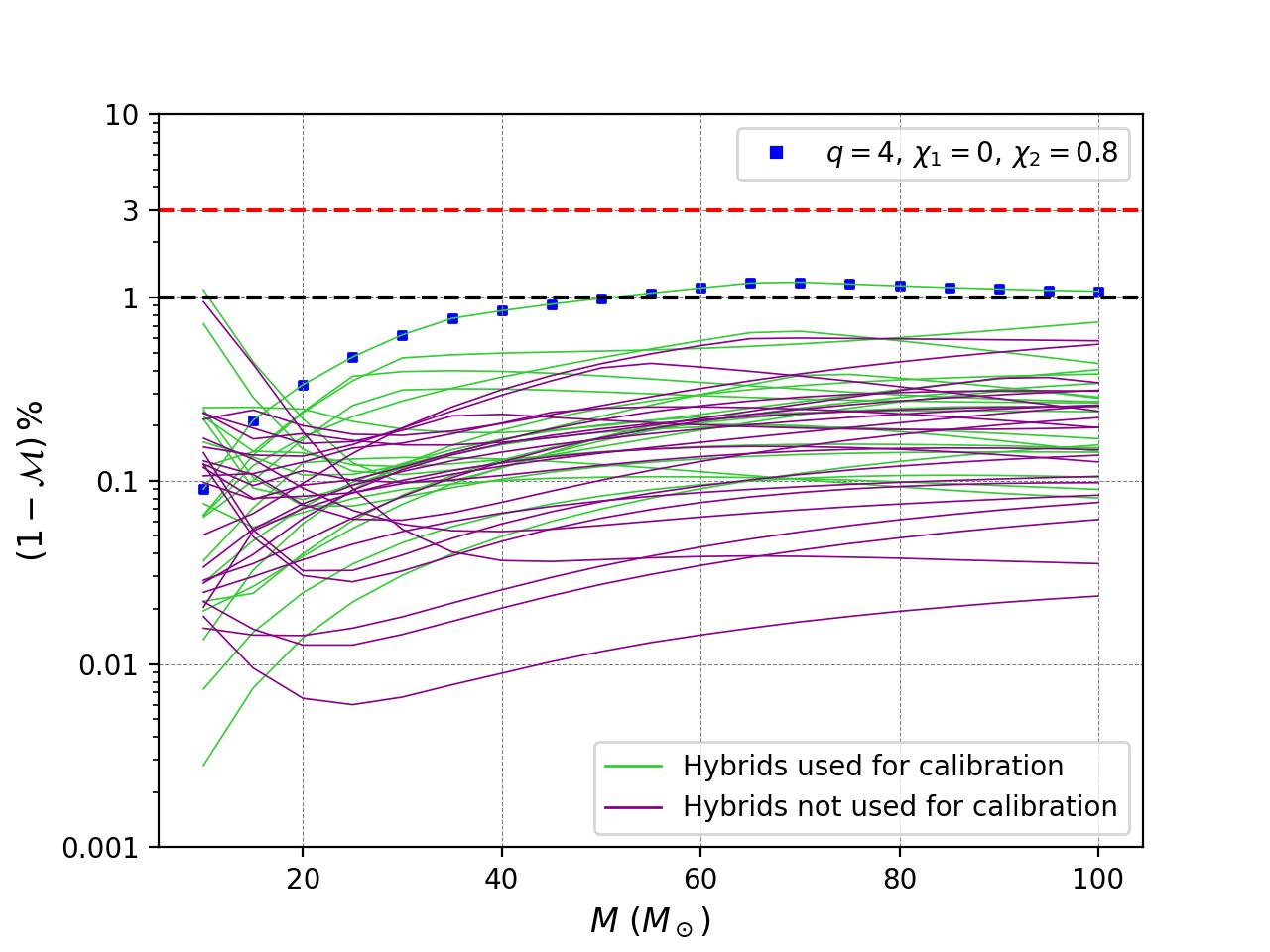}
    \caption{Mismatches (\%) between \texttt{IMRPhenomD\_Horizon} and the hybrid waveforms (Table~\ref{tab:hyb} and Table~\ref{tab:hyb_nc}) in aLIGO ZDHP noise PSD with a lower cutoff of 10 Hz, as a function of the total mass.}
    \label{fig:mismatch}
\end{figure}

A phenomenological model, calibrated with a set of ``target hybrids", should be compared with a larger set of ``test hybrids" to assess its accuracy. The test hybrids should contain both the target hybrids and a new set of hybrids interspersed within the parameter space of calibration. To study the model's behavior beyond the calibration range, test hybrids can be extended further in the parameter space. How well the final model can predict these test hybrids is denoted by its \textit{effectualness} and \textit{faithfulness}~\cite{Ajith:2007kx,Lindblom:2008cm}. A model is effectual if it is accurate enough to predict a GW signal from the detector noise, while to be faithful it also has to have enough accuracy to estimate the binary parameters. A faithful model is also effectual, but the converse may not be true. Given any two signals $h(t)$ and $g(t)$, buried in the noise $n(t)$, one can define a noise-weighted inner product
\begin{equation}
    \braket{h}{g}=2\int_{0}^{\infty}\frac{\Tilde{h}(f)\Tilde{g}^\ast (f)+\Tilde{h}^\ast (f)\Tilde{g}(f)}{S_n(f)}\dd f\,,
\end{equation}
where $\Tilde{h}(f)$ is the Fourier transform of $h(t)$ defined by
\begin{equation}
    \Tilde{h}(f)=\int_{-\infty}^{\infty}h(t)e^{-2\pi i ft}\dd t\,,
\end{equation}
and $S_n(f)$ is the power spectral density (PSD) of the noise.

Equipped with this definition of an inner product between two waveforms, one can define the \textit{match} between them by maximizing the inner product with respect to the constant phase and timeshift freedom between the waveforms $\Tilde{h}(f)$ and $\Tilde{g}(f)$, normalized by their individual norms:
\begin{equation}
   \mathcal{M}=\underset{t_c,\phi_c}{\text{min}}\,\,\frac{\braket{h(\lambda)}{g(\lambda ')}}{\sqrt{\braket{h(\lambda)}{h(\lambda)}\braket{g(\lambda ')}{g(\lambda ')}}}\,.
\end{equation}
Here $\lambda$ and $\lambda '$ denote the intrinsic parameters. The \textit{mismatch}, defined as $1-\mathcal{M}$, quantifies how much the waveform $\Tilde{h}(f)$ morphologically differs from $\Tilde{g}(f)$. The faithfulness of a model is quantified by demanding that the mismatches between the model and the validation dataset (test hybrids) do not exceed a certain threshold. For \texttt{PhenomD}, the threshold was 1\%.

Since the total mass of the binary works as a scaling factor for the gravitational waveforms, it defines how many GW cycles are contained within the sensitive frequency band of a detector. The lower the total mass, the more time the binary spends in its inspiral phase, resulting in a longer waveform given a lower cutoff for the frequency. In Advanced LIGO, complete IMR models become important above a critical value $\sim 12M_\odot$~\cite{Buonanno:2009zt} of the binary mass.
To assess our model's faithfulness, we create a set of test hybrids that are long enough to cover the sensitive frequency band of Advanced LIGO for a total binary mass as low as $10M_\odot$. At 10 Hz, the dimensionless frequency corresponding to a $10M_\odot$ binary is $Mf\approx 0.0005$. 


In Table~\ref{tab:hyb_nc} we list the extra test hybrids created for assessing the model accuracy. Figure~\ref{fig:mismatch} shows the mismatches between the entire set of test hybrids (Table~\ref{tab:hyb} and Table~\ref{tab:hyb_nc}) against the BBH waveform model presented in this chapter. We have used the Advanced LIGO zero-detuned high power~\cite{KAGRA:2013rdx} (ZDHP) noise curve with a lower frequency cutoff of 10 Hz, and upper cutoff of $Mf=0.3$. The mismatches are plotted against the total binary mass. We find that the model accuracy is at par with the accuracy standard of \texttt{PhenomD}, with almost all the hybrids showing mismatches well below 1\%, the majority of them having mismatches around 0.1\% within the mass range $10M_\odot\leqslant M \leqslant 100M_\odot$. The worst mismatch we find is for the $\{q=4,\chi^{}_1=0,\chi^{}_2=0.8\}$ configuration, which crosses the 1\% level above $\sim 50M_\odot$, and reaches a maximum of 1.21\% at $70M_\odot$.

\section{Model Comparison}\label{sec:comparison}

\begin{figure}
     \centering
     \begin{subfigure}[b]{0.65\textwidth}
         \centering
         \includegraphics[width=\textwidth]{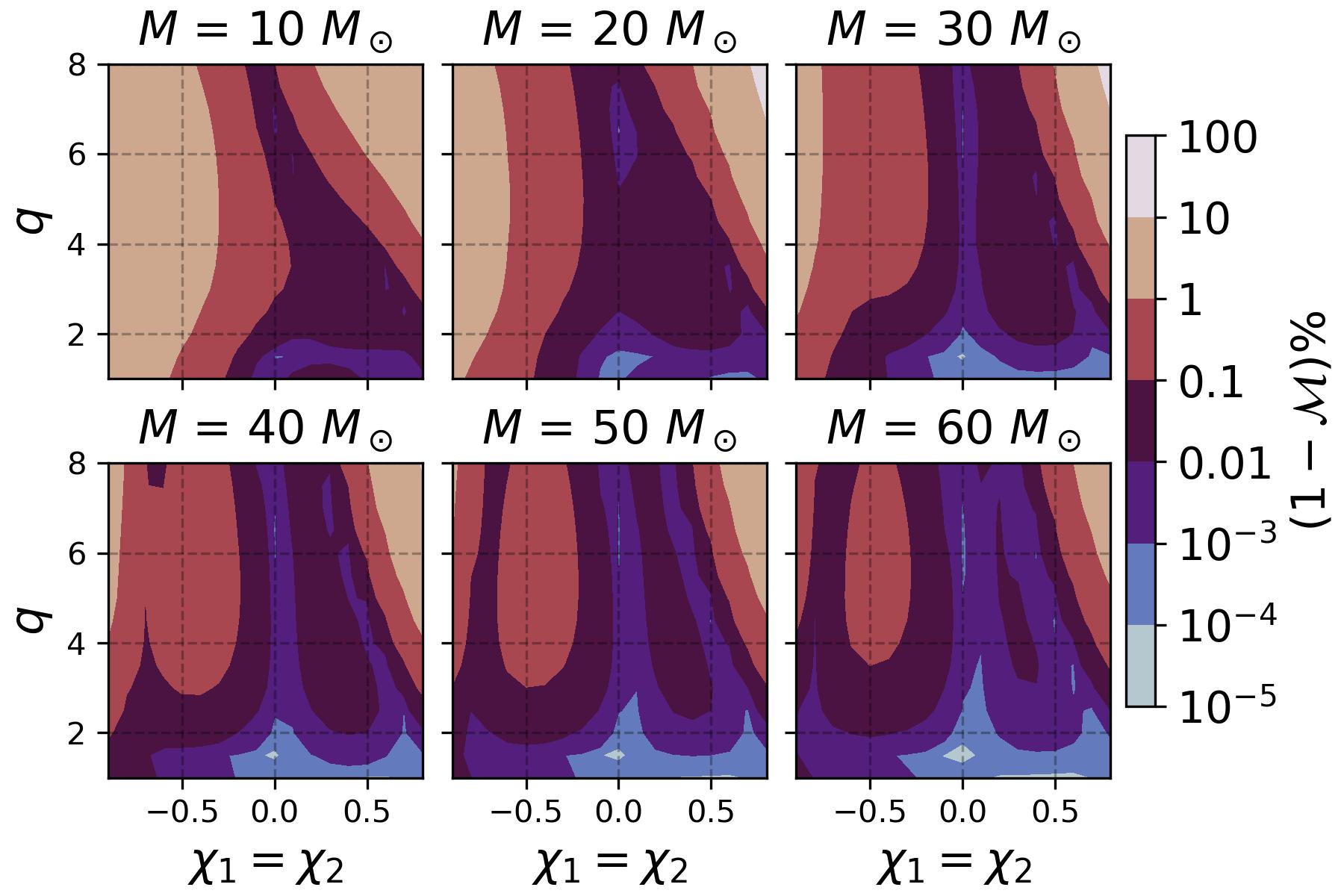}
         \caption{}
         \label{fig:mismatch_D_phaseonly}
     \end{subfigure}
     \hfill
     \begin{subfigure}[b]{0.65\textwidth}
         \centering
         \includegraphics[width=\textwidth]{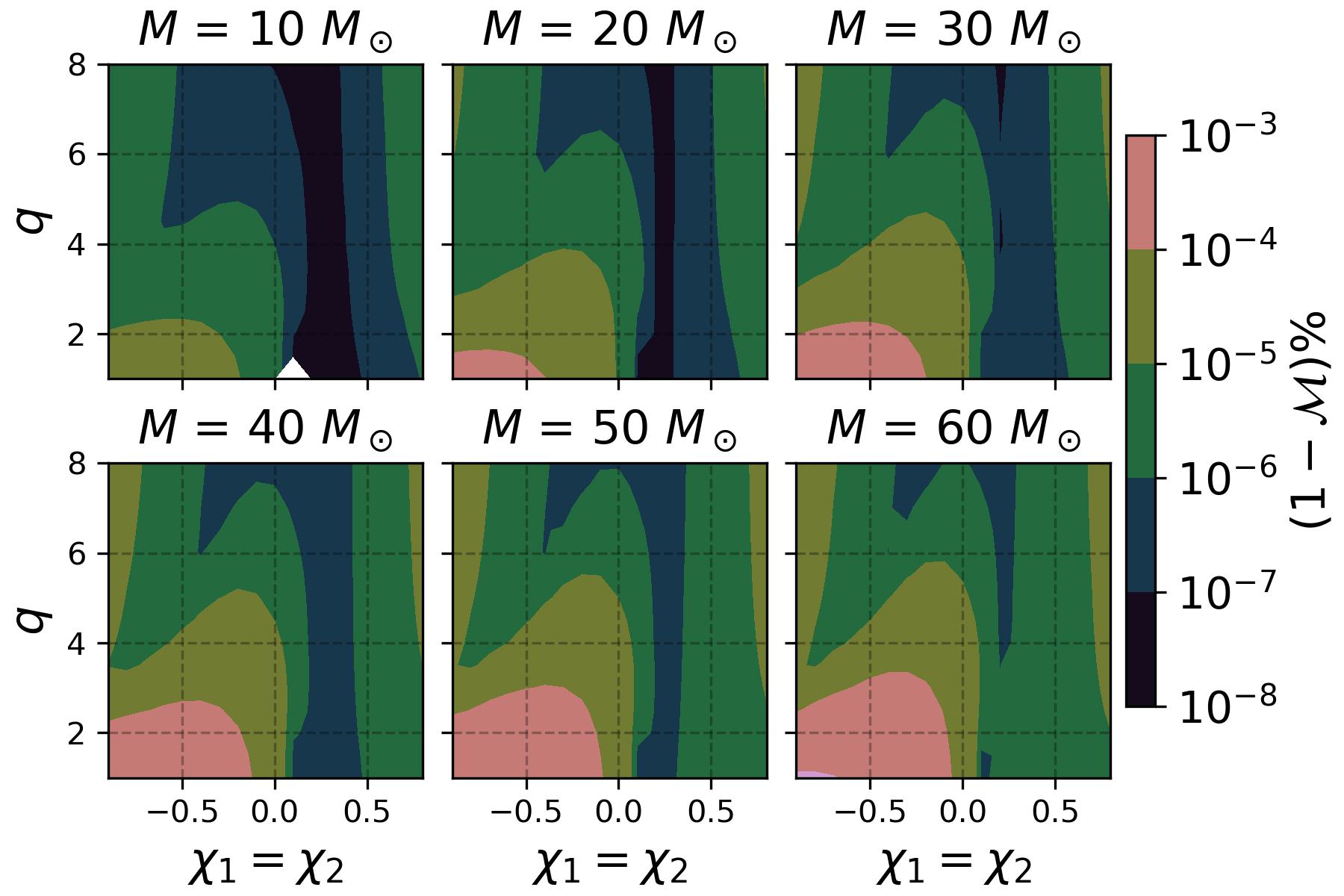}
         \caption{}
         \label{fig:mismatch_D_amponly}
     \end{subfigure}
     \hfill
     \begin{subfigure}[b]{0.65\textwidth}
         \centering
         \includegraphics[width=\textwidth]{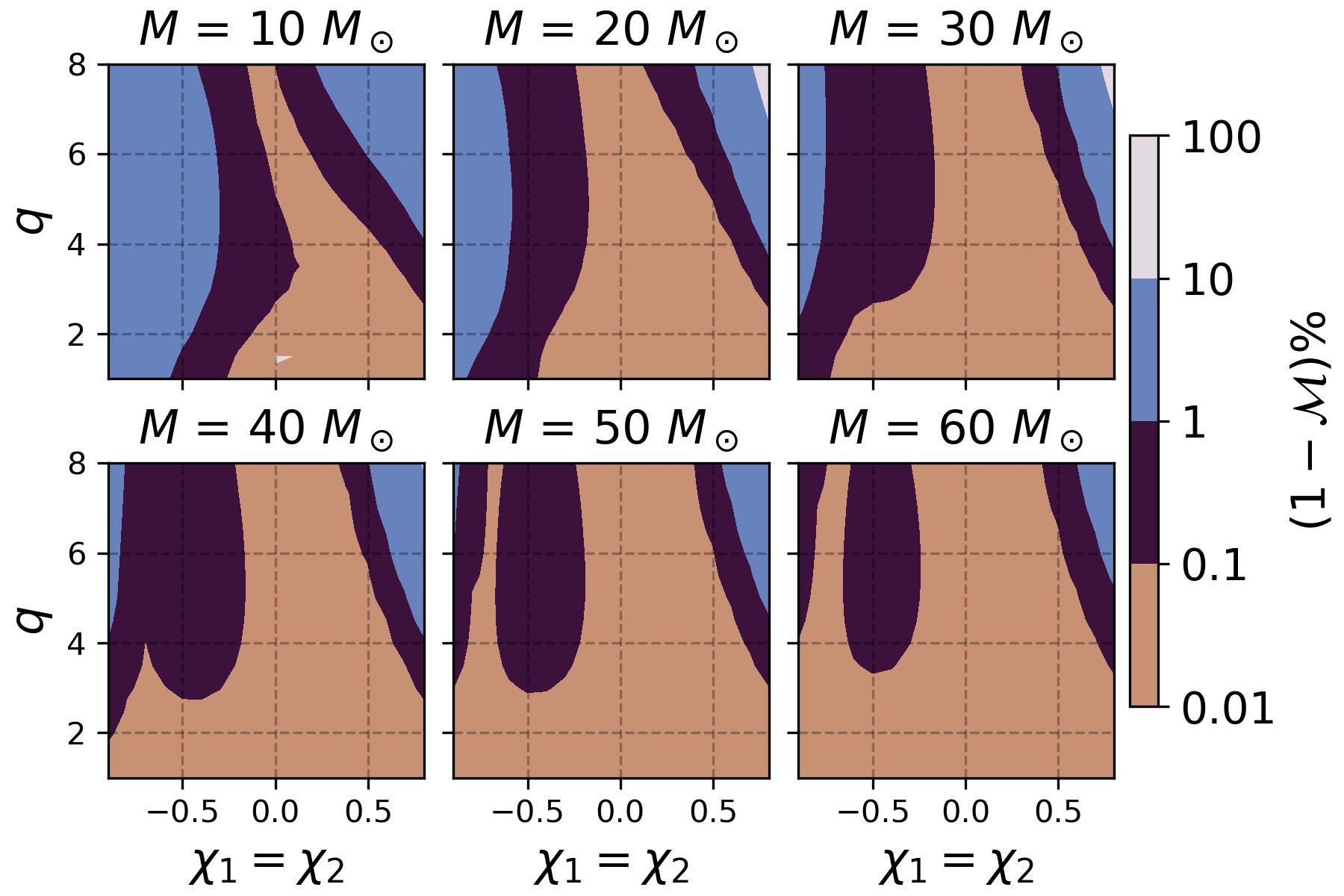}
         \caption{}
         \label{fig:mismatch_D_both}
     \end{subfigure}
        \caption{Mismatches (\%) between \texttt{PhenomD\_Horizon} and \texttt{PhenomD}. (a) Mismatches with only the phase correction but no amplitude correction in \texttt{PhenomD\_Horizon}. (b) Mismatches with only the amplitude correction without any phase correction. (c) Mismatches with corrections in both the phase and amplitude. The plots show that the dominant contributor to these mismatches is the phase correction.}
        \label{fig:mismatch_vs_D}
\end{figure}

Figure~\ref{fig:mismatch_vs_D} shows the mismatches between \texttt{PhenomD\_Horizon} and \texttt{PhenomD}, keeping $\chi^{}_1=\chi^{}_2$. We explore the mismatches within the parameter space $q \in [1,8]$ and $\chi^{}_1,\chi^{}_2 \in [-0.9,0.9]$, for six values of the binary mass, $M/M_\odot \in [10,20,30,40,50,60]$. In Fig.~\ref{fig:mismatch_D_phaseonly}, we show the mismatches including only the phase correction - i.e. using the recalibrated inspiral phase, but keeping the IMR amplitude the same as \texttt{PhenomD}. Since the phase correction due to TH written in Eq.~\eqref{eq:phase correction} is $\propto 1/\eta$ at the leading order, higher mismatches are found for higher mass ratios, for a fixed value of the spins. With increasing spins, mismatches rise due to the increased significance of TH. The mismatch reaches a maximum of $\sim 16\%$ for $30M_\odot$ binaries with $q\gtrsim 7$ and  $\chi^{}_1=\chi^{}_2\gtrsim 0.8$, and a minimum of $\sim 10^{-5}\%$ for heavier binaries. For $M\leqslant 30M_\odot$, we see substantial regions of the parameter space with mismatches between 1-10\% for spin magnitudes of $|\chi^{}_1|=|\chi^{}_2|\gtrsim 0.5$. For $M > 30M_\odot$, the 1\% level is crossed only for $q\gtrsim 4$. 

Figure~\ref{fig:mismatch_D_amponly} shows mismatches when only the amplitude correction is added to the inspiral and the intermediate region is re-evaluated accordingly, but the phase is kept the same as \texttt{PhenomD}. The mismatches, in this case, are orders of magnitude less than the phase-corrected model, and never exceed 0.001\% within this parameter space. This result also vindicates our choice of not recalibrating the amplitude model, as discussed in Sec~\ref{sec:amplitude}. We also note from Eq.~\eqref{eq:amp correction} that at the leading order, $\Tilde{A}_{\rm TH}\propto \sqrt{\eta}$, which implies that the TH corrections are more significant for more symmetric masses. This is in contrast to the behaviour of the phase correction, and renders the amplitude correction ineffective for high mass-ratio binaries with notable TH contribution in the phase. Nevertheless, $\Tilde{A}_{\rm TH}$ still rises with increasing spins. 

In Fig.~\ref{fig:mismatch_D_both}, we use the complete \texttt{IMRPhenomD\_Horizon} model, by including both the phase and amplitude corrections. The contours of mismatches $\geqslant 0.1\%$ are almost the same as in Fig.~\ref{fig:mismatch_D_phaseonly}, but amplitude corrections elevate them above 0.01\% throughout other regions of the parameter space. However, mismatches $<0.1\%$ are below the accuracy level of the model, for which we considered a 1\% tolerance level of modeling errors (Fig.~\ref{fig:mismatch}). This suggests that including the amplitude correction in the TH corrected model has negligible effect compared to the phase correction in GW data analysis. 

\begin{figure}
\vspace{5mm}
    \centering
    \includegraphics[width=0.85\textwidth]{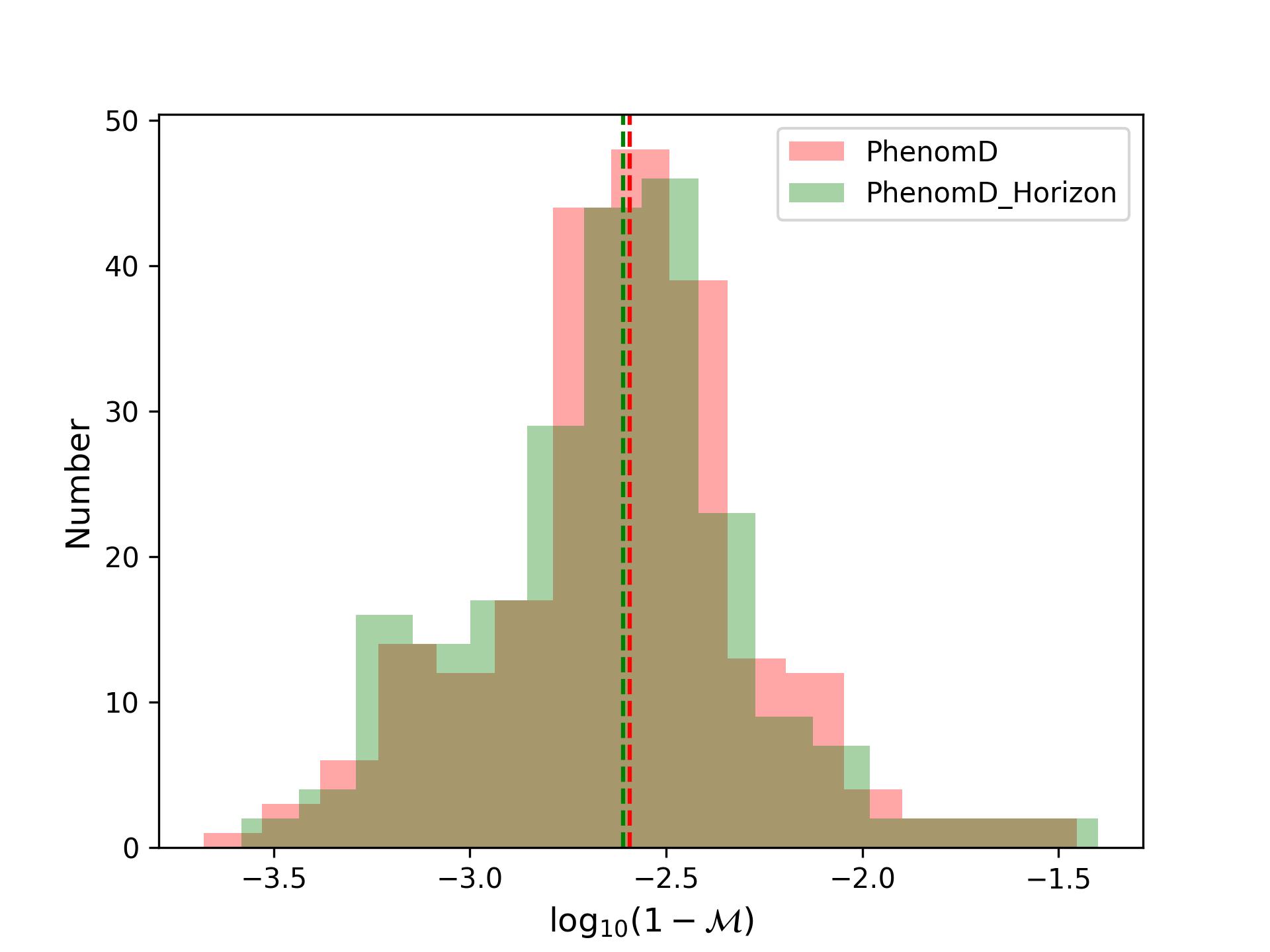}
    \caption{Mismatches of 219 non-precessing noneccentric NR waveforms from SXS with \texttt{PhenomD\_Horizon} and \texttt{PhenomD}, in aLIGO ZDHP noise curve, averaged in the binary mass range $12-100M_\odot$. The dashed lines show the median values of the distributions.}
    \label{fig:mismatchNRzdhp}
\end{figure}

\begin{figure}
\vspace{5mm}
    \centering
    \includegraphics[width=0.85\textwidth]{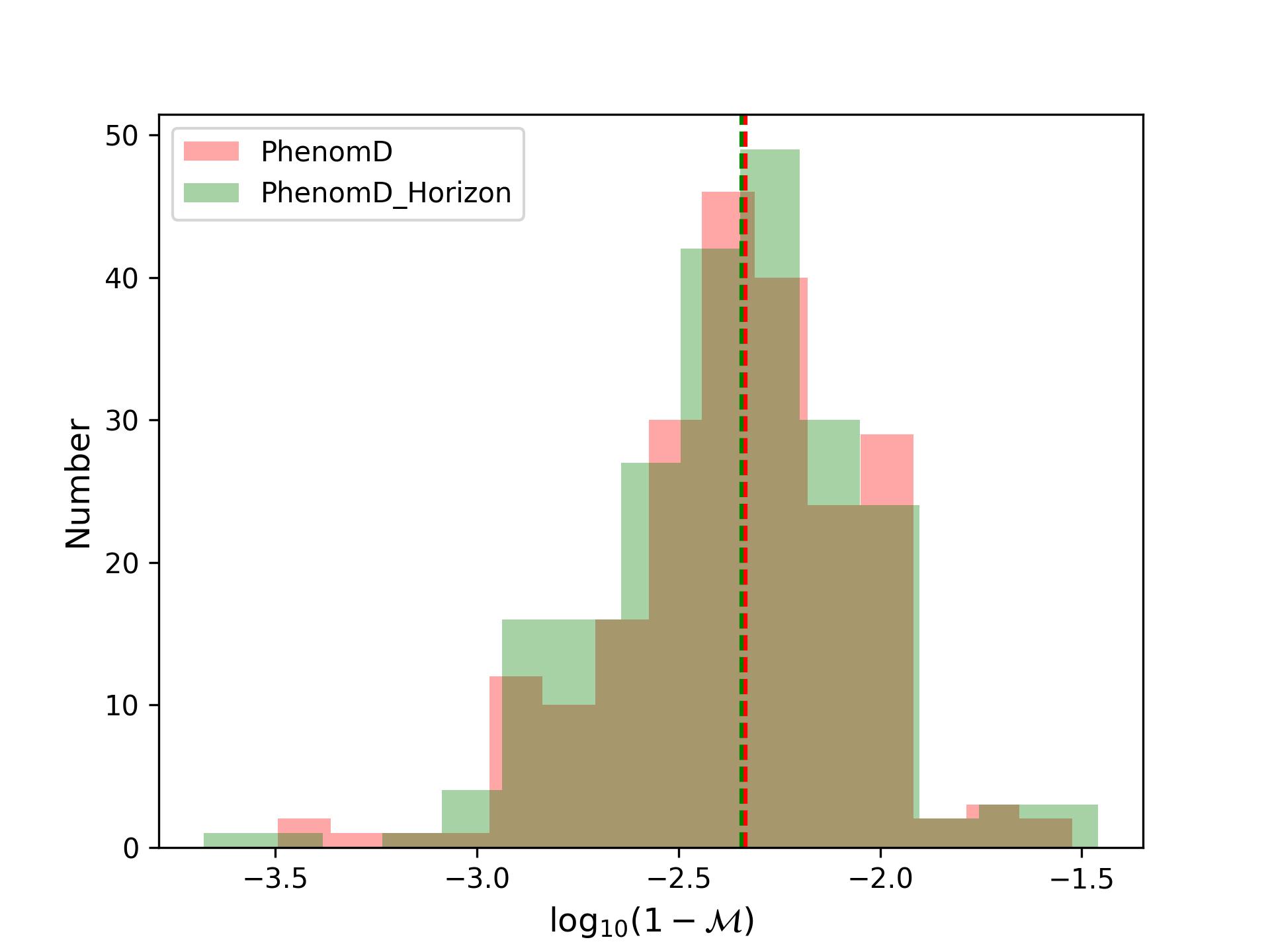}
    \caption{Same as in Fig.~\ref{fig:mismatchNRzdhp}, but with a flat noise curve.}
    \label{fig:mismatchNRflat}
\end{figure}

To demonstrate the improvement in accuracy that our model introduces over \texttt{PhenomD}, we use 219 NR waveforms from the SXS catalog and compare them against \texttt{PhenomD} and \texttt{PhenomD\_Horizon} separately by computing mismatches. In Fig.~\ref{fig:mismatchNRzdhp}, we show the distributions of the mismatches, by using the Advanced LIGO ZDHP noise curve. We plot histograms of $\log_{10}(1-\mathcal{M})$, where $(1-\mathcal{M})$ is the average mismatch in the binary mass range $12M_\odot \leqslant M \leqslant 100M_\odot$. The dashed lines show the medians of the two histogram plots, where a slight overall improvement is seen. The fractional shift in the medians (denoted by an overbar), is
\begin{equation}
    \frac{\overline{(1-\mathcal{M})}_{\rm PhenomD} - \overline{(1-\mathcal{M})}_{\rm PhenomD\_Horizon}}{\overline{(1-\mathcal{M})}_{\rm PhenomD}} \approx 0.04,
\end{equation}
showing an improvement of $\sim 4\%$. Figure~\ref{fig:mismatchNRflat} shows similar histograms with a flat noise curve. In this case, also we see a shift in the median value towards lower mismatches, with an improvement of $\sim 2\%$.


\section{Discussion and Summary}\label{sec:discussion}

We have presented a phenomenological gravitational waveform model for binary black holes that carries the signatures of tidal heating of the holes explicitly throughout the binary evolution, starting from the early inspiral phase and ending at their merger. In the frequency domain, it contains corrections due to this effect both in the phase and amplitude. The entire IMR model for the phase is divided into two parts. The inspiral phase model, defined by the frequency range $0.0035 \leqslant Mf\leqslant 0.018$, is a recalibrated version of \texttt{IMRPhenomD}~\cite{Husa:2015iqa,Khan:2015jqa} to account for the modifications due to tidal heating. The merger-ringdown parts, defined as $Mf\geqslant 0.018$ in terms of the dimensionless frequency, is the same as \texttt{PhenomD} since this region contains purely NR data, which inherently incorporates the dephasing due to tidal heating. The amplitude corrections are added by dividing the model into three parts. The inspiral model, with $0.0035 \leqslant Mf\leqslant 0.014$, is constructed by directly adding the correction term to the amplitude model of \texttt{PhenomD} in this region. The intermediate region, $0.014\leqslant Mf \leqslant Mf_{\rm peak}$, $f_{\rm peak}$ being the frequency corresponding to the peak amplitude, is approximated as a polynomial in $f$ and evaluated by imposing $C^{(1)}$ continuity at the two ends and a collocation point at their mid-frequency. The amplitude model is identical to \texttt{PhenomD} for $f>f_{\rm peak}$. The final model is calibrated within the range in the mass ratio $q\in [1,8]$ and the effective spin parameter $\chi_{\rm PN}\in [-0.79,0.72]$.

We conducted a study of faithfulness, by calculating mismatches between the model and a validation dataset of 40 hybrid waveforms constructed with tidal heating included. We demonstrated that modeling errors do not lead to mismatches higher than 1\% (barring a single outlier), with most of the mismatch values lying around 0.1\%. This standard of faithfulness is comparable to \texttt{PhenomD}, which also considered a maximum tolerance level of 1\% mismatch.

How the new model differs from its parent model \texttt{PhenomD} was assessed by calculating mismatches between them within the calibration range of the intrinsic parameter space, for binary masses ranging from $10M_\odot$ to $60M_\odot$. Phase corrections in the inspiral result in mismatches between these two models as high as $\sim 16\%$, whereas amplitude corrections alone are unable to produce mismatches above $\sim 0.001\%$. With increasing mass asymmetry, the impact of the phase correction increases, but the amplitude correction becomes less effective. Combined with the fact that the former has more significance in GW searches and PE, this reciprocal nature of these two correction terms leaves the latter mostly superfluous for practical purposes. 

We compared the model with a set of 219 NR waveforms from the SXS catalog~\cite{Boyle:2019kee}, and show the mismatch distribution along with the same for \texttt{PhenomD}, averaged over the binary mass range $12M_\odot \leqslant M \leqslant 100M_\odot$. With the Advanced LIGO ZDHP noise curve, we see an improvement of $\sim 4\%$ in the medians of the two distributions. With a flat noise curve, the improvement is relatively modest, about 2\%. These overall improvements point towards a mild yet non-negligible increase in waveform accuracy. 

Tidal heating of black holes, after all, is a weak effect in the context of comparable-mass binaries. Nevertheless, employing complete waveform models deficient in its signatures can 
bias the estimated parameters of a binary. Its absence can also mimic a deviation from GR predictions, raising complications in tests of GR. We leave these studies to future work. Additionally, tidal heating has the important property of varying significantly according to the nature of the component objects, which makes it a potential discriminator for BHs from other objects. However, to utilize the full power of TH as a BH identifier, one needs to model the energy and angular momentum fluxes down the BH horizons also in the late inspiral regime described by NR, where the tidal fields are the strongest. We explore this aspect in Chapter~\ref{Chapter5}.

The current work, in its entirety, is based on the frequency-domain waveform model \texttt{IMRPhenomD}. For this reason, we perform all the comparisons of the model presented here with the said approximant only. More recent and state-of-the-art phenomenological waveform models are available currently, namely, the \texttt{IMRPhenomX}~\cite{Pratten:2020fqn,Pratten:2020ceb} family of waveforms. These waveforms are calibrated with a much larger set of hybrid waveforms, and use superior techniques like hierarchical modeling~\cite{Jimenez-Forteza:2016oae} to reduce modeling errors, and also to generate more accurate waveforms with unequal spins. These models boast 1 to 2 orders of magnitude better faithfulness than \texttt{PhenomD}, and a dramatic shift in the mismatch distribution against NR waveforms towards lower values~\cite{Pratten:2020fqn}. Including the effects of TH in these models by recalibrating the inspiral is beyond the scope of this work, which uses only publicly available NR data from the SXS catalog. We leave them as future possibilities.

\clearpage
\chapter{Horizon Parameters in Inspiral-Merger-Ringdown Waveforms}
\label{Chapter5}

\section{Introduction}

Inspiral-merger-ringdown (IMR) gravitational waveforms, as discussed in Chapter~\ref{Chapter4}, are important for binaries heavier than $\sim 12M_\odot$, above which merger-ringdown parts of a GW signal contribute significantly to the SNR in the second-generation ground-based detectors. Using data from NR simulations, these waveforms feature an accurate description of the late inspiral and merger phases. IMR modeling strategies include phenomenological waveform modeling, waveform models based on NR-calibrated EOB descriptions, and surrogate and reduced-order modeling. The inspiral description of these models comes from post-Newtonian (PN) calculations truncated at an appropriate cutoff frequency or EOB resummations. In the PN framework, the GW generation corresponds to only the point-particle effect at the leading order, where only the ratio of the two masses define the fluxes of GWs. Spin effects start to appear from 1.5PN order. At 2.5PN order, effects of the component objects dissipating energy from the orbits and altering the rate of inspiral come into play. As we saw in previous chapters, this is the tidal heating (TH) effect. For BHs, TH is the strongest -- due to the fact that the horizons of BHs define an one-way membrane that does not allow any mass-energy to escape outward. In the context of BHs, the fluxes of GWs that correspond to TH are often referred to as horizon fluxes. For horizonless compact objects, e.g. the theoretical exotic compact objects (ECOs) that can act as BH-mimickers, TH effects are subdued compared to BHs since the surfaces of these objects feature a nonzero reflectivity. For nonspinning objects, TH first appears at the 4PN order.
 
For any generic inspiralling binary, the amount of TH fluxes lost from the orbit can be expressed as a ratio of that for a BBH~\cite{Datta:2019epe}. In Chapter~\ref{Chapter3}, we have introduced the \textit{horizon parameters}, that essentially represent the ratio of TH fluxes for a generic binary to that of a BBH. These parameters take values between 0 and 1, the latter representing a BH. An object with the horizon parameter close to zero partakes in negligible TH (e.g. neutron stars (NSs)). 

In GW data analysis, strain data received by the GW detectors are compared with an immense bank of template waveforms that span a large region of possible parameter space. Techniques like matched filtering are then used to dig the GW signal out of the detector noise. Accurate detection and prediction of the source parameters largely relies on the accuracy of the template bank.
At the leading order of the phase evolution of a compact binary coalescence (CBC), the masses of the two objects $m_1$ and $m_2$ appear as the combination $\mathcal{M}_c=(m_1m_2)^{3/5}/(m_1+m_2)^{1/5}$, known as the ``chirp mass". At 1PN order, the phase evolution depends sensitively on the reduced mass $\mu=m_1m_2/(m_1+m_2)$, allowing the individual masses to be measured with high accuracy~\cite{Cutler:1994ys}. Once the mass estimates are made from the leading order phase and frequency behaviour of the signal, the binary is generally classified as a BBH if both the masses are higher than $\sim 3M_\odot$. This assumption is motivated by the fact that this mass range is beyond the theoretically allowed maximum mass of NSs~\cite{KAGRA:2021vkt}. However, if one entertains the possibility of exotic objects heavier than NSs mimicking BHs in a binary coalescence~\cite{Cardoso:2019rvt}, their identification may demand probes beyond just the mass estimate. Additionally, objects in the mass-gap (2-5$M_\odot$) pose challenges in their classification since observational and theoretical assurance on whether they are BHs or NSs dwindles in this regime. 

In this scenario, one can utilize the horizon parameters to tell apart BHs from other objects~\cite{datta2020recognizing}. Since TH is a subleading effect in comparable-mass binaries, precise estimation of these parameters from GW data faces challenges. In Chapter~\ref{Chapter3}, we showed that the future GW detectors Einstein Telescope and Cosmic Explorer will be able to estimate them much better than the current detectors. We used an inspiral-only approximant for our analyses in that chapter, truncating it at the innermost stable circular orbit (ISCO). In Chapter~\ref{Chapter4}, we prepared a full IMR gravitational waveform model, based on the approximant \texttt{IMRPhenomD}, that includes the corrections in phase and amplitude in the frequency domain due to the tidal heating of black holes. However, the final model, referred to as \texttt{IMRPhenomD\_Horizon}, is valid exclusively for BBHs since there are no free parameters quantifying the possible presence of horizonless objects. This chapter aims to build upon that work to construct a waveform model with these parameters introduced.

At later stages of the inspiral, effects of TH are more significant due to the stronger tidal fields acting on the compact objects. Unfortunately, PN expansions become increasingly inaccurate in the strong-gravity regime and, consequently, one can no longer use the TH descriptions under this scheme. In this chapter, our effort to build an approximant with horizon parameters is based on modeling the TH fluxes from the horizon data provided by the SXS catalog of NR simulations~\cite{Boyle:2019kee}. The model represents nonspinning binaries only. The resulting approximant contains phenomenological parameters, and can be used on \texttt{IMRPhenomD\_Horizon} to obtain a generic waveform for compact objects with horizon parameters quantifying TH up to the merger frequency. We also describe an analytical approximant, based on the most recent expressions for TH fluxes~\cite{Saketh:2022xjb}, which is valid for arbitrary masses and spins. The latter model, being an analytical one, cannot contain the horizon parameters up to the merger. We truncate it at the ISCO frequency.

\section{Analytical Expressions for Horizon fluxes}

The most recent expressions for the horizon fluxes of BHs are provided in Ref.~\cite{Saketh:2022xjb}, that also agree with the calculations in the test-mass limit. In this section, we briefly summarize the results.

Horizon fluxes of BHs change their mass, spin and angular momentum (spin) during the inspiral with time. In the PN barycentric frame, the rate of change of the BH mass ($m_1$) can be written as
\begin{equation}\label{eq:dmdt}
    \left<\dv{m_1}{t}\right>=\Omega(\Omega_{\rm H} - \Omega)C_v\,,
\end{equation}
where
\begin{equation}
    \Omega_{\rm H}=\frac{\chi_1}{2m_1(1+\kappa_1)}\,
\end{equation}
is the horizon angular velocity of the Kerr black hole (KBH) of mass $m_1$. $\chi_1$ is the dimensionless spin of the KBH, and $\kappa_1=\sqrt{1-\chi_1^2}$. $\Omega$ is the angular velocity of the tidal field of the companion BH with mass $m_2$ and spin $\chi_2$, given by~\cite{Chatziioannou:2016kem}
\begin{equation}
    \Omega = \frac{v^3}{M}\left[1-\eta v^2 + \frac{1}{2}(\Tilde{\chi}-\Bar{\chi_1})v^3+\mathcal{O}(v^4)\right]\,,
\end{equation}
where $M=m_1+m_2$, $v=(\pi Mf)^{1/3}$ is the PN expansion parameter, and $\eta=m_1m_2/M^2$ is the symmetric mass ratio. Defining $X_1=m_1/M$ and $X_2=m_2/M$, we can write
\begin{equation}
\begin{aligned}
      \Tilde{\chi}&=(2X_1^2+3\eta)\chi_1+(2X_2^2+3\eta)\chi_2\,,\\
      \Bar{\chi_1}&=X_1(1+X_1)\chi_1+3\eta \chi_2\,.
\end{aligned}
\end{equation}
The coefficient $C_v$ in Eq.~\eqref{eq:dmdt} is given by
\begin{equation}\label{eq:cx}
\begin{aligned}
    C_v=&-\frac{16}{5}m_1^2X_1^2\eta^2(1 + \kappa_1)v^{12}\bigg[1 + 3\chi_1^2 \\ ~&\left. + \frac{1}{4}\left\{3(2 + \chi_1^2) \right.\left.
    + 2X_1(1 + 3\chi_1^2)(2 + 3X_1)\right\}v^2 \right.\\ ~&\left.
    + \left\{\frac{1}{2}(-4 + 3\chi_1^2)\chi_2 - 2X_1(1 + 3\chi_1^2)\left[X_1(\chi_1+\chi_2)  \right.\right.\right.\\ ~&\left.\left. \left.
    + 4B_2(\chi_1)\right] + X_1\left[-\frac{2}{3}(23 + 30\kappa_1)\chi_1  \right.\right. \right.\\ ~&\left. \left.
    + (7 - 12\kappa_1)\chi_1^3 + 4\chi_2 + \frac{9\chi_1^2\chi_2}{2}\right]\right\}v^3\bigg]\,,
\end{aligned}
\end{equation}
where 
\begin{equation}
    B_2(\chi_1)=\text{Im}\left[\psi^{(0)}\left(3+i\frac{2\chi_1}{\kappa_1}\right)\right]\,,
\end{equation}
$\psi^{(0)}(x)$ being the digamma function. The brackets in Eq.~\eqref{eq:dmdt} denote orbit average.

Expression for the rate of change of $m_2$ can be found from the above equations by exchanging $m_1\leftrightarrow m_2$ and $\chi_1\leftrightarrow \chi_2$.

\section{Modeling Horizon fluxes from NR Simulations}

For each simulation of BBHs, the SXS catalog of NR waveforms provides data for the properties of the apparent horizons, capturing their evolution throughout the inspiral and merger. The data contains evolution of the Christodoulou mass, horizon area and the spin of each BH. Time (in units of the total mass $M$) when the common apparent horizon starts to form is also specified, and defines the moment of merger of the BHs. We utilize these data for modeling the fluxes due to TH of BHs in the late inspiral to the merger regime. 

In this work, we perform modeling of only nonspinning binaries, mainly for building the simplest model. Magnitude of the rate of change of BH mass ($\dd m_i/\dd t$, $i=1,2$) is extremely small at lower frequencies, and the horizon data from the highest resolution available start to resolve $\dd m_i/\dd t$ only after $Mf\sim 0.03$, $f$ being the GW frequency. Even afterwards, there are several glitches in the data from many simulations. Moreover, even for nonspinning binaries, fluxes for the lighter BH are significantly smaller than the heavier one, and ill-resolved even close to the merger. To circumvent this issue, we assume that the two objects are identical in nature so that the horizon parameters are equal, $H_1=H_2=H$. This choice exempts us from modeling $\dd m_1/\dd t$ and $\dd m_2/\dd t$ separately, and allows modeling their sum instead.

\subsection{Modeling the Fluxes}

To model the horizon fluxes for nonspinning binaries, we first identify the leading order behaviour from the analytical PN expression:
\begin{equation}
    \left<\dv{m_1}{t}\right>_{\rm PN}=\frac{32}{5}\eta^2X_1^4v^{18}\,.
\end{equation}
So the total mass change of the two BHs is
\begin{equation}\label{eq:dmdt-PN}
   \left<\dv{M}{t}\right>_{\rm PN}= \left<\dv{m_1}{t}+\dv{m_2}{t}\right>_{\rm PN}=\frac{32}{5}\eta^2(X_1^4+X_2^4)v^{18}\,.
\end{equation}

To obtain a phenomenological fit with NR data, we consider an ansatz augmented with pseudo-PN terms:
\begin{equation}\label{eq:dmdt-NR}
    \left<\dv{M}{t}\right>_{\rm NR}=A(X)\eta^2 v^{18}\frac{1+\eta(a_1v+a_2v^2+a_3v^3)}{1+a_4\eta v}\,,
\end{equation}
with $A(X)=(32/5)(X_1^4+X_2^4)$. $\{a_1,a_2,a_3,a_4\}$ are four phenomenological parameters that capture the fluxes at high frequencies. The ansatz also ensures that for low-frequency expansions, the PN expression in Eq.~\eqref{eq:dmdt-PN} is recovered. However, we note that the ansatz in Eq.~\eqref{eq:dmdt-NR} is not unique. In Table~\ref{tab:NR data} we list the NR simulations used for calibrating the phenomenological parameters. We fit the NR data with Eq.~\eqref{eq:dmdt-NR} in the frequency range $Mf\in [0.03,0.04]$. 

After estimating the values of the aforementioned parameters for each of the NR datasets, we fit the individual parameters with a second-order polynomial in $\eta$:
\begin{equation}\label{eq:aifit}
    a_i=\sum_{j=1}^3\beta_{ij}\eta^{j-1}\,.
\end{equation}

\begin{figure}[h]
    \centering
    \includegraphics[width=\linewidth]{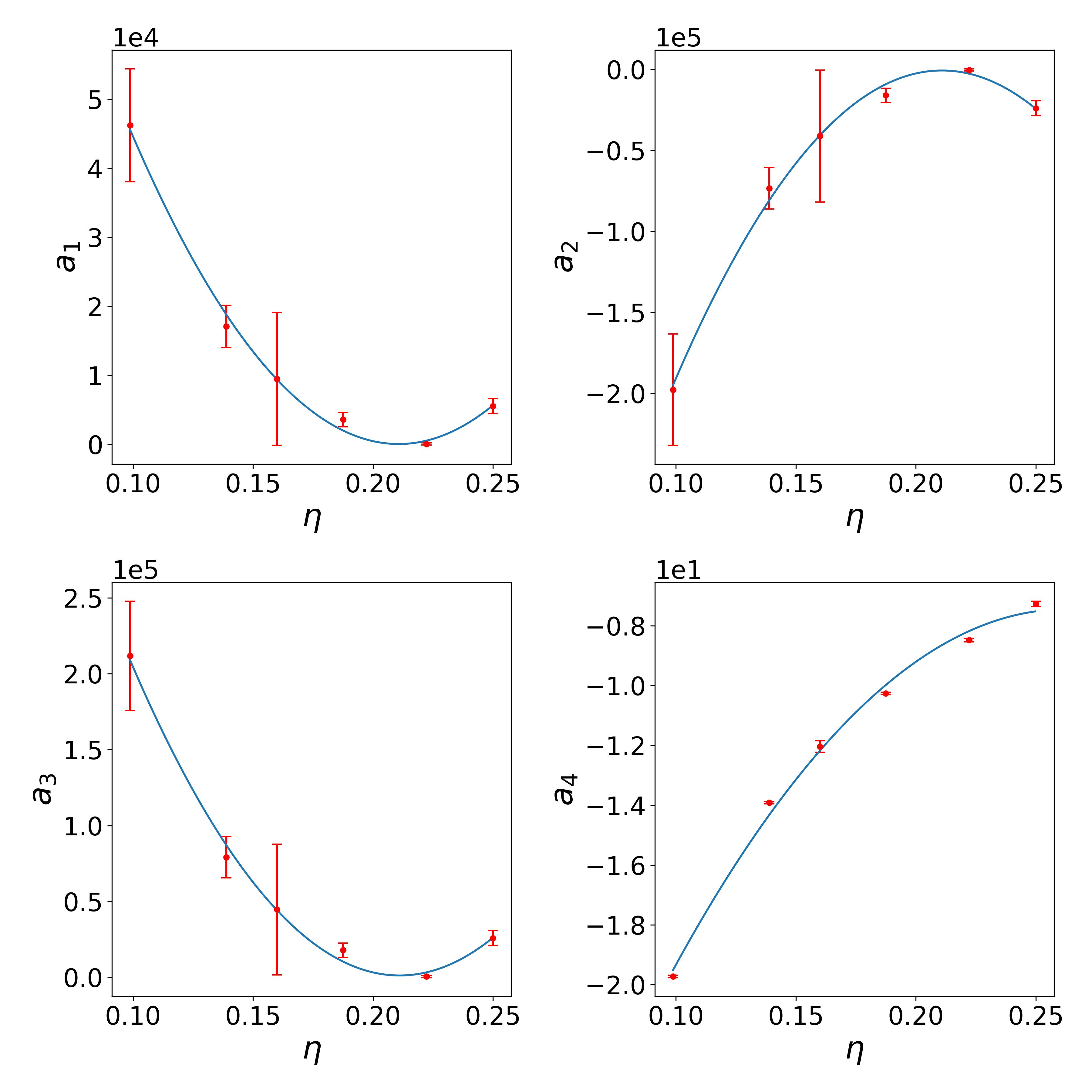}
    \caption{Fits for the phenomenological parameters defined in Eq.~\eqref{eq:dmdt-NR} with the polynomials in Eq.~\eqref{eq:aifit}. The errorbars denote the 1-$\sigma$ errors in estimating each parameter from individual fits.}
    \label{fig:a}
\end{figure}

Figure~\eqref{fig:a} shows the individual fits for the parameters $a_i$ and the fitted polynomials.

\begin{table}[h]
    \centering
    \begin{tabular}{p{0.05\textwidth}p{0.2\textwidth}p{0.15\textwidth}}
        \hline
        \vspace{0.3mm}\\
        \# & Simulation Label & $q=m_1/m_2$ \\
        \vspace{0.3mm}\\
        \hline
        \vspace{0.3mm}\\
        1 & SXS:BBH:0001 & 1\\
        2 & SXS:BBH:0169 & 2\\
        3 & SXS:BBH:0030 & 3\\
        4 & SXS:BBH:0167 & 4\\
        5 & SXS:BBH:0056 & 5\\
        6 & SXS:BBH:0063 & 8\\
        \vspace{0.3mm}\\
        \hline
    \end{tabular}
    \caption{List of NR simulations from the SXS catalog used to calibrate the horizon fluxes.}
    \label{tab:NR data}
\end{table}


The matrix $\beta_{ij}$, with $i$ and $j$ ranging from 1 to 4 and 1 to 3 respectively, is 

\begin{equation}\label{eq:beta}
    \beta_{ij} = \begin{pmatrix}
                161351.05 & -1531185.53 & 3633803.55\\
                -688223.93 & 6520488.6 & -15460193.18\\
                734430.88 & -6946505.25 & 16454233.38\\
                -38.45 & 236.23 & -450
                \end{pmatrix}\,.
\end{equation}


\begin{figure}
     \centering
     \begin{subfigure}[b]{0.49\textwidth}
         \centering
         \includegraphics[width=\textwidth]{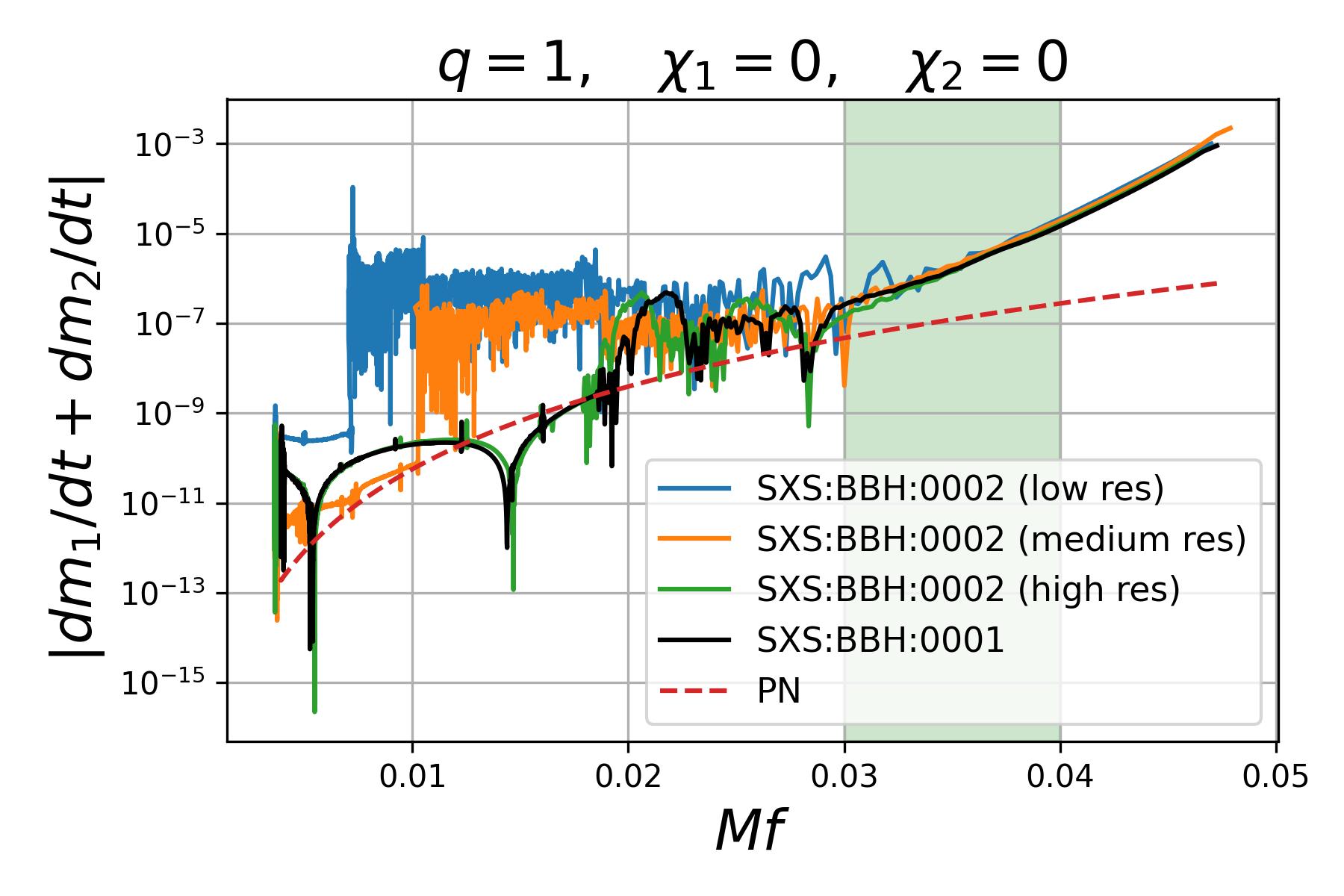}
         \caption{}
     \end{subfigure}
     \hfill
     \begin{subfigure}[b]{0.49\textwidth}
         \centering
         \includegraphics[width=\textwidth]{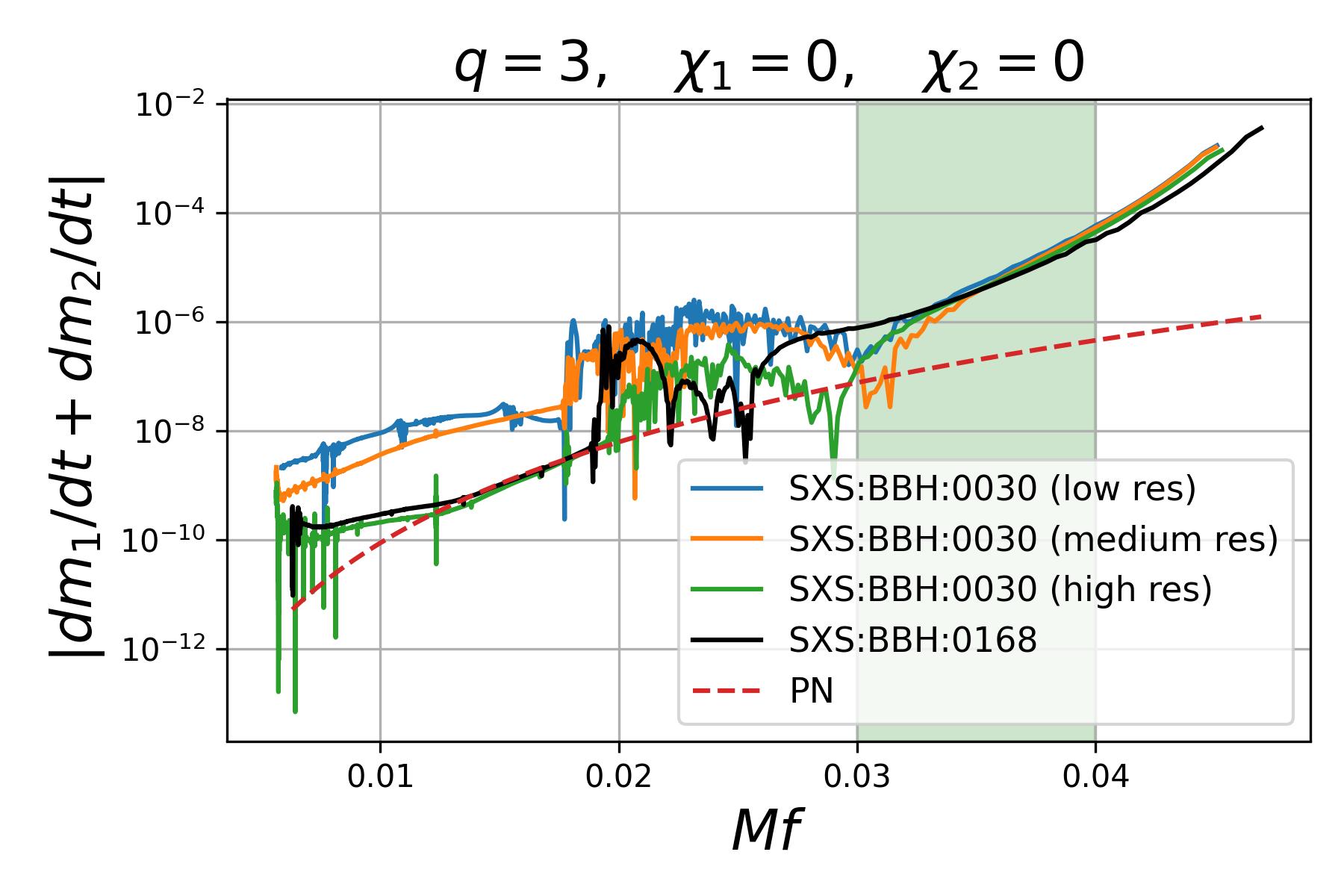}
         \caption{}
     \end{subfigure}
     \hfill
     \begin{subfigure}[b]{0.49\textwidth}
         \centering
         \includegraphics[width=\textwidth]{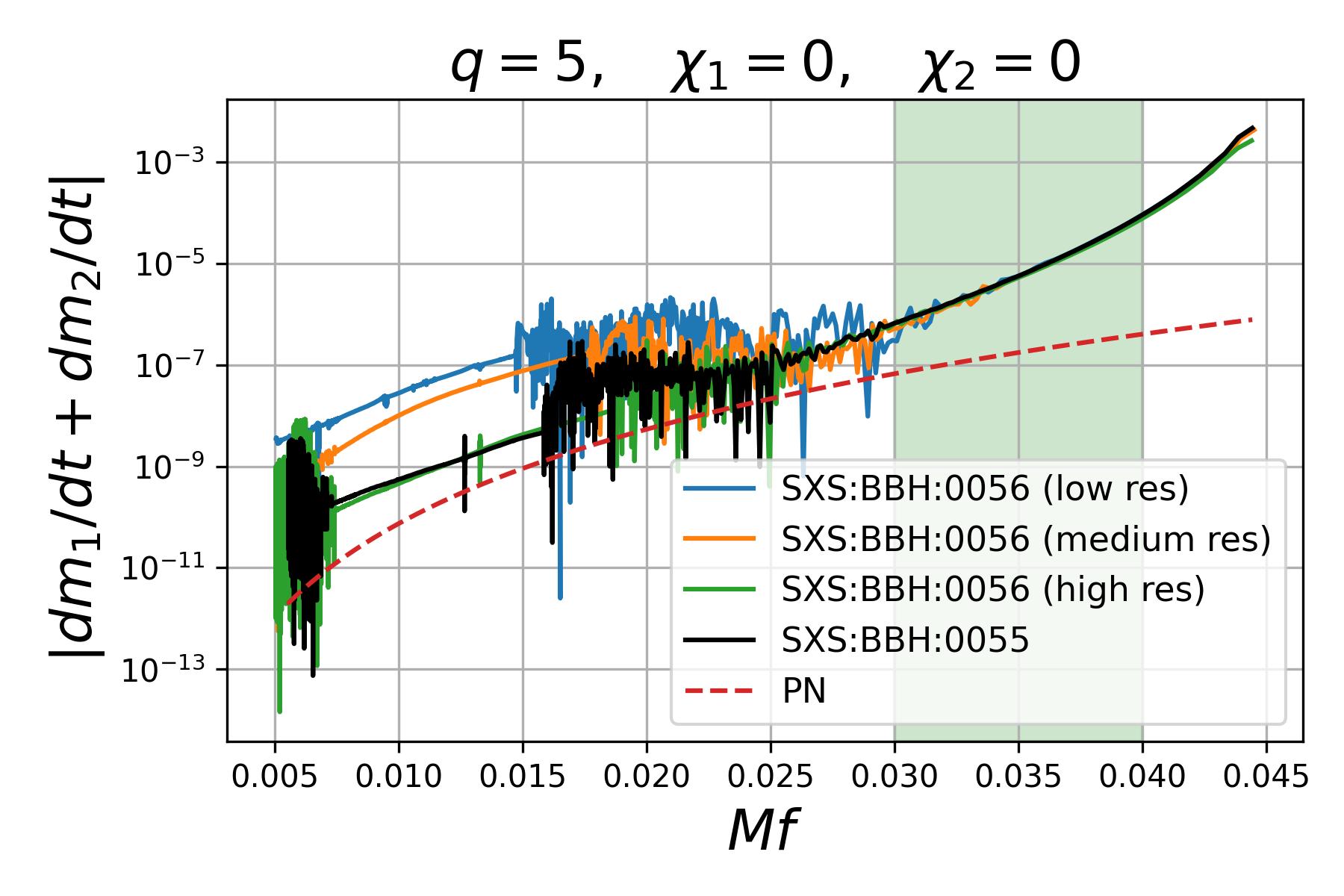}
         \caption{}
     \end{subfigure}
     \caption{Comparison of the accuracy of horizon data from various NR simulations of the SXS catalog. In each figure, we compare three different resolutions of one simulation and the highest available resolution of another simulation with the same intrinsic parameters. PN expressions corresponding to the same configurations are plotted for reference. We show this for three nonspinning binaries with the mass ratios ($q=m_1/m_2$) reported above each plot. The green shaded regions correspond to the frequency interval chosen for fitting the NR data with the ansatz in Eq.~\eqref{eq:dmdt-NR}. Within these regions, data from different simulations are quite close to each other, implying that they are sufficiently resolved. Among these simulations, SXS:BBH:0001, SXS:BBH:0030 (high res) and SXS:BBH:0056 (high res) are used in this chapter for modeling purposes. } 
    \label{fig:NRaccuracy}
\end{figure}

\begin{figure}
     \centering
     \begin{subfigure}[b]{0.49\textwidth}
         \centering
         \includegraphics[width=\textwidth]{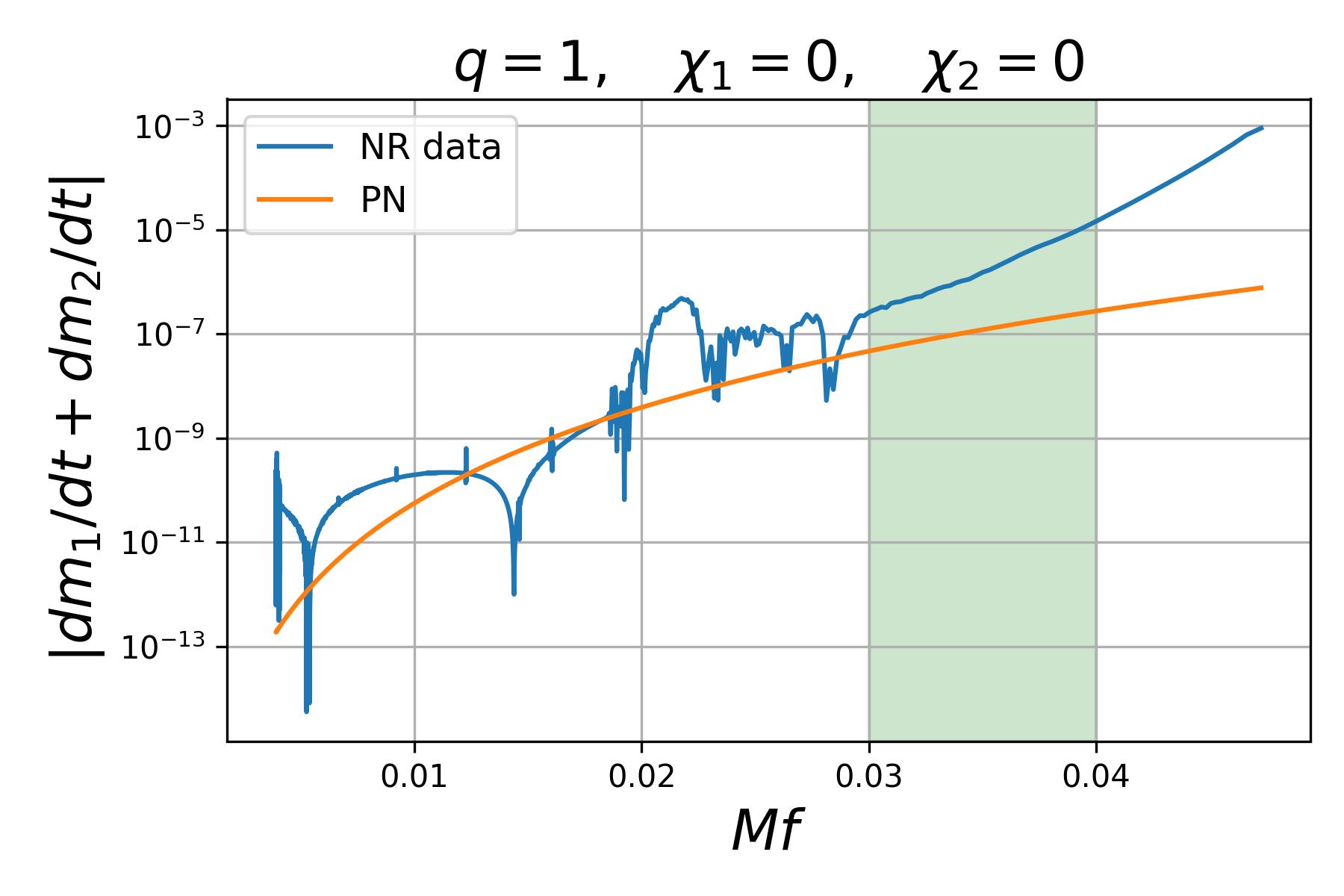}
         \caption{}
     \end{subfigure}
     \hfill
     \begin{subfigure}[b]{0.49\textwidth}
         \centering
         \includegraphics[width=\textwidth]{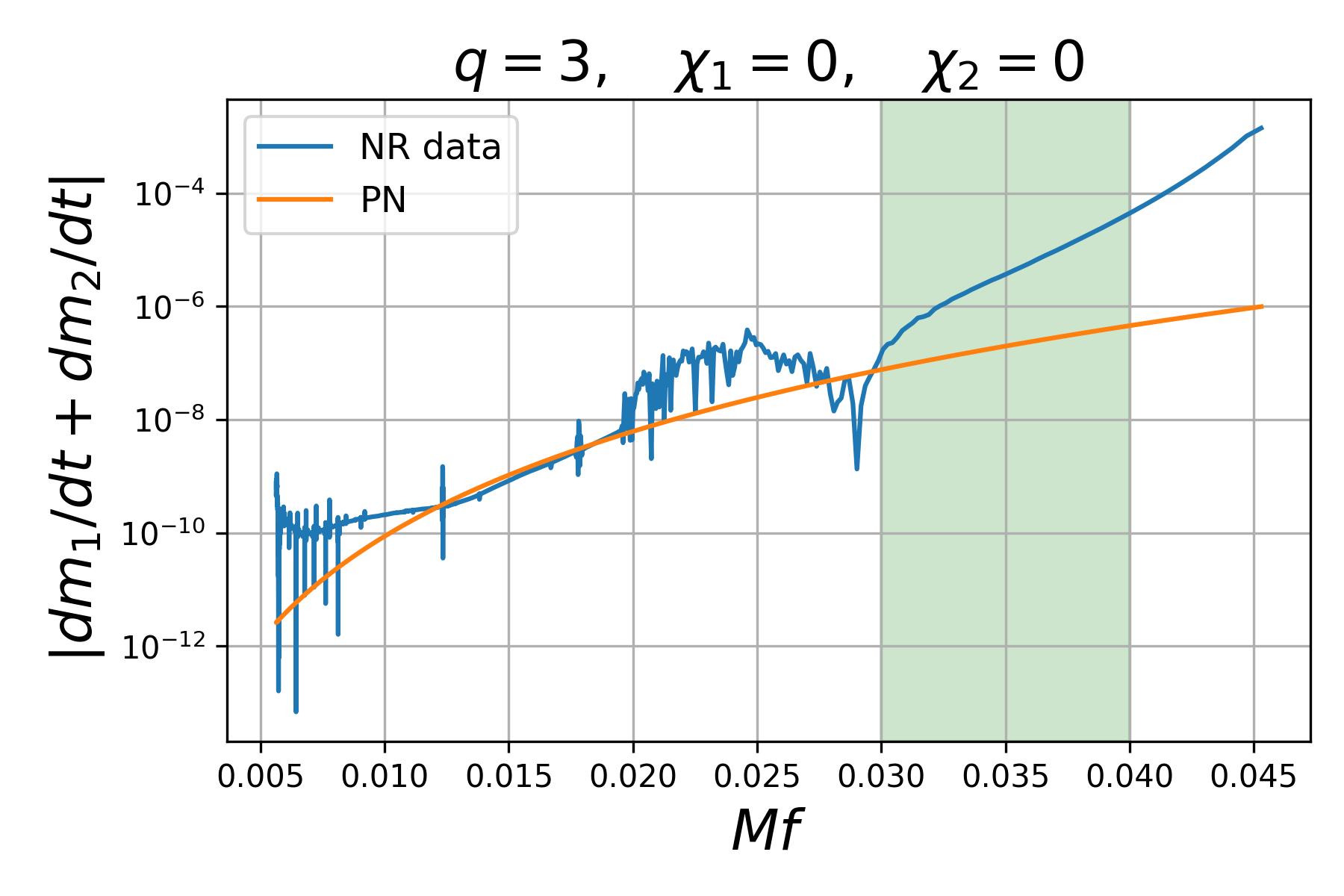}
         \caption{}
     \end{subfigure}
     \hfill
     \begin{subfigure}[b]{0.49\textwidth}
         \centering
         \includegraphics[width=\textwidth]{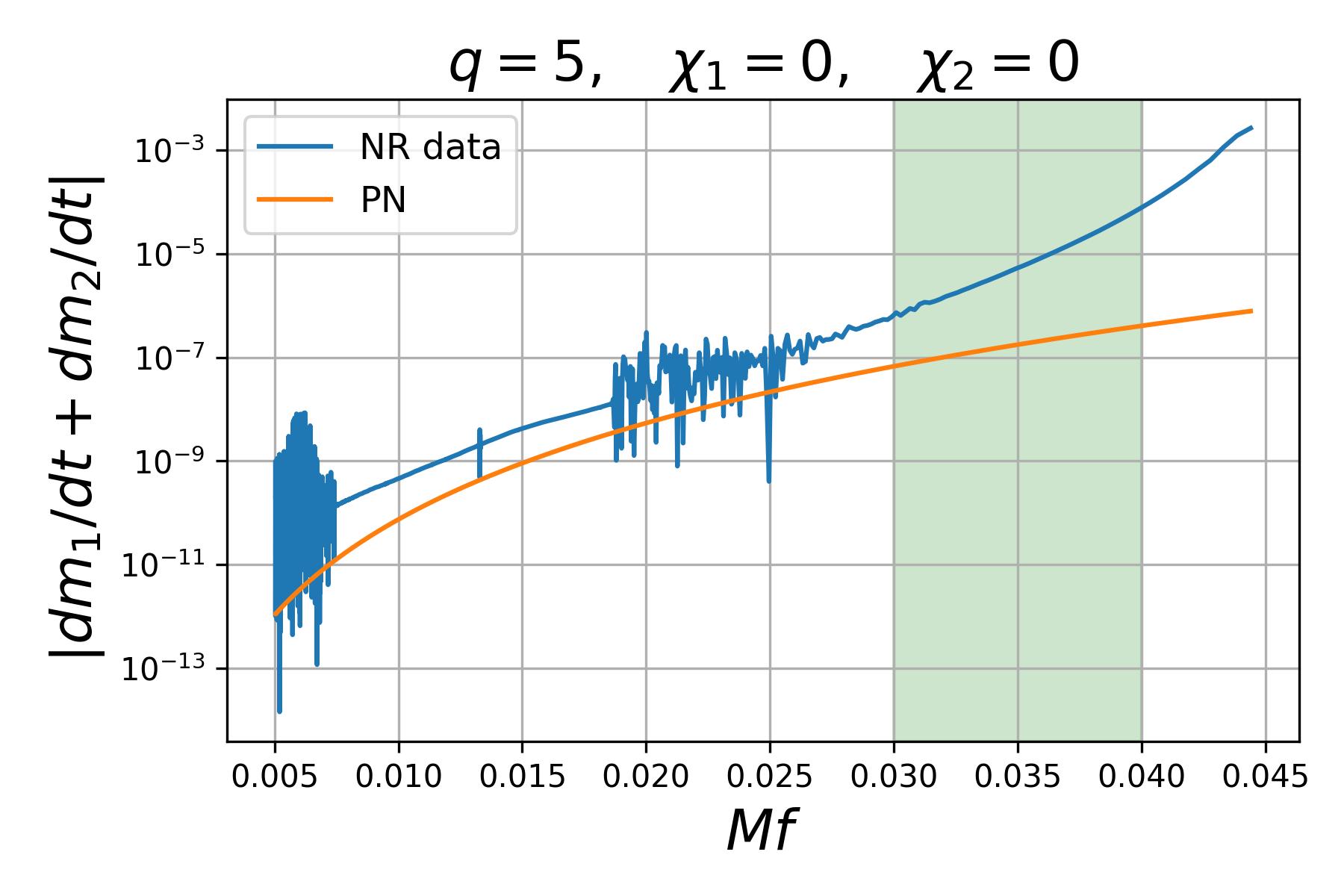}
         \caption{}
     \end{subfigure}
     \hfill
     \begin{subfigure}[b]{0.49\textwidth}
         \centering
         \includegraphics[width=\textwidth]{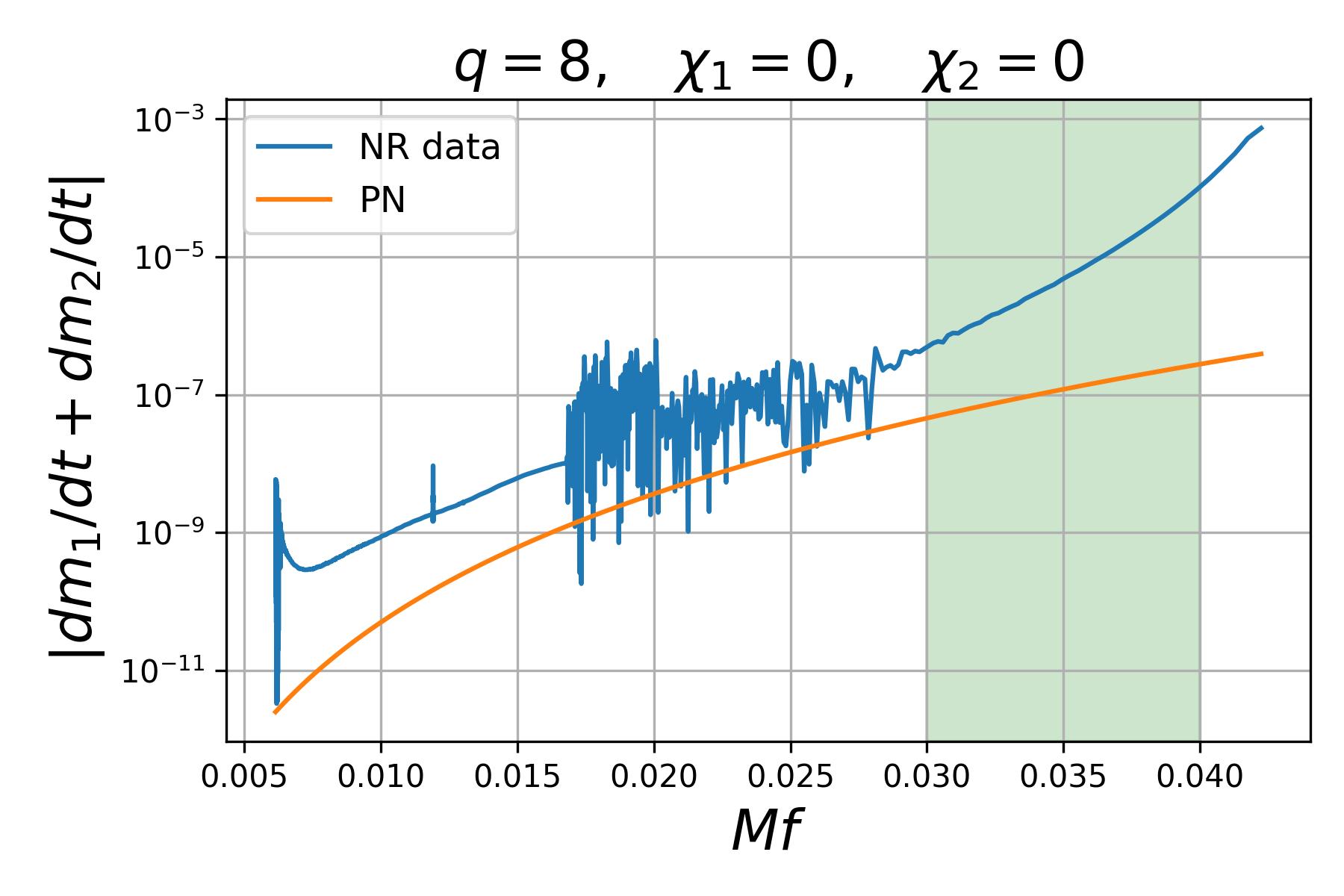}
         \caption{}
     \end{subfigure}
    \caption{This figure shows the total tidal heating fluxes experienced by the binary black holes as demonstrated from NR simulations, and the PN expressions corresponding to the same configurations. We show this for four nonspinning binaries with the mass ratios ($q=m_1/m_2$) reported above each plot. The green shaded regions correspond to the frequency interval chosen for fitting the NR data with the ansatz in Eq.~\eqref{eq:dmdt-NR}.} 
    \label{fig:fluxes}
\end{figure}

\begin{figure}
     \centering
     \begin{subfigure}[b]{0.49\textwidth}
         \centering
         \includegraphics[width=\textwidth]{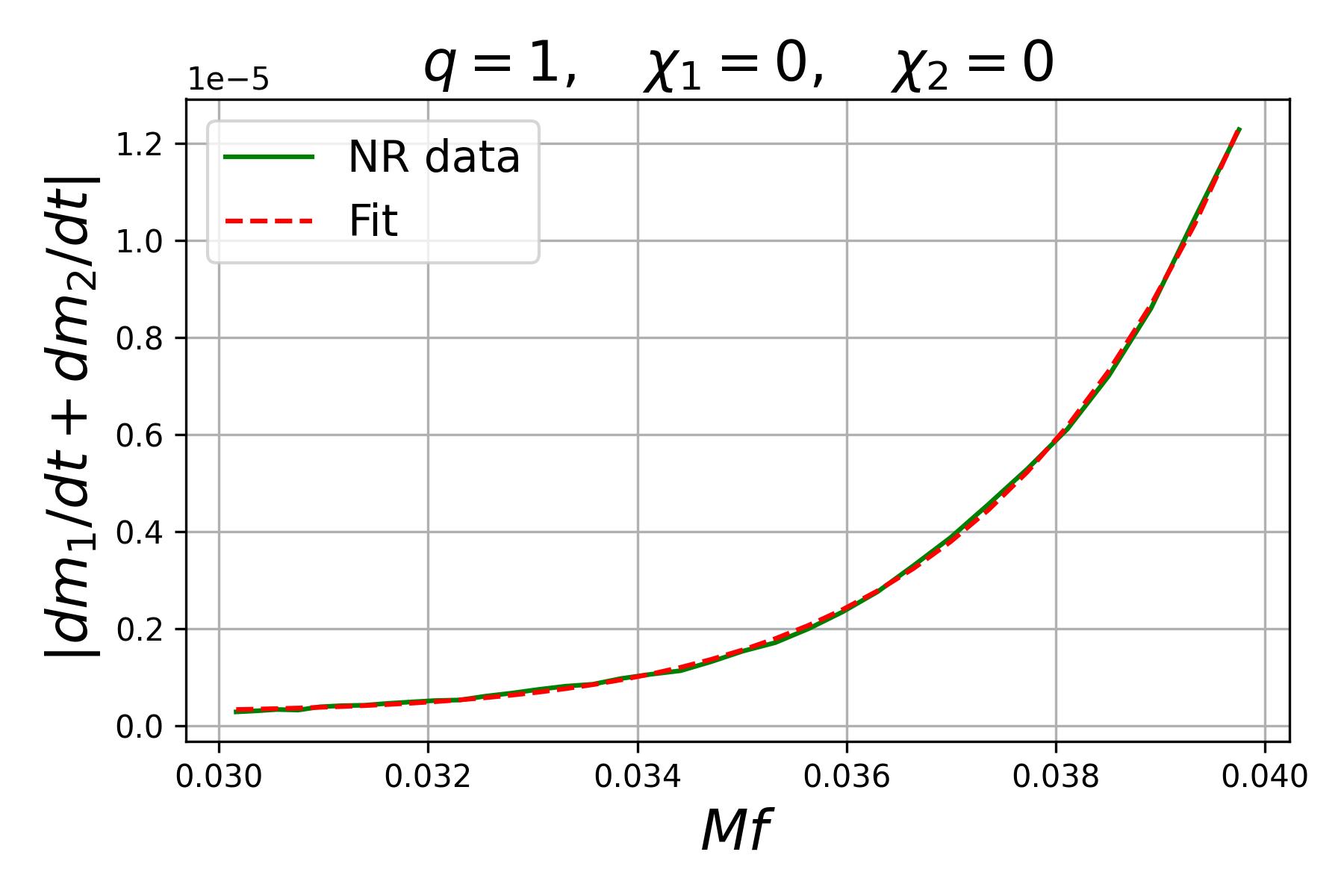}
         \caption{}
     \end{subfigure}
     \hfill
     \begin{subfigure}[b]{0.49\textwidth}
         \centering
         \includegraphics[width=\textwidth]{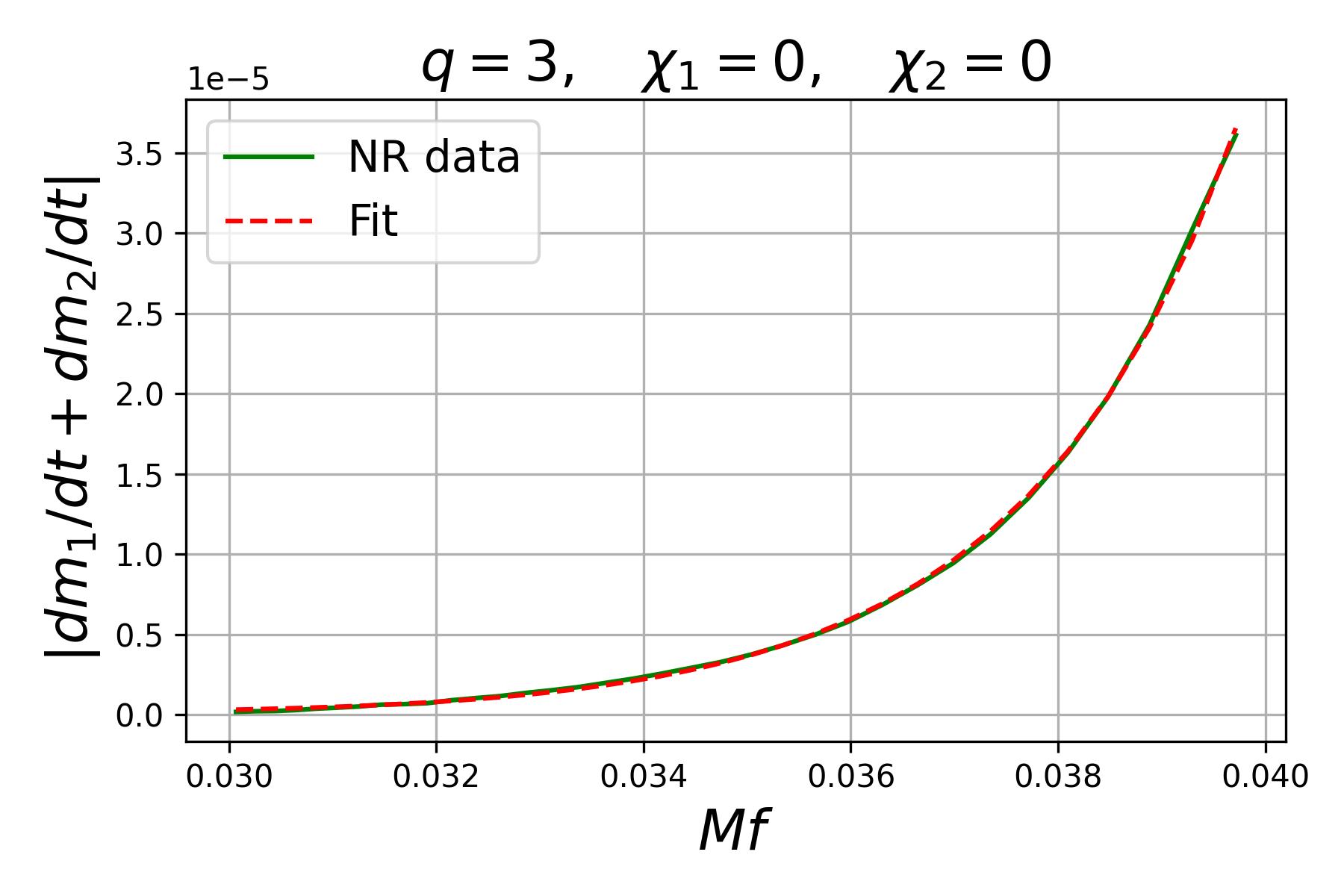}
         \caption{}
     \end{subfigure}
     \hfill
     \begin{subfigure}[b]{0.49\textwidth}
         \centering
         \includegraphics[width=\textwidth]{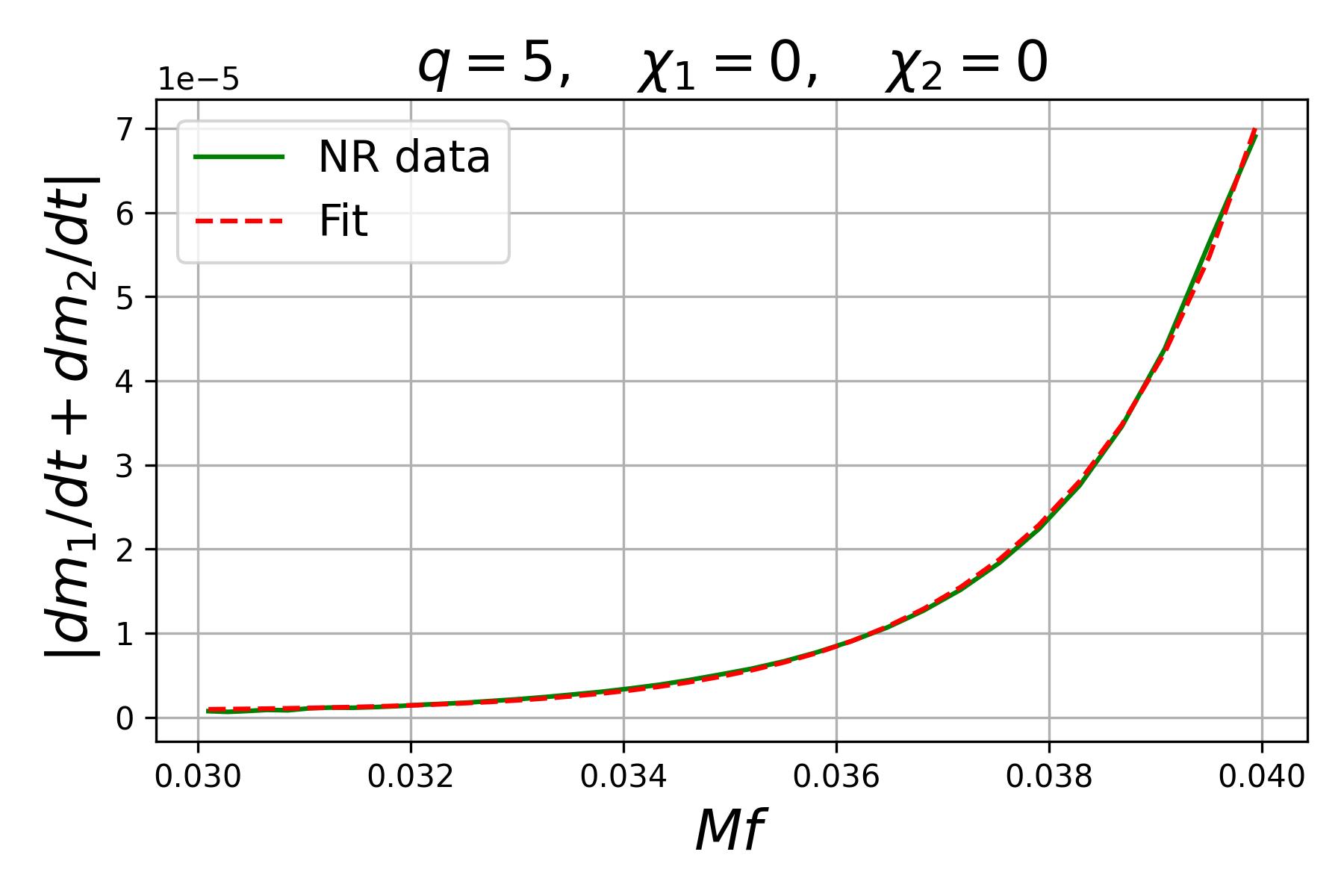}
         \caption{}
     \end{subfigure}
     \hfill
     \begin{subfigure}[b]{0.49\textwidth}
         \centering
         \includegraphics[width=\textwidth]{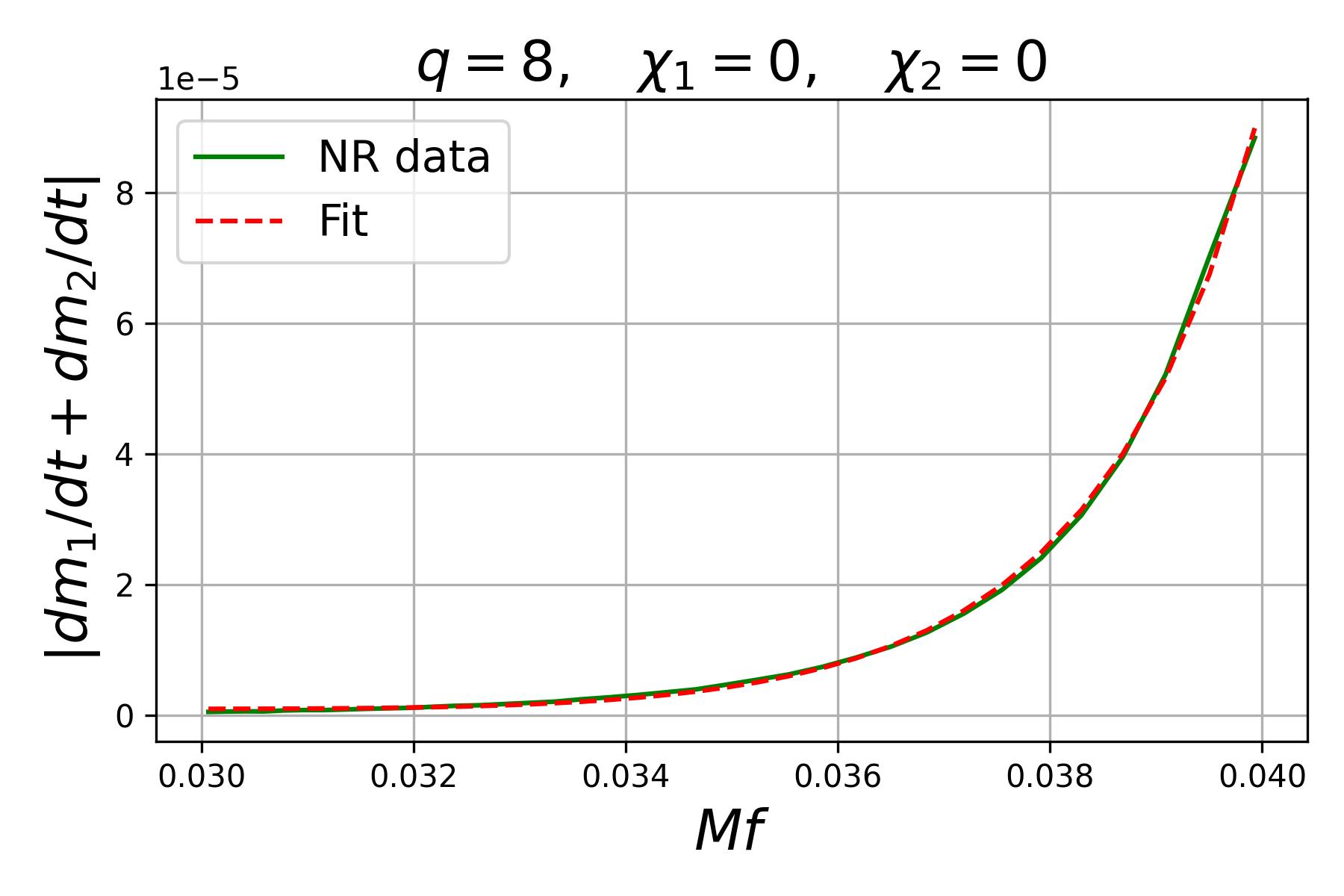}
         \caption{}
     \end{subfigure}
    \caption{The fitted models shown for each of the binaries in Fig.~\ref{fig:fluxes}.} 
    \label{fig:fluxes fit}
\end{figure}

\subsection{Phase Correction in the Frequency Domain}

To find the phase correction due to TH in the frequency domain, we use the energy balance condition
\begin{equation}
    -\Dot{E}_{\rm orb} = \Dot{E}_{\rm \infty} + H\left(\dv{m_1}{t}+\dv{m_2}{t}\right)\,,
\end{equation}
where $\Dot{E}_{\rm orb}$ and $\Dot{E}_{\rm \infty}$ are the rate of change of the orbital energy and the GW flux to the infinity, respectively. Under the stationary phase approximation (SPA), the orbital binding energy $E_{\rm orb}$ can be expressed as an expansion in the PN parameter $v$. We use the expression for $E_{\rm orb}(v)$ and the energy flux to the infinity up to 2PN order~\cite{Blanchet:2013haa}\footnote{In Ref.~\cite{Blanchet:2013haa}, the PN parameter is taken to be $x=v^2$.}:
\begin{equation}
\begin{aligned}
    E_{\rm orb}(v)=&-\frac{1}{2}\eta M v^2\left[ 1 - \left(\frac{3}{4}+\frac{1}{12}\eta\right)v^2 \right.\\& \left.+\left(-\frac{27}{8}+\frac{19}{8}\eta -\frac{1}{24}\eta^2\right)v^4 \right]\,,
\end{aligned}    
\end{equation}
and
\begin{equation}
    \begin{aligned}
        \Dot{E}_{\rm \infty}(v)=&\frac{32}{5}\eta^2 v^{10}\left[1+\left(-\frac{1247}{336}-\frac{35}{12}\eta\right)v^2+4\pi v^3 \right.\\&\left.+ \left(-\frac{44711}{9072}+\frac{9271}{504}\eta+\frac{65}{18}\eta^2\right)v^4\right]\,.
    \end{aligned}
\end{equation}

For nonspinning binaries, the effects of TH first appears at the 4PN order. The NR fits provide higher order terms. Up to 6PN order, the frequency-domain dephasing due to TH can be expressed as
\begin{equation}\label{eq:dpsi}
    \delta\Psi_{\rm TH}(v)=\frac{3}{128\eta v^5}H\left(\sum_{n=8}^{12}\psi_nv^n+\psi_{8,l}v^8\log(v)\right)\,.
\end{equation}
Assuming the form of $\dd M/\dd t$ in Eq.~\eqref{eq:dmdt-NR}, the coefficients $\psi_n$ are
{\allowdisplaybreaks
\begin{align}
    \psi_8=&\frac{25}{36}A(X)v^8\,,\label{eq:psiPN}\\
    \psi_{8,l}=&-3\psi_8\,,\label{eq:psiPNl}\\
    \psi_9=&-\frac{25}{16}A(X)(a_1-a_4)\,,\\
    \psi_{10}=&-\frac{5}{1344}A(X)\{168(a_2-a_1a_4+a_4^2)+995+952\eta\}\,,\\
    \psi_{11}=&-\frac{25}{12096}A(X)[(995+952\eta+168\eta^2)(a_1-a_4)\nonumber\\&+168(a_3-a_2a_4)-1344\pi]\,,\\
    \psi_{12}=&-\frac{25}{113799168}A(X)[30400075-1016064a_3a_4 \nonumber\\&+6017760a_4^2+1016064a_4^4+8128512a_4\pi\nonumber\\&+25702488\eta+5757696a_4^2\eta+17477712\eta^2\nonumber\\&+6048a_2(995+168a_4^2+952\eta)\nonumber\\&-6048a_1(995a_4+168a_4^3+1344\pi+952a_4\eta)]\,.\label{eq:psi}
\end{align}
}

Figure~\ref{fig:fluxes} shows, for four configurations of the mass ratio $q=\{1,3,5,8\}$, how the NR data compare with the PN term. Evidently, the PN expression and the NR results deviate significantly at high frequencies, and at lower frequencies we expect the full model to converge to the PN expressions. For all the datasets of nonspinning binaries, the fluxes start to be resolved after $Mf\sim 0.03$, so we choose this value to be the lowest frequency to start the fitting. The upper frequency $Mf=0.04$ is chosen by identifying the lowest value of the merger frequencies of these binaries. In Fig.~\ref{fig:fluxes fit} we show the fitted models corresponding to the BBHs shown in Fig.~\ref{fig:fluxes}.

Unfortunately, since the modeling of $\dd M/\dd t$ has been done within a short frequency range where the NR simulations are able to resolve the mass change of BHs, the model behaves extremely poorly at lower frequencies. At the zero-frequency limit the model should, in principle, reduce to the PN expression; but within our frequency range of interest there are orders of magnitude difference between the 4PN term and the full model. The complete approximant spanning the inspiral and merger phases can be prepared by dividing the frequency range into parts, and forcing the dephasing to coincide with the PN result below the ISCO frequency. The PN expression is essentially the ansatz in Eq.~\eqref{eq:dmdt-NR} with $a_i=0$. For $Mf< 0.022$, we can express the dephasing by keeping only the PN term in Eq.~\eqref{eq:dmdt-PN}, and for $Mf>0.032$ (starting slightly above the start of the fitting range) we can use the expressions in Eqs.~\eqref{eq:dpsi} to~\eqref{eq:psi}. In the range $0.022\leqslant Mf\leqslant 0.032$, one way to model the dephasing can be to use the NR-fitted formula, with the condition that the coefficients $\{a_i\}$ are linear functions of $Mf$, described as 
\begin{equation}
    a_i(Mf)=a_i^0\left(\frac{Mf-0.022}{0.01}\right)\,,
\end{equation}
or equivalently
\begin{equation}
    a_i(v)=a_i^0\left(\frac{v^3/\pi-0.022}{0.01}\right)\,,
\end{equation}
where $\{a_i^0\}$ are the values from Eqs.~\eqref{eq:aifit} and \eqref{eq:beta}. These functions ensure that $a_i=0$ at $Mf=0.022$, connecting the dephasing with the PN term up to 4PN order given by Eqs.~\eqref{eq:psiPN} and~\eqref{eq:psiPNl}; and $a_i=a_i^0$ at $Mf=0.032$ , so the dephasing coincides with the full model with constant parameters estimated from the fits. The complete phase model in the frequency domain can be built by imposing $C^{(1)}$ continuity at the two interfaces exploiting the overall phase and timeshift freedom. 



\section{Discussion}

In this chapter we have presented a model of the horizon fluxes of black holes, based on NR data, in the late inspiral to the merger regime where they are substantially higher compared to the early inspiral phase described by the PN formalism. The model presented here represents nonspinning binaries, and is calibrated with data from the SXS NR catalog up to mass ratio of $q=8$. We calculate the dephasing in the frequency domain GW waveform due to tidal heating using this model. This dephasing expression can be used directly on a waveform containing TH effects in the phase, e.g. \texttt{IMRPheonmD\_Horizon}, to enable the horizon parameters to act as BH discriminators in binary systems. However, the current modeling of TH only allows the horizon parameters to be identical for both the component objects, thereby making it possible to discriminate systems like BBHs against binary ECOs of the same species. The model presented here, along with the one for BBHs presented in Chapter~\ref{Chapter4}, can be used for estimation of the horizon parameters for heavier binaries, especially using the strong-gravity regime.

An immediate extension of this work would be to extend the model for spinning binaries, allowing for a larger parameter space to probe for BHs against ECOs. In that case, the ansatz for the fluxes has to be modified to accommodate spin-dependent terms. While modeling of spinning binaries would be more involved than the steps detailed here, the upside would be an improvement in the estimation of the horizon parameters due to higher TH fluxes with increasing spins. Additionally, advances in NR simulations in the future have the potential to resolve the TH fluxes more accurately, thereby making it more viable for the techniques presented here to model them throughout a larger frequency range.

\clearpage
\chapter{Conclusion}

Direct detection of gravitational waves by LIGO and Virgo has revolutionized our knowledge of the Cosmos. After decades of efforts and research, the gravitational-wave detectors were successful in detecting numerous events of compact binary coalescence. One of the important possibilities that have been opened up by this field of research is to quantify and test for the existence of horizons. Black holes are a robust prediction of Einstein's General Relativity. Although widely accepted to have a pervasive existence in the Universe, the importance of testing for their evidence quantitatively can hardly be overlooked. There are several species of compact objects hypothesized to exist, from well-motivated theoretical points of view, that can replace black holes or coexist with them within a larger family of dark compact objects. Gravitational-wave observations are a promising way to investigate this quantitatively. The primary objective of this thesis has been to study how well the phenomenon of tidal heating can act as a discriminator for black holes in a binary, and to build a complete inspiral-merger-ringdown model  that can differentiate black holes from other objects using this effect. Deviation from the black-hole scenario can be quantified by introducing two parameters, one for each component of a binary, that account for tidal heating.


In Chapter~\ref{Chapter3} we have explored how well one can measure the two tidal heating parameters in the future ground-based gravitational-wave detectors Einstein Telescope and Cosmic Explorer as well as in the second generation detectors Advanced LIGO and Advanced Virgo. These parameters account for the flux of energy and angular momentum into a black hole associated with tidal heating. The parameters take values of unity for a binary black hole system, and for other objects, they assume fractional positive values. We find the ranges in the binaries' total mass that are most sensitive for estimating these parameters, and show how the measurabilities vary at different regions of the parameter space. This work benchmarks the capabilities of the current and future ground-based gravitational-wave detectors to measure the tidal heating parameters. We also demonstrate that the said parameters have negligible covariances, especially in the third-generation detectors.

Since tidal heating leaves stronger effects as the tidal fields grow in an inspiral, one expects the late inspiral phase to hold the maximum potency to discriminate black holes in comparable-mass binaries. As a step towards building a framework for the goal of utilizing the late inspiral regime, we build an inspiral-merger-ringdown gravitational waveform model in Chapter~\ref{Chapter4} that incorporates the tidal heating effects in a black-hole binary. The work has inherent benefits of alleviating the waveform systematics due to neglecting tidal heating in the early inspiral phase. The model, as we show, is faithful and accurate.

Numerical relativity simulations solve the full Einstein's field equations numerically, and thereby undergo the effects of tidal heating. Data from these simulations predict orders-of-magnitude higher fluxes than the analytical estimates. In Chapter~\ref{Chapter5}, we build a model of tidal heating fluxes from numerical relativity data for nonspinning binaries. We also calculate the dephasing in the frequency domain due to these fluxes. The model presented in this chapter can be used directly on the inspiral-merger-ringdown model of binary black holes that we built with tidal-heating corrections, to search for horizonless alternatives of black holes rigorously.

With the current and future detectors, the field of gravitational-wave astronomy will continue to grow, and many more interesting events with far-reaching implications are yet to be unveiled. The abundance of data from these observations serves as a valuable testbed for astronomy and astrophysics. There are aspects of standard-model physics that still remain elusive, and possible resolutions have been proposed by invoking physics beyond the standard model -- which in turn hints at the existence of exotic compact objects.
The framework developed in this thesis has the potential to make tidal heating a robust discriminator for black holes when the late inspiral phase is accounted for, aiding the quest to confirm or rule out the existence of black-hole mimickers. Regardless of the precise nature of the outcome, exciting new physics can come to light.

\clearpage

\appendix

\renewcommand{\chaptername}{Appendix}
\addtocontents{toc}{\protect\renewcommand{\protect\cftchappresnum}{\chaptername~}}


\chapter{Post-Newtonian inspiral phase}
\label{AppendixA}
In this appendix, we write down the frequency-domain post-Newtonian phase (without tidal heating) used to calibrate the model \texttt{IMRPhenomD\_Horizon}, as discussed in Sec.~\ref{sec:inspiral phase model}.

The 3.5PN frequency-domain phase (without TH) can be written as
\begin{equation}
    \Psi_{\rm TF2}(f) = 2\pi f t_c - \phi_c -\frac{\pi}{4} + \frac{3}{128\eta v^5}\sum_{n=0}^7 \psi_{(n)} v^n\,,
\end{equation}
where $v=(\pi Mf)^{1/3}$.

The coefficients $\{\psi_{(n)}\}$ ($n$=0-7) are given by~\cite{Arun:2008kb,Mishra:2016whh,Wade:2013hoa},
    

\begin{align}
    \psi_{(0)} =~& 1\,,\\
    \psi_{(1)} =~& 0\,,\\
    \psi_{(2)} =~& \frac{3715}{756} + \frac{55}{9}\eta\,,\\
    \psi_{(3)} =~& -16\pi + \frac{113}{3}\delta\chi_a + \left(\frac{113}{3} - \frac{76}{3}\eta\right)\chi_s\,,\\
    \psi_{(4)} =~& \frac{15293365}{508032} + \frac{27145}{504}\eta+ \frac{3085}{72}\eta^2 - \frac{405}{4}\delta\chi_a\chi_s + \left(-\frac{405}{8} + 200\eta\right)\chi_a^2\nonumber \\& + \left(-\frac{405}{8} + \frac{5}{2}\eta\right)\chi_s^2\,, \\
    \psi_{(5)} =~& \left[\frac{38645\pi}{756} - \frac{65\pi}{9}\eta - \left(\frac{732985}{2268} - \frac{24260}{81}\eta - \frac{340}{9}\eta^2\right)\chi_s\nonumber\right. \\& - \left.\left(\frac{732985}{2268} + \frac{140}{9}\eta\right)\delta\chi_a\right]\left(1+3\ln{v}\right)\,,\\
    \psi_{(6)} =~& \frac{11583231236531}{4694215680} - \frac{640\pi^2}{3} - \frac{6848}{21}\gamma_E \nonumber \\& + \eta\left(-\frac{15737765635}{3048192} + \frac{2255\pi^2}{12}\right) + \frac{76055}{1728}\eta^2 - \frac{127825}{1296}\eta^3 \nonumber \\&
         - \frac{6848}{21}\ln{(4v)}  + \pi\left\{\frac{2270}{3}\delta\chi_a + \left(\frac{2270}{3} - 520\eta\right)\chi_s\right\}\nonumber \\& + \left(\frac{75515}{144} - \frac{8225}{18}\eta\right)\delta\chi_a\chi_s 
         + \left(\frac{75515}{288} - \frac{263245}{252}\eta - 480\eta^2\right)\chi_a^2 \nonumber \\&  + \left(\frac{75515}{288} - \frac{232415}{504}\eta + \frac{1255}{9}\eta^2\right)\chi_s^2\,, \\
    \psi_{(7)} =~& \frac{77096675\pi}{254016} + \frac{378515\pi}{1512}\eta- \frac{74045\pi}{756}\eta^2\nonumber \\& + \left\{-\frac{25150083775}{3048192} + \frac{10566655595}{762048}\eta - \frac{1042165}{3024}\eta^2 + \frac{5345}{36}\eta^3\right. \nonumber \\&
         + \left.\left(\frac{14585}{8} - 7270\eta + 80\eta^2\right)\chi_a^2\right\}\chi_s\nonumber \\& + \left\{\left(\frac{-25150083775}{3048192}
         + \frac{26804935}{6048}\eta - \frac{1985}{48}\eta^2\right)\chi_a\right. \nonumber \\& \left. + \left(\frac{14585}{24} - 2380\eta\right)\chi_a^3 + \left(\frac{14585}{8} - \frac{215}{2}\eta\right)\chi_a\chi_s^2\right\}\delta\nonumber \\& + \left(\frac{14585}{24} - \frac{475}{6}\eta + \frac{100}{3}\eta^2\right)\chi_s^3 \,.
\end{align}
Here $\delta = \sqrt{1-4\eta}$, $\chi_s=(\chi_1+\chi_2)/2$, $\chi_a=(\chi_1-\chi_2)/2$. $\gamma_E$ is the Euler's constant, $\gamma_E=0.5772156\cdots$.


\chapter{Ringdown and peak frequency}
\label{AppendixB}

Here we describe the steps to evaluate the ringdown and the peak frequency, from phenomenological relations, that were used in Sec.~\ref{sec:amplitude}.

The quasi-normal mode (QNM) frequency of a BH can be expressed in terms of its oscillatory (real) and damping (imaginary) parts as,
\begin{equation}
    f_{\rm QNM} = f_{\rm RD} - if_{\rm damp}\,.
\end{equation}
$f_{\rm RD}$ can be obtained from the fitting formula~\cite{Nakano:2018vay}
\begin{equation}
    f_{\rm RD} = \frac{1}{2\pi M_f}\left[f_1+f_2(1-\chi^{}_f)^{f_3}\right]\,,
\end{equation}
where $M_f$ and $\chi^{}_f$ are the mass and spin of the merger remnant BH. Evaluation of these quantities is described in Appendix C of Ref~\cite{Favata:2021vhw}.
For the dominant harmonic mode $l=m=2$, the coefficients are given by, $f_1=1.5251,\,f_2=-1.1568,\,f_3=0.1292$.

$f_{\rm damp}$ can be written as
\begin{equation}
    f_{\rm damp} = \frac{f_{\rm RD}}{2Q},
\end{equation}
where the quality factor $Q$ can be expressed by the fitting formula
\begin{equation}
    Q = q_1+q_2(1-\chi^{}_f)^{q_3}.
\end{equation}
For the $l=m=2$ mode, $q_1=0.7,\,q_2=1.4187,\,q_3=-0.499$.

The frequency corresponding to the peak amplitude in the Fourier domain, $f_{\rm peak}$, can then be obtained from the phenomenological relation
\begin{equation}
    f_{\rm peak} = \left| f_{\rm RD} + \frac{f_{\rm damp}\gamma_3(\sqrt{1-\gamma_2^2}-1)}{\gamma_2}  \right|\,,
\end{equation}
$\gamma_2$ and $\gamma_3$ being two phenomenological parameters that can be evaluated for given values of $\{\eta,\chi_1,\chi_2\}$ from the ansatz in Eq.~\eqref{eq:Lambda}. The corresponding coefficients are given in Appendix C of Ref.~\cite{Khan:2015jqa}.

\chapter{Coefficients for the pseudo-PN parameters}
\label{AppendixC}

This appendix lists the numerical values obtained for the phenomenological parameters $\{\sigma_i\}$ as described in Sec.~\ref{sec:inspiral phase model}.

\begin{table}[h]
    \centering
    {\renewcommand{\arraystretch}{1.6}
    \begin{tabular}{p{0.06\linewidth}|p{0.18\linewidth}|p{0.22\linewidth}|p{0.22\linewidth}|p{0.22\linewidth}}
        \hline
         & $\sigma_1$ & $\sigma_2$ & $\sigma_3$ & $\sigma_4$ \\
        \hline
        \hline
        $\lambda_{00}$ & 1650.156 & 14056.472 & $-$86832.697 & 146221.791 \\\hline
        $\lambda_{10}$ & $-$3531.389 & $-$75378.11 & 412796.518 & $-$667838.551 \\\hline
        $\lambda_{01}$ & $-$2661.071 & 137530.39 & $-$623322.321 & 920614.957 \\\hline
        $\lambda_{11}$ & 15385.008 & $-$944130.978 & 4.1416$\times10^6$ & $-$5.959$\times10^6$ \\\hline
        $\lambda_{21}$ & $-$219.668 & 1.7166$\times10^6$ & $-$7.1273$\times10^6$ & 9.8223$\times10^6$ \\\hline
        $\lambda_{02}$ & $-$6845.194 & 169601.655 & $-$737033.591 & 1.0493$\times10^6$ \\\hline
        $\lambda_{12}$ & 38637.011 & $-$1.2802$\times10^6$ & 5.4343$\times10^6$ & $-$7.5726$\times10^6$ \\\hline
        $\lambda_{22}$ & $-$45127.308 & 2.648$\times10^6$ & $-$1.09142$\times 10^7$ & 1.4893$\times10^7$ \\\hline
        $\lambda_{03}$ & $-$2076.335 & 50611.846 & $-$213200.291 & 288184.3 \\\hline
        $\lambda_{13}$ & 17433.08 & $-$468141.986 & 1.9402$\times10^6$ & $-$2.641$\times10^6$ \\\hline
        $\lambda_{23}$ & $-$31696.712 & 1.10645$\times10^6$ & $-$4.5663$\times10^6$ & 6.2489$\times10^6$ \\
        
        \hline
    \end{tabular}}
    \caption{Best-fit values of $\lambda^i_{jk}$ parameters in Eq.~\eqref{eq:Lambda} for the pseudo-PN parameters $\{\sigma_i\}$ in the inspiral phase.}
    \label{tab:lambda}
\end{table}
\newpage

\setstretch{1}

\chapter*{Publications}

\addcontentsline{toc}{chapter}{Publications}
\section*{Limited author papers}

\begin{enumerate}
    \item \textbf{``Toward establishing the presence or absence of horizons in coalescing binaries of compact objects by using their gravitational wave signals"}\\
        \vspace{0.5mm}\\
        \textbf{S. Mukherjee}, S. Datta, S. Tiwari, K. S. Phukon, and S. Bose.\\
        \vspace{0.5mm}\\
        arXiv: \href{https://arxiv.org/abs/2202.08661}{2202.08661  [gr-qc]}\\
        \vspace{0.5mm}\\
        Phys. Rev. D 106, 104032 (2022)\\
        \vspace{.5mm}\\
        DOI: \href{https://doi.org/10.1103/PhysRevD.106.104032}{10.1103/PhysRevD.106.104032}
    \vspace{1mm}\\
    \item \textbf{``A phenomenological gravitational waveform model of binary black holes incorporating horizon fluxes"}\\
         \vspace{0.5mm}\\
        \textbf{S. Mukherjee}, S. Datta, K. S. Phukon, and S. Bose.\\
        \vspace{0.5mm}\\
        arXiv: \href{https://arxiv.org/abs/2311.17554}{2311.17554  [gr-qc]}
    \vspace{1mm}\\
    \item \textbf{``Modeling horizon absorption of black holes from numerical relativity data"}\\
        \vspace{.5mm}\\
        \textbf{S. Mukherjee}, S. Datta, K. S. Phukon, and S. Bose.\\
        \vspace{.5mm}\\
        (Under preparation)
\end{enumerate}
\newpage\null\newpage
\let\cleardoublepage\clearpage

\renewcommand{\bibname}{References}
\fancyhead[LO,RE]{\bfseries \bibname}
\printbibliography[heading=bibintoc]

\end{document}